\documentclass[11pt,a4paper]{article}
\usepackage{graphicx}
\usepackage{url}
\usepackage{wrapfig}
\usepackage{multicol}
\usepackage{multirow}
\usepackage{authblk}
\usepackage{lscape}
\usepackage{lipsum}
\usepackage{esint}
\usepackage{floatflt}
\usepackage[normalem]{ulem}
\usepackage{caption}
\usepackage{subcaption}

\usepackage{cite}
\usepackage{array}

\usepackage{type1cm}
\usepackage{eso-pic}
\usepackage{color}
\usepackage{setspace}
\usepackage[bottom]{footmisc}

\newcommand{\forceindent}{\leavevmode{\parindent=1em\indent}}

\usepackage{array}
\newcolumntype{C}[1]{>{\centering\arraybackslash}p{#1}}

\title{\bf Simulation of live-cell imaging system reveals hidden uncertainties in cooperative binding measurements}
\author{\small Masaki Watabe,$^{1}$ Satya N. V. Arjunan,$^{1,2}$ Wei Xiang Chew,$^{1,3}$ Kazunari Kaizu,$^{1}$ and Koichi Takahashi$^{1,4,5,\ast}$}
\affil{\small
$^{1}$ Laboratory for Biologically Inspired Computing, RIKEN Center for \\
Biosystems Dynamics Research, Suita, Osaka 565-0874, Japan\\
$^{2}$ Lowy Cancer Research Centre, The University of New South Wales, Sydney, Australia\\
$^{3}$ Physics Department, Faculty of Science, University of Malaya, Kuala Lumpur 50603, Malaysia\\
$^{4}$ Institute for Advanced Biosciences, Keio University, Fujisawa, Kanagawa 252-8520, Japan\\
$^{5}$ Department of Biosciences and Informatics, Keio University, Yokohama, Kanagawa 223-8522, Japan}
\date{\small Published by Physical Review E: 3 July, 2019}

\begin{document}
\maketitle

\noindent {We propose a computational method to quantitatively evaluate the systematic uncertainties that arise from undetectable sources in biological measurements using live-cell imaging techniques. We then demonstrate this method in measuring biological cooperativity of molecular binding networks: in particular, ligand molecules binding to cell surface receptor proteins. Our results show how the non-statistical uncertainties lead to invalid identification of the measured cooperativity. Through this computational scheme, the biological interpretation can be more objectively evaluated and understood under a specific experimental configuration of interest.}

\vfill

\noindent {\small {\it URL}: \url{https://link.aps.org/doi/10.1103/PhysRevE.100.010402}} \\
{\small {\it DOI}: 10.1103/PhysRevE.100.010402} \\
{\small {\it E-mail}: masaki@riken.jp}\\

\newpage

\section*{Introduction}
\paragraph{}
Recent progress in robotic and automated techniques for sampling and analyzing complex biological data can reduce statistical uncertainties, increasing precisions in measuring biological and physical properties in living cells \cite{king2009, marx2013, jordan2015, yachie2017}. However, despite advances in computational techniques, the measurement process has resisted quantitative evaluation of systematic uncertainties that arise from inaccuracies in experimental data acquisition and analysis \cite{amrhein2019, geris2016}. An absence and ignorance of evaluation of such systematic uncertainties often lead to excessive interpretations of the measured properties.

\paragraph{}
A key challenge to evaluating the systematic uncertainties is finding meaningful and nonintuitive variance. There exist a large number of systematic sources in the measurement processes, but no well-defined procedure to evaluate systematic variance. Although experienced experimental biophysicists are able to anticipate some systematic sources (e.g., the faulty calibration of measurement equipment) and ensure that most systematic uncertainties are much less than the required precision, other sources cannot even be detected by empirical approaches. The systematic variance that arises from undetectable sources in the measurement process can cause erroneous identification and interpretation of measured properties \cite{taylor1997, bevington2003, geris2016, amrhein2019}. For example, structural uncertainties that arise from various model assumptions in biological network topologies cannot be directly extracted from experimental data, introducing errors into analysis and potentially giving rise to misleading conclusions \cite{babtie2014, kirk2015}. Thus, the measurement processes require a non-experimental evaluation method that allows biophysicists to objectively interpret the measured outputs and draw proper conclusions.

\paragraph{}
To better understand the origin of these hidden uncertainties, we consider a process of data-driven (or inductive) modeling in bioimaging. For the sake of simplicity, we assume modeling the steady state behavior of a specific type of protein (e.g., cell-membrane receptors) in a living cell of interest. In this modeling, we also assume that a fluorescence microscopy system measures the concentrations of the proteins tagged with fluorescent emitters (e.g., green fluorescence proteins) within the focal plane of the optics onto a digital camera placed at the conjugated focal plane. Technical details of extracting the protein concentrations from microscopy images are also omitted in this simplification. 

% as well as extended to general nonequilibrium modeling cases with arbitrary bioimaging systems.
%
%All of the subsequent considerations in the evaluation procedure, however, can be straightforwardly applied to other parameters.
%All of the subsequent considerations in the evaluation procedure can be straightforwardly applied to other model parameters in the measurement process as well as extended to general nonequilibrium modeling cases with arbitrary bioimaging systems.

\paragraph{}
Here, as an example, we evaluate the systematic uncertainties associated with a specific parameter (e.g., image acquisition periods) in the simplified measurement process. We first consider that the measured concentration of the observed proteins is a function of time at the $i$-th image-frame $C{\scriptstyle (t_i)}$. For $m$ cell-samples, the experimenters ensure the stability of the concentration changes in an image acquisition period, $T$. The observed rate in the protein concentrations converges near 0 as $t_i \to T$,
\begin{equation}
\frac{\Delta C_i}{\Delta t_i} \to\ 0 \pm \sigma_{\scriptscriptstyle C'}% \left({\rm stat.}\right)
\label{eqn01;rate}
\end{equation}
where $\Delta C_i = C{\scriptstyle (t_{i})} - C{\scriptstyle (t_{i-1})}$ and $\Delta t_i = t_{i} - t_{i-1}$ are the concentration and time difference at the $i$-th image frame, respectively. $\sigma_{\scriptscriptstyle C'}$ represents the statistical deviations in the observed rates. An ``apparent" steady state of the protein concentrations is then computed by the time-average integration $\overline{C} \pm \sigma_{\overline{\scriptscriptstyle C}}$.

\paragraph{}
In the data-driven approach, the measured protein concentrations are typically fitted to the network model constructed under the data-interpretation that the measured concentrations fully converge to equilibrium within the acquisition period. The rate of concentration changes in the observed proteins interacting with $N$ variables can be theoretically modeled in the form of an ordinary differential equation. In equilibrium modeling, the rate fully converges to $0$ as $t \to \infty$,
\begin{equation}
\frac{d R{\scriptstyle (t)}}{dt} = f{\scriptstyle (R(t),{\bf x}(t),t;\ \theta)} \to 0
\label{eqn02;rate}
\end{equation}
where $R{\scriptstyle (t)}$ is the observed protein concentration as a function of time and model-parameters $\theta$ (e.g., dissociation constants), ${\bf x}{\scriptstyle (t)} = [x_{\scriptscriptstyle 1}{\scriptstyle (t)}, ... , x_{\scriptscriptstyle N}{\scriptstyle (t)}]$ and $x_{\scriptscriptstyle i}{\scriptstyle (t)}$ is the value of the $i$-th variable at $t$ and $\theta$. In the equilibrium modeling, the rate of changes in the $i$-th variable over time also fully converges to $0$ as $t \to \infty$, $d x_i{\scriptstyle (t)}/dt = f_i{\scriptstyle (R(t),{\bf x}(t),t;\ \theta)} \to 0$. The full-equilibrium concentration in the observed proteins is then computed by the time-average integration, $\widetilde{R}{\scriptstyle (\theta)} = {\displaystyle \lim_{\scriptscriptstyle \tau \to \infty}} \frac{1}{\tau} \int^{\tau}_{0} R{\scriptstyle (t; \theta)}dt$. Fitting procedures (e.g., maximum-likelihood method) directly compare the measured concentration ($\overline{C}$) to the theoretical full-equilibrium concentration ($\widetilde{R}{\scriptstyle (\theta)}$) binned into each histogram. The best-fit model parameters ($\overline{\theta}$) are those which minimize the discrepancies defined in an optimization function ${\bf \mathcal{M}}{\scriptstyle( \widetilde{R}{\scriptstyle (\theta)}, \overline{C}, \sigma_{\overline{\scriptscriptstyle C}})}$. Confidence levels and correlations are also estimated in each fitting parameter.

\paragraph{}
Although the experimenters ensure the stability of the measured protein concentration changes in time and show the goodness of fit to the equilibrium model, actual biological cells generally operate out of equilibrium. In Eq. (\ref{eqn01;rate}), the apparent state stability can be not only interpreted as the complete convergence of the protein responses to full-equilibrium but also the incomplete convergence if the responses are so slow that the protein concentration remains nonequilibrium during the acquisition period. To split this double data-interpretations, the experimenters, however, must contend with the experimental configuration and the state-transition speed in the observed proteins. The data-interpretation that gives rise to the complete convergence requires not only maximizing event samples in a shorter image acquisition period but also fasten the observed rates of converging to the full-equilibrium. The configuration to reduce statistical uncertainties constrains the limitation and sensitivity in measuring the state-transition time. Because of these tradeoffs, it is unclear when or whether the measured protein concentration actually converges to the full-equilibrium within the acquisition period $T$. 

\paragraph{}
To see the effects of the double data-interpretations in the apparent steady states, we consider a Taylor expansion of Eq. (\ref{eqn02;rate}) at the model-true equilibrium concentration $\overline{R} = \widetilde{R}{\scriptstyle(\overline{\theta})}$,
\begin{equation}
f{\scriptstyle (R{\scriptscriptstyle(t; \overline{\theta})})} \approx \left( R{\scriptstyle (t; \overline{\theta})} - \overline{R} \right) \frac{\partial f}{\partial R} \bigg|_{\scriptstyle R{\scriptscriptstyle (t; \overline{\theta})} = \overline{R}} +\ \cdots
\label{eqn02;taylor}
\end{equation}
where zeroth-order term vanishes at the model-true equilibrium $f{\scriptstyle (\overline{R})} = 0$. At $t = T$, we evaluate the systematic variance that arises from the following model assumptions. If full-equilibrium is proper as a model assumption, then $f{\scriptstyle (R{\scriptscriptstyle (T; \overline{\theta})})}$ can fully converge to $0$. No systematic variance can be generated in the fitting results, $\left| R{\scriptstyle (T; \overline{\theta})} - \overline{R}\right| < \sigma_{\scriptscriptstyle \overline{C}}$, implying successful restoration of the model-true equilibrium concentrations. However, if nonequilibrium is the proper assumption, then the $f{\scriptstyle (R{\scriptscriptstyle (T; \overline{\theta})})}$ can converge to a finite value near $0$. Under this assumption, measuring the model-true equilibrium requires a limit excess of the experimental configuration, thereby generating an undetectable gap in the fitting results, $\left| R{\scriptstyle (T; \overline{\theta})} - \overline{R} \right| \gg \sigma_{\scriptscriptstyle \overline{C}}$. This implies the reconstruction failure of the model-true equilibrium concentration at $T$, leading to excessive interpretations of the measured protein concentration. Thus, the systematic variance that arises from the double data-interpretations cannot be evaluated without computing the dynamical behavior of the equilibrium models. 

\paragraph{}
A computational modeling approach for a whole experimental system is more relevant to extract the impact of various systematic uncertainties \cite{incerti2009, incerti2018, agostinelli2003, allison2006, allison2016}. In this work, we introduce a comprehensive method to quantitatively evaluate the systematic variance computed not only from a computer simulation of live-cell imaging systems but also image processing and pattern recognition algorithms for biological images. In particular, we construct a bioimage simulation module for an oblique illumination fluorescence microscopy system configured to observe receptor proteins binding to ligands on an apical surface of biological cells. We then show how the non-statistical variance leads to misidentification of cooperativity in the binding system.

\section*{Computational method}
\paragraph{}
A key insight into evaluating systematic variance is to estimate how well the ground-true model properties can be restored through analytical procedures, influencing the interpretation of biological properties reconstructed (or extracted) from actual biological images \cite{taylor1997, bevington2003, geris2016}. Such model-driven evaluation allows us to quantify the restoration efficiency and defects (or failure) in the reconstruction processes. We cannot fully know the true-model; otherwise, there would not be any uncertainty in the biological measurements. However, we can approximately know what to expect either from earlier experimental results or from the biological models derived from experimental knowledge. Such approximations can function as a guide to compute systematic variance in an organized manner from a specific experimental configuration of interest.

\paragraph{}
Systematic variance generally causes the reconstructed properties to be shifted in one direction from the ground-true model properties \cite{taylor1997, bevington2003, geris2016}. Of particular importance is the quantifying of systematic variance that arises from undetectable sources in the measurement processes. If the ground-true property is well-restored through the reconstruction process, then it is unchanged and has weak influence on the biological interpretation. However, if the reconstruction process poorly restores the true property, it may change significantly, affecting the biological interpretation. For example, geometric uncertainties imposed by the unobservable sources such as complex and irregular shapes of membrane-enclosed cellular compartments (e.g., endoplasmic reticulum) in FRAP experiments lead to invalid measurement of molecular mobility and compartmental connectivity \cite{sbalzarini2013, mai2013}.

\paragraph{}
A realistic ground-true model simulation that approximately represents a whole experimental system can be constructed for the model-driven evaluation. Biological measurements using live-cell imaging techniques are generally governed by various natural laws and principles of biochemistry and physics. Models of each imaging process can be simulated within the limited range and dimensions of model-parameter space. Various model-simulation studies exist in physics simulations for molecular fluorescence and optical apparatus \cite{lakowicz2006, pawley2008, mansuripur2009, valeur2012} and systems biology simulations aimed at explaining and predicting biological phenomenon \cite{kaneko2006, alon2007a, tomita1999}. We integrate these model-simulations into a unified model corresponding approximately to a whole experimental system, thus helping to explore the extended dimensions of model-parameter space that can affect potential imaging and analytical outcomes \cite{boulanger2009, rezatofighi2013, sbalzarini2013, angiolini2015, watabe2015, venkataramani2016, linden2016, weigert2018}. In particular, we have developed the bioimage simulation platform for handling a large range of biochemical and physical parameters that govern image-based measurement systems, generating computational photomicrographs that arise from the various systematic sources: spatiotemporal model of biological cells, photophysics and imaging apparatus \cite{watabe2015}. Through such simulation platforms, the biological interpretation of measured properties can be more objectively evaluated and understood under a specific experimental configuration of interest.

\section*{Result}
\paragraph{}
The computational method presented here enables us to evaluate the systematic variances that arise from inaccuracy in the cooperative binding measurement using fluorescence microscopy. In particular, we construct a bioimage simulation module for the oblique illumination fluorescence microscopy system configured to observe biochemical reactions and aggregation of the HRG ligand induced ErbB receptors on an apical region of the cell membrane [see sections A and B in the supporting information (SI)]. We then evaluate the cooperativity in two cell-models: simple binding of the single ligands to the single receptors (see SI section C.1) and dimer formation of ligand-enhanced receptors (see SI section C.2). The result for the dimer formation is as follows. 

\paragraph{}
M. Hiroshima and his colleagues have used the oblique illumination technique to acquire single-molecule images in an equilibrium region, and then identified apparent (or observed) cooperativity to be "negative" for the dimer formation \cite{hiroshima2012}. In their experiment, the ErbB receptors on a cell-surface that expresses the HRG ligands tagged with tetramethylrhodamine (TMR) fluorescent molecules appear as spots (or blobs) in the single-molecule images. Spot-detection algorithm was applied to the images not only for extracting spot-properties (e.g., intensity, size), but also reconstructing the spot area-density (${\rm spots/\mu m^2}$) on the cell-surface. Converting the reconstructed area-density to the equilibrium binding curve and Scatchard plot, the cooperativity was finally identified in the dimer formation.

\paragraph{}
Using the bioimage simulation module, we generated the single-molecule images simulated under the model-assumption that true-cooperativity is negative (see SI section B). We then appied the analytical procedure to the computational images in order to quantify the differences between the ground-true properties and the reconstructed ones (see SI section C.2.2). 

\subsection*{Computational photomicrography}
\paragraph{}
First, we programed the simulation module for the single-molecule experimental system using fluorescence microscopy. This simulation module can generate computational single-molecule images of the three dimensional cell-model structures represented by the dimer and higher-order oligomer (e.g., trimer, tetramer, pentamer) formations on the cell-surface \cite{hiroshima2012}. There are two major simulation components; the optical system and the cell-model.

\begin{figure}[!h]
  \centering
  \leftline{\bf \hspace{1.0cm} (A)}
      \includegraphics[width=8.0cm]{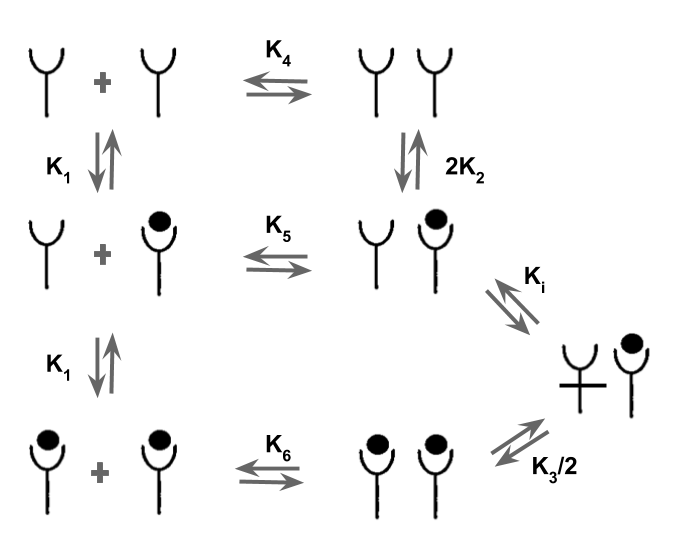}
      \hspace{8.0cm}
      
  \vspace{0.4cm}
  \centering
  \leftline{\bf \hspace{1.0cm} (B)}
      \includegraphics[width=16.0cm]{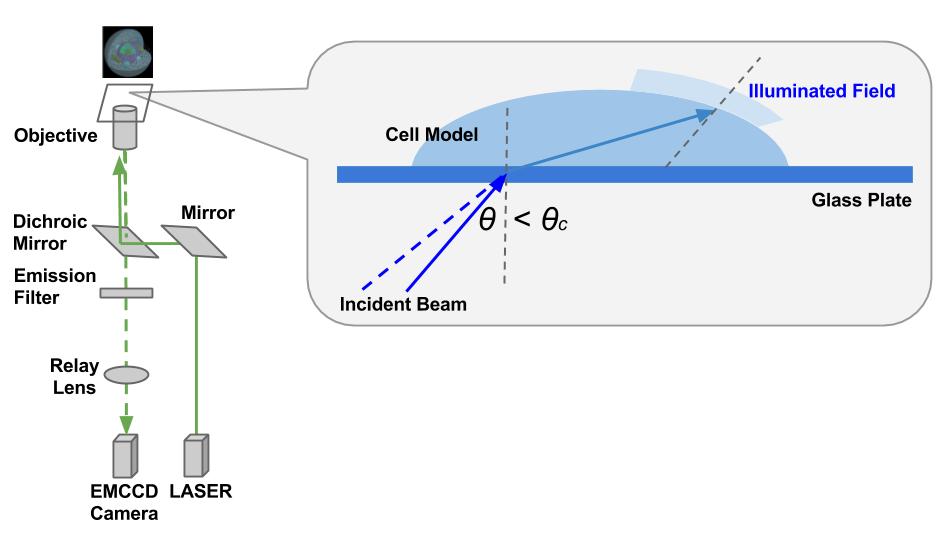}

  \vspace{0.4cm}
  \caption{{\bf Cell-model and microscopy configuration}. (A) Reaction network for the dimer formations of HRG ligand induced ErbB receptors \cite{hiroshima2012}. Y-shaped object and black filled-in circle represent ErbB receptor and HRG ligand. Receptors randomly walk from voxel to voxel, and slowly diffuse at $0.015\ {\rm \mu m^2/sec}$. (B) Optical configuration of fluorescence microscope. The magnified panel shows a schematic side view of oblique illumination to the spatial cell-model. The hemi-elliptical cell was placed on the glass surface. Dashed and solid blue lines represent critical angle ($\theta_c$) and incident beam angle ($\theta$). An incident angle less than the critical angle leads to the oblique illumination at the apical cell surface. Molecules were distributed in the cell-surface compartment.}
  \label{fig;figure_01}
\end{figure}

\begin{enumerate}
\item [(1)] Optical system (see SI section A): The simulation for the optical system is composed of three parts; (a) An illumination system transfers the photon flux from a light source to the cell-model, to generate a prescribed photon distribution and maximize the flux delivered to the cell-model. An incident beam of excitation wavelength that passed through the objective lens is assumed to uniformly illuminate specimen. In particular, the optics simulation has been implemented for a selective visualization of apical and basal surface regions of the cell-model. Various illuminations can be configured: epifluorescence illumination, oblique illumination and total internal reflection fluorescence illumination. (b) Fluorophores defined in the cell-model absorb photons from the photon distribution, and are quantum-mechanically excited to higher energy states. Molecular fluorescence is the result of photophysical and biochemical processes in which the fluorophores emit photons in the excited state. In particular, Monte Carlo simulations have been implemented for the fluorescence processes of the Beer-Lambertz law and photobleaching/photoblinking effects. (c) Finally, an image-forming system relays a nearly exact image of the cell-model to the light-sensitive detector. In particular, we have programmed for the optics simulation for the formation and convolution of point spreading function (PSF), and the Monte Carlo simulation of the electron-multiplying charge-coupled device (EMCCD) camera detection process.
\item [(2)] Cell-model (see SI section B): For the dimer formation, the Spatiocyte cell simulation method can provide the spatiotemporal cell-model of biological fluctuation that arises from stochastic changes in the cell surface geometry, the number of ErbB receptors, HRG ligand binding and unbinding, receptor states (e.g., monomer and dimer), and the translational diffusion of each receptors \cite{arjunan2010, takahashi2005}. In particular, negative cooperativity has been incorporated in the model by introducing an intermediate state transition. Figure \ref{fig;figure_01}A illustrates the biochemical reaction network of the dimer formation. Values of model parameters are shown in Table \ref{tab;simple_oligomer_model_parameters}. \\
\forceindent In addition, we extended the dimerization model to higher-order oligomer formation: trimer, tetramer and pentamer. In this trial extension, we introduced four additional equilibrium constants for the molecular reactions: two interactive constants are $K_{7}$ and $K_{8}$ that characterize respectively the trimer formation of monomer and dimer, and the tetramer formation of two dimers, and others are $K_{9}$ and $K_{10}$ that characterize respectively the $(n+1)$-th and $(n+2)$-th order of oligomer formations ($3 \le n \le 12$). The higher-order oligomers were assumed to be immobile on the cell membrane. The right panel of Figure \ref{fig;simple_oligomer_network} illustrates the biochemical reaction network of the higher-order oligomer formation. Values of additional model parameters are shown in Table \ref{tab;simple_oligomer_model_parameters}.
\end{enumerate}

\begin{figure}[!h]
  \centering
  \leftline{\bf \hspace{0.5cm} (A) \hspace{8.0cm} (B)}
      \includegraphics[width=18.0cm]{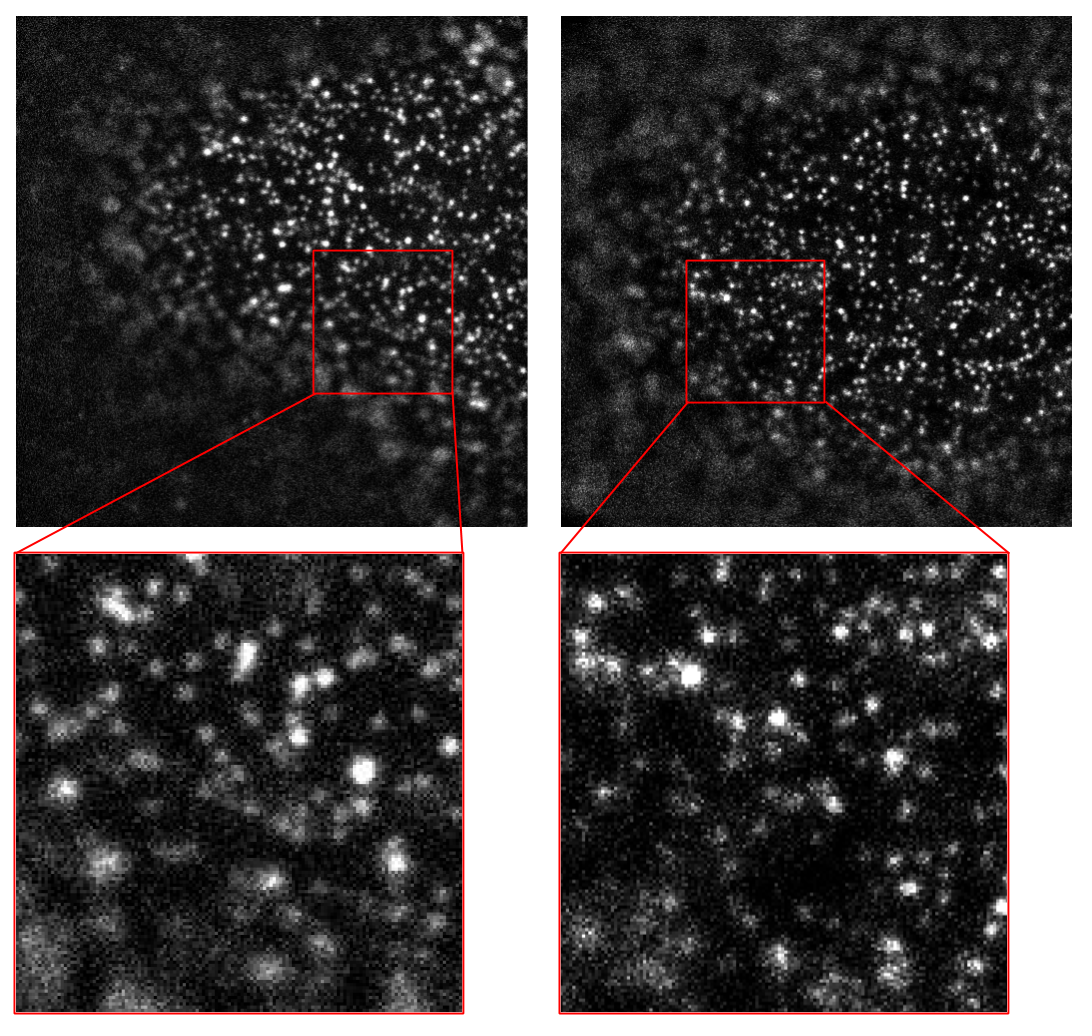}

  \vspace{0.4cm}
  \caption{{\bf Image comparison}. (A) A snapshot of an actual image obtained by single-molecule experiment using fluorescence microscopy \cite{hiroshima2012}. (B) A snapshot of the single-molecule images obtained by the simulation of fluorescence microscopy. Image size of each snapshots is $34 \times 34\ {\rm \mu m^2}$. The bottom panels show the magnified images for the selected cell regions. Actual minimum and maximum values of the intensity histogram are $2,000$ and $3,000$ ADC counts. The intensity is rescaled in the range of $0$ to $255$.}
  \label{fig;figure_02}
\end{figure}

\paragraph{}
Figures \ref{fig;figure_01}B represent optical arrangement and cell-surface geometry of the cell-model illuminated with an incident beam angle less than the critical angle. Microscopy specifications and the operating condition are shown in Table \ref{tab;specification3}. Single-molecule images obtained by the simulation of oblique illumination microscopy can be accordingly compared with actual single-molecule images at the level of photon-counting unit. The image comparison is shown in Figures \ref{fig;figure_02}. A movie for the image comparison is available online at \url{https://youtu.be/Q_RkZ3E4fAM}. However, the image comparisons are insufficient to check the validity of the microscopy simulation module. A more elaborate set of verification and calibration is required in the future.

\subsection*{Analysis}
\begin{figure}[!h]
  \centering
  \leftline{\bf \hspace{1.0cm} (A) \hspace{7.0cm} (B)}
      \includegraphics[width=8.0cm]{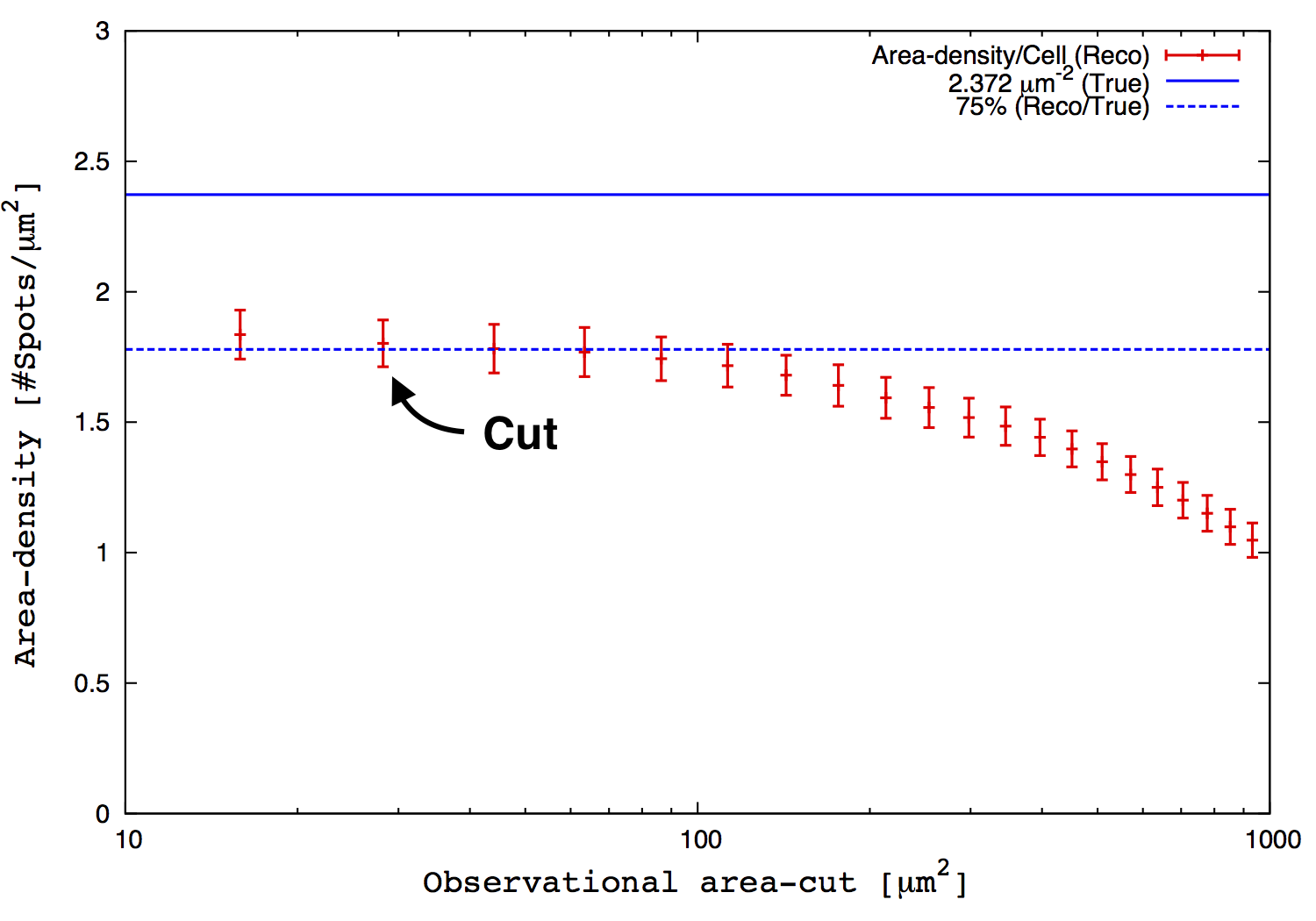}
      \includegraphics[width=8.0cm]{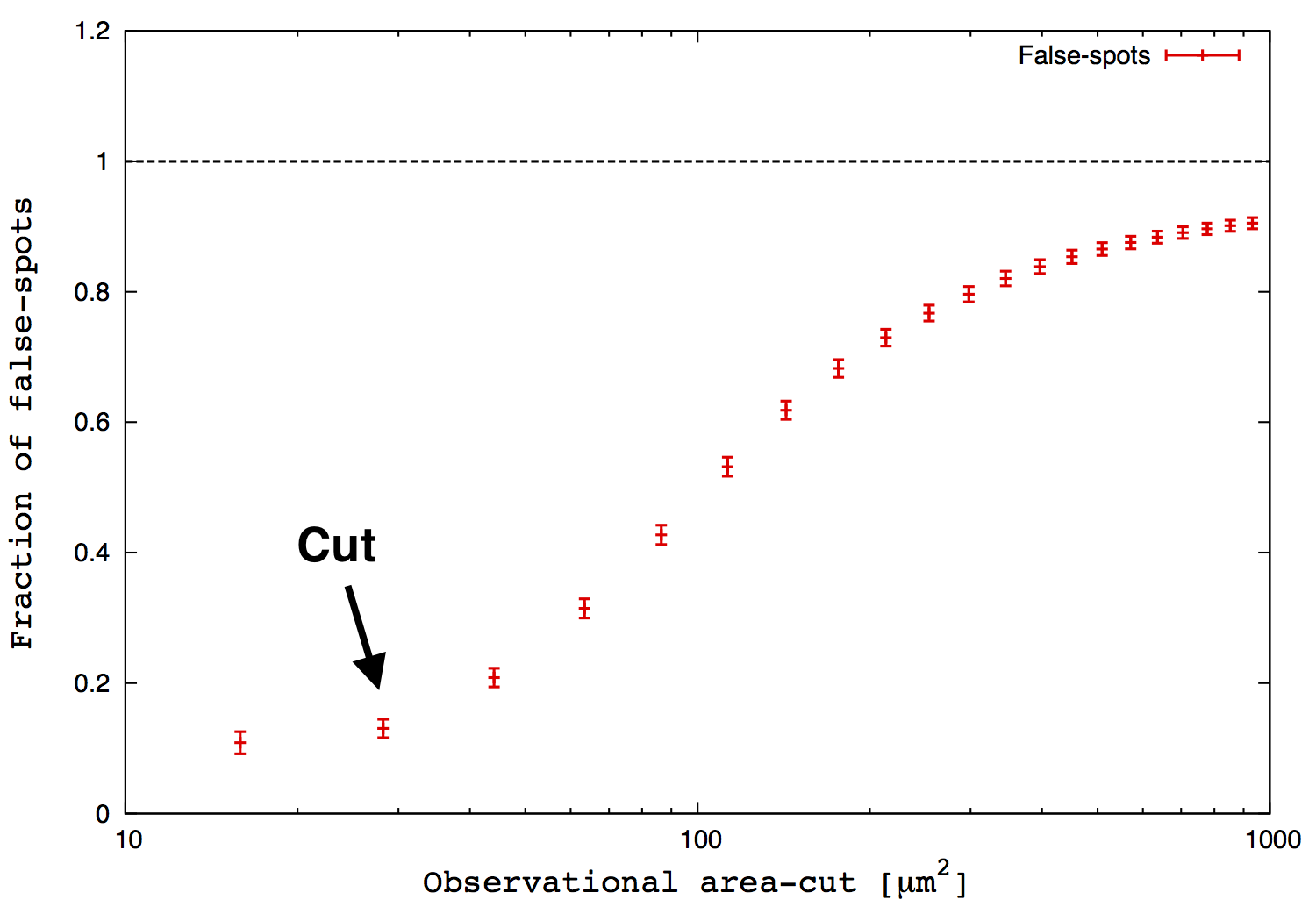}

  \vspace{0.5cm}
  \centering
  \leftline{\bf \hspace{1.0cm} (C)}
      \includegraphics[width=8.0cm]{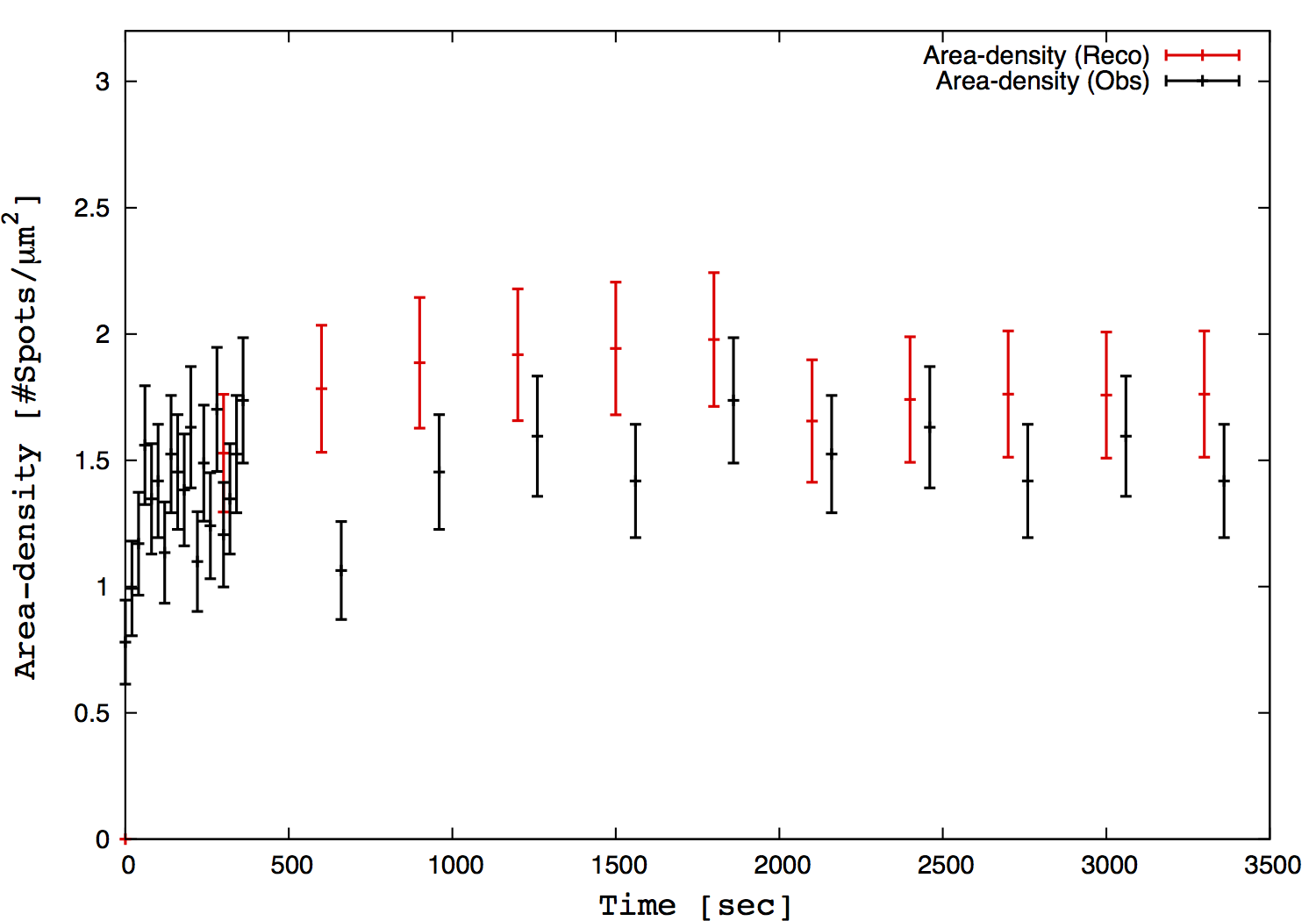}
      \hspace{8.0cm}

  \vspace{0.5cm}
  \caption{{\bf Spot area-density}. $120$ cell-samples are stimulated with $1.0\ {\rm nM}$ ligand. We assume that the ground-true spot area-density is $2.372\ {\rm spots/\mu m^2}$. (A) Reconstruction efficiency for various area-cuts. (B) Fractional occupancy of false-spots for various area-cuts. (C) For the $28\ {\rm \mu m^2}$ area-cut, the reconstructed time-course data of ligand binding to receptor is directly compared with actual data. Black and red lines represent the reconstructed  and the actual area-density.}
  \label{fig;figure_03}
\end{figure}

%\subsubsection*{Reconstructing spot area-density}
\paragraph{}
The Laplacian of the Gaussian (LoG) method \cite{vanderwalt2014} was applied to detect spot-like features in the computational single-molecule images. We then counted the number of the detected spots for various observational area-cuts. Each area-cuts are fixed at an image center. The spot area-density is reconstructed for  the area-size varying from $10\ {\rm \mu m^2}$ to $1000\ {\rm \mu m^2}$. For $10$ cell-samples stimulated to the $1.0\ {\rm nM}$ ligand, Figures \ref{fig;figure_03}A and B show the comparison of ground-true spot area-density ($2.372\ {\rm spots/\mu m^2}$) to the reconstructed one. Figure \ref{fig;figure_03}A shows the efficiency of the density reconstruction for various area-cuts. While the reconstruction efficiency is limited to $75\%$ below $300\ {\rm \mu m^2}$ area-cut, the efficiency is underestimated above the area-cut due to including the defocused cell-surface regions. Figure \ref{fig;figure_03}B shows the fractional occupancy of defected-spots for the various area-cuts. The detected spots are either one of true-molecule spots or the defects that arise from inaccuracy in the spot-detection algorithm. Such defects are one of the systematic sources that can misidentify the true-molecule spots. For the specific area-cut ($28\ {\rm spots/\mu m^2}$), Figure \ref{fig;figure_03}C shows the direct comparison of the reconstructed area-density varying in time to the actual one. Binding sites are saturated near $1.7\ {\rm spots/\mu m^2}$.

\begin{figure}[!h]
  \centering
      \includegraphics[width=16.0cm]{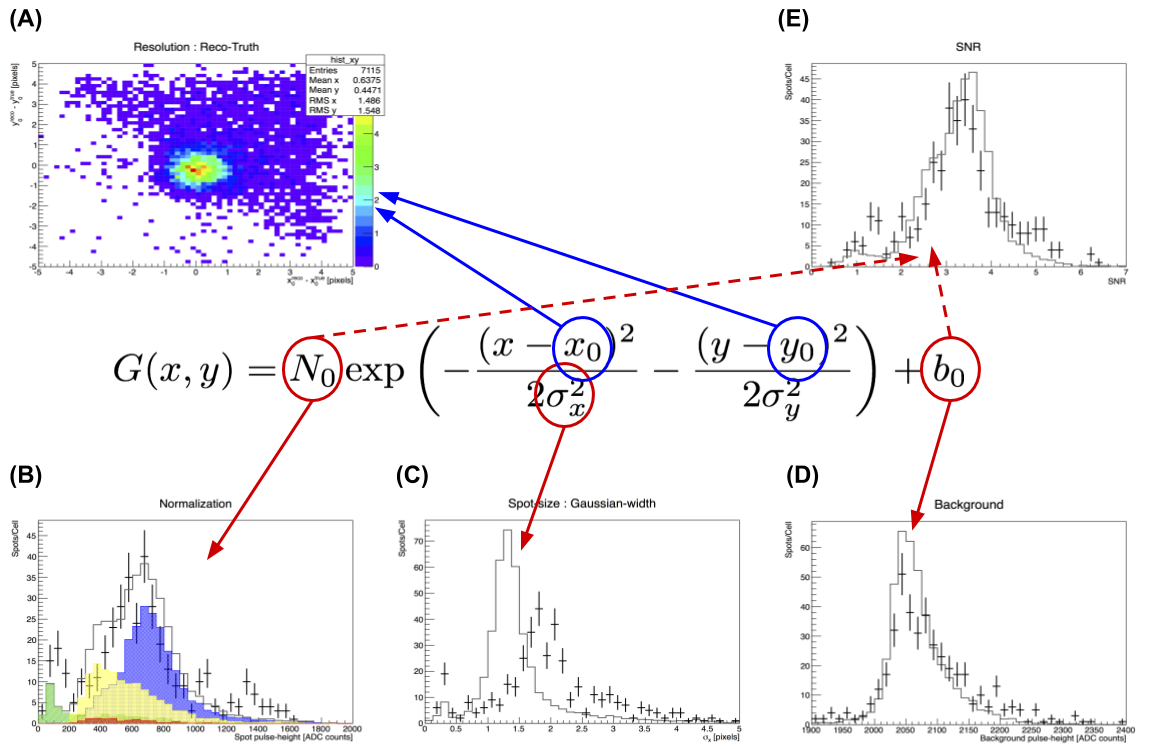}

  \vspace{0.5cm}
  \caption{{\bf Spot-properties}. For $10$ cell-samples stimulated with $1.0\ {\rm nM}$ ligand and $28\ {\rm \mu m^2}$ area-cut at the image center, the spot-properties are reconstructed as the Gaussian function: spot pulse-height ($N_0$), spot-position ($x_0, y_0$), spot-size ($\sigma_x, \sigma_y$), and background pulse-height ($b_0$). Figure A shows the 2-dimensional distribution of the difference between the ground-true spot-position and the reconstructed one. Figures B to E show the direct comparison of the reconstructed to the actual spot-properties. Black lines and crosses represent the reconstructed and actual distributions of the detected spots. Red, yellow and blue colored distributions represent true-molecule spots for monomer ({\bf R}), monomer-like dimer ({\bf rR} or {\bf pR}) and dimer ({\bf RR}). Green represents the distribution for the defect-spots.  All simulated spectra are normalized by the total number of the detected spots.}
  \label{fig;figure_04}
\end{figure}

%\subsubsection*{Extracting spot-properties}
\paragraph{}
Properties of the detected spots are characterized as a Gaussian function of six parameters: spot pulse-height or intensity ($N_0$), spot-positions ($x_0, y_0$), spot pulse-widths ($\sigma_x, \sigma_y$) and background pulse-height ($b_0$). All detected spots are fitted to the Gaussian function. For the $10$ cell-samples stimulated to the $1.0\ {\rm nM}$ ligand and the $28\ {\rm spots/\mu m^2}$ area-cut, Figure \ref{fig;figure_04}A shows the difference between the ground-true spot-position ($x_0^{\ true}, y_0^{\ true}$) and the reconstructed one ($x_0^{\ reco}, y_0^{\ reco}$). In the 2-dimensional distribution of the localization errors: ($x_0^{\ reco} - x_0^{\ true}, y_0^{\ reco} - y_0^{\ true}$), we confirmed that a peak is located near zero on each axes, and nearly formed as Gaussian function. Root mean squared (RMS) value represents the positional resolution to $0.78\ {\rm pixels}$ ($58\ {\rm nm}$). However, the tails of each distribution appear to be asymmetric. One of the most probable explanations for the asymmetry is because of the z-axis. In our analysis, we assumed that spots are characterized as a 2-dimensional Gaussian function, ignoring the z-axis. The 3-diemnsional Gaussian fitting may be able to resolve the asymmetry of each distribution.

\paragraph{}
Figures \ref{fig;figure_04}B to E show the direct comparison between the spot-properties reconstructed from the simulated and the actual single-molecule images . So far, no significant discrepancy is found in the the spot pulse-height ($N_0$), background pulse-height ($b_0$) and signal-to-noise (SNR) distributions. However, we found a large difference in the comparison of the spot-size ($\sigma_x$) distribution. While the averaged size of the reconstructed spots is about $1.35\ {\rm pixels}$ ($89.63\ {\rm nm}$), the actual spot-size is relatively large: $2.00\ {\rm pixels}$ ($133\ {\rm nm}$). There exist many systematic sources that cause such discrepancy: light-scattering, autofluorescence, and optical aberration. More detailed implementation of the simulation module is required to resolve the discrepancies. Furthermore, such direct comparisons for the spot-properties are insufficient to check the validity of the microscopy simulation module. A more elaborate set of verification and calibration is required in the future. 

\begin{figure}[!h]
  \centering
  \leftline{\bf \hspace{1.0cm} (A) \hspace{7.0cm} (B)}
      \includegraphics[width=8.0cm]{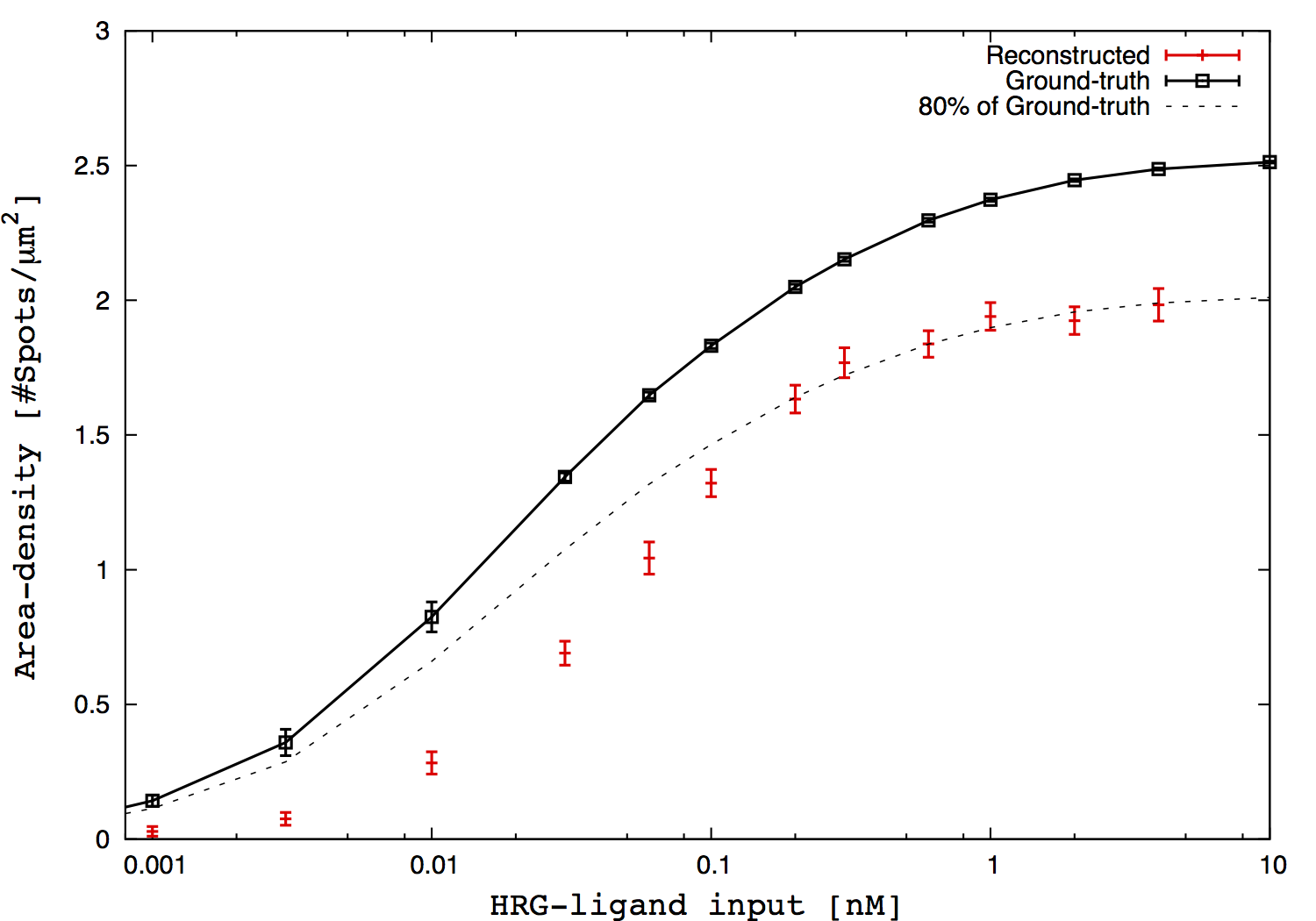}
      \includegraphics[width=8.0cm]{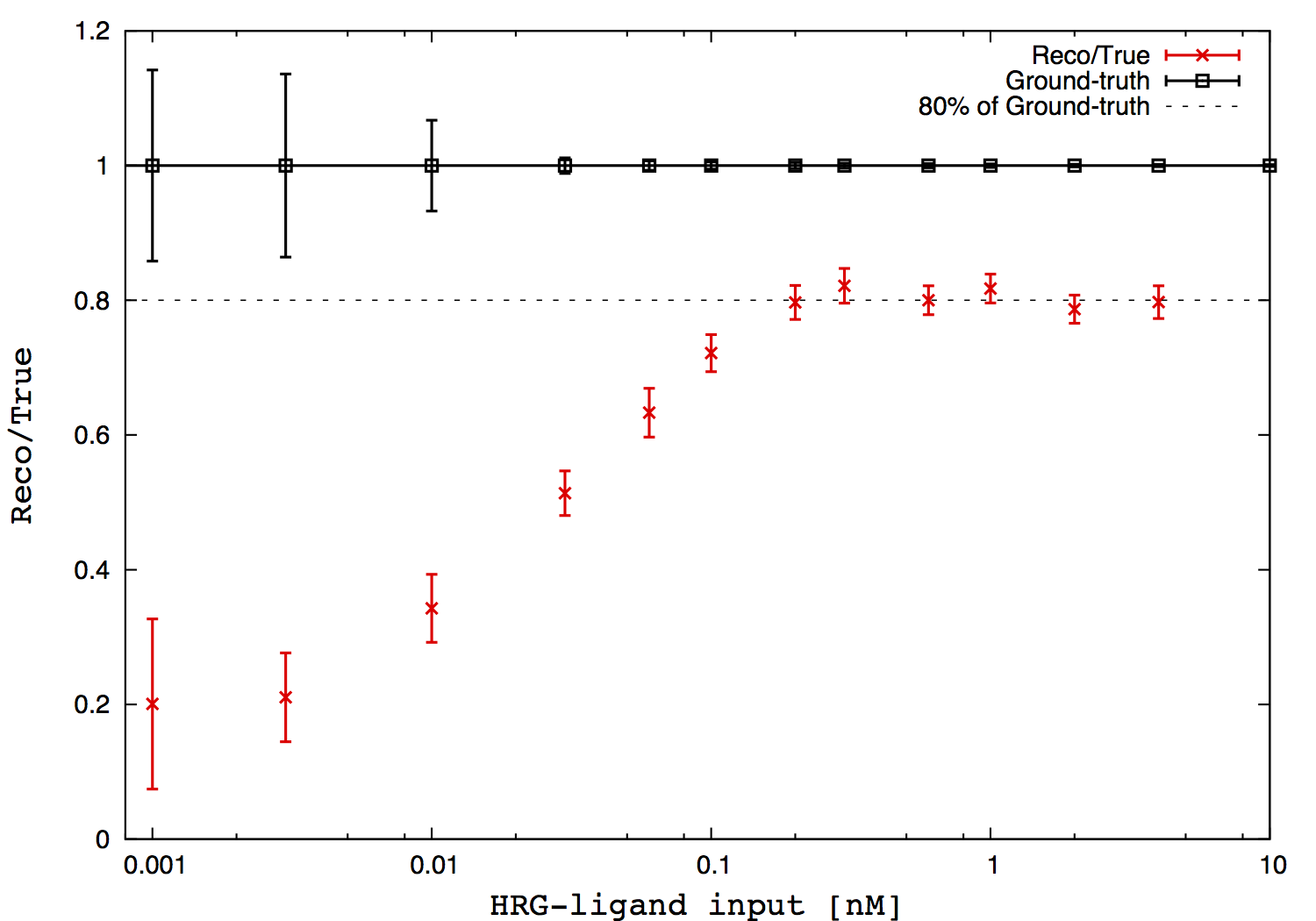}

  \vspace{0.5cm}
  \centering
  \leftline{\bf \hspace{1.0cm} (C) \hspace{7.0cm} (D)}
      \includegraphics[width=8.0cm]{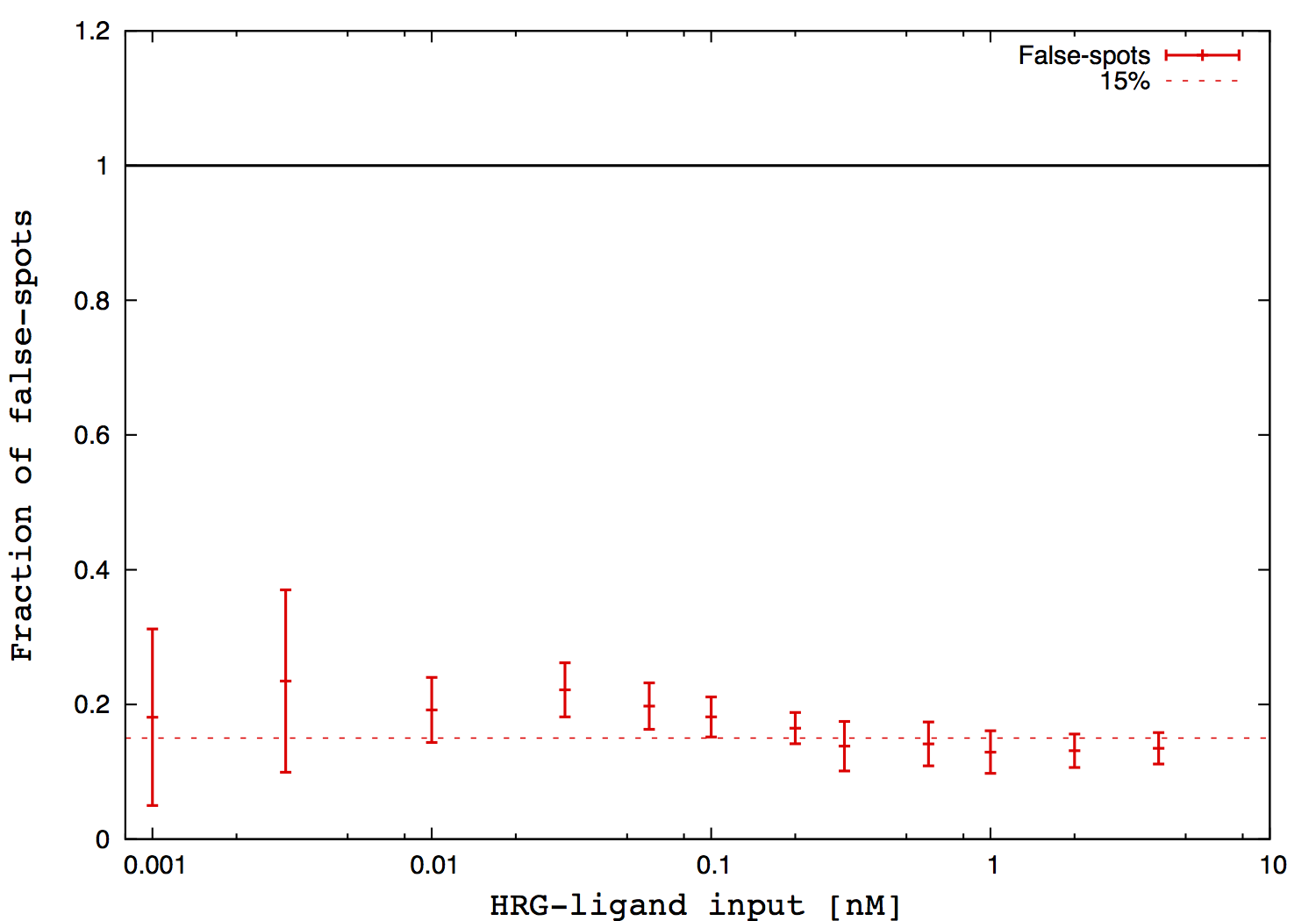}
      \includegraphics[width=8.0cm]{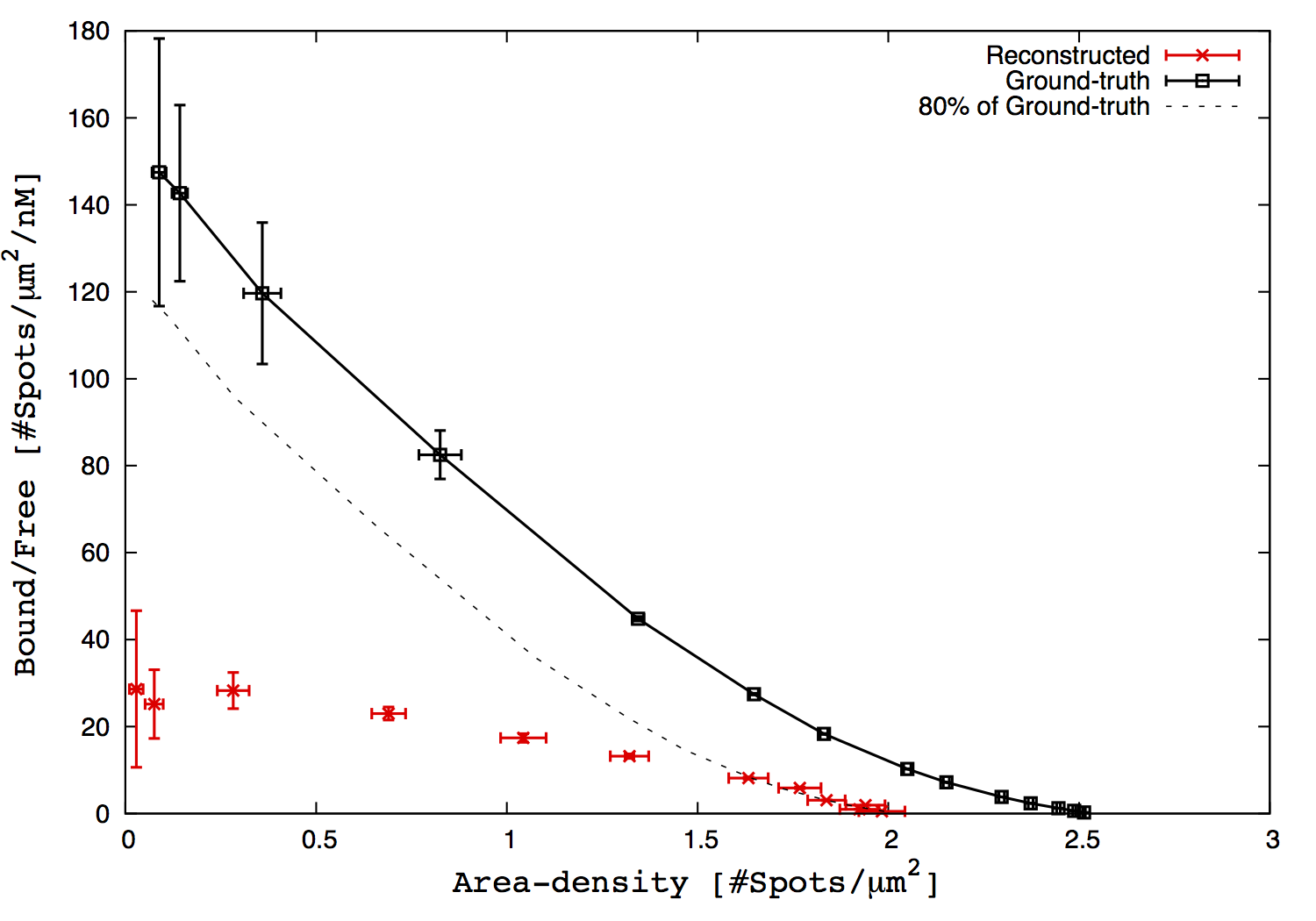}

  \vspace{0.5cm}
  \caption{{\bf Biological properties}. $120$ cell-samples are stimulated with the ligands in the concentration range of $1.0\ {\rm pM}$ to $4.0\ {\rm nM}$. Solid and dashed black lines represent the ground-true model properties and $80\%$ of the true properties. Reconstructed properties are represented with red crosses. (A) Ligand-receptor binding curve. (B) The ratio of the reconstructed curve to the ground-true model curve. (C) Fractional occupancy of false-spots. Red dashed line represents $15\%$ of the occupancy. (D) Scatchard plot.}
  \label{fig;figure_05}
\end{figure}

%\subsection*{Interpreting cooperativity}
\paragraph{}
Finally, we reconstructed the equilibrium binding curve and Scatchard plot for $120$ cell-samples stimulated with HRG ligands in the concentration range of $1.0\ {\rm pM}$ to $4.0\ {\rm nM}$. Cooperativity generally appears in the shape and concavity of the binding curve and Scatchard plot. Figures \ref{fig;figure_05}A and B show the binding curve and the ratio of the reconstructed curve to ground-true curve. These curves clearly show that the shape of the true binding curve is partially restored in the reconstructed one. While reconstruction efficiency is about $80\%$ and steady at relatively higher ligand concentration inputs ($> 0.2\ {\rm nM}$), the efficiency significantly reduces at lower ligand concentration region ($< 0.2\ {\rm nM}$). Figure\ref{fig;figure_03}C shows fractional occupancy of the defected-spots. Approximately $13\%$ of the detected spots are defected.

\begin{table}[b]
\centering
\begin{tabular}{|c|c|c|c|c|}
\hline
& $B_0\ [{\rm spots/\mu m^2}]$ & $K_A\ [{\rm nM}]$ & $n$ & $\chi_0^2 / d.o.f$ \\ \hline
Ground-true curve & \hspace{0.3cm} $2.551 \pm 0.007$ \hspace{0.3cm} & \hspace{0.2cm} $0.027 \pm 0.001$ \hspace{0.2cm} & \hspace{0.2cm} ${\bf\color{black} 0.722} \pm 0.017$ \hspace{0.2cm} & \hspace{0.1cm}  $2.883$ \hspace{0.1cm} \\ \hline
Reconstructed curve & $1.989 \pm 0.014$ & $0.053 \pm 0.002$ & ${\bf\color{red} 1.109} \pm 0.026$ & $1.616$ \\ \hline
Reconstructed ($0.2$-$4.0\ {\rm nM}$) & $1.983 \pm 0.043$ & $0.046 \pm 0.021$ & ${\bf\color{red} 1.068} \pm {\bf\color{blue} 0.360}$ & $0.345$ \\ \hline
%Observed & $ \pm $ & $\bf  \pm $ & $ \pm $ & $/9$\\ \hline
\end{tabular}
\caption{{\bf Results of fitting to the Hill equation}. The best fit values and statistical errors of each parameters are listed. The ground-true curve and the reconstructed one exhibit negative ($n<1$) and positive cooperativity ($n>1$). $\chi_0^2 / d.o.f$ is the reduced minimum.}
\label{tab;table_01}
\end{table}

\paragraph{}
In a standard approach of biological science, the Hill function can be fitted to the binding curves to quantify cooperative characteristics in the ligand-receptor binding system. The Hill function can be written in the form of
\begin{eqnarray}
B(L) & = & \frac{B_0 L^n}{K^n_A + L^n}
\end{eqnarray}
where $L$, $B_0$, $K_A$ and $n$ represent ligand concentration, maximum area-density of ligand binding, ligands occupying half of the binding sites and the Hill coefficient. The fitting results are shown in Table \ref{tab;table_01}. The ground-true curve of the binding system exhibits negative cooperativity ($n<1$), but positive cooperativity ($n>1$) appears in the reconstructed curves.

\paragraph{}
The systematic shift clearly appears in the Scatchard plot. Figure \ref{fig;figure_05}D shows the comparison of the true Scatchard plot to the reconstructed one. The plot shows that the concavity of the true plot is not well-restored in the reconstructed one. While the model-truth is well-characterized as a concave-up curve that represents negative cooperativity, the reconstructed one exhibits a concave-down curve (or a straight line) that represents positive cooperativity (or no cooperativity). 
 
\subsection*{What causes the systematic shift?}
\begin{figure}[!h]
  \centering
  \leftline{\bf \hspace{1.0cm} (A) \hspace{7.0cm} (B)}
      \includegraphics[width=8.0cm]{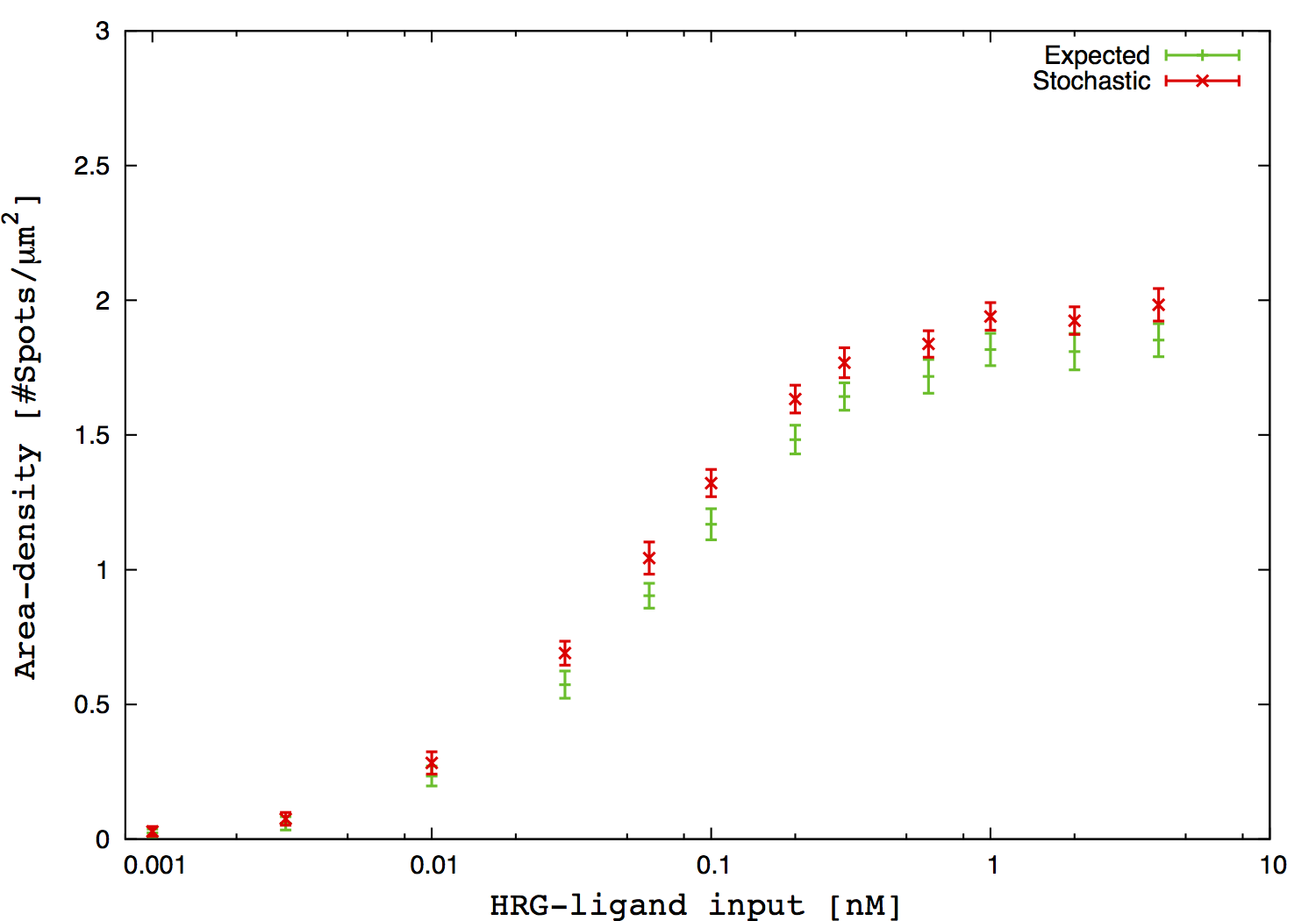}
      \includegraphics[width=8.0cm]{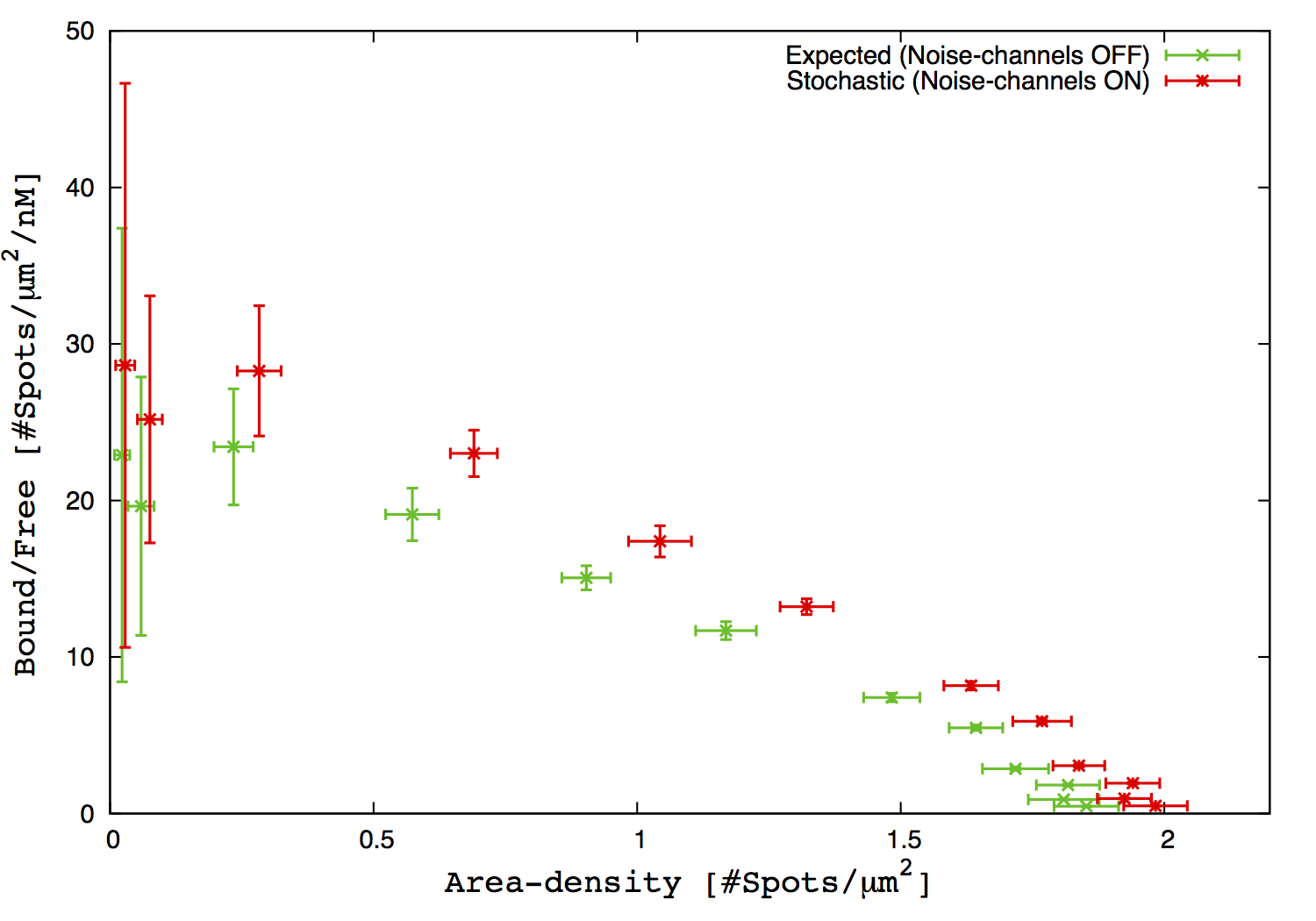}

  \leftline{\bf \hspace{1.0cm} (C) Ligand : $0.600\ {\rm nM}$ \hspace{3.5cm} (D) Ligand : $0.300\ {\rm nM}$ }
  \centering
      \includegraphics[width=8.0cm]{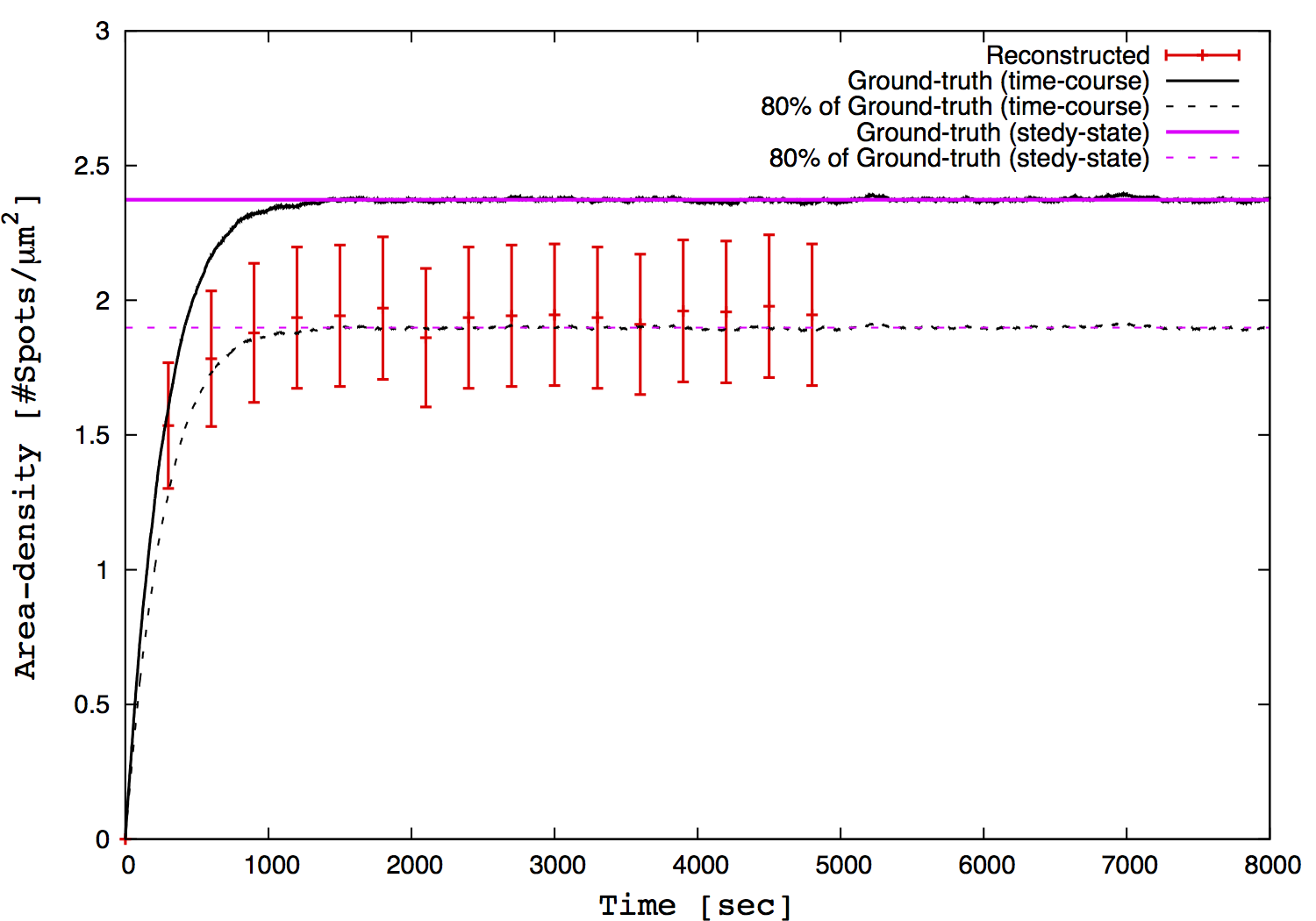}
      \includegraphics[width=8.0cm]{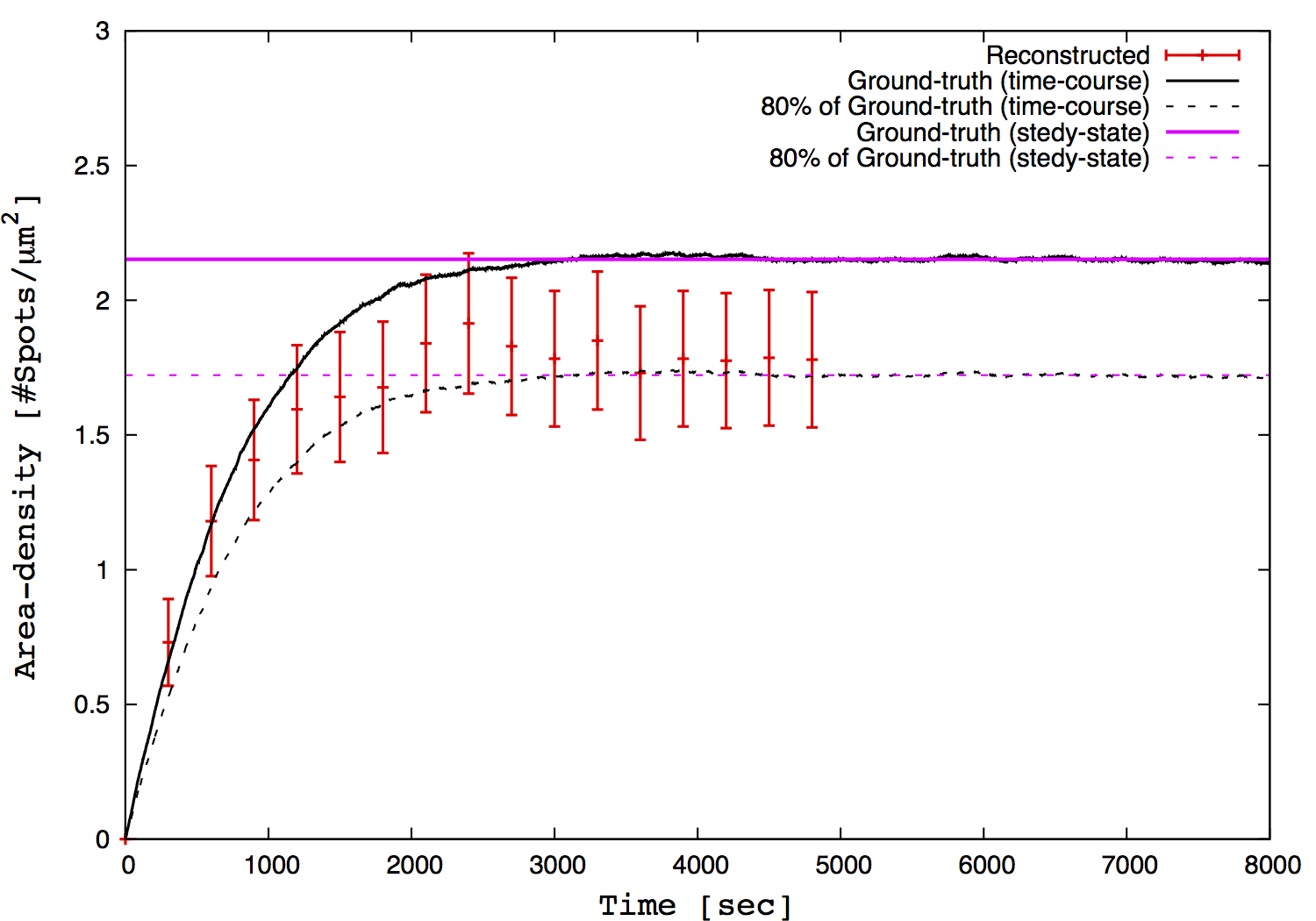}

  \leftline{\bf \hspace{1.0cm} (E) Ligand : $0.030\ {\rm nM}$ \hspace{3.5cm} (F) Ligand : $0.003\ {\rm nM}$ }
  \centering
      \includegraphics[width=8.0cm]{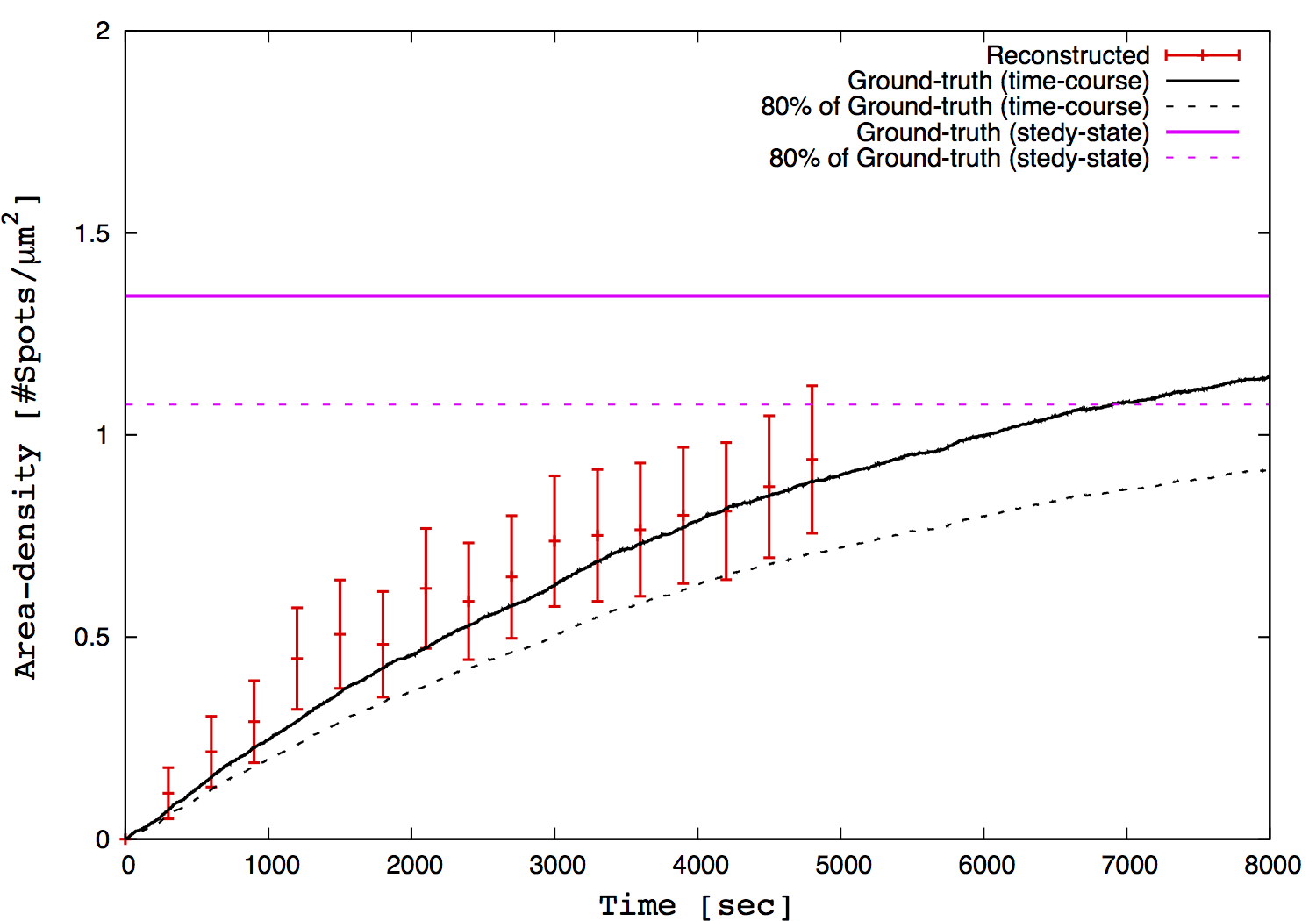}
      \includegraphics[width=8.0cm]{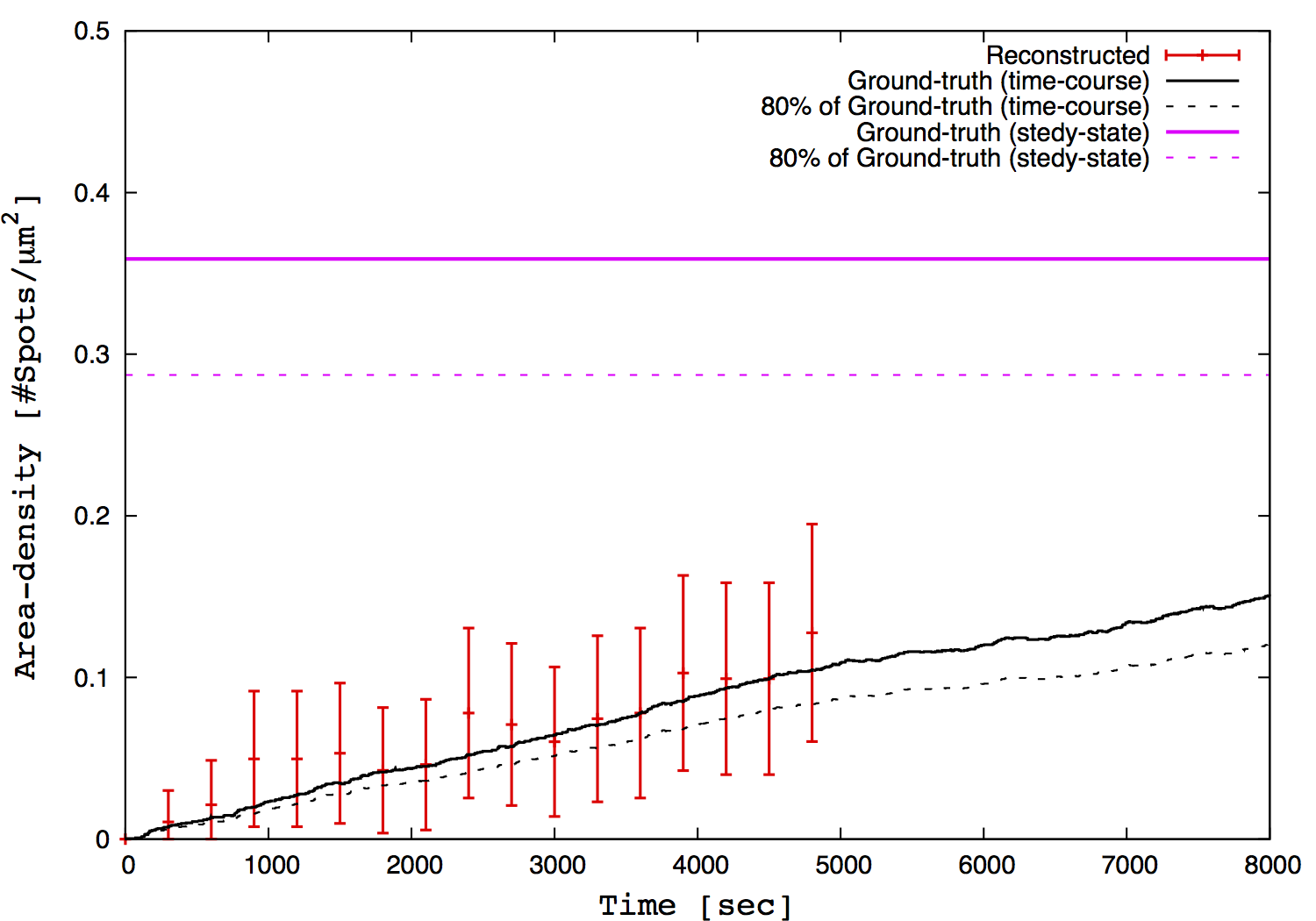}

  \vspace{0.5cm}
  \caption{{\bf Systematic sources}. (A-B) Comparisons of the reconstructed properties: the equilibrium binding curve and Scatchard plot. Green and red lines represent the reconstructed biological properties obtained from the expected and stochastic images. (C-D) Time-course data for $0.600\ {\rm nM}$ and $0.300\ {\rm nM}$ ligand inputs. Solid and dashed pink lines represent the ground-true equilibrium state and $80\%$ of the true equilibrium state. The ground-true response of binding state is represented with black solid line. The $80\%$ restoration of the true response is colored with dashed black lines. (E-F) Time-course data for $0.030\ {\rm nM}$ and $0.003\ {\rm nM}$ ligand inputs.}
  \label{fig;figure_06}
\end{figure}

\paragraph{}
There are three major sources that can generate such systematic shifts: (1) The stochastic process of photon-detection is one of the systematic sources that may influence on the reconstruction procedures, changing the shape and concavity of the equilibrium binding curve and Scatchard plot. We generate the stochastic and expected images by turning on and off the noise channels in the simulation module. We then analyze those images to evaluate the systematic influence to the biological properties. Figures \ref{fig;figure_06}A and B show that all reconstructed properties are aligned in parallel, finding no significant change in the shape and concavity. (2) Imperfect performance of the spot-detection algorithm may influence on our biological interpretation of cooperativity. Figures \ref{fig;figure_06}C and D show examples of the time course data for the ligand concentration inputs more than $0.2\ {\rm nM}$. In this concentration range, the reconstructed responses (presented in the red cross in the Figures) successfully converge to the $80\%$ of the true-equilibrium state (pink dashed lines), reaching the $80\%$ restoration of the true-equilibrium state. As we showed in Figure \ref{fig;figure_05}B, the shape and concavity of the reconstructed curves are unchanged from the ground-truth in the concentration range, implying weak influence on the identification of cooperativity. (3) Quasi-static responses may be a critical issue for the identification of cooperativity. This process is a slow transition of the binding system to come to an equilibrium state. If the binding system is quite sensitive to such slow response, then the system cannot converge to the equilibrium state within the acquisition period. Figures \ref{fig;figure_06}E and F show examples of the time course data for the ligand concentration inputs less than $0.2\ {\rm nM}$. In this concentration range, the reconstructed responses still remain in the nonequilibrium state within the image acquisition period: $0$ to $5,000\ {\rm sec}$. Although the true-nonequilibrium binding  state (black lines) is well-restored through the reconstruction procedure, the reconstructed binding state (red cross) is failed converging to the true-equilibrium states (presented in the pink lines), generating a gap between the reconstructed binding state and the true-equilibrium one. Such a gap cannot even be detected during the image acquisition period, thus leading to the misidentification of cooperativity. 

\subsection*{Confidence in the fitting results}
\begin{figure}[!h]
  \centering
  \leftline{\bf \hspace{1.0cm} (A) \hspace{7.0cm} (B)}
      \includegraphics[width=8.0cm]{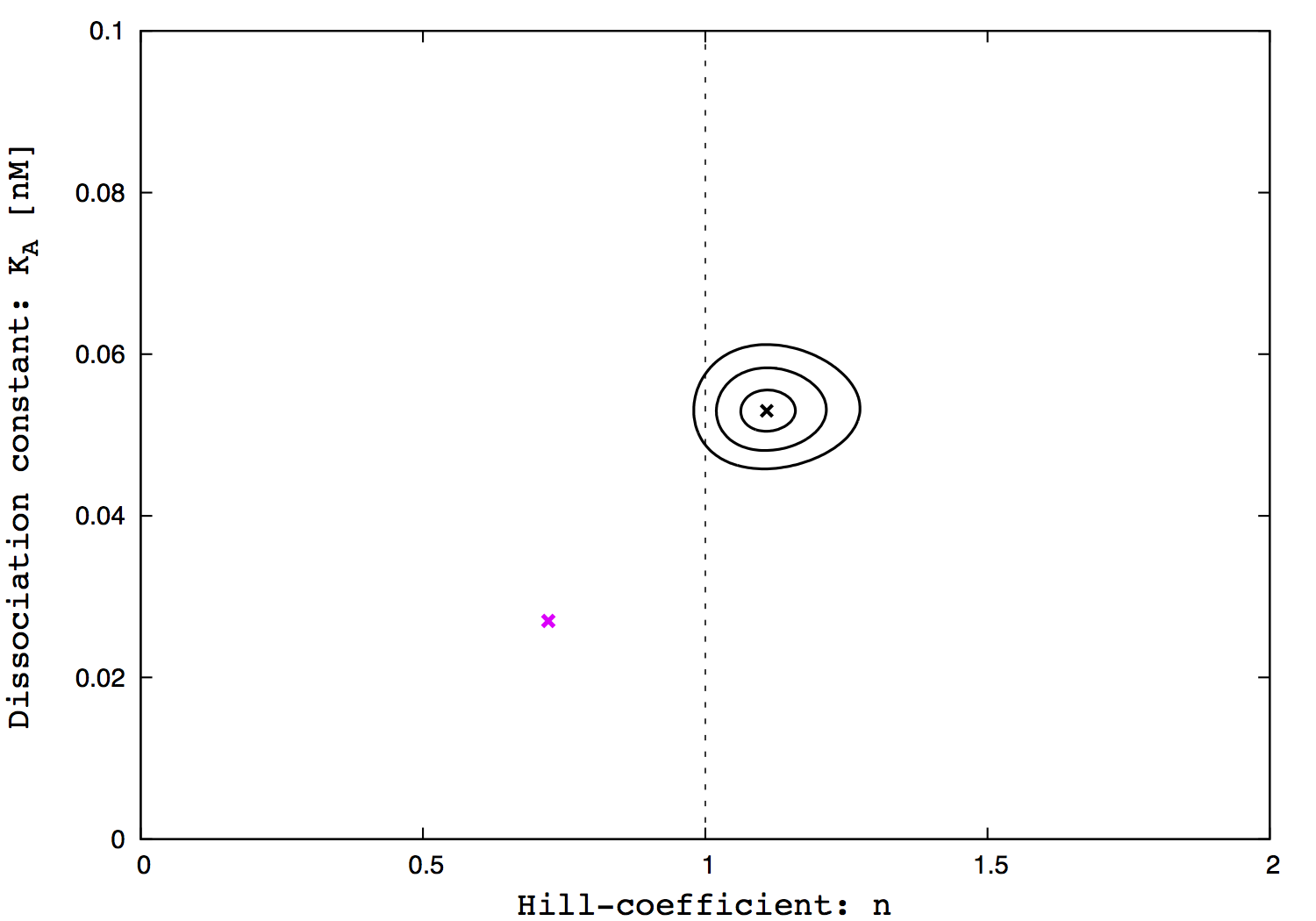}
      \includegraphics[width=8.0cm]{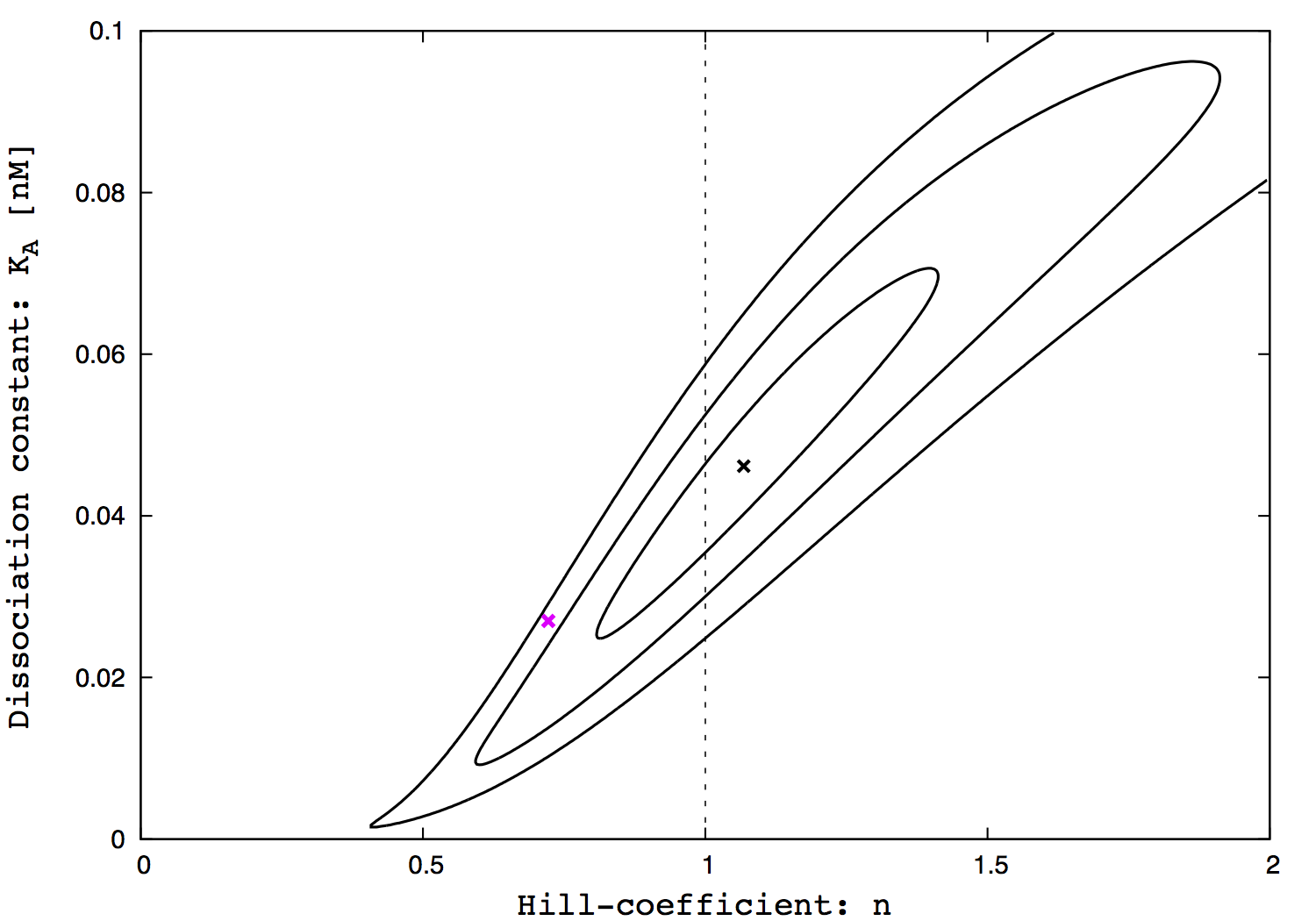}

  \vspace{0.5cm}
  \caption{{\bf Confidence interval}. (A) The $\Delta \chi^2$-contour plot for the image analysis in the full concentration range. The ground-truth and the best fit are represented in pink and black stars. Black contour lines represent $1\sigma$ ($68\%$), $2\sigma$ ($95\%$) and $3\sigma$ ($99\%$) confidence intervals. (B) The $\Delta \chi^2$-contour plot for the image analysis limited in the high concentration range.}
  \label{fig;figure_08}
\end{figure}

\paragraph{}
We estimated statistical uncertainties in the fitting parameters to indicate numerically our confidence in our fitting results \cite{taylor1997, bevington2003}. Figure \ref{fig;figure_08}A shows the $\Delta \chi^2$-contour plot for the results of the parameter fitting in the full concentration range: the ground-truth (pink point) is located out of the $3\sigma$ confidence contour line, implying the restoration failure of the true parameter values. In this analysis, the quasi-static response in the low-concentration range is the major systematic source that can generate the gap between the reconstructed binding state and the true-equilibrium state, thus causing the restoration failure of the ground-true Hill coefficient. Such a systematic gap cannot even be closed by increasing the number of cell-samples. 

\begin{figure}[!h]
  \centering
  \leftline{\bf \hspace{1.0cm} (A) \hspace{7.0cm} (B)}
      \includegraphics[width=8.0cm]{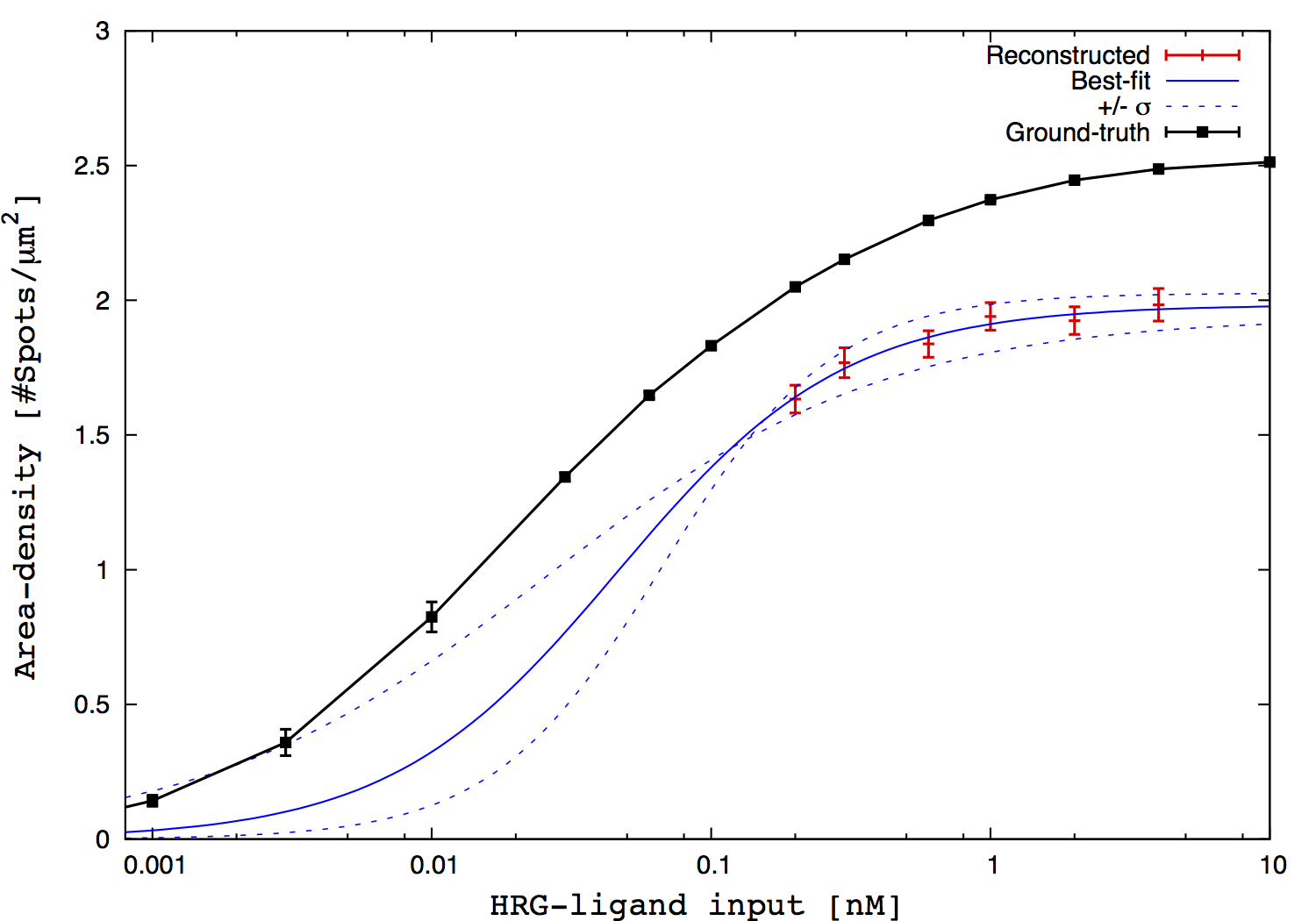}
      \includegraphics[width=8.0cm]{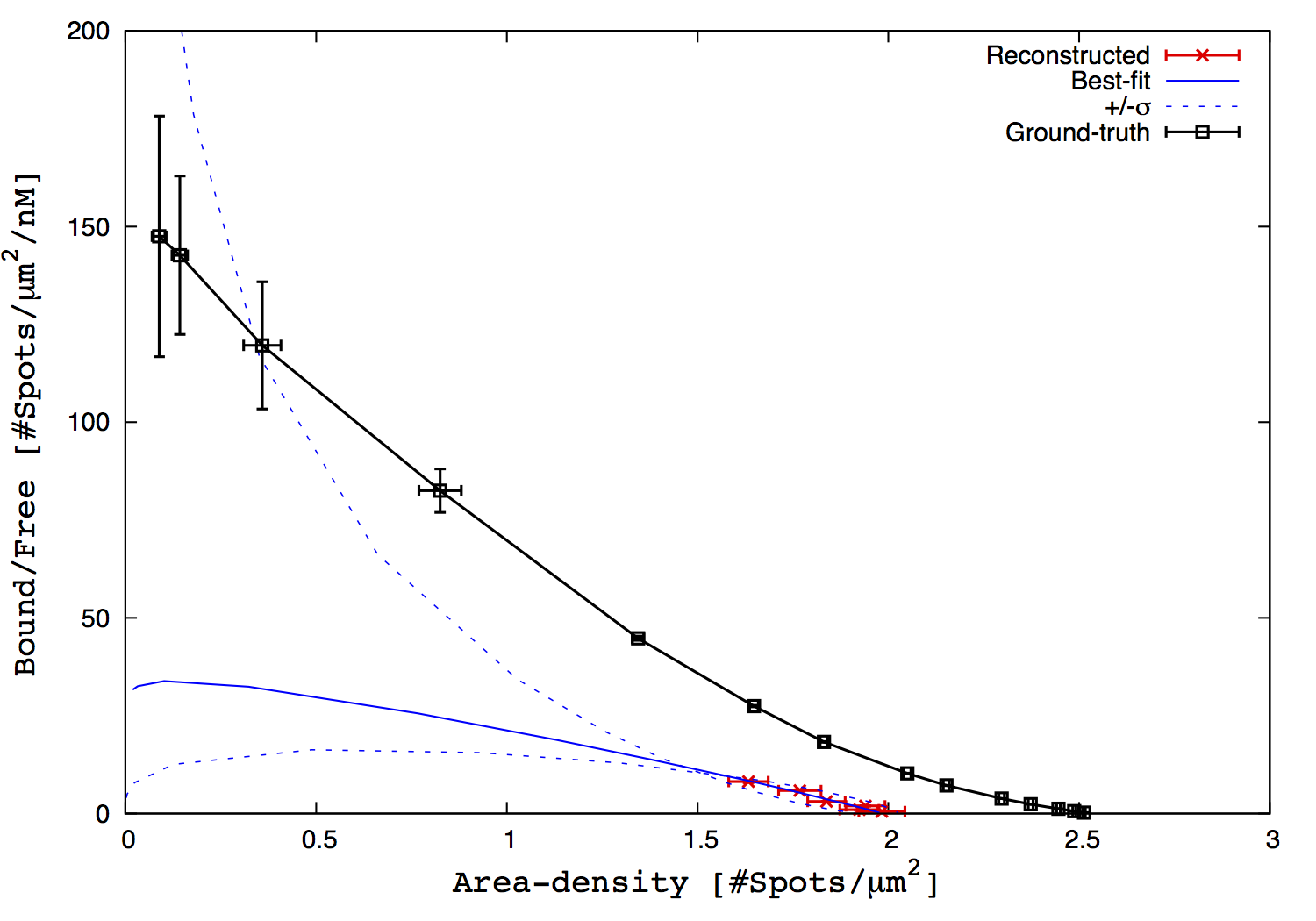}

  \vspace{0.5cm}
  \caption{{\bf Results of the parameter fitting limited in the high concentration range}. $60$ cell-samples are stimulated with the concentration range of $0.2$ to $4.0\ {\rm nM}$.ligand. Black and red lines represent the reconstructed and the ground-true biological properties. The best fit properties and the properties shifted $\pm 1\sigma$ from the best fit are presented in blue solid and dashed lines. (A) Equilibrium binding curve. (B) Scatchard plot.}
  \label{fig;figure_07}
\end{figure}

\paragraph{}
Alternatively, we performed the parameter fitting in the high concentration: $0.2$ to $4.0\ {\rm nM}$ (see SI section C.2.3). The fitting result are shown in the last row of Table \ref{tab;table_01} and in Figures \ref{fig;figure_07}. Figure \ref{fig;figure_08}B shows the $\Delta \chi^2$-contour plot for the fitting results: the ground-truth (pink point) is located within the $3\sigma$ confidence contour line. While the ground-truth of the receptor system exhibits the negative cooperativity, the reconstructed Hill coefficients largely fluctuate around a unity. The cooperative characteristics is thus less determinable in this analysis, but the contour line covers the ground-truth, displaying better results.

\section*{Conclusion}
\paragraph{}
Major scientific activities in biological sciences are dedicated not only to extracting laws and patterns from experimental data but also experimentally validating the biological models derived from experimental knowledge \cite{kitano2002_sci, kitano2002_nat, godfrey2003}. Various model candidates can be either confirmed or refuted by repeating these activities. The key to successful reporting of experimental results is to provide an objective evaluation and representation of the uncertainties that arise from imprecision and inaccuracies in the experimental processes. The study and estimation of the experimental uncertainties have been generally known as error analysis, its main function being to allow biophysicists to numerically indicate the validity and confidence of their experimental results \cite{taylor1997, bevington2003, geris2016}.

\paragraph{}
In the error analysis, statistical analysis of the experimental data is only half the story. The other half is the computation of the systematic uncertainties that affect the sensitivity and limitation of a given experimental configuration to the model parameters and, even more significantly, to new advancements in biophysics and biology. An identification of the new findings in the experimental approaches must always contend with those error estimations. For this reason, we proposed the computational method to evaluate the impact of various systematic uncertainties in the biological measurements using live-cell imaging techniques. We then presented the first examples of not only estimating the systematic uncertainties in model assumptions and parameters (e.g., image acquisition periods and spot-detection efficiency) that can affect the cooperative binding measurements but also reducing them to levels allowing for proper conclusions. In the near future, we believe our computational scheme can help bridges the gap between theory and experiment in biological sciences.

\section*{Acknowledgements}
We would like to thank Yasushi Okada, Tomonobu M. Watanabe, Jun Kozuka, Michio Hiroshima, Kozo Nishida, Kazunari Iwamoto, Yuki Shindo, Hanae Shimo, Yin Fai Chin, Suguru Kato, Toru Niina, Shin Moriga, Taku Tsuzuki, Koji Ochiai, Keiko Itano, Sibel Ozer, Kotone Itaya and Kaoru Ikegami for their guidance and support throughout this research work. We would also like to thank Yasushi Sako and Kylius Wilkins for their critical reading of the manuscript. This research work is supported by the JSPS (Japanese Society for the Promotion of Science) KAKENHI Grants-in-Aid program for Challenging Exploratory Research (15K12146).

\section*{Author Contributions}
MW conceived and designed the problem. MW programmed the fluorescence microscopy simulation module. SNVA, WXC and KK develop E-Cell simulator. MW constructed cell-models. MW performed analyses. MW wrote manuscript. KK and KT provided support and guidance.

\newpage

\newpage

\newpage

\newpage

\newpage

\newpage

\newpage

\newpage

%\bibliography{ref2017}

\newpage

\appendix

\setcounter{figure}{0} \renewcommand{\thefigure}{S\arabic{figure}}
\setcounter{table}{0} \renewcommand{\thetable}{S\arabic{table}}

\leftline{\bf\Large Supporting Information}
\vspace{0.3cm}
\hrule
\vspace{0.3cm}

\paragraph{}
This supporting information not only provides updates of new implementation for the fluorescence microscopy simulation module, but also details for the model construction and the reconstruction of physical and biological properties.\\
\\
\\

\tableofcontents
\addtocontents{toc}{\protect\rule{\textwidth}{.2pt}\par}

\newpage

\section{Implementation updates}

\begin{wrapfigure}{r}{9.3cm}
\vspace{-20pt}
  \centering
      \includegraphics[width=9cm]{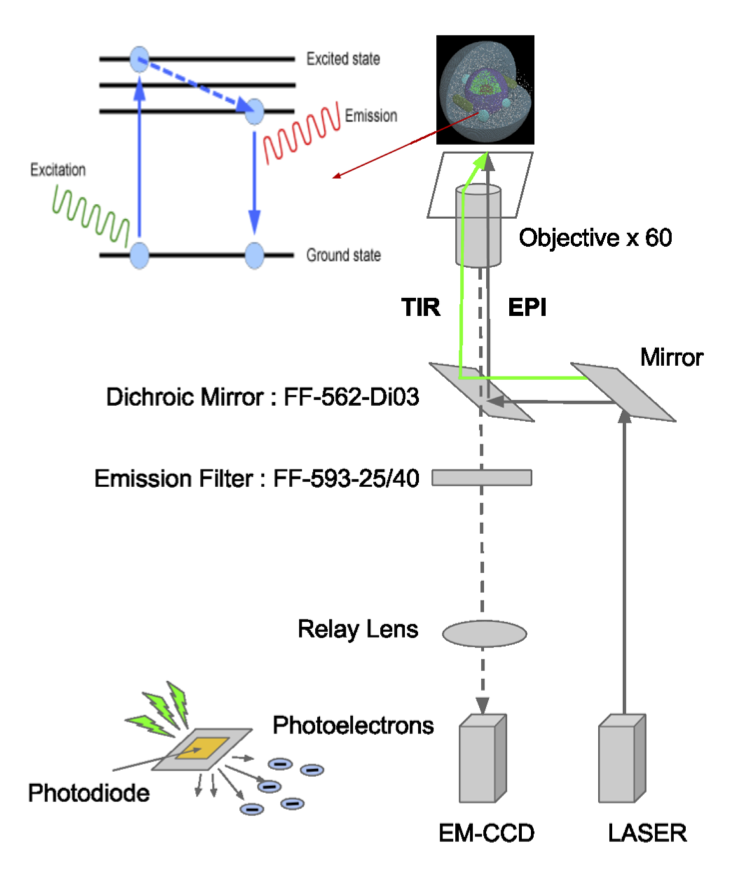}
  \caption[Optical configurations]{Optical configurations of the TIRFM simulation module \cite{watabe2015}.}
  \label{fig;tirfm_config}
\vspace{-30pt}
\end{wrapfigure}

\paragraph{}
We have implemented the simulation modules of total internal reflection fluorescence microscopy (TIRFM) and laser-scanning confocal microscopy (LSCM) \cite{watabe2015}. Those simulation modules were designed to generate digital images that closely represent the actual digital images obtained using actual fluorescence microscopy systems. In this section, we provide updates of new developments to extend the TIRFM simulation module characteristics since the first publication. There are two new features in this update. (1) Monte Carlo method is applied to simulate Beer-Lambert law and the stochastic processes of photobleaching and photoblinking effects. (2) Fluorescence microscopy enables selective illumination of the apical regions of cell-model. Epi-illumination and oblique illumination techniques are implemented. 

%In addition, the complete source code of the simulation module was written in Python and released as an open-source framework at \url{https://github.com/ecell/bioimaging}. The package is freely available for Linux and Mac OS X.

\subsection{TIRFM simulation module}
\paragraph{}
The TIRFM simulation module enables a selective visualization of basal surface regions of a cell-model. Its optical configuration is shown in Figure \ref{fig;tirfm_config} \cite{watabe2015}. Implementation assumptions are summarized in Table \ref{tab;char_tirfm}. The illumination system transfers the photon flux from a light source to a cell-model, to generate a prescribed photon distribution and maximize the flux delivered to the cell-model. Fluorophores defined in the cell-model absorb photons from the distribution, and are quantum-mechanically excited to higher energy states. Molecular fluorescence is the result of physical and chemical processes in which the fluorophores emit photons in the excited state. Finally, the image-forming system relays a nearly exact image of the cell-model to the light-sensitive detector. \\
\\

\begin{table}[!h]
    \centering
    \begin{tabular}{|c|c|}
    \hline
    \multirow{2}{*}{Principle} & Photon-counting \\
    & 1st-order paraxial approximation (Linear term) \\ \hline
    \multirow{2}{*}{Illumination} & Evanescent fields \\
    & Continuous / Uniform / Linearly-polarized \\ \hline
    \multirow{2}{*}{Fluorescence} & Beer-Lamberts law \\
    & Photophysics (photobleaching ... etc)\\ \hline
    \multirow{2}{*}{Image-forming} & 3-D PSF Models (Unpolarized analytical form) \\
    & EMCCD camera \\ \hline
    \end{tabular}
    \caption{Implementation assumptions for the TIRFM simulation module. The detection process for the cameras is performed with Monte Carlo simulation, where EMCCD stand for electron-multiplying charge-coupled device }
    \label{tab;char_tirfm}
\end{table}

\newpage

\subsubsection{Molecular fluorescence}
\paragraph{}
Molecular fluorescence is the result of physical and chemical processes in which fluorophores emit photons from electronically excited states \cite{valeur2012, pawley2008, lakowicz2006}. The Monte Carlo simulation of the overall fluorescence process includes a statistical model of the systematic effects that are influenced by the absorption and emission spectra, quantum yield, lifetime, quenching, photobleaching and blinking, anisotropy, energy transfer, solvent effect, diffusion, complex formation, and a host of environmental variables. In this study, we particularly implement for the fluorescence processes of Beer-Lambertz law, photobleaching and photobliking. Implementation details are described as follows;

\begin{enumerate}
\item[(1)] {\it Beer-Lambert law} \cite{valeur2012, pawley2008, lakowicz2006} : A fundamental aspect of molecular fluorescence is the attenuation of a photon to the properties of materials through which the photon is traveling. The left panel of Figure \ref{fig;beer_lamberts} shows the schematic view of the molecular fluorescence. The relationship of the number of photons entering volume ($n_0$) to the number of photons leaving the volume ($n_1$) is written in the form of 
\begin{eqnarray}
n_{1} & = & n_{0} \times 10^{-A} \\
& & {\rm where}\ \ \ n_{0} = \frac{\sigma\ \delta T}{4 \pi} \left| {\bf A}_{T} \right|^2 \nonumber
\end{eqnarray}
where $\Phi_{QY}$, $\left| {\bf A}_{T} \right|^2$, and $\delta T$ are quantum yield, the transmitted beam flux density, and detection time. The absorption cross-section is given by $\sigma = \frac{\ln(10)}{N_A}\ \epsilon$ where $N_A$ is Avogadro's number. The detector is located in a specific direction. We expect to observe the number of photons devided by an unit surface area of a sphere ($4\pi$). The amplitude of the transmitted beam flux density depends on the index of refraction, and the incident beam angle, amplitude and polarization. The absorption coefficient ($A$) is given by
\begin{equation}
A = \ \log{\left( \frac{I}{I_0} \right)} = \epsilon\ c\ l
\end{equation}
where $\epsilon$, $c$ and $l$ are molar absorption coefficient (or cross-section), volume concentration and path length. In our simulation, we assume that the volume concentration, penetration depth and detection time are given by Spatiocyte voxel-volume, voxel-diameter, and time interval. Finally the expected number of photons emitted from a single fluorophore is given by
\begin{eqnarray}
n_{emit} & = & \Phi_{QY}\ n_{abs} \nonumber \\
& = & \Phi_{QY} \left( n_{0} - n_{1} \right) \nonumber \\
& = & \Phi_{QY}\ n_{0} \left( 1 - 10^{-A} \right)
%& = & \Phi_{QY}\ \frac{\sigma\ \delta T}{4 \pi} \left| {\bf A}_{T} \right|^2 \left( 1 - 10^{-A} \right) 
%\label{eqn;beer_lamberts}
\end{eqnarray}
In previous implementations of bioimaging simulation \cite{watabe2015}, we have assumed that the fluorescence molecules subsequently emit a single photon of longer wavelength while they absorb one million photons of excitation wavelength, and the cross-section of photon-molecule interaction is roughly $10^{-14}\ {\rm cm^2}$. Such approximation is given by
\begin{eqnarray}
n_{emit} & \approx & \Phi_{QY}\ n_{0} \ \ln(10)\ A \nonumber \\
& \sim & \Phi_{QY}\ n_{0} \times 10^{-6}
%\label{eqn;approx}
\end{eqnarray}
The right panel of Figure \ref{fig;beer_lamberts} represents the number of emitted photons as a function of absorption coefficient for various flux densities. Red and black lines represent Beer-Lamberts law and its approximation for various incident beam flux densities. The intersection of the black and red lines is found in the typical range of the absorption coefficients; $10^4$ to $10^6\ {\rm [1/(M\ cm)]}$. 

\item[(2)] {\it Aging Mechanism} \cite{valeur2012, zhao2011, thompson2010, pawley2008, didier2005, zondervan2004} : Statistical aging process, such as photobleaching effect, involves photodynamic interactions between excited fluorophores and molecular oxygen ($O_2$) in its triplet state, dissolved in the sample media. The aging process is governed by Levy statistics. In a simple model, we assume a limited number of photons that each fluorophore can emit (photon budget), and the presence of thermally activated barriers in the state transition between a bright state and a permanently photobleached dark state. Such transition leads to the number of photons decreasing along the time, preventing long exposure time experiments. We assume that the probability distribution associated to time scales of the photobleaching effects, is given by a power-law, and written in the form of
\begin{equation}
P(\tau) = \frac{\alpha}{\tau_0} \left( \frac{\tau_0}{\tau} \right)^{1 + \alpha}
\label{eqn;prob_power}
\end{equation}
where $\alpha$ and $\tau_0$ are dimenssionless constants ($0 < \alpha < 1$) and a characteristics time. In addition, the number of photons emitted before photobleaching depends on photochemical interactions of intracellular singlet oxygen cencentration and on the distance between the fluorophores and intracellular components such as protein and lipids.

\begin{figure}[t]
  \centering
	\includegraphics[width=7.6cm]{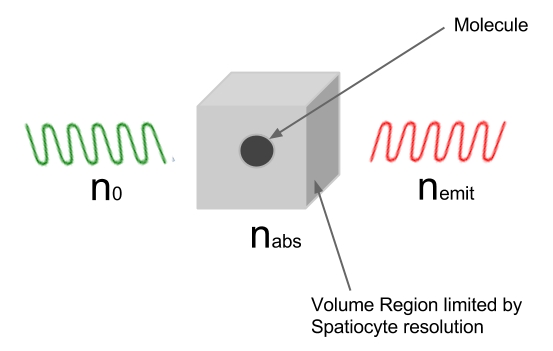}
	\includegraphics[width=7.6cm]{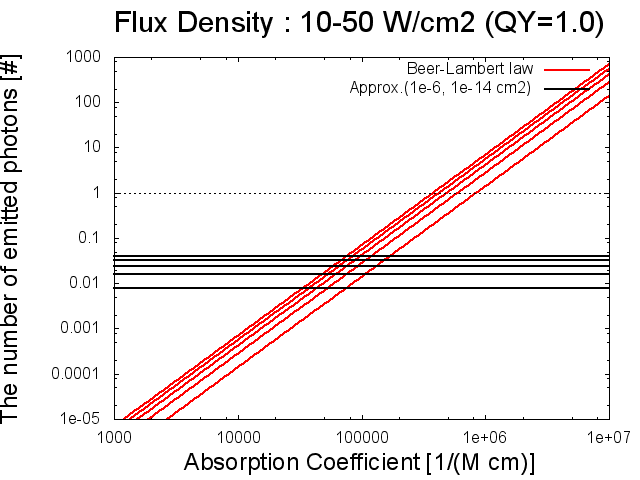}

  \caption{(Left) Schematic view of Beer-Lamberts law. While $n_0$ photons enters the volume region given by Spatiocyte resolution, a fluorescent molecule absobs $n_{abs}$ photons and leaves $n_{emit}$ photons. (Right) The number of emitted photons as a function of absorption coefficient. Red and black lines represent Beer-Lambert's law and its approximation for various incident beam flux densities: $10, 20, 30, 40$, and $50\ {\rm W/cm^2}$, bottom to top.}
  \label{fig;beer_lamberts}
\end{figure}

\item[(3)] {\it Photoblinking} \cite{hoogenboom2007, margolin2005, brokmann2003, shimizu2001} : Time series of each fluorophore often exhibit fluorescence intermittency or blinking, where at random times the fluorophores display the transition between a bright state (ON state in which it stays a time $\tau_{on}$) and a reversible dark state (OFF state in which it stays a time $\tau_{off}$). The process is characterized based on a time sequence of ON and OFF $\{ \tau^{(1)}_{on}, \tau^{(1)}_{off}, \tau^{(2)}_{\tiny on}, \tau^{(2)}_{\tiny off}, \tau^{(3)}_{on}, \tau^{(3)}_{off}, ..... \tau^{(n)}_{on}, \tau^{(n)}_{off} \}$.  The time $\tau^{(i)}_{on/off}$ are drawn at random from the probability distribution given by the equation (\ref{eqn;prob_power}).
%\item[(4)] {\it Photoactivation} \cite{thompson2010} : Effective turn-on ratio including signal from preactivated molecules in dark measurement, is defined as follows;
%\begin{equation}
%R_{eff} = \frac{p R}{1 + q R}
%\label{eqn;r_eff_preact}
%\end{equation}
%where $p$ and $q$ are the overall activation yield, and the fraction of pre-activated molecules. $R = n_{off}$ is the number of dark molecules measured in bulk experiment.
\end{enumerate}

\subsubsection{Examples of single-molecule images}
\begin{figure}[!h]
  \centering
      \includegraphics[width=10cm]{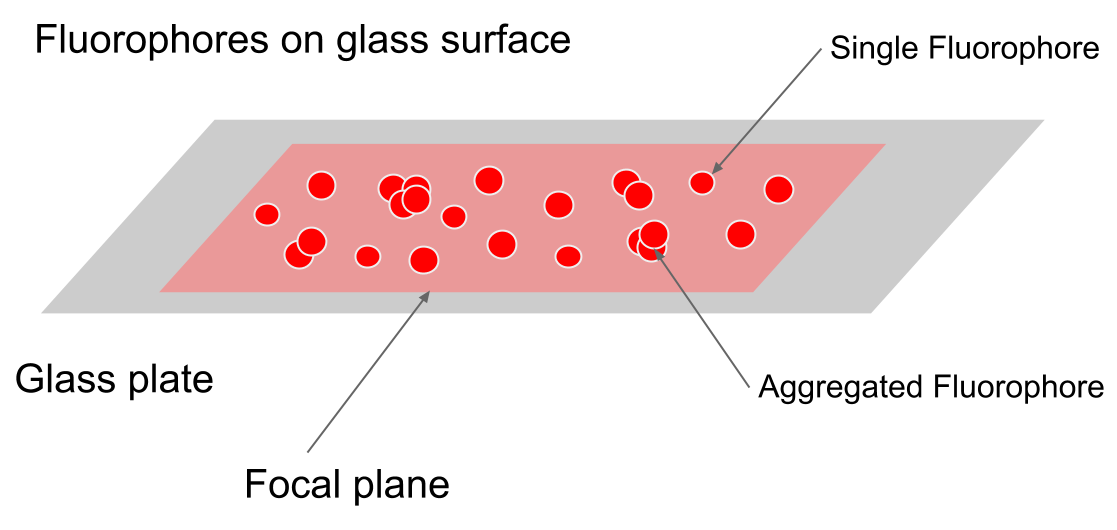}

  \caption{Realistic particle model of TMR-tagged proteins located on glass surface. The aggregation process is included in the model.}
  \label{fig;aggregation_model}
\end{figure}

\begin{figure}[!h]
  \vspace{0.1cm}
  \centerline{\bf 1st Frame ($0.0\ {\rm sec}$) \hspace{3.5cm} 51th Frame ($5.0\ {\rm sec}$)}
  \vspace{0.1cm}
  \centering
	\includegraphics[width=3.4cm]{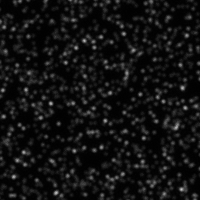}
	\includegraphics[width=4.0cm]{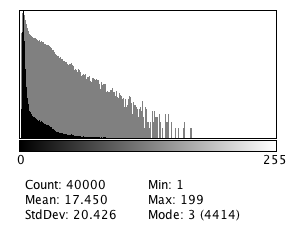}
	\includegraphics[width=3.4cm]{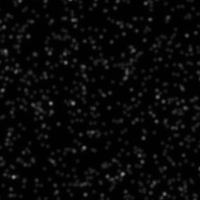}
	\includegraphics[width=4.0cm]{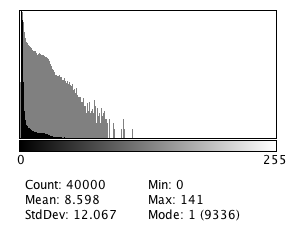}
		
  \vspace{0.1cm}
  \centerline{\bf 11th Frame ($1.0\ {\rm sec}$) \hspace{3.5cm} 61th Frame ($6.0\ {\rm sec}$)}
  \vspace{0.1cm}
  \centering
	\includegraphics[width=3.4cm]{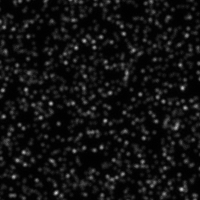}
	\includegraphics[width=4.0cm]{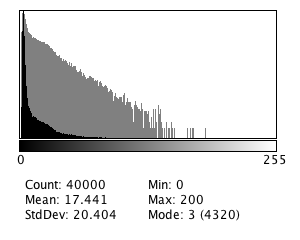}
	\includegraphics[width=3.4cm]{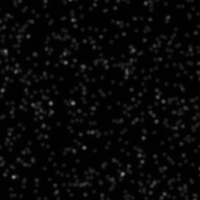}
	\includegraphics[width=4.0cm]{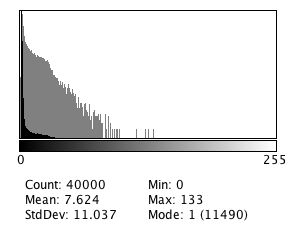}
		
  \vspace{0.1cm}
  \centerline{\bf 21th Frame ($2.0\ {\rm sec}$) \hspace{3.5cm} 71th Frame ($7.0\ {\rm sec}$)}
  \vspace{0.1cm}
  \centering
	\includegraphics[width=3.4cm]{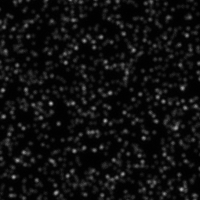}
	\includegraphics[width=4.0cm]{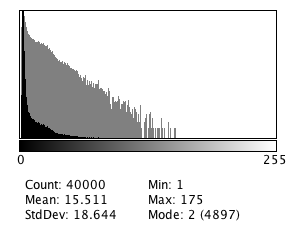}
	\includegraphics[width=3.4cm]{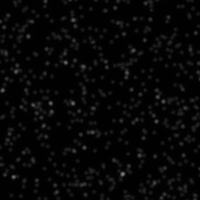}
	\includegraphics[width=4.0cm]{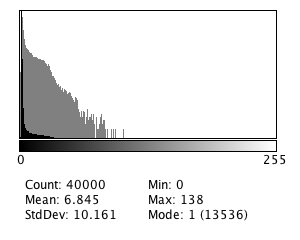}
		
  \vspace{0.1cm}
  \centerline{\bf 31th Frame ($3.0\ {\rm sec}$) \hspace{3.5cm} 81th Frame ($8.0\ {\rm sec}$)}
  \vspace{0.1cm}
  \centering
	\includegraphics[width=3.4cm]{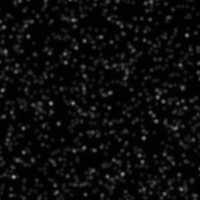}
	\includegraphics[width=4.0cm]{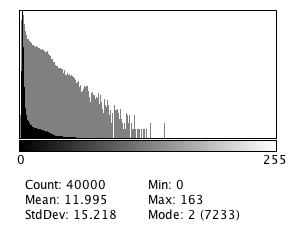}
	\includegraphics[width=3.4cm]{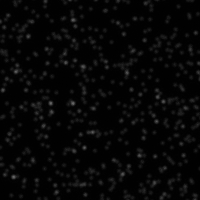}
	\includegraphics[width=4.0cm]{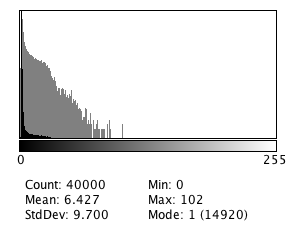}
		
  \vspace{0.1cm}
  \centerline{\bf 41th Frame ($4.0\ {\rm sec}$) \hspace{3.5cm} 91th Frame ($9.0\ {\rm sec}$)}
  \vspace{0.1cm}
  \centering
	\includegraphics[width=3.4cm]{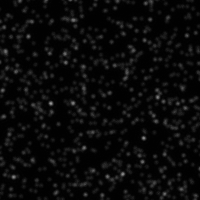}
	\includegraphics[width=4.0cm]{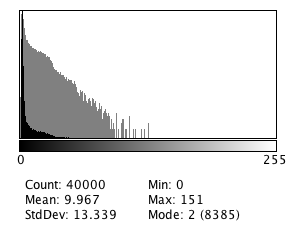}
	\includegraphics[width=3.4cm]{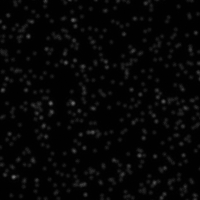}
	\includegraphics[width=4.0cm]{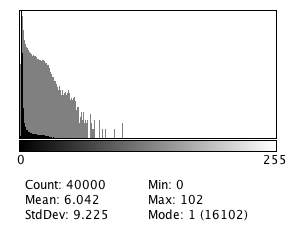}
				
  \caption{{\bf Examples of single-molecule images (1).} 10 frames of the aggregation model that includes photobleaching effects. Each image size is $200 \times 200$ pixels. Actual minimum and maximum values of the intensity histogram are $1,900$ and $9,500$ ADC counts. The intensity is rescaled in the range of $0$ to $255$. }
  \label{fig;example_images_plane_aggr05_bleach}
\end{figure}

\begin{figure}[!h]
  \vspace{0.1cm}
  \centerline{\bf 1st Frame ($0.0\ {\rm sec}$) \hspace{3.5cm} 51th Frame ($5.0\ {\rm sec}$)}
  \vspace{0.1cm}
  \centering
	\includegraphics[width=3.4cm]{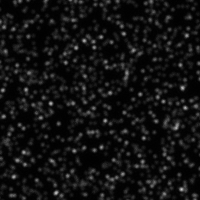}
	\includegraphics[width=4.0cm]{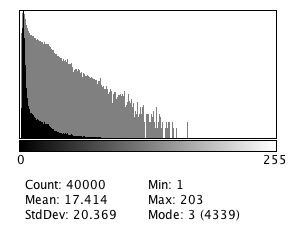}
	\includegraphics[width=3.4cm]{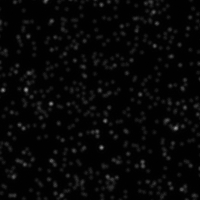}
	\includegraphics[width=4.0cm]{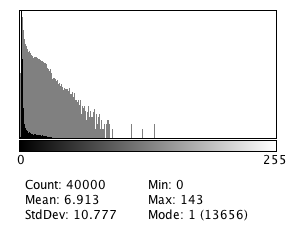}
		
  \vspace{0.1cm}
  \centerline{\bf 11th Frame ($1.0\ {\rm sec}$) \hspace{3.5cm} 61th Frame ($6.0\ {\rm sec}$)}
  \vspace{0.1cm}
  \centering
	\includegraphics[width=3.4cm]{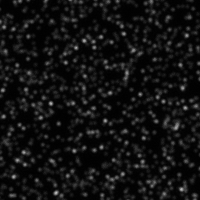}
	\includegraphics[width=4.0cm]{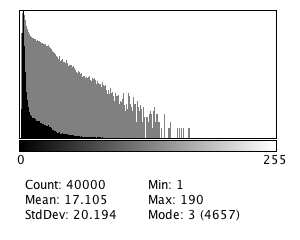}
	\includegraphics[width=3.4cm]{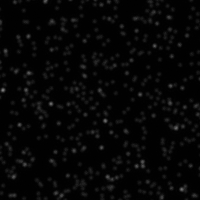}
	\includegraphics[width=4.0cm]{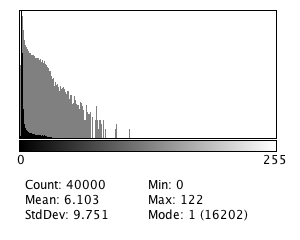}
		
  \vspace{0.1cm}
  \centerline{\bf 21tn Frame ($2.0\ {\rm sec}$) \hspace{3.5cm} 71th Frame ($7.0\ {\rm sec}$)}
  \vspace{0.1cm}
  \centering
	\includegraphics[width=3.4cm]{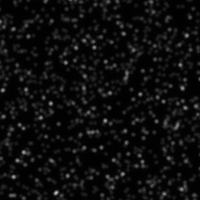}
	\includegraphics[width=4.0cm]{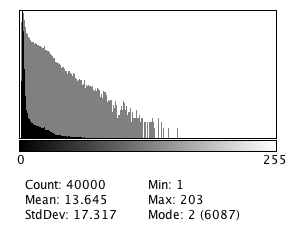}
	\includegraphics[width=3.4cm]{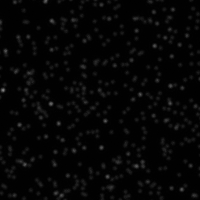}
	\includegraphics[width=4.0cm]{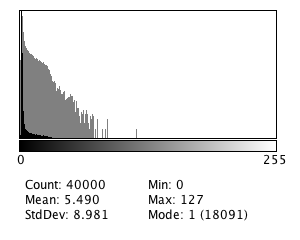}
		
  \vspace{0.1cm}
  \centerline{\bf 31th Frame ($3.0\ {\rm sec}$) \hspace{3.5cm} 81th Frame ($8.0\ {\rm sec}$)}
  \vspace{0.1cm}
  \centering
	\includegraphics[width=3.4cm]{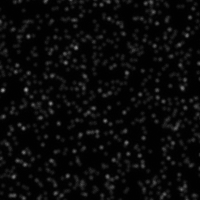}
	\includegraphics[width=4.0cm]{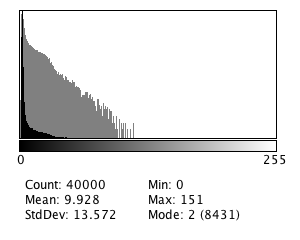}
	\includegraphics[width=3.4cm]{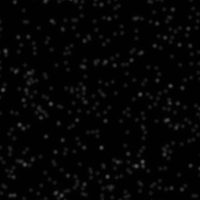}
	\includegraphics[width=4.0cm]{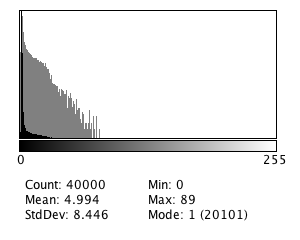}
		
  \vspace{0.1cm}
  \centerline{\bf 41th Frame ($4.0\ {\rm sec}$) \hspace{3.5cm} 91th Frame ($9.0\ {\rm sec}$)}
  \vspace{0.1cm}
  \centering
	\includegraphics[width=3.4cm]{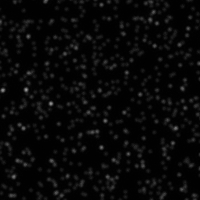}
	\includegraphics[width=4.0cm]{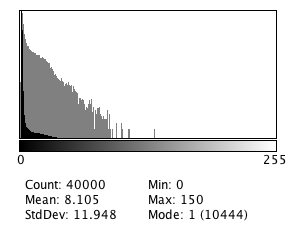}
	\includegraphics[width=3.4cm]{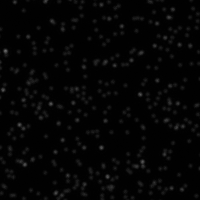}
	\includegraphics[width=4.0cm]{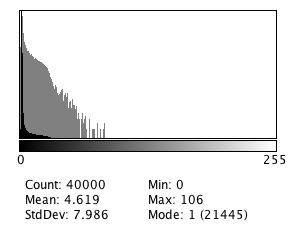}
				
  \caption{{\bf Examples of single-molecule images (2).} 10 frames of the aggregation model that includes photobleaching and photoblinking effects. Each image size is $200 \times 200$ pixels. Actual minimum and maximum values of the intensity histogram are $1,900$ and $9,500$ ADC counts. The intensity is rescaled in the range of $0$ to $255$. }
  \label{fig;example_images_plane_aggr05_blink}
\end{figure}

\paragraph{}
We constructed realistic particle model of TMR-tagged proteins on glass surface as shown in Figure \ref{fig;aggregation_model}. We assumed that $10,000$ fluorescent molecules are stationary, and randomly distributed on the surface ($30 \times 30\ {\rm \mu m^2}$). An aggregation process where many fluorophores form clusters in a colloidal suspension, is included in the model. Images are simulated for the optical specification and condition of the TIRFM simulation module shown in Table \ref{tab;specification1}. Incident beam angle is $65.7^{\circ}$ (Evanescent fields). Photophysical parameterizations of fluorophores are also shown in the Table. Results are shown in Figures \ref{fig;example_images_plane_aggr05_bleach} and \ref{fig;example_images_plane_aggr05_blink}. These Figures show snapshot images captured with TIRFM imaging configuration.

\begin{figure}[!h]
  \centering
	\includegraphics[width=7.0cm]{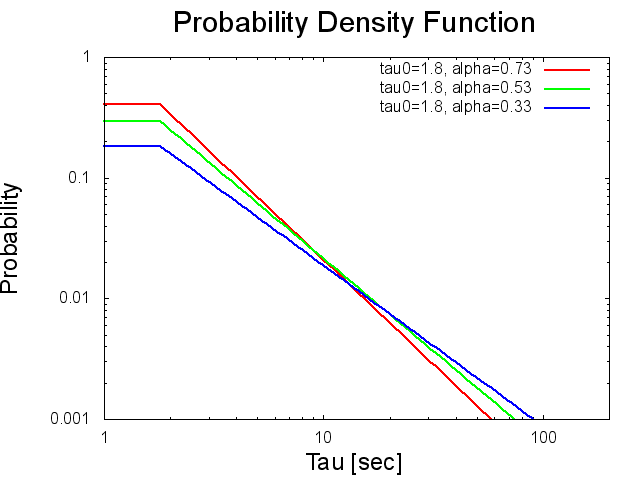}
	\includegraphics[width=7.0cm]{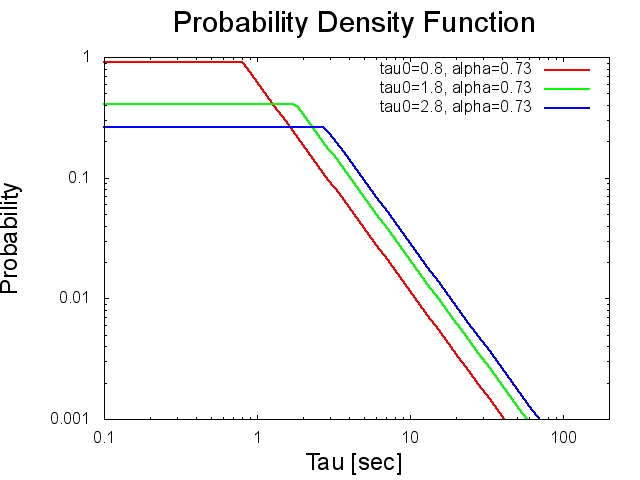}

  \caption{(Left) Probability distribution for fixed $\tau_0$ ($= 1.8\ {\rm sec}$) and various $\alpha$ ($= 0.73, 0.53, 0.33$). (Right) Probability distribution for various $\tau_0$ ($= 0.8, 1.8, 2.8\ {\rm sec}$) and fixed $\alpha$ ($= 0.73$). }
  \label{fig;beer_lamberts}
%\end{figure}
%
%\begin{figure}[!h]
  \centering
	\includegraphics[width=7.0cm]{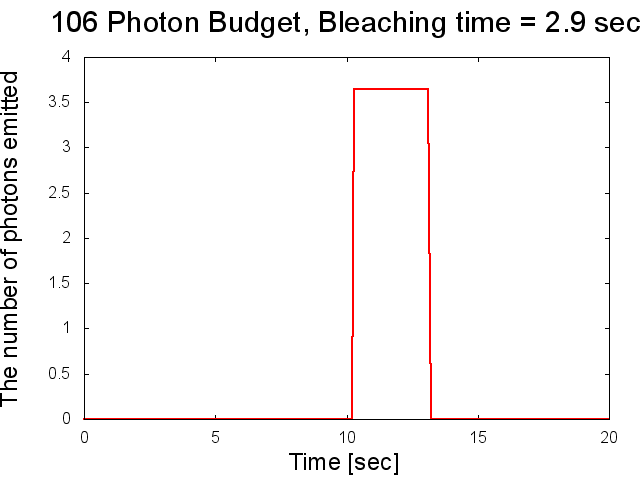}
	\includegraphics[width=7.0cm]{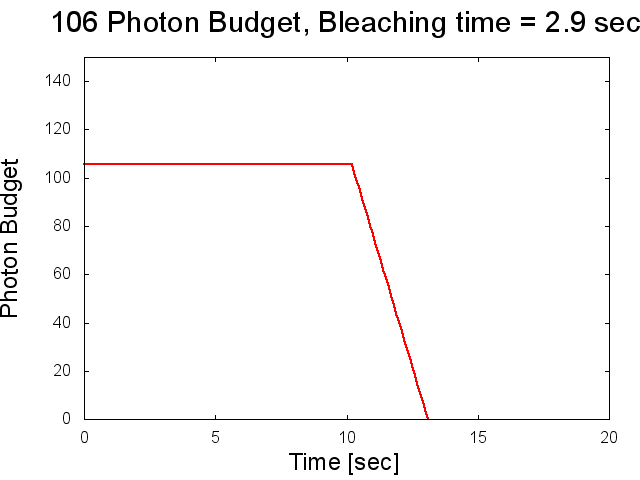}

  \caption{(Left) The behavior of photobleaching effect is presented as the number of photons emitted per single-molecule in exposure time. A fluorophore is activated around $10\ {\rm sec}$, and emits $3.6$ photons per cycle. (Right) Photon budget reducing with time. No photon is emitted after photobleaching at $\sim 13\ {\rm sec}$.}
  \label{fig;bleaching}
%\end{figure}
%
%\begin{figure}[!h]
  \centering
	\includegraphics[width=7.0cm]{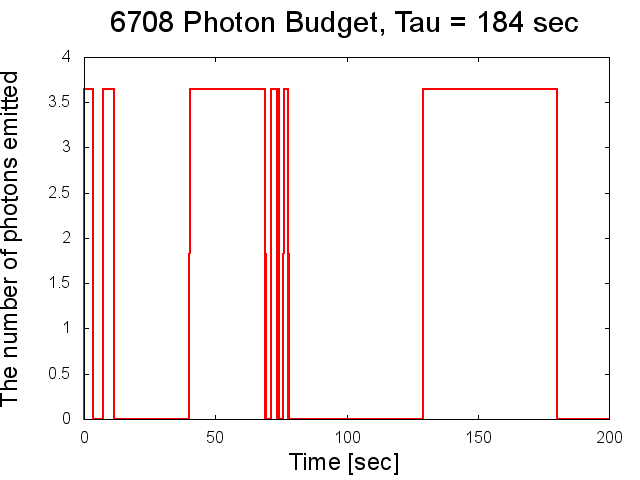}
	\includegraphics[width=7.0cm]{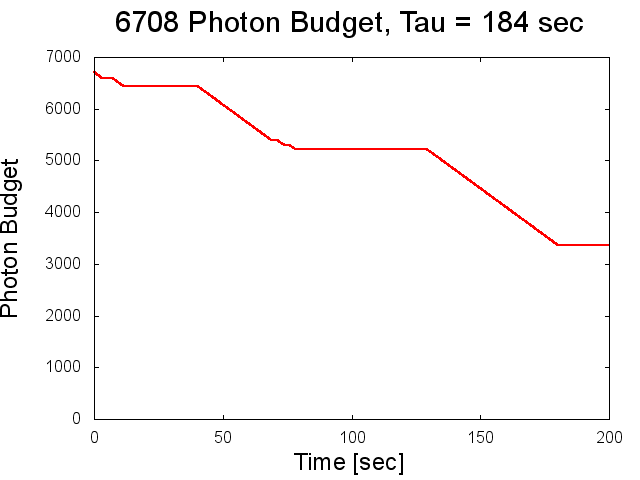}

  \caption{(Left) The binary behavior of photoblinking effect is shown. The time sequence of ON and OFF states (photoblinking process) is presented in the number of photons emitted per single-molecule over exposure time. A fluorophore is activated around $0\ {\rm sec}$, and emits $3.6$ photons per cycle. (Right) Photon budget reducing in time. No photon is emitted when the molecule is on the OFF state. }
  \label{fig;blinking}
\end{figure}

\newpage

\begin{table}[!h]
\centering
\begin{tabular}{|p{7cm}|c|}
\hline
\multicolumn{2}{|c|}{\bf Excitation Beam} \\ \hline
Flux density & $20\ {\rm W/cm^2}$ \\ \hline
Wavelength & $488\ {\rm nm}$ \\ \hline
Refraction index & $1.47\ ({\rm glass})$, $1.33\ ({\rm water})$ \\ \hline
Critical angle & $65.6^{\circ}$ \\ \hline
\multicolumn{2}{|c|}{\bf Optical elements}\\ \hline
Objective & $\times\ 60\ /\ {\rm N.A.}\ 1.49$ \\ \hline
Dichroic mirror & Semrok FF-562-Di03 \\ \hline
Emission filter & Semrok FF-593-25/40 \\ \hline
Tube lens & $\times\ 4.17$ \\ \hline
Optical magnification & $\times\ 250$ \\ \hline
Optical background & $0.0\ {\rm photons/pixel}$ \\ \hline
\multicolumn{2}{|c|}{\bf EMCCD Camera (Hamamatsu model)}\\ \hline
Image size & $512 \times 512$ \\ \hline
Pixel length & $16\ {\rm \mu m}$ \\ \hline
Quantum efficiency & $92\ \%$ \\ \hline
EM Gain & $\times 300$ \\ \hline
Exposure time & $100\ {\rm msec}$ \\ \hline
Readout noise & $100\ {\rm electrons}$ \\ \hline
Full well & $370,000\ {\rm electrons}$ \\ \hline
Dynamic range & $71.3\ {\rm dB}$ \\ \hline
Excess noise & $\sqrt{2}$ \\ \hline
A/D Converter & $16$-bit \\ \hline
Gain & $5.82\ {\rm electrons/count}$ \\ \hline
Offset  & $2000\ {\rm counts}$ \\ \hline
\multicolumn{2}{|c|}{\bf Photophysics} \\ \hline
Fluorophore & \hspace{1.0cm}  TMR\ (${\rm Abs.}\ 548\ {\rm nm}\ /\ {\rm Em.}\ 608\ {\rm nm}$) \hspace{1.0cm} \\ \hline
Fluorescence quantum yield & $61\ \%$ \\ \hline
Absorption coefficient (cross-section) & $83400\ {\rm M^{-1} cm^{-1}}$ ($\sigma = 3.19 \times 10^{-16}\ {\rm cm^2}$) \\ \hline
Photobleaching (assuming power-law) & $\tau_0 = 1.8\ {\rm sec}$, $\alpha = 0.73$ \\ \hline
\multirow{2}{*}{Photoblinking (assuming power-law)} & ON\ \ ($\tau_0 = 1.0\ {\rm sec},\ \alpha = 0.58$) \\
& OFF\ ($\tau_0 = 10\ {\rm \mu sec},\ \alpha = 0.48$) \\ \hline
%\multirow{2}{*}{Photoactivation} & $1000$ (Turn on ratio) \\
%& $10\ \%$ (Activation yield) \\
%& $0.00\ \%$ (Fraction of preactivation) \\ \hline
\end{tabular}
%\vspace{0.3cm}
\caption{Microscopy specifications and operating conditions for imaging the particle models.}
\label{tab;specification1}
\end{table}

\newpage

\newpage

\newpage

\subsection{Fluorescence microscopy simulation module}
\paragraph{}
We implemented the simulation module for fluorescence microscopy. The simulation module enables a selective visualization of apical and basal surface regions of a cell-model. Its optical configuration is shown in Figure \ref{fig;tirfm_config}. Implementation assumptions are summarized in Table \ref{tab;char_oim}. 

\begin{table}[!h]
    \centering
    \begin{tabular}{|c|c|}
    \hline
    \multirow{2}{*}{Principle} & Photon-counting \\
    & 1st-order paraxial approximation (Linear term) \\ \hline
    \multirow{2}{*}{Illumination} & Epi-illumination / Oblique illumination\\
    & Continuous / Uniform / Linearly-polarized \\ \hline
    \multirow{2}{*}{Fluorescence} & Beer-Lamberts law \\
    & Photophysics (photobleaching ... etc)\\ \hline
    \multirow{2}{*}{Image-forming} & 3-D PSF Models (Unpolarized analytical form) \\
    & EMCCD camera \\ \hline
    \end{tabular}
    \caption{Implementation assumptions for the epifluorescence microscopy. The detection process for the cameras is performed with Monte Carlo simulation. }
    \label{tab;char_oim}
\end{table}

\subsubsection{Epi-illumination / oblique illumination}
\begin{figure}[!h]
  \centering
      \includegraphics[width=16cm]{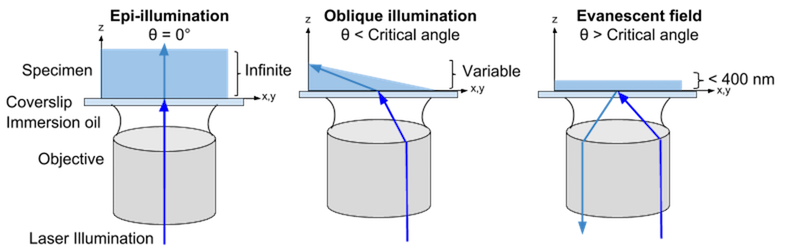}

  \caption{Epi-illumination (left), Oblique illumination (middle) and Evanescent field (right).}
  \label{fig;illumination_schematics}
\end{figure}

\begin{figure}[!h]
  \centering
      \includegraphics[width=8cm]{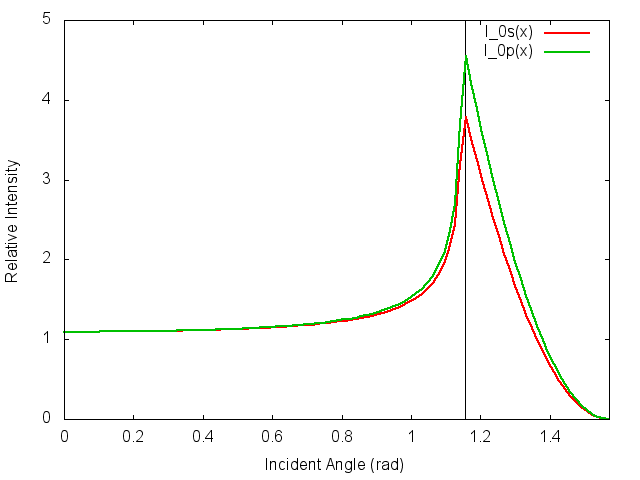}
      \includegraphics[width=8cm]{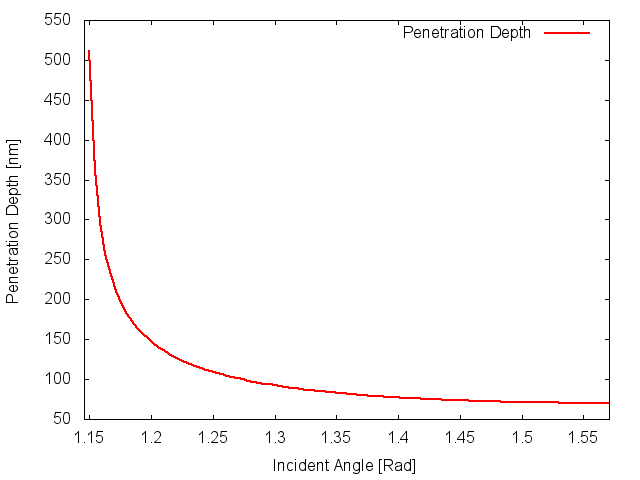}
      
  \caption{Intensity transition of epi-illumination to the evanescent field (left) and the penetration depth of the evanescent field as a function of the incident beam angle (right). }
  \label{fig;illumination_intensity}
\end{figure}

\paragraph{}
An incident beam of excitation wavelength ($\lambda$) that passed through the objective lens is assumed to uniformly illuminate the specimen. The surviving photons through the use of excitation filters interact with the fluorophores in the cell-model, and excite the fluorophores to the electrically excited state. The optics simulations for the focusing of the incident photons through the objective lens include a statistical model of systematic parameter ruled by specifications including numerical aperture (NA), magnification, working distance, degree of aberration, correction of refracting surface radius, thickness, refractive index and details of each lens element. Figure \ref{fig;illumination_schematics} illustrates schematic views of epi-illumination, oblique illumination and Evanescent field. The left panel of Figure \ref{fig;illumination_intensity} shows the relative intensity of incident electric fields as a function of incident beam angle for s- and p-polarization. The polarization of the illumination field depends on the incident beam polarization, which can be either p-pol (polarized in the plane of the incidence formed by the incident and reflected rays, denoted here as the x-z plane) or s-pol (polarized normal to the plane of incidence; here, in the y-direction). 

\newpage

\paragraph{}
If the incidence beam angles are less than the critical angles given by $\sin \theta_{c} = n_2/n_1$ where medias are fused silica ($n_1 = 1.46$) and water ($n_2=1.33$), then most of the incidence beam propagates through the interface into the lower index material with a refraction angle given by Snell's Law. If the incidence angle is at $0$ degree, then the epi-illumination is generated along the z-axis. If the angle is at $0 < \theta \ll \theta_{c}$, the oblique illumination is generated along the refracted beam axis. Finally, if the angle is $\theta > \theta_{c}$, then the incidence beam undergoes total internal refraction (TIR). The evanescent field is generated along the z-axis, perpendicular to the TIR surface, and is capable of exciting the fluorescent molecules near the surface. The intensity of the evanescent field at any position exponentially decays with z. The right panel of Figure \ref{fig;illumination_intensity} shows the penetration depth of the evanescent field as a function of the incident beam angle. More details are described in reference \cite{miyanaga2009, axelrod2008, wazawa2005, axelrod2003}. 

\begin{figure}[!h]
  \centering
      \includegraphics[width=16cm]{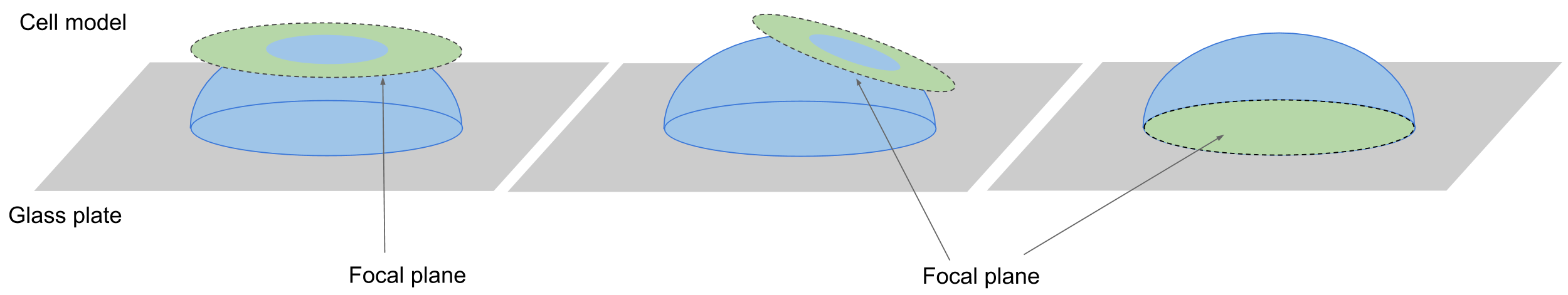}

  \caption{Epi-illumination (left), oblique illumination (midddle) and Evanescent field (right).}
  \label{fig;cell_model_illumination}
\end{figure}

\begin{figure}[!h]
  \leftline{\bf 1st Frame ($0.0\ {\rm sec}$)}
  \vspace{0.1cm}
  \centering
	\includegraphics[width=4.1cm]{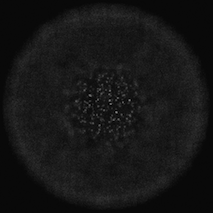}
	\includegraphics[width=4.1cm]{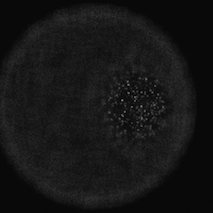}
	\includegraphics[width=4.1cm]{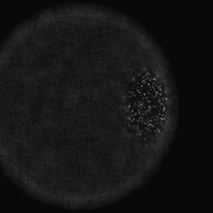}
	\includegraphics[width=4.1cm]{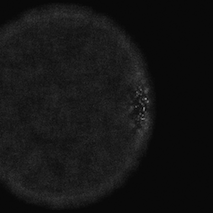}
	
  \leftline{\bf 11th Frame ($1.0\ {\rm sec}$)}
  \vspace{0.1cm}
  \centering
	\includegraphics[width=4.1cm]{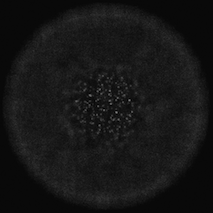}
	\includegraphics[width=4.1cm]{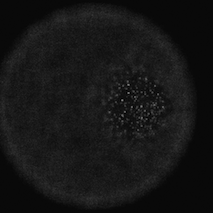}
	\includegraphics[width=4.1cm]{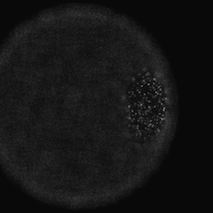}
	\includegraphics[width=4.1cm]{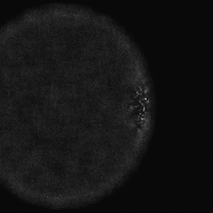}
	
  \leftline{\bf 21st Frame ($2.0\ {\rm sec}$)}
  \vspace{0.1cm}
  \centering
	\includegraphics[width=4.1cm]{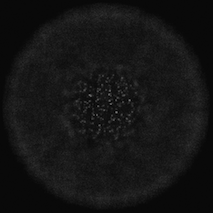}
	\includegraphics[width=4.1cm]{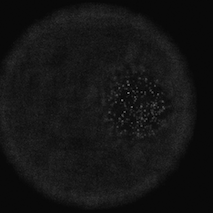}
	\includegraphics[width=4.1cm]{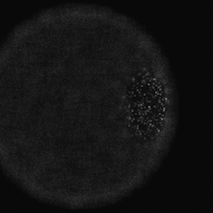}
	\includegraphics[width=4.1cm]{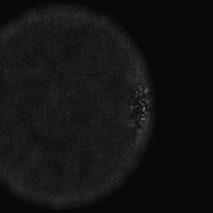}
	
  \leftline{\bf 31st Frame ($3.0\ {\rm sec}$)}
  \vspace{0.1cm}
  \centering
	\includegraphics[width=4.1cm]{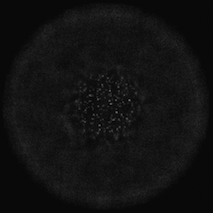}
	\includegraphics[width=4.1cm]{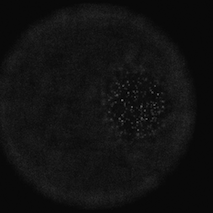}
	\includegraphics[width=4.1cm]{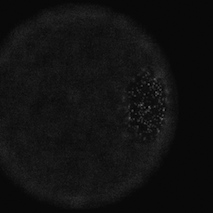}
	\includegraphics[width=4.1cm]{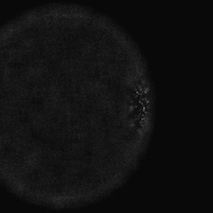}
	
  \leftline{\bf Epi-illumination ($\theta = 0^{\circ}$) \hspace{2.0cm} Oblique illumination ($\theta = 8^{\circ},\ 16^{\circ},\ 24^{\circ}$)}
  \vspace{0.1cm}
  \caption{{\bf Examples of single-molecule images (1).} 8 frames of the simple cell-model that includes particle aggregation and photobleaching effects. Each image size is $512 \times 512$ pixels. Actual minimum and maximum values of the intensity histogram are $1,900$ and $4,800$ ADC counts. The intensity is rescaled in the range of $0$ to $255$. }
  \label{fig;example_images_cell_model_bleach_0}
\end{figure}

\begin{figure}[!h]
  \leftline{\bf 41st Frame ($4.0\ {\rm sec}$)}
  \vspace{0.1cm}
  \centering
	\includegraphics[width=4.1cm]{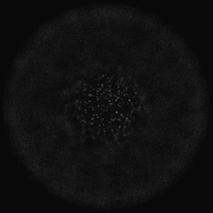}
	\includegraphics[width=4.1cm]{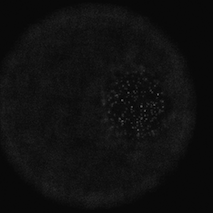}
	\includegraphics[width=4.1cm]{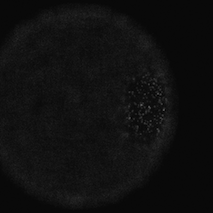}
	\includegraphics[width=4.1cm]{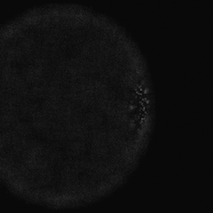}
	
  \leftline{\bf 51st Frame ($5.0\ {\rm sec}$)}
  \vspace{0.1cm}
  \centering
	\includegraphics[width=4.1cm]{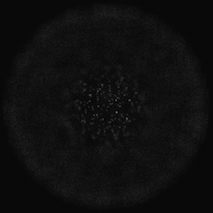}
	\includegraphics[width=4.1cm]{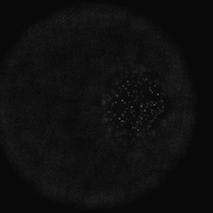}
	\includegraphics[width=4.1cm]{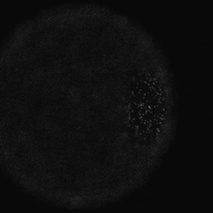}
	\includegraphics[width=4.1cm]{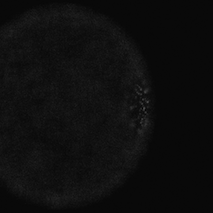}
	
  \leftline{\bf 61st Frame ($6.0\ {\rm sec}$)}
  \vspace{0.1cm}
  \centering
	\includegraphics[width=4.1cm]{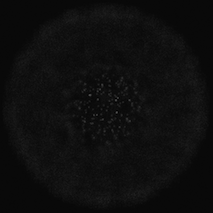}
	\includegraphics[width=4.1cm]{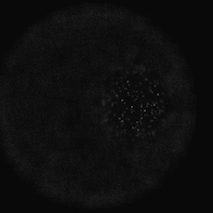}
	\includegraphics[width=4.1cm]{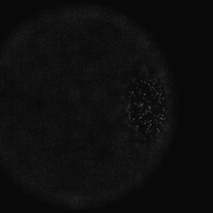}
	\includegraphics[width=4.1cm]{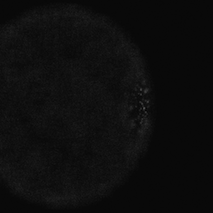}
	
  \leftline{\bf 71st Frame ($7.0\ {\rm sec}$)}
  \vspace{0.1cm}
  \centering
	\includegraphics[width=4.1cm]{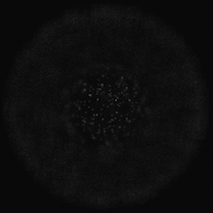}
	\includegraphics[width=4.1cm]{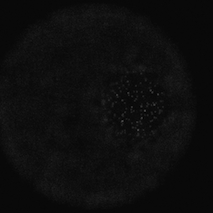}
	\includegraphics[width=4.1cm]{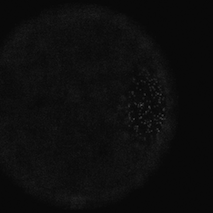}
	\includegraphics[width=4.1cm]{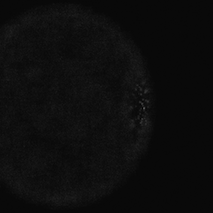}
	
  \leftline{\bf Epi-illumination ($\theta = 0^{\circ}$) \hspace{2.0cm} Oblique illumination ($\theta = 8^{\circ},\ 16^{\circ},\ 24^{\circ}$)}
  \vspace{0.1cm}
  \caption{{\bf Examples of single-molecule images (2).} 8 frames of the cell-model that includes particle aggregation and photobleaching effects. Each image size is $512 \times 512$ pixels. Actual minimum and maximum values of the intensity histogram are $1,900$ and $4,800$ ADC counts. The intensity is rescaled in the range of $0$ to $255$. }
  \label{fig;example_images_cell_model_bleach_1}
\end{figure}

\paragraph{Examples of single-molecule images :}
We constructed the simple cell-model of TMR-tagged proteins distributed on membrane surface. We assumed that $10,000$ fluorescent molecules are stationary, and randomly distributed on the cell surface ($30 \times 30 \times 10 \ {\rm \mu m^3}$). The aggregation process where many fluorophores form clusters in a colloidal suspension, is included in the model. Single-molecule images are simulated for the optical specifications and operating conditions of the fluorescence miroscopy simulation module shown in Table \ref{tab;specification1}. Figure \ref{fig;cell_model_illumination} shows the epi-illumination ($\theta = 0^{\circ}$), the oblique illumination ($\theta = 8^{\circ},\ 16^{\circ},\ 24^{\circ}$) and evanescent field ($\theta = 65.7^{\circ}$). Photophysical parameterizations of fluorophores are also shown in the Table. Results are shown in Figures \ref{fig;example_images_cell_model_bleach_0}-\ref{fig;example_images_cell_model_bleach_1}. Each figures show snapshot images captured with fluorescence miroscopy imaging configuration.

\newpage

\newpage

\newpage

\subsubsection{Oblique illumination near the critical angle}
\begin{figure}[!h]
  \centering
      \includegraphics[width=8.0cm]{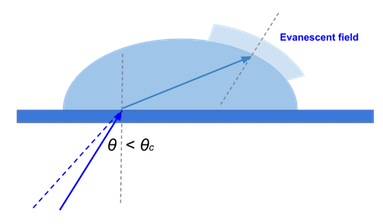}
      \includegraphics[width=8.0cm]{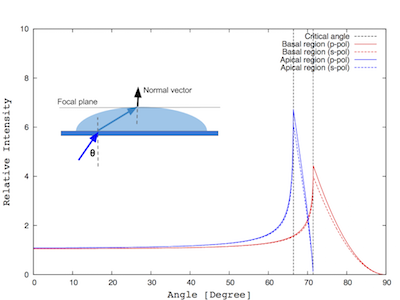}

  \caption{(Left) Schematic view of oblique illumination near the critical angle. Dashed blue line represents the critical angle. (Right) Relative intensity as a function of incident beam angles.}
  \label{fig;low_angle_oblique_illumination1}
\end{figure}

\paragraph{}
Oblique illumination is designed to drastically reduce the background of scattering light originating from solution, dust and optical elements \cite{hiroshima2012, teramura2006, uyemura2005, sako2002, sako2000, tokunaga1997}. This illumination technique allows us to investigate the dynamics of individual molecules at an apical cell surface. The left panel of Figure \ref{fig;low_angle_oblique_illumination1} illustrates schematic views of the oblique illumination near the critical angle, generating evanescent field at the apical cell surface. An incident beam of excitation wavelength ($\lambda$) that passed through the objective lens is assumed to uniformly illuminate the specimen. The surviving photons through the use of excitation filters interact with the fluorophores in the cell-model, and excite the fluorophores to the electrically excited state. If the incidence beam angles are less than the critical angles given by $\sin \theta_{c} = n_{i}/n_{j}$ where media are fused silica ($n_1 = 1.460$), cell ($n_2=1.384$) and culture medium ($n_3 = 1.337$) \cite{jena2008}, then most of the incidence beam propagates through the interface into the lower index material with a refraction angle given by Snell's Law. If the incident beam angle is near the critical angle ($0 \ll \theta \leq \theta_{c}$), then the incidence beam can be reflected between cell cytoplasm and culture medium, and undergoes total internal refraction (TIR) at the apical cell surface. The evanescent field is generated along the axis as perpendicular to the apical surface, and is capable of exciting the fluorescent molecules near the apical surface. The intensity of the evanescent field at any position depends on cell geometry, and exponentially decays with the surface normal axis. 

\begin{figure}[!h]
  \centering
      \includegraphics[width=9cm]{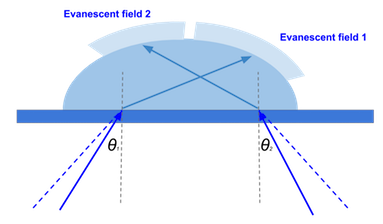}

  \caption{Schematic view of oblique illumination using two excitation beams. Dashed blue lines represent critical angles.}
  \label{fig;low_angle_oblique_illumination2}
\end{figure}

\begin{figure}[!h]
  \leftline{\bf \hspace{3.2cm} $60^{\circ}$ \hspace{4.0cm} $61^{\circ}$ \hspace{4.0cm} $62^{\circ}$}
  \centering
	\includegraphics[width=4.6cm]{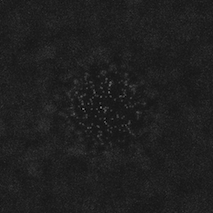}
	\includegraphics[width=4.6cm]{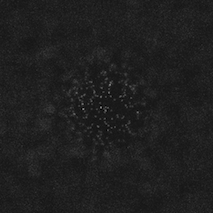}
	\includegraphics[width=4.6cm]{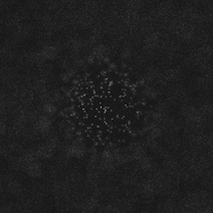}

  \leftline{\bf \hspace{3.2cm} $63^{\circ}$ \hspace{4.0cm} $64^{\circ}$ \hspace{4.0cm} $65^{\circ}$}
  \centering
	\includegraphics[width=4.6cm]{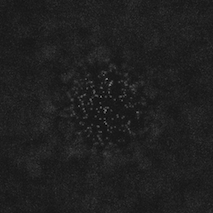}
	\includegraphics[width=4.6cm]{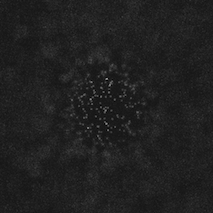}
	\includegraphics[width=4.6cm]{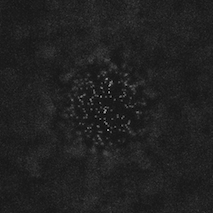}

  \leftline{\bf \hspace{3.2cm} $66^{\circ}$ \hspace{4.0cm} $67^{\circ}$ \hspace{4.0cm} $68^{\circ}$}
  \centering
	\includegraphics[width=4.6cm]{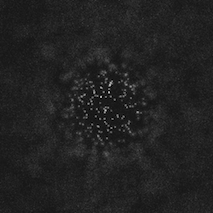}
	\includegraphics[width=4.6cm]{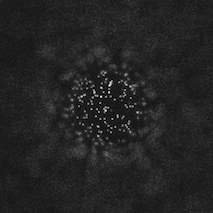}
	\includegraphics[width=4.6cm]{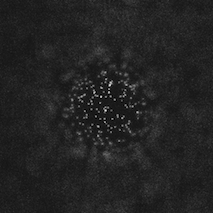}

  \leftline{\bf \hspace{3.2cm} $69^{\circ}$ \hspace{4.0cm} $70^{\circ}$ \hspace{4.0cm} $71^{\circ}$}
  \centering
	\includegraphics[width=4.6cm]{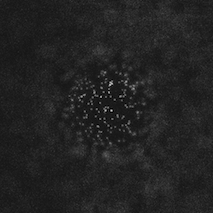}
	\includegraphics[width=4.6cm]{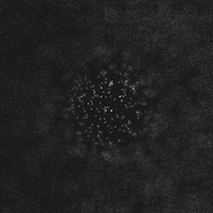}
	\includegraphics[width=4.6cm]{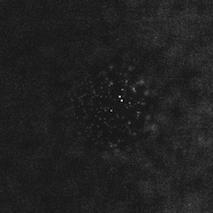}

  \caption{{\bf Examples of single-molecule images (1).} Each image size is $512 \times 512$ pixels. Actual minimum and maximum values of the intensity histogram are $1,900$ and $3,000$ ADC counts. The intensity is rescaled in the range of $0$ to $255$. }
  \label{fig;example_images_cell_model_low_angle}
\end{figure}

\begin{figure}[!h]
  \leftline{\bf \hspace{2.6cm} $60^{\circ}$ \hspace{4.6cm} $61^{\circ}$ \hspace{4.8cm} $62^{\circ}$}
  \centering
	\includegraphics[width=5.2cm]{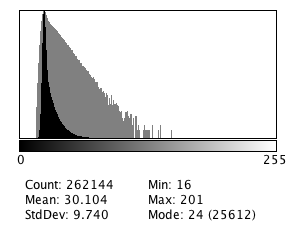}
	\includegraphics[width=5.2cm]{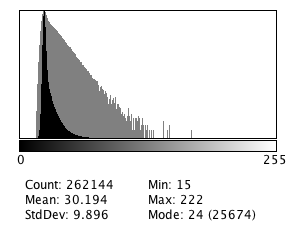}
	\includegraphics[width=5.2cm]{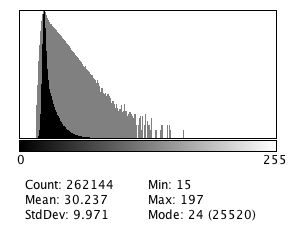}

  \leftline{\bf \hspace{2.6cm} $63^{\circ}$ \hspace{4.6cm} $64^{\circ}$ \hspace{4.8cm} $65^{\circ}$}
  \centering
	\includegraphics[width=5.2cm]{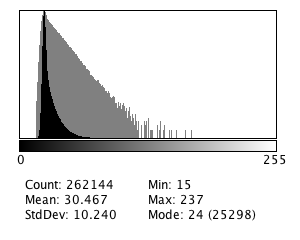}
	\includegraphics[width=5.2cm]{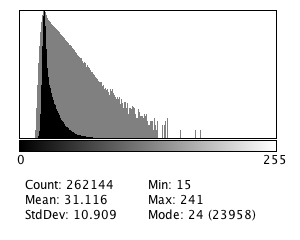}
	\includegraphics[width=5.2cm]{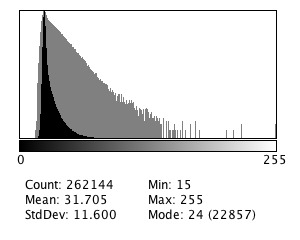}

  \leftline{\bf \hspace{2.6cm} $66^{\circ}$ \hspace{4.6cm} $67^{\circ}$ \hspace{4.8cm} $68^{\circ}$}
  \centering
	\includegraphics[width=5.2cm]{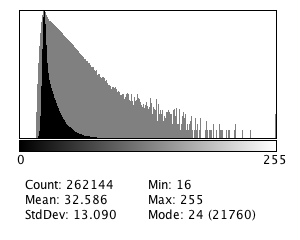}
	\includegraphics[width=5.2cm]{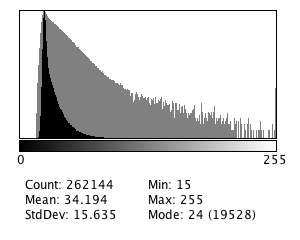}
	\includegraphics[width=5.2cm]{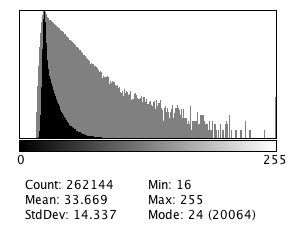}

  \leftline{\bf \hspace{2.6cm} $69^{\circ}$ \hspace{4.6cm} $70^{\circ}$ \hspace{4.8cm} $71^{\circ}$}
  \centering
	\includegraphics[width=5.2cm]{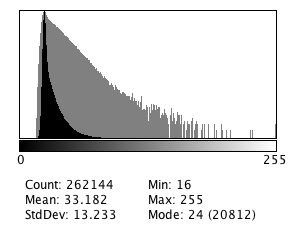}
	\includegraphics[width=5.2cm]{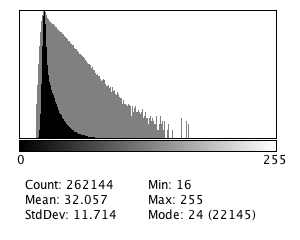}
	\includegraphics[width=5.2cm]{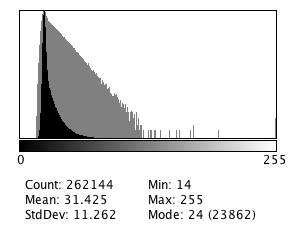}

  \caption{{\bf Histograms of single-molecule images (1).} Each image size is $512 \times 512$ pixels. Actual minimum and maximum values of the intensity histogram are $1,900$ and $3,000$ ADC counts. The intensity is rescaled in the range of $0$ to $255$. }
  \label{fig;example_histograms_cell_model_low_angle}
\end{figure}

\begin{figure}[!h]
  \leftline{\hspace{1.0cm} $z_0 = 10\ {\rm \mu m}$ (Top)}
  \centering
	\includegraphics[width=4.6cm]{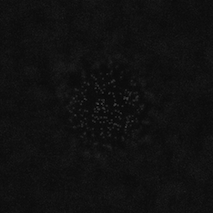}
	\includegraphics[width=4.6cm]{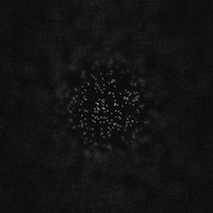}
	\includegraphics[width=4.6cm]{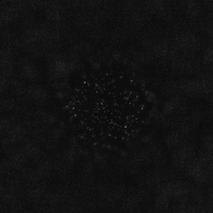}

  \leftline{\hspace{1.0cm} $z_0 = 9.6\ {\rm \mu m}$}
  \centering
	\includegraphics[width=4.6cm]{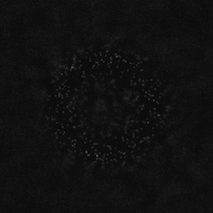}
	\includegraphics[width=4.6cm]{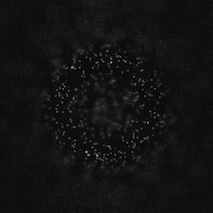}
	\includegraphics[width=4.6cm]{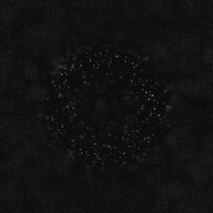}

  \leftline{\hspace{1.0cm} $z_0 = 8.4\ {\rm \mu m}$}
  \centering
	\includegraphics[width=4.6cm]{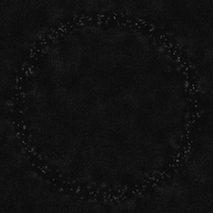}
	\includegraphics[width=4.6cm]{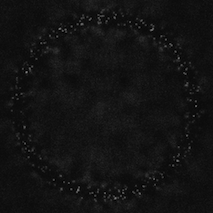}
	\includegraphics[width=4.6cm]{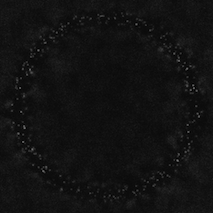}

  \leftline{\hspace{1.0cm} $z_0 = 7.2\ {\rm \mu m}$}
  \centering
	\includegraphics[width=4.6cm]{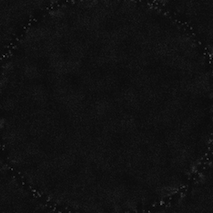}
	\includegraphics[width=4.6cm]{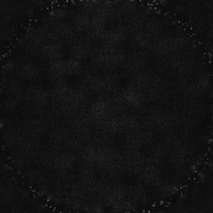}
	\includegraphics[width=4.6cm]{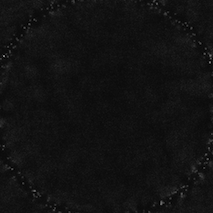}

  \leftline{\bf \hspace{2.0cm} $\theta_1 = -\theta_2 = 60^{\circ}$ \hspace{2.0cm} $\theta_1 = -\theta_2 = 66^{\circ}$ \hspace{2.0cm} $\theta_1 = -\theta_2 = 71^{\circ}$}
  \caption{{\bf Examples of single-molecule images (2).} Two excitation beams are used to generate these images. Each image size is $512 \times 512$ pixels. $\theta_1$, $\theta_2$ and $z_0$ represent the beam angles and the focal distance from the glass surface. Actual minimum and maximum values of the intensity histogram are $1,900$ and $4,200$ ADC counts. The intensity is rescaled in the range of $0$ to $255$. }
  \label{fig;example_images_cell_model_low_angle2}
\end{figure}

\begin{figure}[!h]
  \leftline{\hspace{0.4cm} $z_0 = 10\ {\rm \mu m}$ (Top)}
  \centering
	\includegraphics[width=5.2cm]{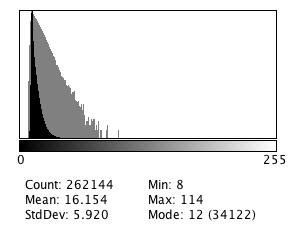}
	\includegraphics[width=5.2cm]{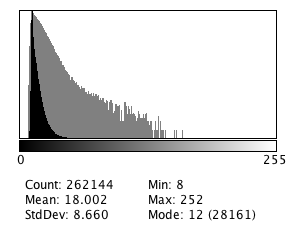}
	\includegraphics[width=5.2cm]{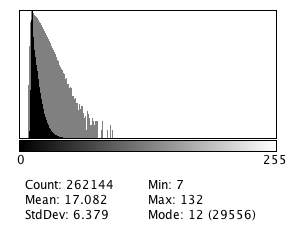}

  \leftline{\hspace{0.4cm} $z_0 = 9.6\ {\rm \mu m}$}
  \centering
	\includegraphics[width=5.2cm]{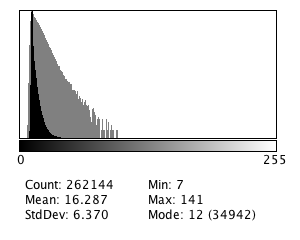}
	\includegraphics[width=5.2cm]{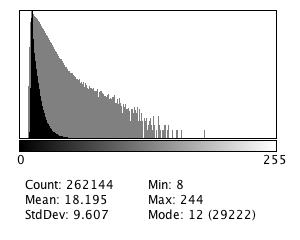}
	\includegraphics[width=5.2cm]{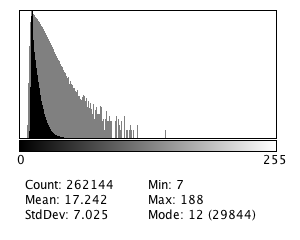}

  \leftline{\hspace{0.4cm} $z_0 = 8.4\ {\rm \mu m}$}
  \centering
	\includegraphics[width=5.2cm]{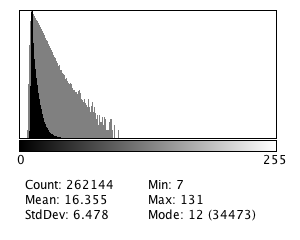}
	\includegraphics[width=5.2cm]{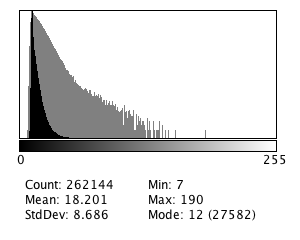}
	\includegraphics[width=5.2cm]{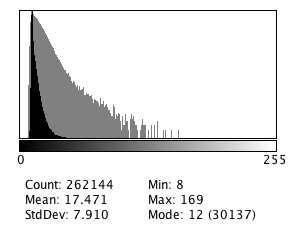}

  \leftline{\hspace{0.4cm} $z_0 = 7.2\ {\rm \mu m}$}
  \centering
	\includegraphics[width=5.2cm]{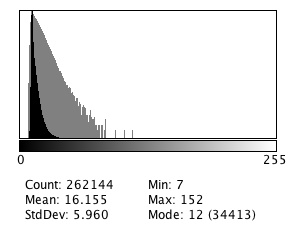}
	\includegraphics[width=5.2cm]{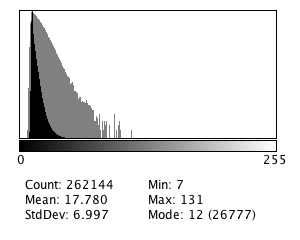}
	\includegraphics[width=5.2cm]{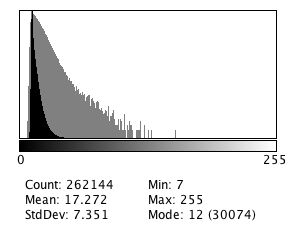}

  \leftline{\bf \hspace{1.6cm} $\theta_1 = - \theta_2 = 60^{\circ}$ \hspace{2.8cm} $\theta_1 = - \theta_2 = 66^{\circ}$ \hspace{2.8cm} $\theta_1 = - \theta_2 = 71^{\circ}$}
  \caption{{\bf Histograms of single-molecule images (2).} Two excitation beams are used to generate these images. Each image size is $512 \times 512$ pixels. $\theta_1$, $\theta_2$ and $z_0$ represent the incident beam angles and the focal distance from the glass surface. Actual minimum and maximum values of the intensity histogram are $1,900$ and $4,200$ ADC counts. The intensity is rescaled in the range of $0$ to $255$. }
  \label{fig;example_histograms_cell_model_low_angle2}
\end{figure}

\paragraph{Examples of single-molecule images :}
We assumed a hemispherical cell-model that represents the fluorescence molecules tagged with tetramethylrhodamine (TMR). The cell measuring $50\ {\rm \mu m}$ in diameter, and $6\ {\rm \mu m}$ in height was placed on the glass surface. Approximately $8,000$ molecules are uniformly distributed on the cell surface. We then simulated single-molecule images for the optical specification and condition of the fluorescence microscopy simulation module shown in Table \ref{tab;specification2}. Photophysical parameterizations of fluorophores are also shown in the Table. Photobleaching and blinking effects are not included in the example images. The right panel of Figure \ref{fig;low_angle_oblique_illumination1} shows relative intensity of incident electric fields as a function of incident beam angle. Results are shown in Figures \ref{fig;example_images_cell_model_low_angle} and \ref{fig;example_histograms_cell_model_low_angle}. Each Figure shows snapshot images captured with fluorescence microscopy imaging configuration. In addition, the use of two excitation beams allows us to observe the dynamics of single-molecules for a given focal height \cite{uyemura2005}. Figure \ref{fig;low_angle_oblique_illumination2} llustrates schematic views of the oblique illumination using two incident beams. Assuming the simple cell-model introduced above, we simulated single-molecule images for the microscopy specifications and operating conditions shown in Table \ref{tab;specification2}. Results are shown in Figures \ref{fig;example_images_cell_model_low_angle2} and \ref{fig;example_histograms_cell_model_low_angle2}. Each Figure shows snapshot images changing focal heights from top regions of the cell-model. 

\newpage

\begin{table}[!h]
\centering
\begin{tabular}{|p{7cm}|c|}
\hline
\multicolumn{2}{|c|}{\bf Excitation Beam} \\ \hline
Flux density & $20\ {\rm W/cm^2}$ \\ \hline
Wavelength & $488\ {\rm nm}$ \\ \hline
Refraction index & $1.46\ (glass)$, $1.384\ (cell)$, $1.337\ (culture\ medium)$ \\ \hline
Critical angle & $71.43^{\circ}\ (glass \to cell)$, $75.02^{\circ}\ (cell \to medium)$ \\ \hline
Incident beam angle & $60 \sim 71^{\circ}$ \\ \hline
\multicolumn{2}{|c|}{\bf Optical elements}\\ \hline
Objective & $\times\ 60\ /\ {\rm N.A.}\ 1.40$ \\ \hline
Dichroic mirror & Semrok FF-562-Di03 \\ \hline
Emission filter & Semrok FF-593-25/40 \\ \hline
Tube lens & $\times\ 4.17$ \\ \hline
Optical magnification & $\times\ 250$ \\ \hline
Optical background & $0.0\ {\rm photons/pixel}$ \\ \hline
\multicolumn{2}{|c|}{\bf EMCCD Camera (Hamamatsu model)}\\ \hline
Image size & $512 \times 512$ \\ \hline
Pixel length & $16\ {\rm \mu m}$ \\ \hline
Quantum efficiency & $92\ \%$ \\ \hline
EM Gain & $\times 300$ \\ \hline
Exposure time & $100\ {\rm msec}$ \\ \hline
Readout noise & $100\ {\rm electrons}$ \\ \hline
Full well & $370,000\ {\rm electrons}$ \\ \hline
Dynamic range & $71.3\ {\rm dB}$ \\ \hline
Excess noise & $\sqrt{2}$ \\ \hline
A/D Converter & $16$-bit \\ \hline
Gain & $5.82\ {\rm electrons/count}$ \\ \hline
Offset  & $2000\ {\rm counts}$ \\ \hline
\multicolumn{2}{|c|}{\bf Photophysics} \\ \hline
Fluorophore & \hspace{1.0cm}  TMR\ (${\rm Abs.}\ 548\ {\rm nm}\ /\ {\rm Em.}\ 608\ {\rm nm}$) \hspace{1.0cm} \\ \hline
Fluorescence quantum yield & $61\ \%$ \\ \hline
Absorption coefficient (cross-section) & $83400\ {\rm M^{-1} cm^{-1}}$ ($\sigma = 3.19 \times 10^{-16}\ {\rm cm^2}$) \\ \hline
%Photobleaching (Power-law) & $\tau_0 = 1.8\ {\rm sec}$, $\alpha = 0.73$ \\ \hline
%\multirow{2}{*}{Photoblinking (Power-law)} & ON\ \ ($\tau_0 = 1.0\ {\rm sec},\ \alpha = 0.58$) \\
%& OFF\ ($\tau_0 = 10\ {\rm \mu sec},\ \alpha = 0.48$) \\ \hline
%\multirow{2}{*}{Photoactivation} & $1000$ (Turn on ratio) \\
%& $10\ \%$ (Activation yield) \\
%& $0.00\ \%$ (Fraction of preactivation) \\ \hline
\end{tabular}
%\vspace{0.3cm}
\caption{Fluorescence microscopy specifications and its operating conditions for imaging the cell-models. Photobleaching and blinking effects are not included in this microscopy configuration. }
\label{tab;specification2}
\end{table}

\newpage

\newpage

\newpage

\newpage

\newpage

\section{Model construction}
\subsection{Hiroshima's model of dimer formation}
\begin{figure}[!h]
  \centering
	\includegraphics[width=7.4cm]{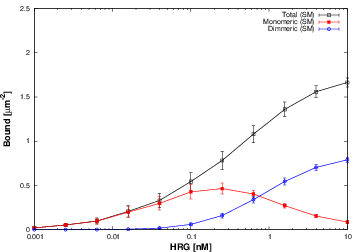}
	\hspace{0.4cm}
	\includegraphics[width=7.4cm]{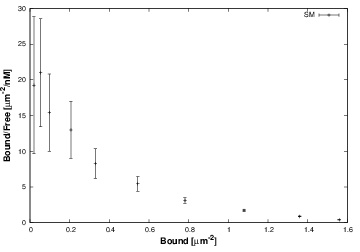}

  \caption{{\bf Simulation outputs of the dimerization model}. (Left) Equilibrium binding curve and (Right) Scatchard plot.}
  \label{fig;dimer_model_simulation}
\end{figure}

\paragraph{}
We constructed a mathematical cell-model of HRG ligand induced ErbB receptor dimerization in a biological cell that represents the HRG ligands tagged with tetramethylrhodamine (TMR). A cell simulation method with Spatiocyte was used to construct the cell-model of the dimer formation. Spatiocyte provides the spatial cell-model of biological fluctuation that arises from stochastic changes in the cell surface geometry, number of ErbB receptors, HRG ligand binding, molecular states (monomer or dimer), and translational diffusion of each receptors \cite{arjunan2010}. The left panel of Figure \ref{fig;simple_oligomer_network} illustrates biochemical reaction network of the dimer formation. An intermediate state transition of the proteins is included in this model. The hemi-elliptical cell measuring $65\ {\rm \mu m}$ and $45\ {\rm \mu m}$ in minor and major axes, and $2.0\ {\rm \mu m}$ in height was placed on the glass surface. Average observational surface area of the cell-model is $4,731\ {\rm \mu m^2}$. Monomer and dimer of ErbB receptors are presented as solid spheres where voxel radius is $20\ {\rm nm}$. Each receptors is uniformly distributed in the surface compartment, and randomly walk from voxel to voxel, slowly diffusing at $0.015\ {\rm \mu m^2/sec}$. Molecular collisions occur between walks. We defined seven equilibrium constants for the molecular collisions: $K_1$, $K_2$, and $K_3$ are the binding constants of HRG to an isolated receptor, to a receptor in a dimer where one site is free and one bound. $K_4$, $K_5$, and $K_6$ are the dimerization constants of two free receptors, one bound and one free receptor, and two bound receptors. $K_i$ that characterizes the intermediate state transition constant in the dimer where one site is free and one bound. Values of each model parameters are shown in Table \ref{tab;simple_oligomer_model_parameters}. We simulated the dimerization model without the systematic effects that arise from microscopy specification and its operating condition. Simulation results are shown in Figure \ref{fig;dimer_model_simulation}. The left and the right panels show the equilibrium binding curve and the Scatchard plot. The Scatchard plot shows a concave-up curve that represents negative cooperativity of the binding system. We consequently reproduced the properties obtained by Hiroshima's analysis \cite{hiroshima2012}. 

\subsection{Toy model of higher-order oligomer formation}
\begin{figure}[!h]
  \centering
      \includegraphics[width=14.0cm]{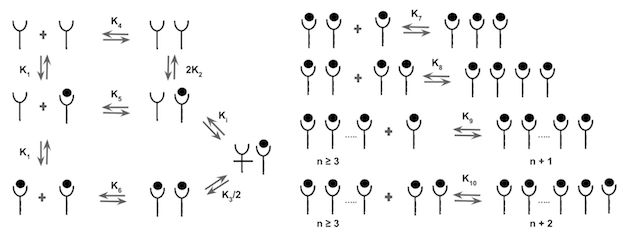}
      
  \caption{{\bf Cell-model}. (Left) Reaction network for the model of HRG ligand induced ErbB receptors \cite{hiroshima2012}. The black dot and Y-shaped object represent the TMR-HRG ligand and ErbB protein. (Right) Additional reaction networks for simple model extension of the dimerization to $n$-th order of oligomer formation (e.g., trimer, tetramer and pentamer).}
  \label{fig;simple_oligomer_network}
\end{figure}

\paragraph{}
In order to extend the dimerization model to higher-order oligomer formation, we defined four additional equilibrium constants for the molecular collisions: two interactive constants are $K_{7}$ and $K_{8}$ that characterize respectively the trimer formation of monomer and dimer, and the tetramer formation of two dimers; and others are $K_{9}$ and $K_{10}$ that characterize respectively the $(n+1)$-th and $(n+2)$-th order of oligomer formations ($3 \le n \le 12$). We assumed that the monomers and dimers are the subunits which polymerize on the cell membrane, and each higher-order oligomers are presented as immobile polymerized chains. Figure \ref{fig;simple_oligomer_network}B illustrates an additional reaction network of the higher-order oligomer formation. Values of each parameter are shown in Table \ref{tab;simple_oligomer_model_parameters}. In fact, average observational area on the cell surface ($43^2 = 1,930\ {\rm \mu m^2}$) is larger than actual image size ($34^2\ {\rm \mu m^2}$) \cite{hiroshima2012}. The observational area needed to be adjusted to $661\ {\rm \mu m^2}$. Such parameter adjustment consequently leads to changes in the total area density ($1.70 \to 4.96\ {\rm \mu m^{-2}}$) and the equilibrium constants ($K_{4,5,6}$). 

\begin{table}[!h]
\centering
\begin{tabular}{|p{4.0cm}|p{12cm}|}
\hline
\multicolumn{2}{|c|}{\bf Area density of ErbB proteins} \\ \hline
Total (Hiroshima-2012) & $1.70\ {\rm \mu m^{-2}}$ (Monomer : $1.32$, Dimer : $0.193$) \\ \hline
Total (Extended model) & $4.96\ {\rm \mu m^{-2}}$ (Monomer : $4.96$, Dimer : $0.00$, Others : $0.00$) \\ \hline
\multicolumn{2}{|c|}{\bf Equilibrium constants and reaction rates} \\ \hline
$K_{1}=d_1/k_1$ & $3.63\ {\rm nM}$ \hspace{0.3cm} $(k_1 = 0.00193\ {\rm nM^{-1}\ sec^{-1}},\ d_1 = 0.00700\ {\rm sec^{-1}})$ \\ \hline
$K_{2}=d_2/k_2$ & $0.0155\ {\rm nM}$ \hspace{0.3cm} $(k_2 = 0.00255\ {\rm nM^{-1}\ sec^{-1}},\ d_2 = 3.95 \times 10^{-5}\ {\rm sec^{-1}})$ \\ \hline
%$K'_{3} = d'_3/k'_3$ & $0.1022\ {\rm nM}$ \hspace{0.3cm} $(k'_3 = 5.09\ {\rm nM^{-1}\ sec^{-1}},\ d'_3 = 0.52\ {\rm sec^{-1}})$ \\ \hline
$K_{3} = d_3/k_3$ & $0.553\ {\rm nM}$ \hspace{0.3cm} $(k_3 = 4.09\ {\rm nM^{-1}\ sec^{-1}},\ d_3 = 2.26\ {\rm sec^{-1}})$ \\ \hline
$K_{4} = d_4/k_4$ & $9.01\ {\rm \mu m^{-2}} \hspace{0.6cm} \rightarrow\ 26.316$ \hspace{0.3cm} (Assume $d_4 = 0.01\ {\rm sec^{-1}} \rightarrow\ 1.00$)\\ \hline
$K_{5} = d_5/k_5$ & $0.0770\ {\rm \mu m^{-2}} \hspace{0.2cm} \rightarrow\ 0.2247$ \hspace{0.3cm} (Assume $d_5 = 0.01\ {\rm sec^{-1}} \rightarrow\ 0.10$)\\ \hline
$K_{6} = d_6/k_6$ & $0.000818\ {\rm \mu m^{-2}}\ \rightarrow\ 0.00238$ \hspace{0.3cm} (Assume $d_6 = 0.03\ {\rm sec^{-1}}$)\\ \hline
$K_{i} = d_i/k_i$ & $0.139 \hspace{0.3cm} (k_i = 4.51\ {\rm sec^{-1}},\ d_i = 0.629\ {\rm sec^{-1}})$ \\ \hline
$K_{7}=d_7/k_7$ & $4.831\ {\rm \mu m^{-2}}$ \hspace{0.3cm} (Assume $k_7 = 0.0207\ {\rm \mu m^2/sec}$, $d_7 = 0.10\ {\rm sec^{-1}}$) \\ \hline
$K_{8}=d_8/k_8$ & $9.615\ {\rm \mu m^{-2}}$ \hspace{0.3cm} (Assume $k_8 = 0.0104\ {\rm \mu m^2/sec}$, $d_8 = 0.10\ {\rm sec^{-1}}$) \\ \hline
$K_{9}=d_9/k_9$ & $9.615\ {\rm \mu m^{-2}}$ \hspace{0.3cm} (Assume $d_9 = 0.10\ {\rm sec^{-1}}$) \\ \hline
$K_{10}=d_{10}/k_{10}$ & $9.615\ {\rm \mu m^{-2}}$ \hspace{0.3cm} (Assume $d_{10} = 0.10\ {\rm sec^{-1}}$) \\ \hline
\end{tabular}
\caption{{\bf Model parameters for oligomer formation}. Hiroshima has provided two sets of parameter values; primary and secondary parameters for the interactions between HRG and ErbB molecules \cite{hiroshima2012}. We use the primary set of reaction parameter values. Monomers and dimers are properly selected for analyzing each reaction rate. However, the monomer and dimers are not properly selected for obtaining the secondary set of parameter values. The secondary set is obtained by fitting of the dimerization model to all observed molecules, including higher-order oligomers ($\sim 10\%$).}
\label{tab;simple_oligomer_model_parameters}
\end{table}

\newpage

\begin{figure}[!h]
  \leftline{\hspace{0.5cm} 1-st frame ($20\ {\rm sec}$)}
  \centering
	\includegraphics[width=4.8cm]{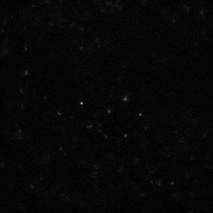}
	\includegraphics[width=4.8cm]{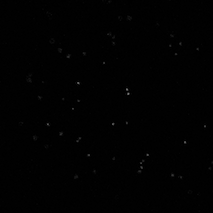}
	\includegraphics[width=4.8cm]{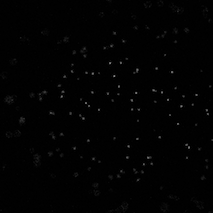}

  \leftline{\hspace{0.5cm} 4-th frame ($80\ {\rm sec}$)}
  \centering
	\includegraphics[width=4.8cm]{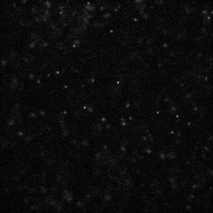}
	\includegraphics[width=4.8cm]{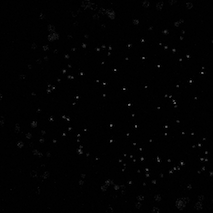}
	\includegraphics[width=4.8cm]{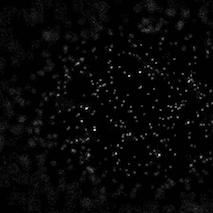}

  \leftline{\hspace{0.5cm} 7-th frame ($140\ {\rm sec}$)}
  \centering
	\includegraphics[width=4.8cm]{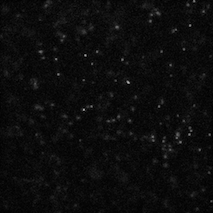}
	\includegraphics[width=4.8cm]{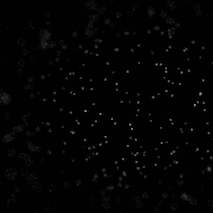}
	\includegraphics[width=4.8cm]{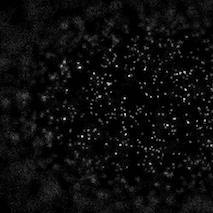}

  \leftline{\hspace{0.5cm} 10-th frame ($200\ {\rm sec}$)}
  \centering
	\includegraphics[width=4.8cm]{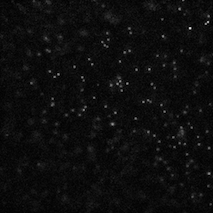}
	\includegraphics[width=4.8cm]{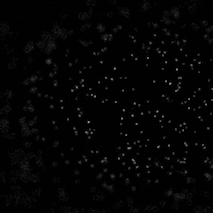}
	\includegraphics[width=4.8cm]{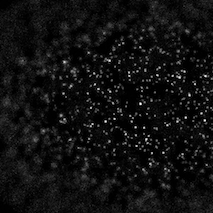}

  \caption{{\bf Image comparison (1)}. Each image size is $512 \times 512$ pixels. Actual minimum and maximum values of the intensity histogram are $2,000$ and $3,000$ ADC counts. The intensity is rescaled in the range of $0$ to $255$.}
  \label{fig;image_comparison_0}
\end{figure}

\begin{figure}[!h]
  \leftline{\hspace{0.5cm} 13-th frame ($260\ {\rm sec}$)}
  \centering
	\includegraphics[width=4.8cm]{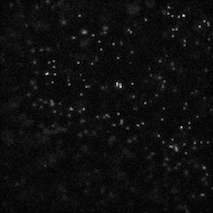}
	\includegraphics[width=4.8cm]{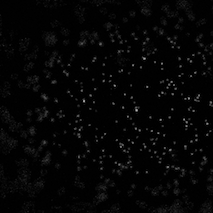}
	\includegraphics[width=4.8cm]{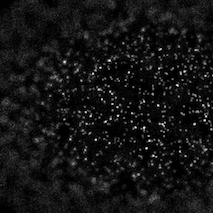}

  \leftline{\hspace{0.5cm} 16-th frame ($320\ {\rm sec}$)}
  \centering
	\includegraphics[width=4.8cm]{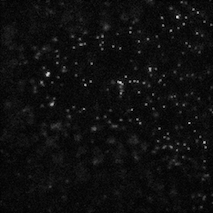}
	\includegraphics[width=4.8cm]{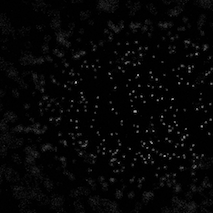}
	\includegraphics[width=4.8cm]{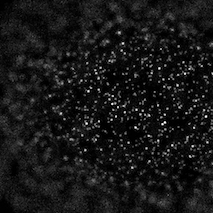}

  \leftline{\hspace{0.5cm} 19-th frame ($960\ {\rm sec}$)}
  \centering
	\includegraphics[width=4.8cm]{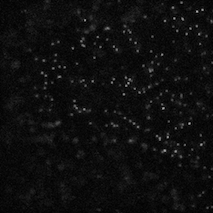}
	\includegraphics[width=4.8cm]{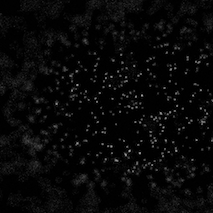}
	\includegraphics[width=4.8cm]{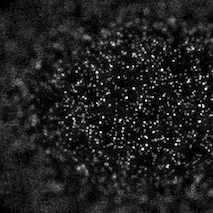}

  \leftline{\hspace{0.5cm} 22-th frame ($1,560\ {\rm sec}$)}
  \centering
	\includegraphics[width=4.8cm]{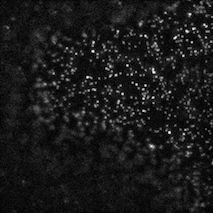}
	\includegraphics[width=4.8cm]{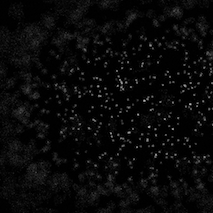}
	\includegraphics[width=4.8cm]{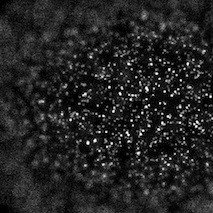}

  \caption{{\bf Image comparison (2)}. Each image size is $512 \times 512$ pixels. Actual minimum and maximum values of the intensity histogram are $2,000$ and $3,000$ ADC counts. The intensity is rescaled in the range of $0$ to $255$.}
  \label{fig;image_comparison_1}
\end{figure}

\begin{figure}[!h]
  \leftline{\hspace{0.5cm} 24-th frame ($2,160\ {\rm sec}$)}
  \centering
	\includegraphics[width=4.8cm]{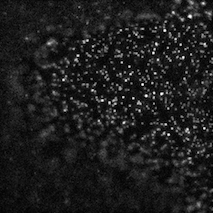}
	\includegraphics[width=4.8cm]{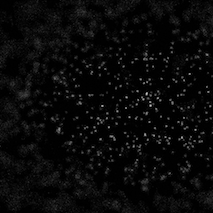}
	\includegraphics[width=4.8cm]{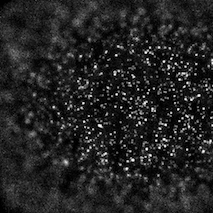}

  \leftline{\hspace{0.5cm} 26-th frame ($2,760\ {\rm sec}$)}
  \centering
	\includegraphics[width=4.8cm]{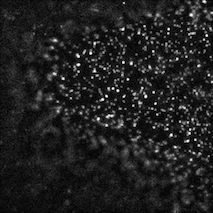}
	\includegraphics[width=4.8cm]{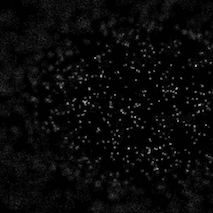}
	\includegraphics[width=4.8cm]{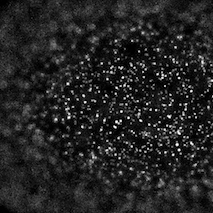}

  \leftline{\hspace{0.5cm} 28-th frame ($3,360\ {\rm sec}$)}
  \centering
	\includegraphics[width=4.8cm]{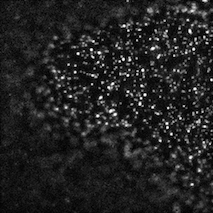}
	\includegraphics[width=4.8cm]{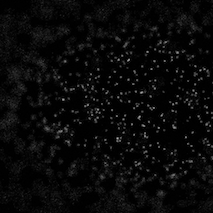}
	\includegraphics[width=4.8cm]{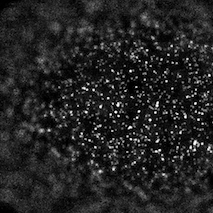}

  \leftline{\hspace{0.5cm} 30-th frame ($3,960\ {\rm sec}$)}
  \centering
	\includegraphics[width=4.8cm]{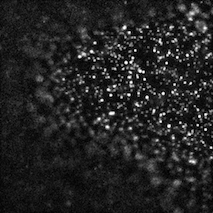}
	\includegraphics[width=4.8cm]{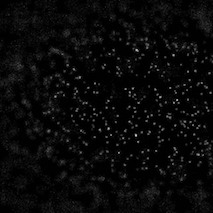}
	\includegraphics[width=4.8cm]{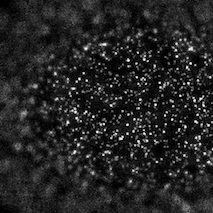}

  \caption{{\bf Image comparison (3)}. Each image size is $512 \times 512$ pixels. Actual minimum and maximum values of the intensity histogram are $2,000$ and $3,000$ ADC counts. The intensity is rescaled in the range of $0$ to $255$.}
  \label{fig;image_comparison_2}
\end{figure}

\paragraph{Image comparison :}
We simulated single-molecule imaging of the apical surface region of each model for the optical specification and condition of the fluorescence microscopy simulation module shown in Table \ref{tab;specification3}. Comparisons of actual single-molecule images to the simulated single-molecule images for $1.0\ {\rm nM}$ ligand input are shown in Figure \ref{fig;image_comparison_0}-\ref{fig;image_comparison_2}. The left column shows the actual images obtained by the single-molecule experiment (Movie S1 in Ref. \cite{hiroshima2012}). Cell size in the actual image is relatively large, but not known. Middle and right columns show the single-molecule images obtained by the bioimaging simulations for the dimerization model and higher-order oligomerization. The photobleaching effects are included in each simulated images. 
The simulated images of the higher-order oligomer formation model are visually similar to the corresponding real images at steady state. Thus, the simulated images were properly compared with the images obtained using the actual fluorescence microscopy system at the level of photon-counting units. However, there still remains differences in the resulting images. A more elaborate set for calibration is required for further understanding of the receptor system. 

\begin{table}[!h]
\centering
\begin{tabular}{|p{7cm}|c|}
\hline
\multicolumn{2}{|c|}{\bf Excitation Beam} \\ \hline
Flux density & $50\ {\rm W/cm^2}$ (Assumed) \\ \hline
Wavelength & $532\ {\rm nm}$ \\ \hline
Refraction index & $1.46\ ({\rm glass})$, $1.384\ ({\rm cell})$, $1.337\ ({\rm culture\ medium})$ \\ \hline
Critical angle & $71.43^{\circ}\ (glass \to cell)$, $75.02^{\circ}\ (cell \to medium)$ \\ \hline
Incident beam angle & $60^{\circ}$ (Oblique illumination) \\ \hline
\multicolumn{2}{|c|}{\bf Optical elements}\\ \hline
Objective & $\times\ 60\ /\ {\rm N.A.}\ 1.49$ \\ \hline
Dichroic mirror & Semrok FF-562-Di03 \\ \hline
Emission filter & not available \\ \hline
Tube lens & $\times\ 4.02$ \\ \hline
Optical magnification & $\times\ 241$ \\ \hline
Optical background & $0.01\ {\rm photons/pixel}$ (Assumed) \\ \hline
\multicolumn{2}{|c|}{\bf EMCCD Camera (Hamamatsu model)}\\ \hline
Image size & $512 \times 512$ \\ \hline
Pixel length & $16\ {\rm \mu m}$ \\ \hline
Quantum efficiency & $92\ \%$ \\ \hline
EM Gain & $\times 300$ \\ \hline
Exposure time & $150\ {\rm msec}$ \\ \hline
Readout noise & $100\ {\rm electrons}$ \\ \hline
Full well & $370,000\ {\rm electrons}$ \\ \hline
Dynamic range & $71.3\ {\rm dB}$ \\ \hline
Excess noise & $\sqrt{2}$ \\ \hline
A/D Converter & $16$-bit \\ \hline
Gain & $5.82\ {\rm electrons/count}$ \\ \hline
Offset  & $2000\ {\rm counts}$ \\ \hline
\multicolumn{2}{|c|}{\bf Photophysics} \\ \hline
PSF normalization factor & $1.0$ (Assumed) \\ \hline
Fluorophore & \hspace{1.0cm}  TMR\ (${\rm Abs.}\ 548\ {\rm nm}\ /\ {\rm Em.}\ 608\ {\rm nm}$) \hspace{1.0cm} \\ \hline
Fluorescence quantum yield & $61\ \%$ \\ \hline
Absorption coefficient (cross-section) & $83400\ {\rm M^{-1} cm^{-1}}$ ($\sigma = 3.19 \times 10^{-16}\ {\rm cm^2}$) \\ \hline
Photobleaching (assuming power-law) & $\tau_0 = 2.27\ {\rm sec}$, $\alpha = 0.73$ \\ \hline
%\multirow{2}{*}{Photoblinking (Power-law)} & ON\ \ ($\tau_0 = 1.0\ {\rm sec},\ \alpha = 0.58$) \\
%& OFF\ ($\tau_0 = 10\ {\rm \mu sec},\ \alpha = 0.48$) \\ \hline
%\multirow{2}{*}{Photoactivation} & $1000$ (Turn on ratio) \\
%& $10\ \%$ (Activation yield) \\
%& $0.00\ \%$ (Fraction of preactivation) \\ \hline
\end{tabular}
%\vspace{0.3cm}
\caption{Fluorescence microscopy specification and its operating condition to obtain single-molecule images of the oligomer formation.}
\label{tab;specification3}
\end{table}

\newpage

\newpage

\newpage

\newpage

\subsection{PSF-verification}
\paragraph{}
Verification is the process of confirming the simulation modules are correctly implemented with respect to conceptual description and analytical solutions \cite{watabe2015}. During the verification process, the simulation modules must be tested to find and estimate numerical errors in the implementations. The simulation module presented here is designed to count the number of photons that passed through the optical configuration. A wrong estimation of the numerical errors that arise from the photon-counting principle can provide a wrong intensity of the final images. 
 
\paragraph{}
In our optics simulation, an image-forming system enables the formation and convolution of the Born-Wolf form of point spreading functions (PSF). We assume that single fluorophore (voxel-radius $= 20\ {\rm nm}$, path-length $= 40\ {\rm nm}$) is placed on glass surface. We then evaluate how well the exact-from of the Born-Wolf PSF are correctly implemented for various lateral distance (z-axis) from the focal plane. In particular, two constraints are applied to optimize the PSF simulation processes: r-cut ($r < 1\ {\rm nm}$) and z-cut ($z < 1\ {\rm nm}$). Microscopy configuration is shown in Table \ref{tab;specification3}. The results of the PSF verification are presented below.

\begin{figure}[!b]
  \centerline{\bf Oblique illumination ($60^{\circ}$) \hspace{4.0cm} Evanescence ($72^{\circ}$)}
  \centering
      \includegraphics[width=7.4cm]{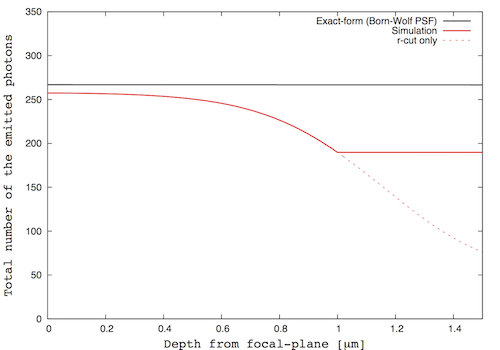}
      \hspace{0.4cm}
      \includegraphics[width=7.4cm]{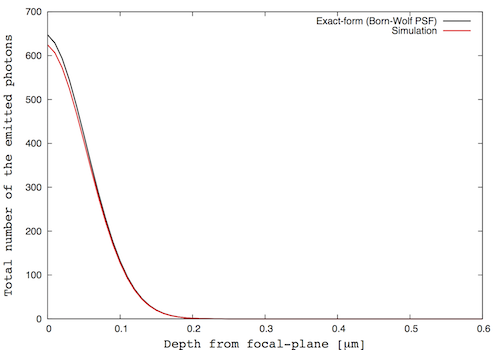}

  \centering
      \includegraphics[width=7.4cm]{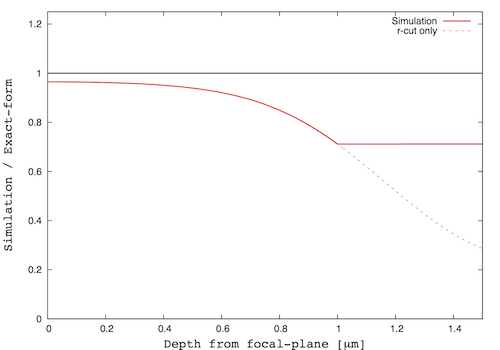}
      \hspace{0.4cm}
      \includegraphics[width=7.4cm]{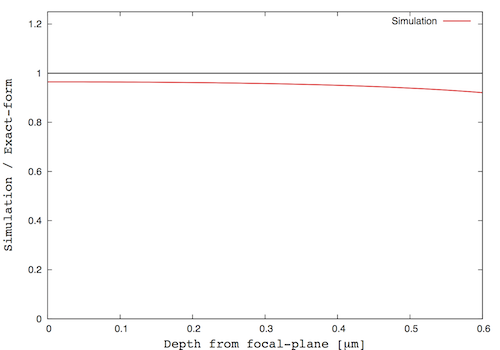}

  \vspace{0.5cm}
  \caption{{\bf Total number of the emitted photons}. The top and the bottom panels show the total number of the emitted photons  as a function of depth from the focal plane, and the ratio of the total photon number of the exact-form to the simulated-form. The black and red lines represent the exact-form and the simulated-form of PSF. The z-cut is omitted in the dashed red lines.}
  \label{fig;photons_verification}
\end{figure}

\paragraph{}
We computed the total number of the emitted photons from the exact-form and the simulated-form of the PSF for two illumination configurations: oblique illumination ($\theta_{inc} = 60^{\circ}$) and evanescent field ($\theta_{inc} = 72^{\circ}$). The top panels of Figures \ref{fig;photons_verification} show the comparison of the simulated-form (red lines) to the exact-form (black lines) of the PSF with respect to the lateral distance from the focal plane. While the total number of the emitted photons from the exact-form of the PSF is constant in the oblique illumination configuration, an exponential decrease is shown in the evanescent (or TIR) configuration. The bottom panels show the ratio of the total number from the exact-form to the simulated-form of the PSF, implying that the total number is well-conserved within $4$-$6\%$ differences near the focal plane (below the $600\ {\rm nm}$ lateral depth). The differences, however, increase up to $30\%$ at defocused region or background (above the $600\ {\rm nm}$ depth). If the z-cut is omitted from the simulation processes, then the total photon number can be violated with respect to the lateral distance (red dashed lines). The additional z-cut can, however, help recovering the total photon number up to $70\%$ levels.

\begin{figure}[!b]
  \centerline{\bf Oblique illumination ($60^{\circ}$) \hspace{4.0cm} Evanescence ($72^{\circ}$)}
  \centering
      \includegraphics[width=7.4cm]{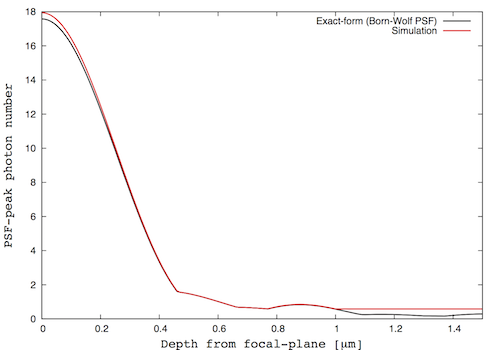}
      \hspace{0.4cm}
      \includegraphics[width=7.4cm]{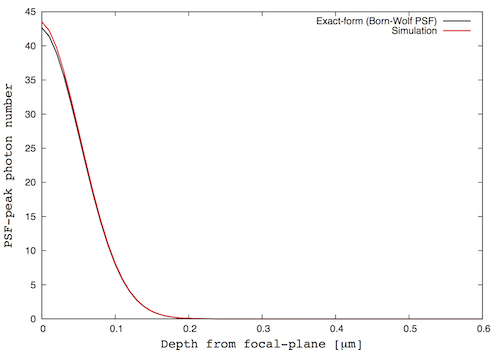}

  \centering
      \includegraphics[width=7.4cm]{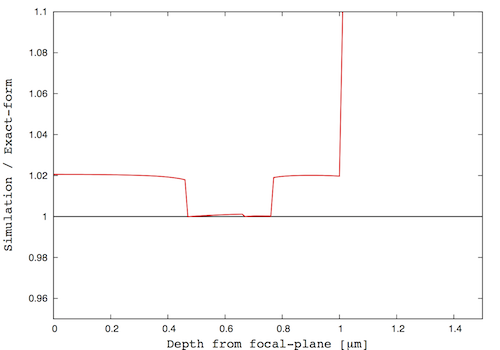}
      \hspace{0.4cm}
      \includegraphics[width=7.4cm]{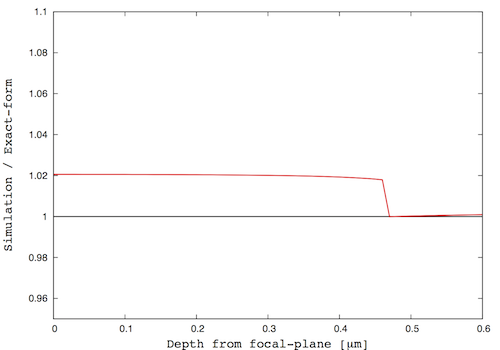}

  \vspace{0.5cm}
  \caption{{\bf PSF-peak (or -maximum) photon number}. The top and the bottom panels show the PSF-peak (or -maximum) photons as a function of depth from the focal plane, and the ratio of the peak photon number of the exact-form to the the simulated-form. The black and red lines represent the exact-form and the simulated-form.}
  \label{fig;peaks_verification}
\end{figure}

\begin{figure}[!h]
  \leftline{\bf\hspace{0.2cm} $z = 0\ {\rm nm}$}
  \centering
      \includegraphics[width=2.4cm]{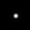}
      \includegraphics[width=2.4cm]{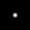}
      \includegraphics[width=3.0cm]{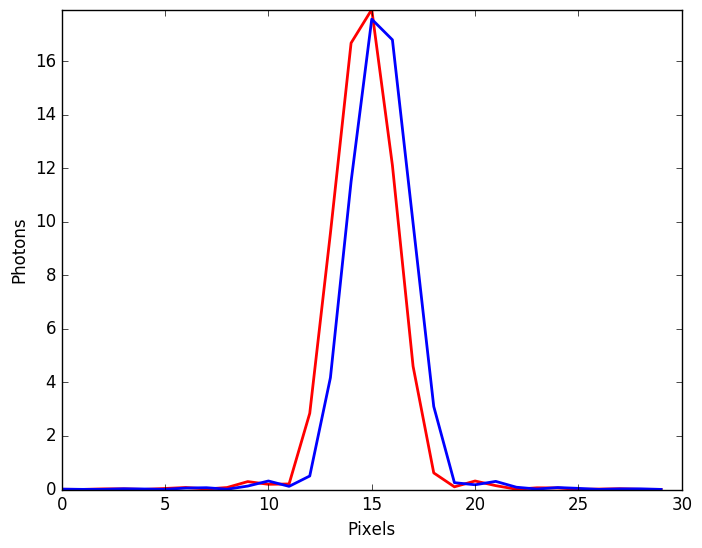}
      \hspace{0.1cm}
      \includegraphics[width=2.4cm]{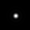}
      \includegraphics[width=2.4cm]{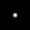}
      \includegraphics[width=3.0cm]{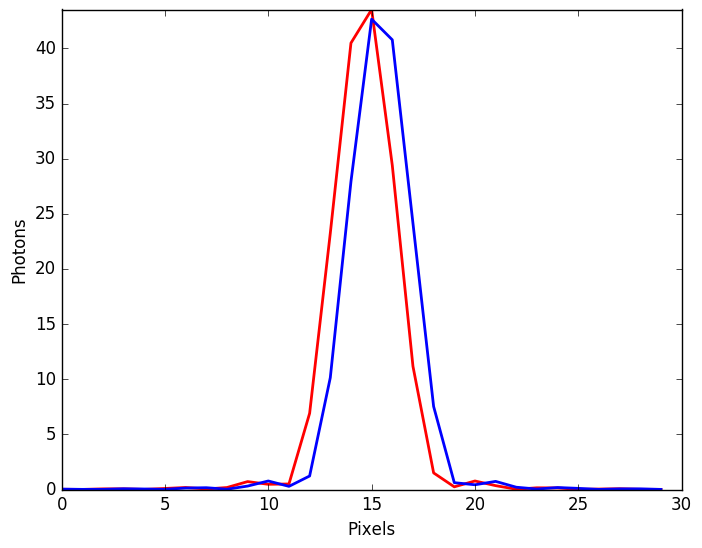}

  \leftline{\bf\hspace{0.2cm} $z = 100\ {\rm nm}$}
  \centering
      \includegraphics[width=2.4cm]{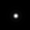}
      \includegraphics[width=2.4cm]{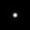}
      \includegraphics[width=3.0cm]{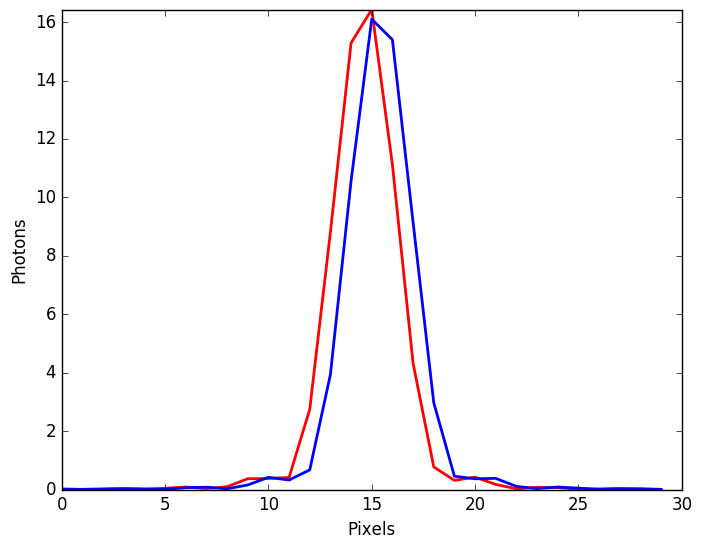}
      \hspace{0.1cm}
      \includegraphics[width=2.4cm]{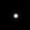}
      \includegraphics[width=2.4cm]{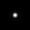}
      \includegraphics[width=3.0cm]{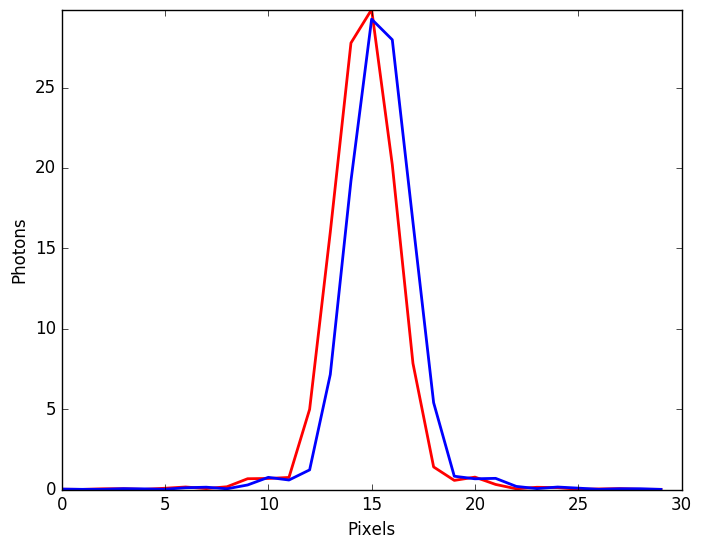}

  \leftline{\bf\hspace{0.2cm} $z = 200\ {\rm nm}$}
  \centering
      \includegraphics[width=2.4cm]{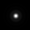}
      \includegraphics[width=2.4cm]{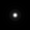}
      \includegraphics[width=3.0cm]{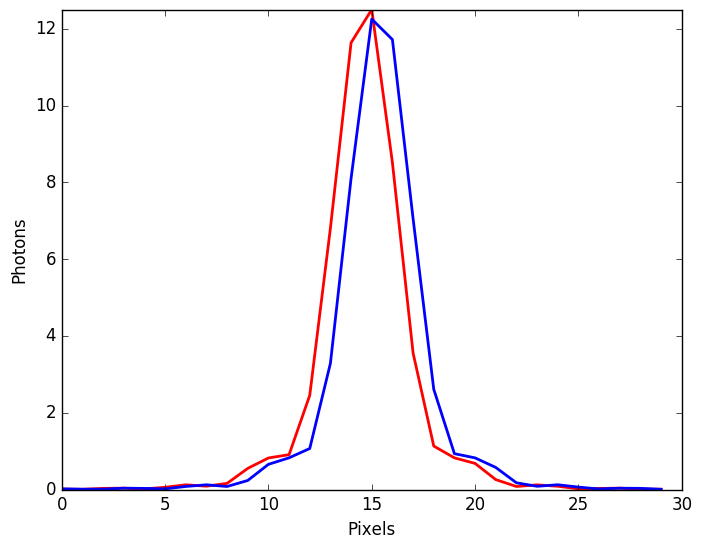}
      \hspace{0.1cm}
      \includegraphics[width=2.4cm]{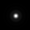}
      \includegraphics[width=2.4cm]{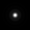}
      \includegraphics[width=3.0cm]{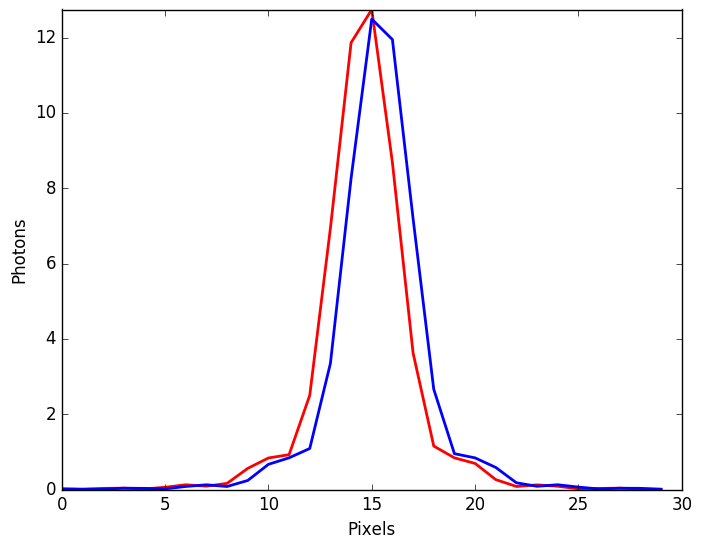}

  \leftline{\bf\hspace{0.2cm} $z = 300\ {\rm nm}$}
  \centering
      \includegraphics[width=2.4cm]{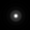}
      \includegraphics[width=2.4cm]{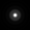}
      \includegraphics[width=3.0cm]{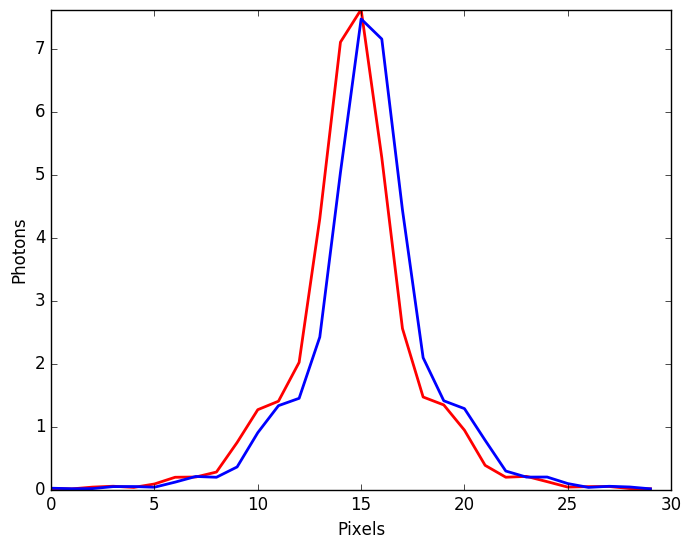}
      \hspace{0.1cm}
      \includegraphics[width=2.4cm]{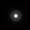}
      \includegraphics[width=2.4cm]{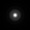}
      \includegraphics[width=3.0cm]{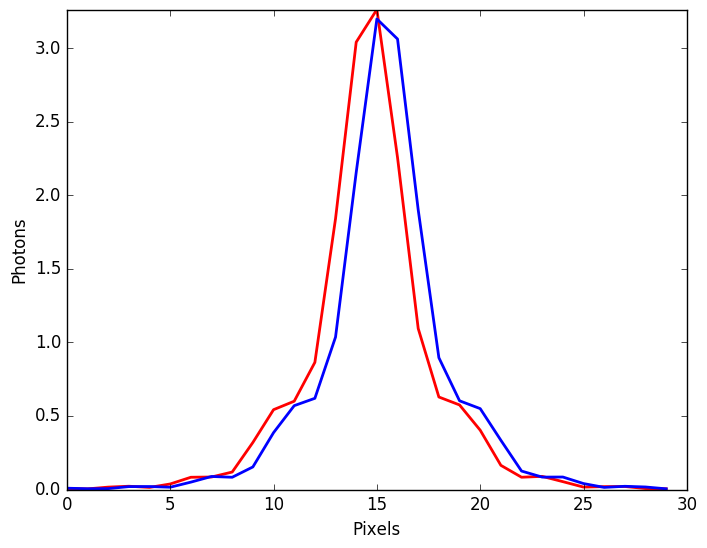}

  \leftline{\bf\hspace{0.2cm} $z = 400\ {\rm nm}$}
  \centering
      \includegraphics[width=2.4cm]{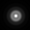}
      \includegraphics[width=2.4cm]{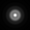}
      \includegraphics[width=3.0cm]{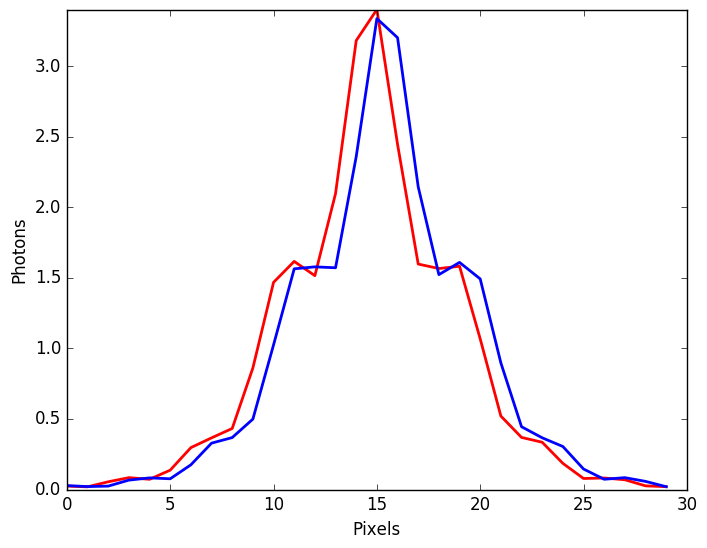}
      \hspace{0.1cm}
      \includegraphics[width=2.4cm]{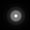}
      \includegraphics[width=2.4cm]{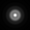}
      \includegraphics[width=3.0cm]{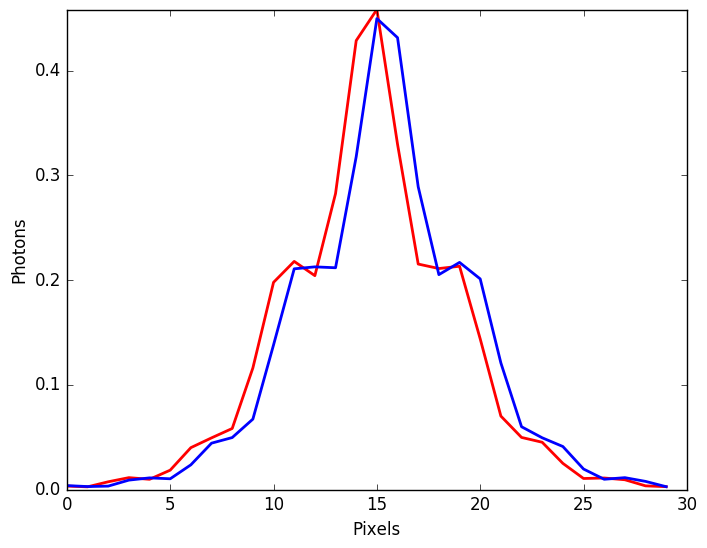}

  \leftline{\bf\hspace{0.2cm} $z = 500\ {\rm nm}$}
  \centering
      \includegraphics[width=2.4cm]{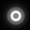}
      \includegraphics[width=2.4cm]{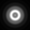}
      \includegraphics[width=3.0cm]{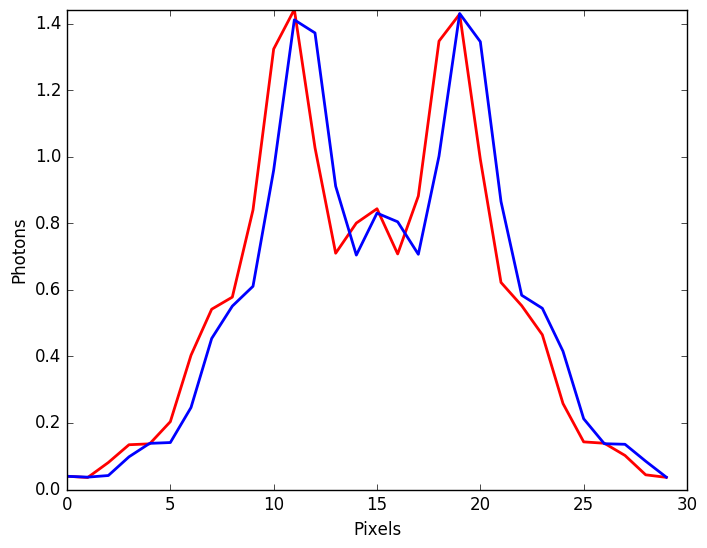}
      \hspace{0.1cm}
      \includegraphics[width=2.4cm]{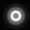}
      \includegraphics[width=2.4cm]{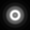}
      \includegraphics[width=3.0cm]{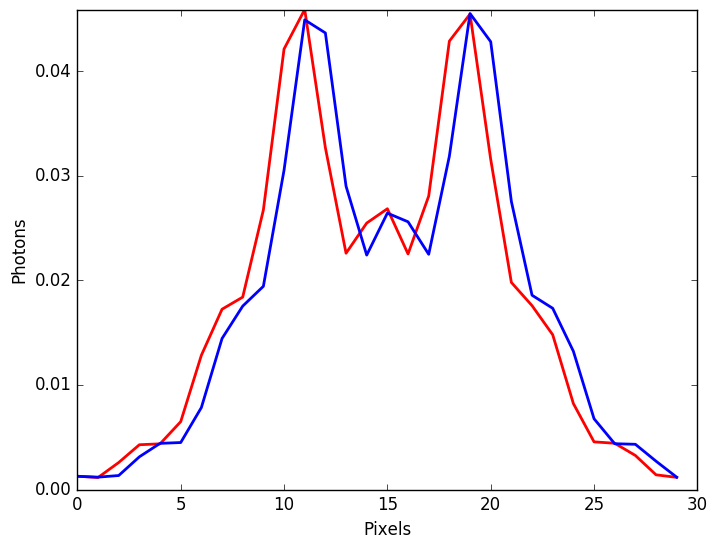}

  \leftline{\bf\hspace{0.2cm} $z = 600\ {\rm nm}$}
  \centering
      \includegraphics[width=2.4cm]{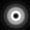}
      \includegraphics[width=2.4cm]{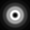}
      \includegraphics[width=3.0cm]{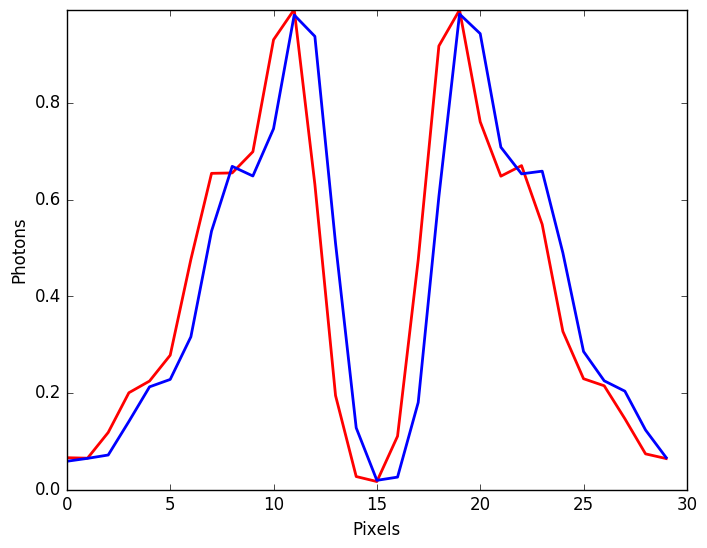}
      \hspace{0.1cm}
      \includegraphics[width=2.4cm]{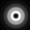}
      \includegraphics[width=2.4cm]{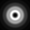}
      \includegraphics[width=3.0cm]{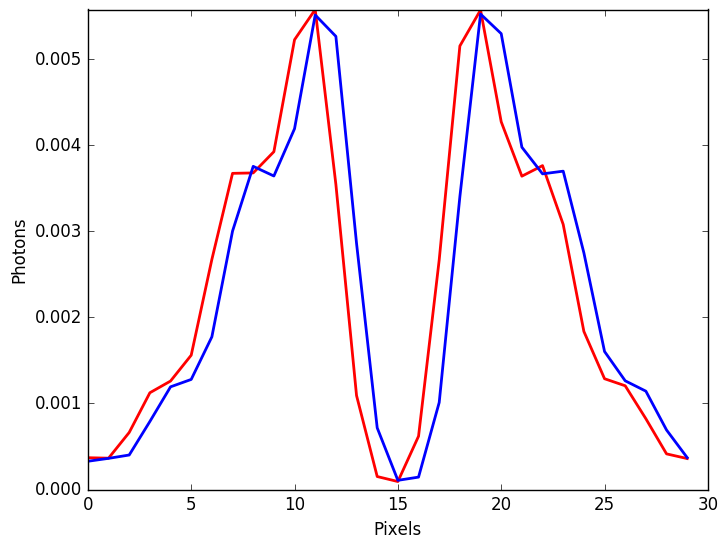}

  \centerline{\bf Oblique illumination \hspace{4.0cm} Evanescence}
  \vspace{0.5cm}
  \caption{{\bf PSF-verification 1}. For each illumination, the right and middle figures show the exact-form and the simulated-form of PSF. The right panels show the comparison of the photon distributions (PSF) along x-axis. The blue and red lines represent the exact-form and the simulated-form.}
  \label{fig;psf_verification_1}
\end{figure}

\begin{figure}[!h]
  \leftline{\bf\hspace{0.2cm} $z = 700\ {\rm nm}$}
  \centering
      \includegraphics[width=2.4cm]{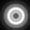}
      \includegraphics[width=2.4cm]{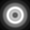}
      \includegraphics[width=3.0cm]{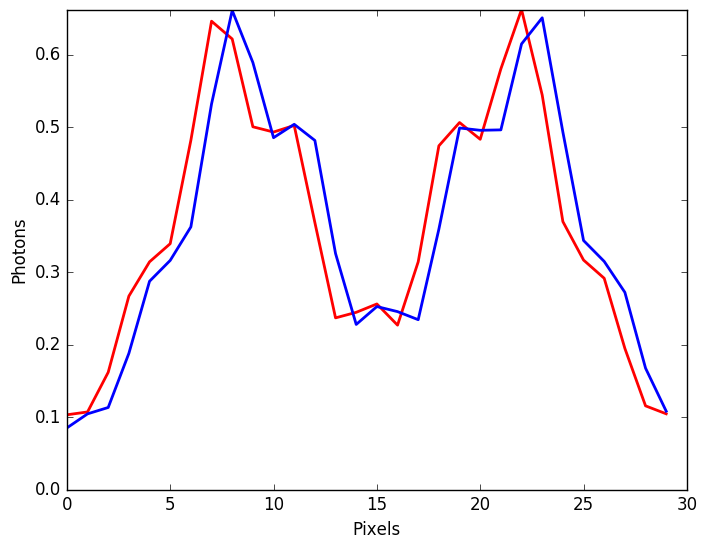}
      \hspace{0.1cm}
      \includegraphics[width=2.4cm]{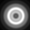}
      \includegraphics[width=2.4cm]{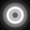}
      \includegraphics[width=3.0cm]{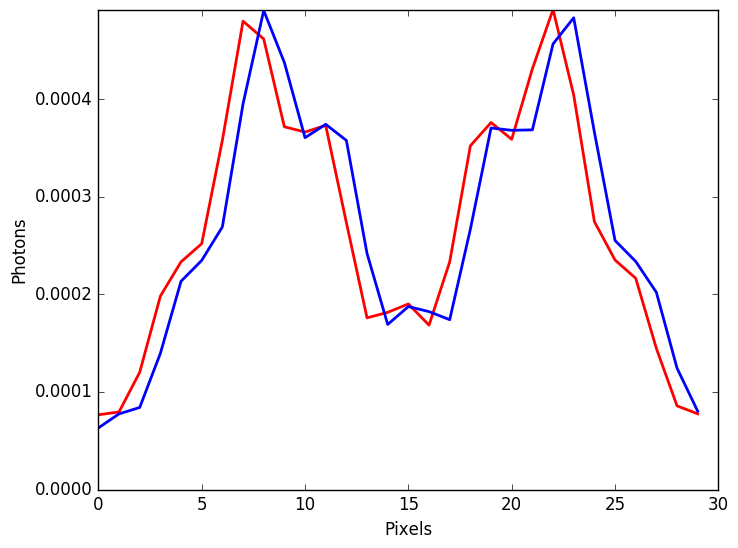}

  \leftline{\bf\hspace{0.2cm} $z = 800\ {\rm nm}$}
  \centering
      \includegraphics[width=2.4cm]{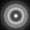}
      \includegraphics[width=2.4cm]{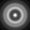}
      \includegraphics[width=3.0cm]{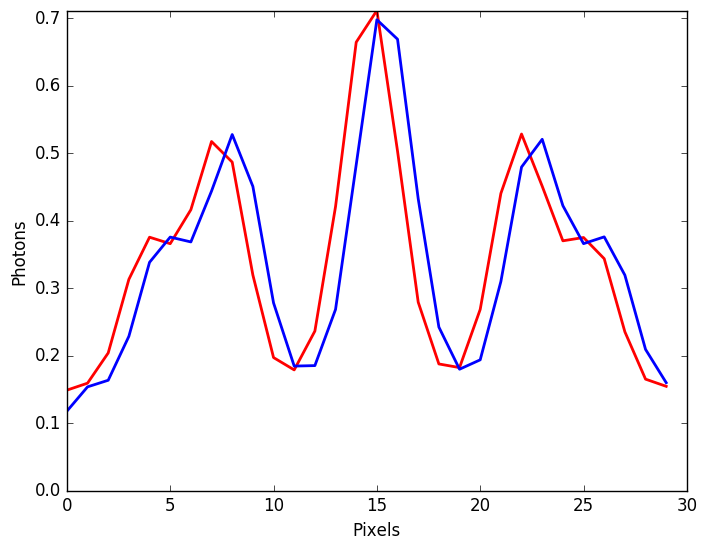}
      \hspace{0.1cm}
      \includegraphics[width=2.4cm]{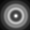}
      \includegraphics[width=2.4cm]{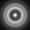}
      \includegraphics[width=3.0cm]{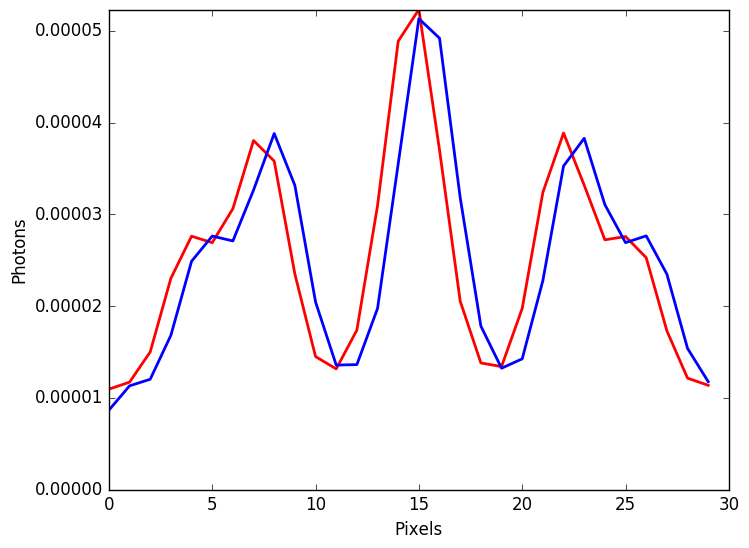}

  \leftline{\bf\hspace{0.2cm} $z = 900\ {\rm nm}$}
  \centering
      \includegraphics[width=2.4cm]{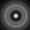}
      \includegraphics[width=2.4cm]{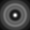}
      \includegraphics[width=3.0cm]{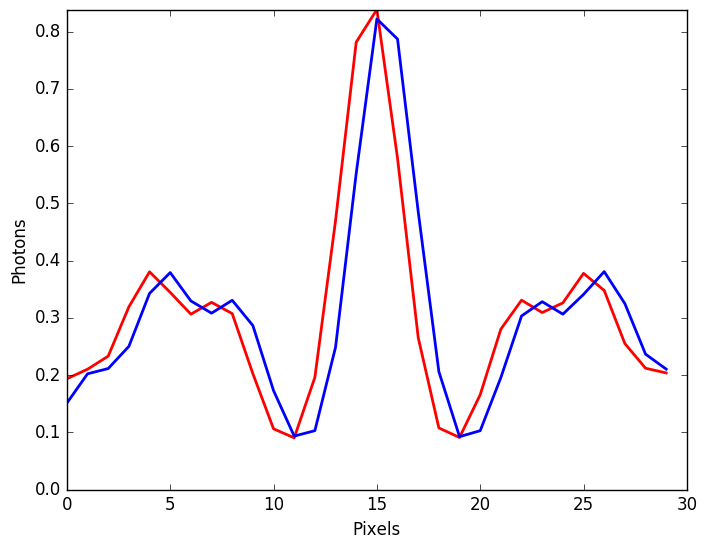}
      \hspace{0.1cm}
      \includegraphics[width=2.4cm]{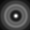}
      \includegraphics[width=2.4cm]{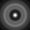}
      \includegraphics[width=3.0cm]{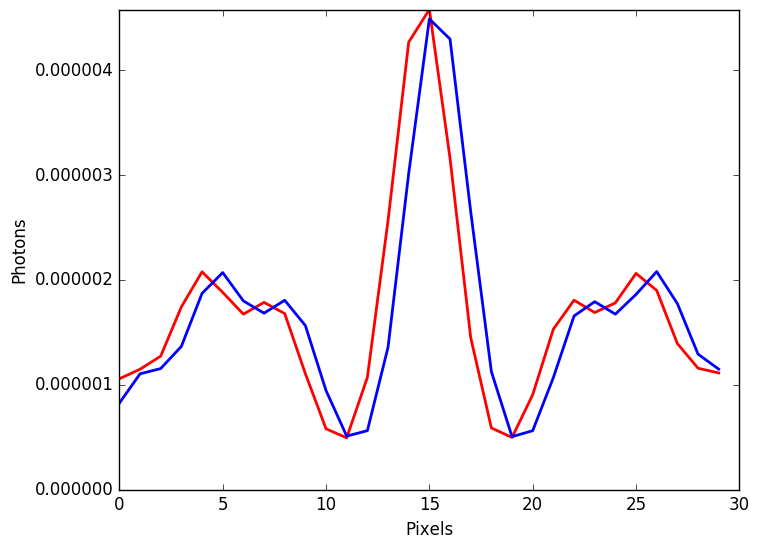}

  \leftline{\bf\hspace{0.2cm} $z = 1000\ {\rm nm}$}
  \centering
      \includegraphics[width=2.4cm]{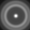}
      \includegraphics[width=2.4cm]{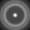}
      \includegraphics[width=3.0cm]{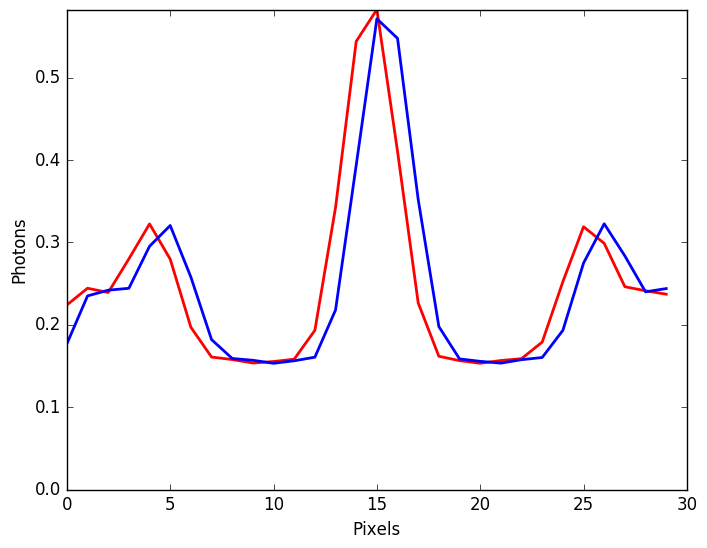}
      \hspace{0.1cm}
      \includegraphics[width=2.4cm]{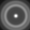}
      \includegraphics[width=2.4cm]{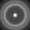}
      \includegraphics[width=3.0cm]{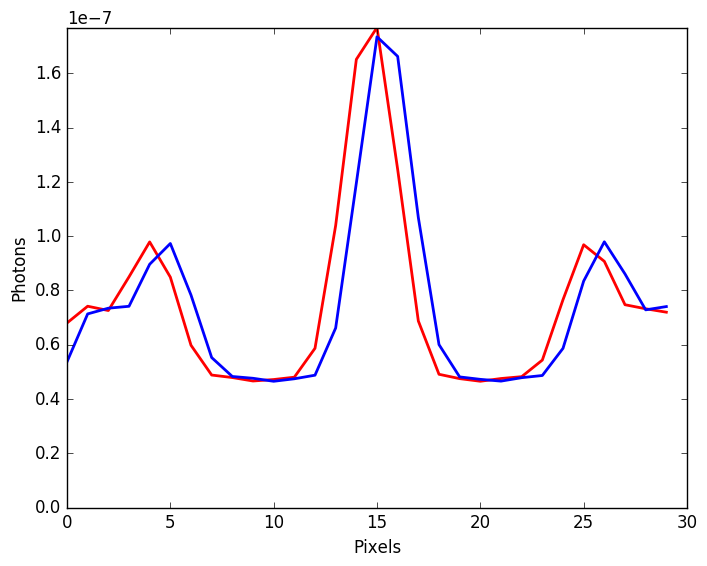}

  \leftline{\bf\hspace{0.2cm} $z = 1100\ {\rm nm}$}
  \centering
      \includegraphics[width=2.4cm]{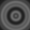}
      \includegraphics[width=2.4cm]{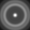}
      \includegraphics[width=3.0cm]{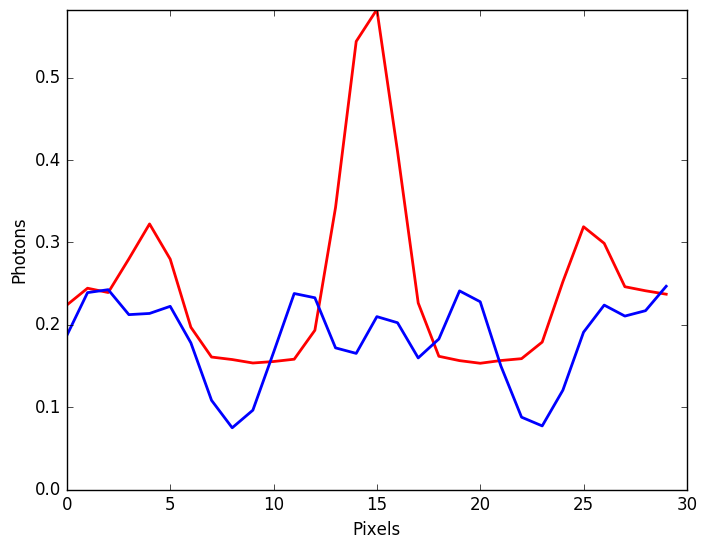}
      \hspace{0.1cm}
      \includegraphics[width=2.4cm]{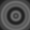}
      \includegraphics[width=2.4cm]{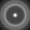}
      \includegraphics[width=3.0cm]{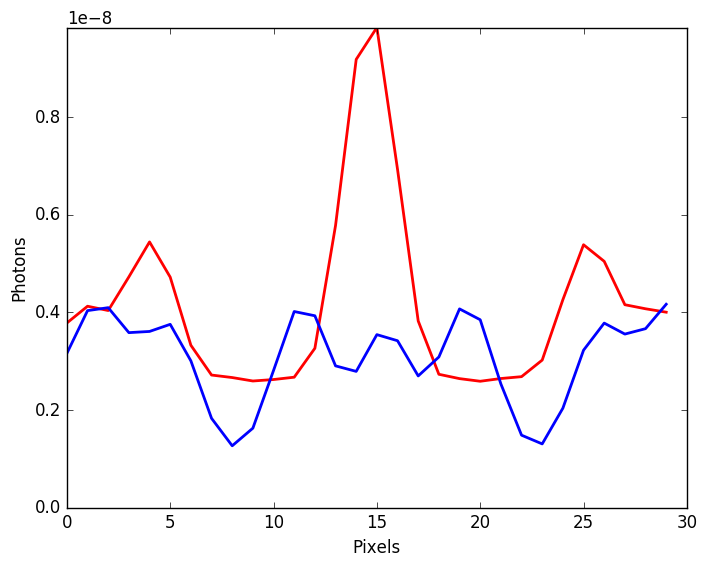}

  \leftline{\bf\hspace{0.2cm} $z = 1200\ {\rm nm}$}
  \centering
      \includegraphics[width=2.4cm]{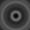}
      \includegraphics[width=2.4cm]{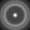}
      \includegraphics[width=3.0cm]{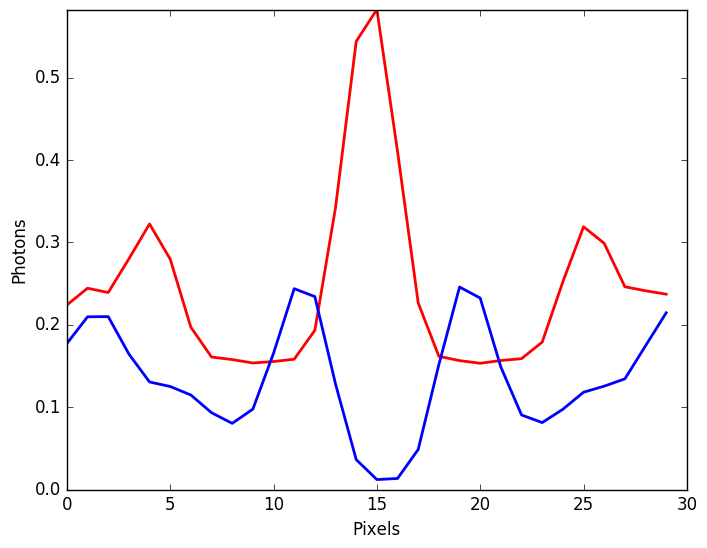}
      \hspace{0.1cm}
      \includegraphics[width=2.4cm]{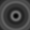}
      \includegraphics[width=2.4cm]{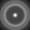}
      \includegraphics[width=3.0cm]{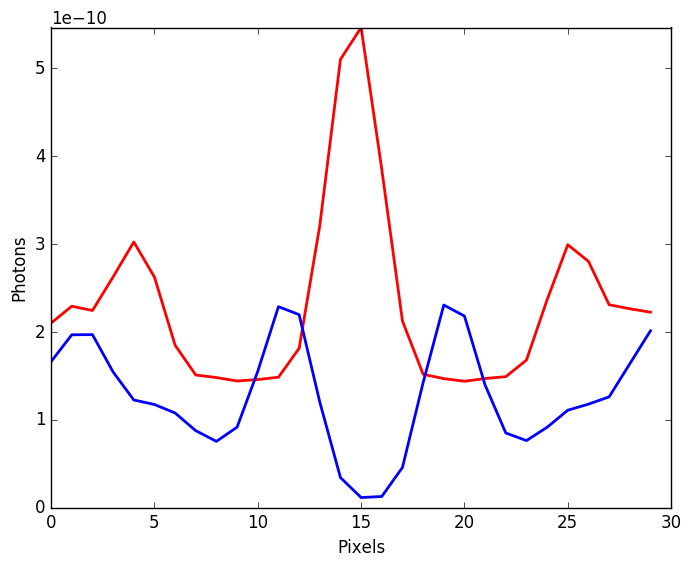}

  \leftline{\bf\hspace{0.2cm} $z = 1300\ {\rm nm}$}
  \centering
      \includegraphics[width=2.4cm]{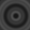}
      \includegraphics[width=2.4cm]{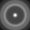}
      \includegraphics[width=3.0cm]{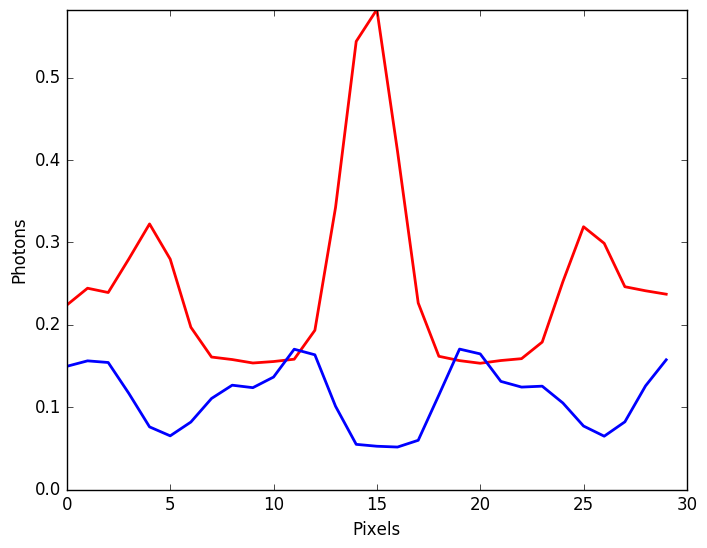}
      \hspace{0.1cm}
      \includegraphics[width=2.4cm]{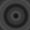}
      \includegraphics[width=2.4cm]{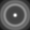}
      \includegraphics[width=3.0cm]{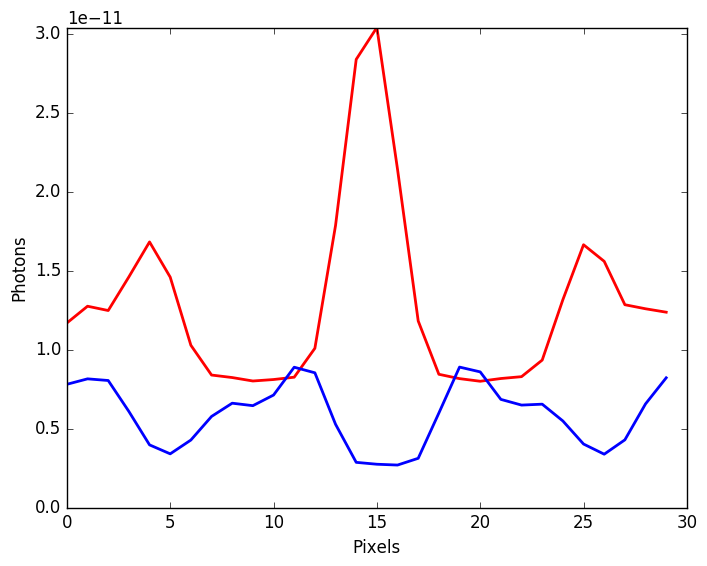}

  \centerline{\bf Oblique illumination \hspace{4.0cm} Evanescence}
  \vspace{0.5cm}
  \caption{{\bf PSF-verification 2}. For each illumination, the right and middle figures show the exact-form and the simulated-form of PSF. The right panels show the comparison of the photon distributions (PSF) along x-axis. The blue and red lines represent the exact-form and the simulated-form.}
  \label{fig;psf_verification_2}
\end{figure}

\paragraph{}
We also evaluated the peak (or -maximum) photon number from the exact-from and the simulated-form of the PSF. The top panels of Figures \ref{fig;peaks_verification} show the comparison of the PSF-peak photon number as a function of the lateral distance from the focal plane. The bottom panels show the ratio of the peak photon number from the exact-form to the simulated-form of the PSF, implying that the peak photon number is well-conserved within $2\%$ differences below the z-cut limit. Although the difference is overflow above the limit, the peak photon number is relatively low compared with the value on the focal plane. This may weakly influence on the image-forming processes. In addition, Figures \ref{fig;psf_verification_1} and \ref{fig;psf_verification_2} show the PSF-image comparison for various lateral distances. We found no significant difference near the focal region. The differences, however, clearly appear above the z-cut limit.

\newpage

\newpage

\newpage

\section{Reconstruction of physical and biological properties}
\paragraph{}
A primary task for quantitative image analysis is to accurately reconstruct physical properties of single-molecules such as molecular localization and diffusion coefficient. Our collaborators strongly rely on the reconstruction method (or single particle tracking) constructed by T. M. Watanabe in 2005 (unpublished work). In this study, we modified spot-scanning process in his reconstruction method, and computationally evaluate the performance of the modified method. There are four major steps in the modified reconstruction procedure. The steps are shortly described as follows.

\begin{enumerate}
\item[(1)] Spot-detection : We applied the Laplacian of the Gaussian (LoG) method to identify spots (or blobs) in single-molecule images \cite{vanderwalt2014}\footnote{Also see \url{https://en.wikipedia.org/wiki/Blob_detection} and \url{http://scikit-image.org/docs/dev/auto_examples/features_detection/plot_blob.html}.}. The method detects spot-like features by searching for scale space extrema of a scale-normalized LoG :
\begin{eqnarray}
\nabla^2_{norm} L(x, y; \sigma) & = & \sigma^2 \left( L_{xx} + L_{yy}\right)
\end{eqnarray}
where $L(x, y; \sigma) = g(x,y;\sigma) \ast f(x,y)$ is a scale space representation at a certain scale $\sigma$. $g(x,y;\sigma)$ and $f(x,y)$ are Gaussian kernel and input image. There are five input parameters; (a) minimum and maximum standard deviation of the Gaussian kernel, (b) the number of intermediate values in the range of minimum to maximum standard deviations, (c) local maxima smaller than threshold are ignored, reducing detection of blobs with less intensities, and (d) an overlap value between 0 and 1, which can eliminate the smaller blobs if the area of two blobs overlaps by a fraction greater than the threshold. In addition, more accurate methods are discussed in ref. \cite{basset2015, ruusuvuori2010, smal2010}.
\item[(2)] Spot-property : Spot-properties are characterized as a Gaussian function of six parameters. Each detected spots can be fitted to the function written in the form of
\begin{eqnarray}
G(x,y) & = & N_0 \exp{\left( - \frac{(x - x_0)^2}{2 \sigma^2_x} - \frac{(y - y_0)^2}{2 \sigma^2_y}\right)} + bg
\end{eqnarray}
where $N_0$, ($x_0, y_0$), $\sigma_{x,y}$ and $bg$ are spot pulse-height (or normalization factor), central spot position, spot pulse-widths, and background pulse-height. 
\item[(3)] Spot-tracking  (or event-identification) : We assume two conditions in linking two spots. The primary condition is the distance between $i$-th spot in $k$-th image-frame and $j$-th spot in ($k+1$)-th image frame must be less than average of each spot sizes. The condition can be written in the form of
\begin{eqnarray}
\left| \vec{r}_{j,k+1} - \vec{r}_{i,k} \right| & < & \frac{\sigma_{r_{i,k}} + \sigma_{r_{j,k+1}}}{2}
\end{eqnarray}
where $i$ and $j$ represent spot indexes at $k$-th and ($k+1$)-th image-frames. The secondary condition is the intensity difference between $i$-th spot in $k$-th image-frame and $j$-th spot in ($k+1$)-th image frame must be less than sum of each spot shot-noise. The condition is given as follows.
\begin{eqnarray}
\left| n(\vec{r}_{j,k+1}) - n(\vec{r}_{i,k}) \right| & < & \sqrt{n(\vec{r}_{i,k}) + n(\vec{r}_{j,k+1})}
\end{eqnarray}
where $n$ represents the spot pulse-height in photoelectron unit. If there is more than one spot that satisfies the two conditions, then the spot in the shortest distance can be linked. Finally, all linked spots are presented as space-time series of spots, and identified as an event. 
\item[(4)] Event-property : The physical properties of each spot-series are reconstructed: diffusion coefficient, event-length, event pulse-height and event-vertex. \\
\end{enumerate}

\newpage

\subsection{Simple cell-models}
\subsubsection{No binding}
\paragraph{}
The use of simulated single-molecule images allows us to quantitatively check the performance of the spot and event reconstructions. The results are shown as follows.

\paragraph{Cell-model :}
We constructed a relatively simple cell-model where monomer and dimers slowly diffuse on a cellular membrane. The GEOMETRY and LENGTH variables of the Spatiocyte cell compartment are set to the ellipsoidal shape and $45 \times 35 \times 2.0\ {\rm \mu m^3}$ ($Y \times Z \times X$). The radius of the HCP lattice voxels is set to $20\ {\rm nm}$. The cell-surface area is $2576.72$ ${\rm \mu m^2}$ with $715$ monomers and $357$ dimers uniformly distributed on the cell-surface. Those molecules slowly diffuse with $0.015\ {\rm \mu m^2/sec}$. No molecular reaction is defined in this simple model.

\begin{figure}[!h]
  \centering
        \includegraphics[width=7.4cm]{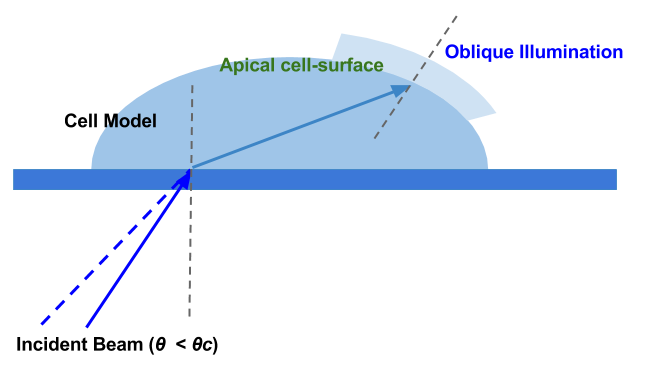}
        \hspace{0.4cm}
        \includegraphics[width=7.4cm]{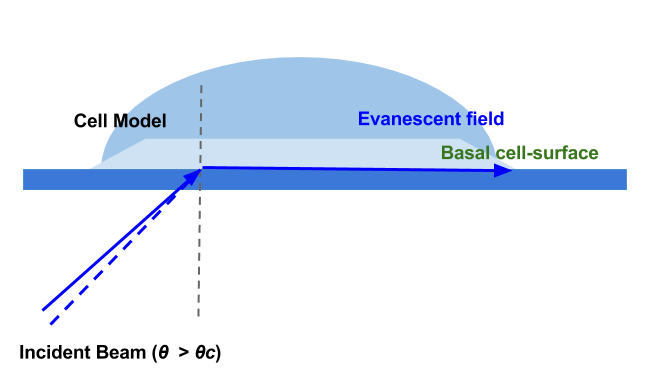}        

  \caption{{\bf Illumination configuration on apical and basal cell-surfaces}.}
  \label{fig;cell_illumination}
\end{figure}

\paragraph{Microscopy configuration :}
Single-molecule imaging of apical and basal regions of the simple cell-model is simulated for the optical specification and its operating condition of the fluorescence microscopy simulation module shown in Table \ref{tab;specification3}. Photobleaching is included in the simulated single-molecule images. Incident beam angle is set to $72^{\circ}$ to observe the basal region of the cell-model. In order to observe the apical cell-area, we also set the incident beam angle to $60^{\circ}$ to generate oblique illumination. While exposure time is configured to $0.150\ {\rm sec}$, image acquisition period is $10\ {\rm sec}$ ($66$ frames). The illumination configuration on the cell-surfaces is shown in Figure \ref{fig;cell_illumination}.

\begin{figure}[!h]
  \centering
        \includegraphics[width=6.6cm]{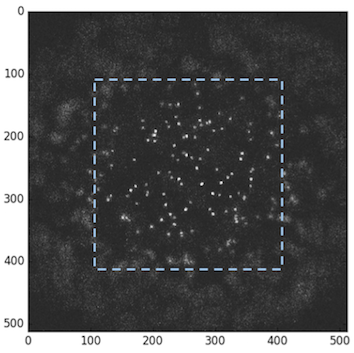}
        \hspace{1.2cm}
        \includegraphics[width=6.6cm]{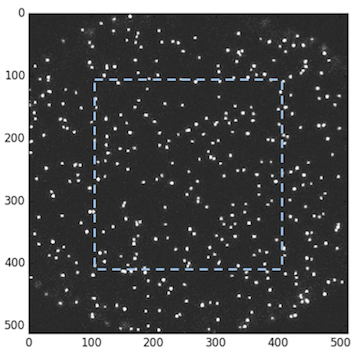}        

  \vspace{0.1cm}
  \centering
        \includegraphics[width=6.6cm]{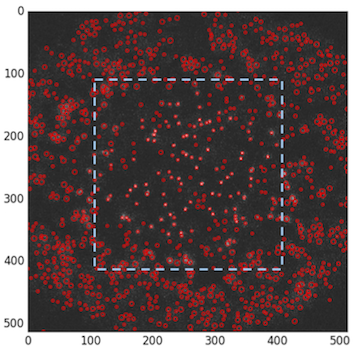}
        \hspace{1.2cm}
        \includegraphics[width=6.6cm]{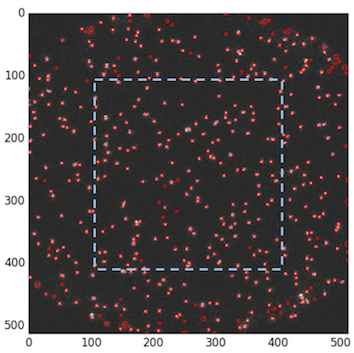}        

  \centerline{\bf Apical cell-region \hspace{4.2cm} Basal cell-region}
  \caption{{\bf Example images of the spot detection on the apical and basal cell-regions}. Single-molecule images before and after spot-detection are shown in the top and the bottom panels.}
  \label{fig;spot_images_1}
\end{figure}

\paragraph{(1) Spot-detection :}
In this test configuration, the parameter values for the spot-detection algorithm are configured to $\sigma_{min} = 2.0\ {\rm pixels}$, $\sigma_{max} = 4.0\ {\rm pixels}$, $20$ intermediate values in the deviation range, threshold $= 15$, and overlap $= 0.5$. Figure \ref{fig;spot_images_1} shows example images of spot-detection on the apical and basal cell-regions. The observational area cut to $300 \times 300\ {\rm pixels}$ ($397\ {\rm \mu m^2}$) is shown with dashed lines. Red circles represent molecular-spots detected by using the LoG algorithm. Each image size is $512 \times 512\ {\rm pixels}$. Actual minimum and maximum values of the image intensity are $1,900$ and $2,500$ ADC counts. The image intensity is rescaled in the range of $0$ to $255$. The molecular-spots are clear and well detected in the basal cell-region. In the apical cell-region, the microscopy is configured to focused on the image center. While molecular-spots are clear and properly picked up at the focused area, the spots are blur and not well detected near the curved-region on the cellular membrane.

\begin{figure}[t]
  \centering
        \includegraphics[width=7.4cm]{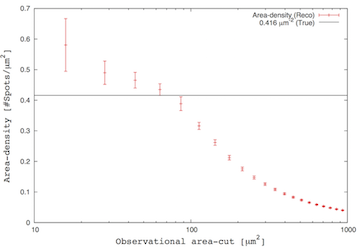}
        \hspace{0.4cm}
        \includegraphics[width=7.4cm]{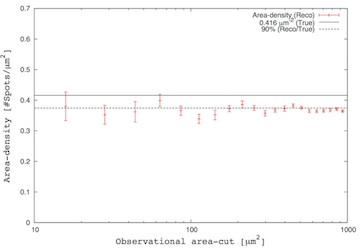}

  \centerline{\bf Apical cell-region \hspace{4.0cm} Basal cell-region}
  \caption{{\bf Reconstructed area-density for various area-cuts}. The reconstruction efficiency is shown in the apical and basal cell-regions.}
  \label{fig;reco_density_1}
\end{figure}

\paragraph{}
In addition, using the simulated images not including photobleaching effects, we reconstruct spot area-density and estimate reconstruction efficiencies for various observational cuts. The area is fixed at the image center, and the area-size varies from $10\ {\rm \mu m^2}$ to $1000\ {\rm \mu m^2}$. Figure \ref{fig;reco_density_1} shows the reconstruction efficiency of the area-density in apical and basal cell-regions. Ground-true spot area-density is set to $0.416\ {\rm spots/\mu m^2}$. While the reconstruction efficiency in the basal cell-region is constant at $90\%$ for the various area-cuts, the efficiency is largely varied by the area-cuts in the apical cell-regions. The reconstructed area-density is over- and under-estimated below and above $100\ {\rm \mu m^2}$ area-cut.

\begin{figure}[!h]
  \centering
        \includegraphics[width=7.4cm]{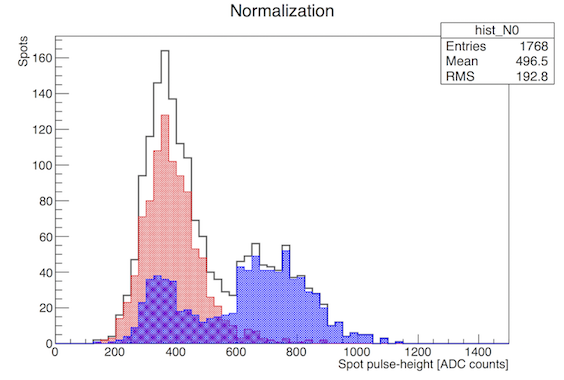}
        \hspace{0.4cm}
        \includegraphics[width=7.4cm]{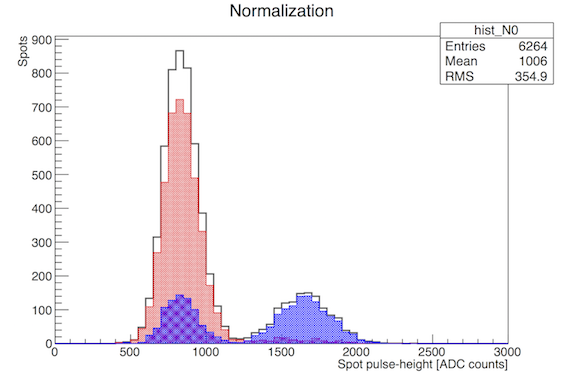}

  \centering
        \includegraphics[width=7.4cm]{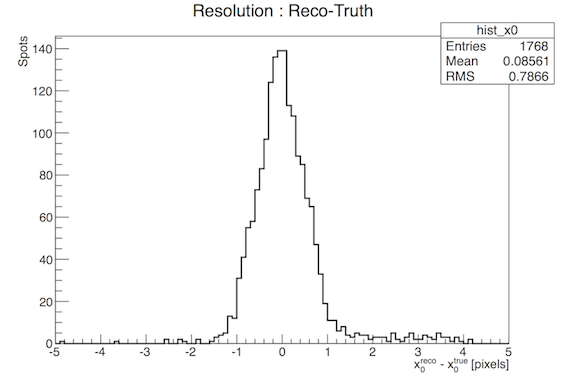}
        \hspace{0.4cm}
        \includegraphics[width=7.4cm]{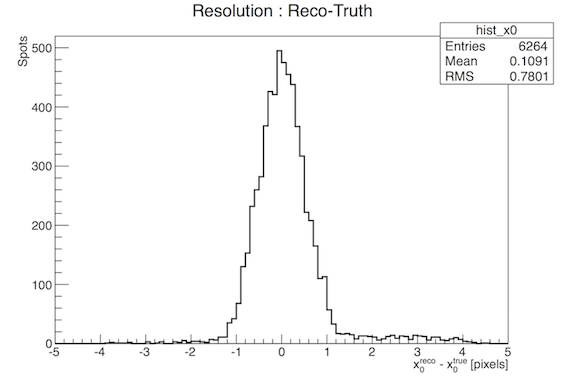}

  \centering
        \includegraphics[width=7.4cm]{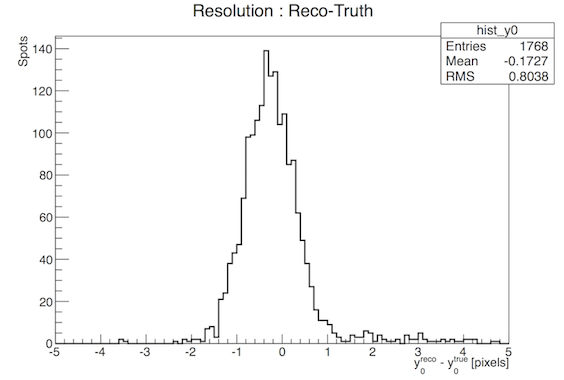}
        \hspace{0.4cm}
        \includegraphics[width=7.4cm]{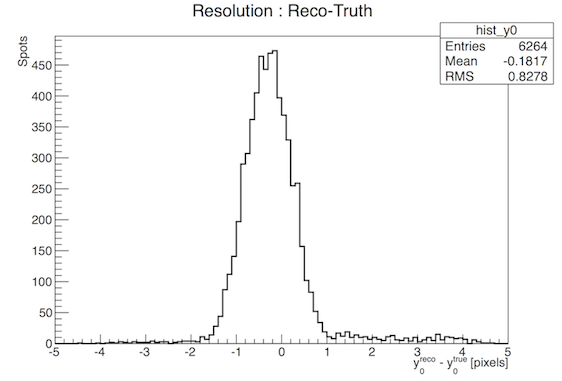}
 
  \centering
        \includegraphics[width=7.4cm]{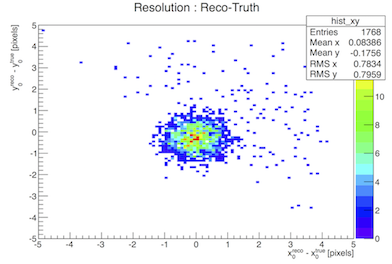}
        \hspace{0.4cm}
        \includegraphics[width=7.4cm]{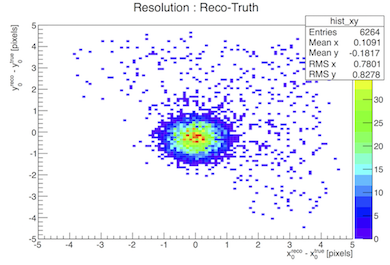}
 
  \centerline{\bf Apical cell-region \hspace{4.0cm} Basal cell-region}

  \caption{{\bf Spot-properties 1}. The top panels show distributions of spot pulse-height. Distributions and correlation of positional resolution (or localization error) in x-y axes are shown in the bottom three panels.}
  \label{fig;spot_property_1}
\end{figure}

\begin{figure}[!h]
  \centering
        \includegraphics[width=7.4cm]{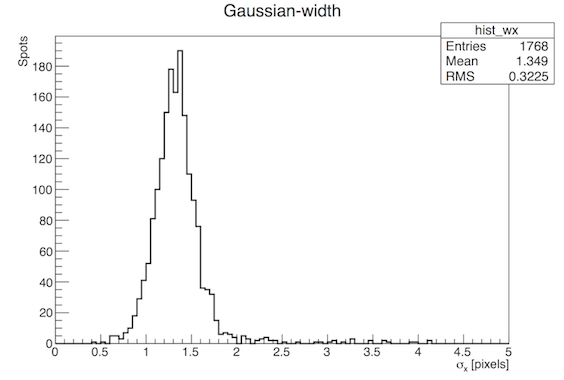}
        \hspace{0.4cm}
        \includegraphics[width=7.4cm]{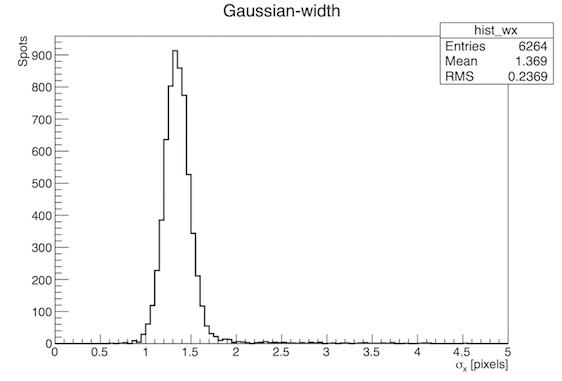}

  \centering
        \includegraphics[width=7.4cm]{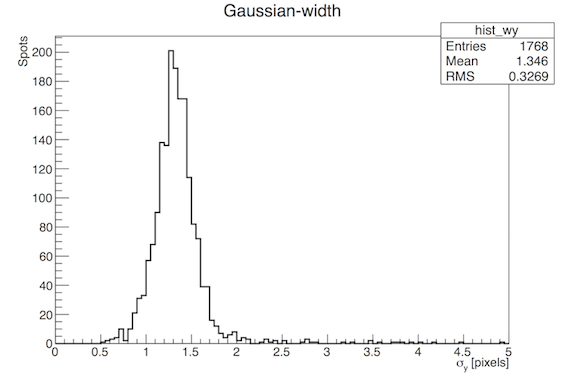}
        \hspace{0.4cm}
        \includegraphics[width=7.4cm]{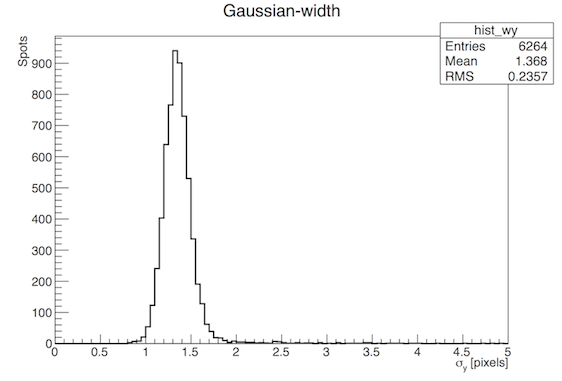}
 
  \centering
        \includegraphics[width=7.4cm]{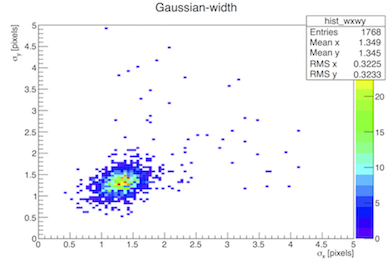}
        \hspace{0.4cm}
        \includegraphics[width=7.4cm]{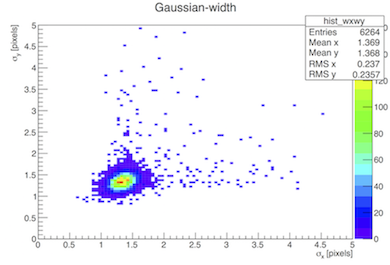} 

  \centering
        \includegraphics[width=7.4cm]{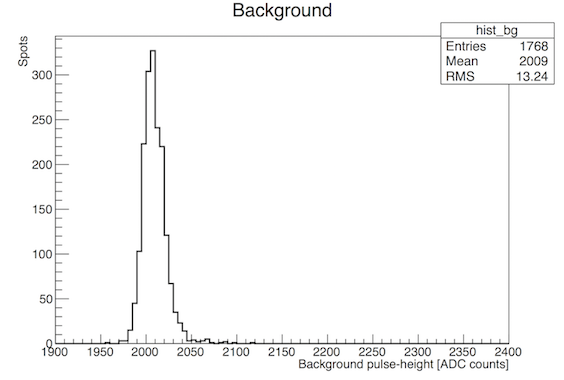}
        \hspace{0.4cm}
        \includegraphics[width=7.4cm]{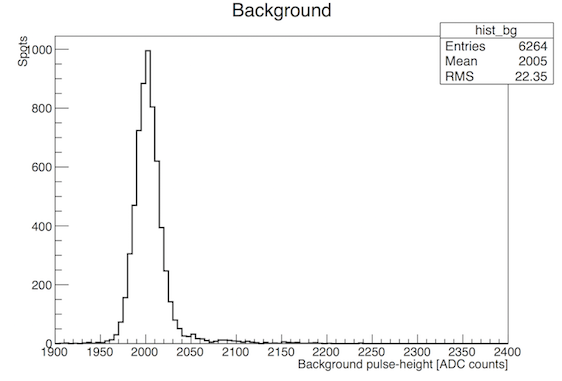} 

  \centerline{\bf Apical cell-region \hspace{4.0cm} Basal cell-region}
  \caption{{\bf Spot-properties 2}. Distributions and correlations of reconstructed pulse-width (or Gaussian-witdth) in x-axis and y-axis are shown in top three panels. Distribution of background pulse-height is shown in the bottom panels.}
  \label{fig;spot_property_2}
\end{figure}

\begin{figure}[!h]
  \centering
        \includegraphics[width=7.4cm]{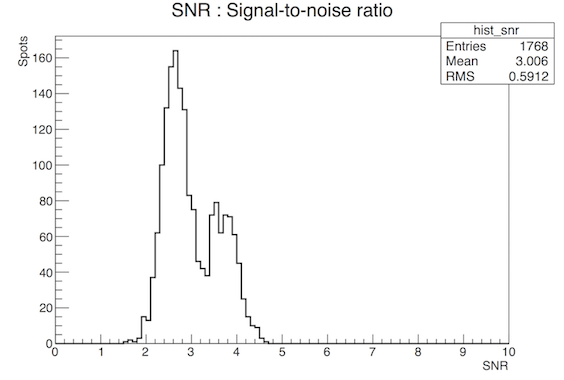}
        \hspace{0.4cm}
        \includegraphics[width=7.4cm]{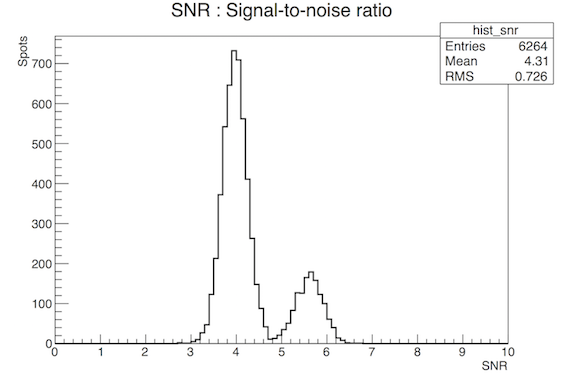} 

  \centering
        \includegraphics[width=7.4cm]{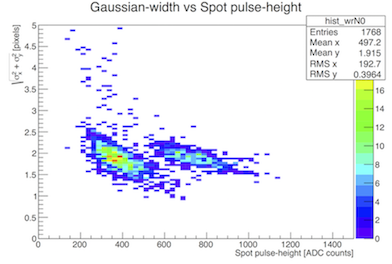}
        \hspace{0.4cm}
        \includegraphics[width=7.4cm]{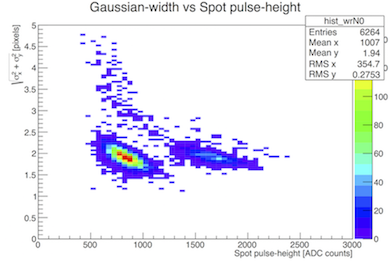} 

  \centering
        \includegraphics[width=7.4cm]{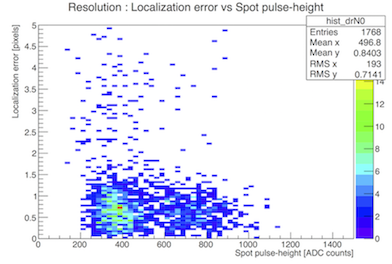}
        \hspace{0.4cm}
        \includegraphics[width=7.4cm]{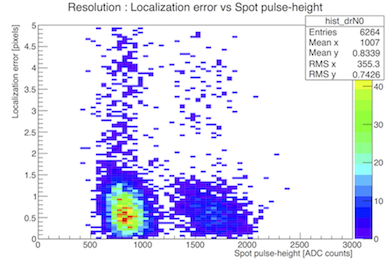} 

  \centerline{\bf Apical cell-region \hspace{4.0cm} Basal cell-region}
  \caption{{\bf Spot-properties 3}. The top panels show SNR distribution. Bottom two panels show correlation distributions of spot pulse-height to spot-size and localization error.}
  \label{fig;spot_property_3}
\end{figure}

\paragraph{(2) Spot-property :}
Using the simulated images including photobleaching effects, we reconstruct spot-properties. The reconstructed spot-properties are presented as six parameters of the Gaussian function; spot pulse-height (or normalization factor), central position, spot pulse-width and background pulse-height. Distributions and correlations of each parameters are shown in Figures \ref{fig;spot_property_1}, \ref{fig;spot_property_2} and \ref{fig;spot_property_3}.
\begin{enumerate}
\item[(a)] Spot pulse-height ($N_0$) : Distributions of the spot pulse-height are shown in the top panels of Figure \ref{fig;spot_property_1}. This is also known as cluster size in Hiroshima's work \cite{hiroshima2012}. The black solid line represents observed distribution of reconstructed spots (monomer and dimer together).  Two peaks are observed in these histograms. First and second peaks correspond to monomer-like and dimer-like spot distributions. Red and blue filled histograms represent ground-truth distribution of monomer ({\bf R}) and dimer ({\bf RR}) spots. Some of dimer spots photobleaches and forms monomer-like spots. The histograms show that the photobleached-dimer spots (blue) are identified as monomer-like spots (red).
\item[(b)] Localization errors ($\vec{r}^{\ reco} - \vec{r}^{\ true}$) : Positional resolutions (or localization error) of reconstructed spot-positions are shown in the bottom three panels of Figure \ref{fig;spot_property_1}. In those Figures, we confirmed that peaks of each resolution distribution are located near zeros, and are nearly formed as Gaussian distributions. Root mean squared (RMS) value represents the positional resolution to $0.78\ {\rm pixels}$ ($58\ {\rm nm}$). \\
\forceindent However, the tails of each distribution appear to be asymmetric. One of the possible explanations of the asymmetry is because of the z-axis. In our analysis, we assumed that spots are characterized as 2-D Gaussian distributions, ignoring the z-axis. 3-D Gaussian fitting may be able to resolve the asymmetry of each distribution.
\item[(c)] Spot-size ($\sigma_{x}, \sigma_{y}$ and $\sigma_{x}$ vs $\sigma_{y}$) : Distributions of spot-size (or Gaussian-widths) are shown in top three panels of Figure \ref{fig;spot_property_2}. Those histograms show that the spots are nearly symmetric in x-y axes and their-size is about $1.35\ {\rm pixels}$ ($89.63\ {\rm nm}$).
\item[(d)] Background pulse-height ($bg$) : Distribution of background pulse-height is shown in the bottom panels of Figures \ref{fig;spot_property_2}. The background pulse-heights are distributed around the offset value, and nearly form a Gaussian distribution. 
\item[(e)] SNR : The top panels of Figures \ref{fig;spot_property_3} show SNR distributions. We simply assume that the SNR is the ratio of the spot pulse-height to the spot and background noise. It can be written in the form of $SNR = \frac{s}{\sqrt{s + b}}$ where $s$ and $b$ are spot and background pulse-height in photoelectron unit. The overall SNR distribution of the basal cell-images is shifted to relatively larger SNR values than that of the apical cell-images. 
\item[(f)] Correlations : Correlation distributions of the spot pulse-height to spot-size and localization error are shown in the middle and the bottom panels of Figures \ref{fig;spot_property_3}. Two peaks are observed in those correlations. First and second peaks correspond to monomer-like and dimer-like spots. 
\end{enumerate}

%\begin{eqnarray}
%SNR & = & \frac{s}{\sqrt{s + b}}
%\end{eqnarray}

\newpage

\newpage

\newpage

\newpage

\begin{figure}[!h]
  \leftline{\bf \hspace{0.5cm} Event-trajectories in observation}
  \centering
        \includegraphics[width=7.4cm]{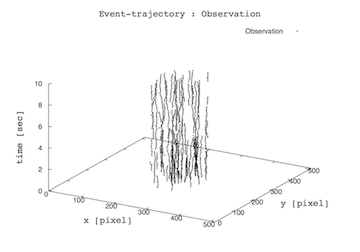}
        \hspace{0.4cm}
        \includegraphics[width=7.4cm]{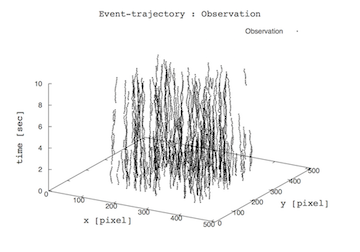}

  \leftline{\hspace{0.5cm} \bf Event-trajectories in ground-truth}
  \centering
        \includegraphics[width=7.4cm]{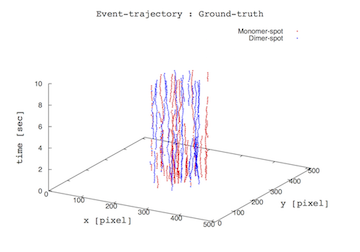}
        \hspace{0.4cm}
        \includegraphics[width=7.4cm]{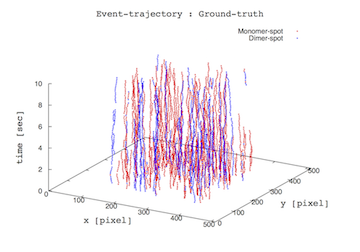}

  \centerline{\bf Apical cell-region \hspace{4.0cm} Basal cell-region}
  \caption{{\bf Event-trajectories 1}. Image pixel-coordinates in x-y axes is shown in the horizontal plane. The vertical axis represents time.}
  \label{fig;event_trajectory_1}
\end{figure}

\begin{figure}[!h]
  \leftline{\bf \hspace{0.5cm} Trajectories for each event-types :}
  \centering
        \includegraphics[width=7.4cm]{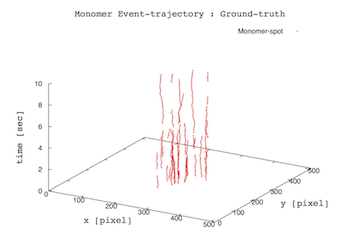}
        \hspace{0.4cm}
        \includegraphics[width=7.4cm]{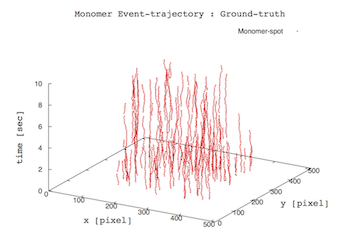}

  \centering
        \includegraphics[width=7.4cm]{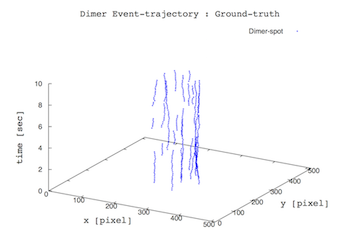}
        \hspace{0.4cm}
        \includegraphics[width=7.4cm]{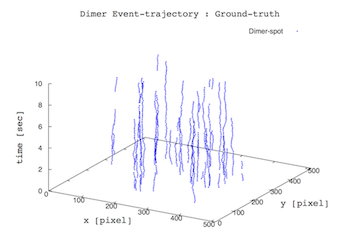}
        
  \centering
        \includegraphics[width=7.4cm]{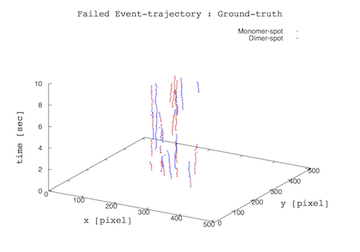}
        \hspace{0.4cm}
        \includegraphics[width=7.4cm]{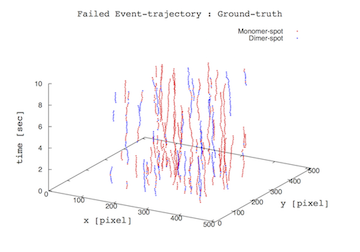}

  \leftline{\bf \hspace{0.5cm} Examples :}
  \centering
        \includegraphics[width=7.4cm]{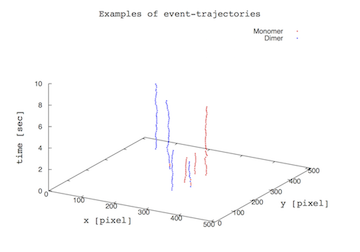}
        \hspace{0.4cm}
        \includegraphics[width=7.4cm]{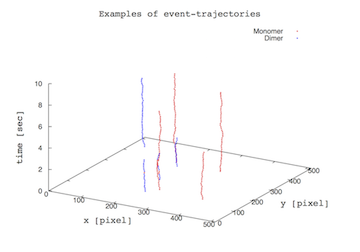}

  \centerline{\bf Apical cell-region \hspace{4.0cm} Basal cell-region}
  \caption{{\bf Event-trajectories 2}.}
  \label{fig;event_trajectory_2}
\end{figure}

\begin{figure}[!h]
  \leftline{\bf \hspace{0.5cm} (a) True-monomer events :}
  \centering
        \includegraphics[width=7.4cm]{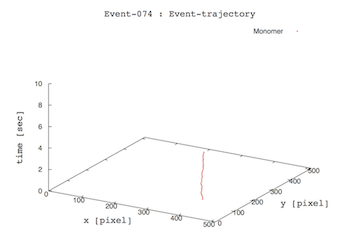}
        \hspace{0.4cm}
        \includegraphics[width=7.4cm]{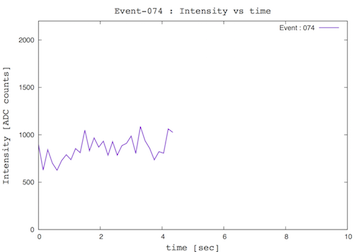}

  \centering
        \includegraphics[width=7.4cm]{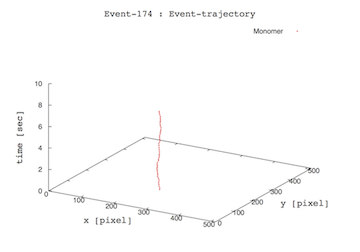}
        \hspace{0.4cm}
        \includegraphics[width=7.4cm]{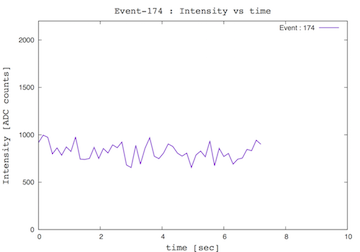}
        
  \leftline{\bf \hspace{0.5cm} (b) True-dimer events :}
  \centering
        \includegraphics[width=7.4cm]{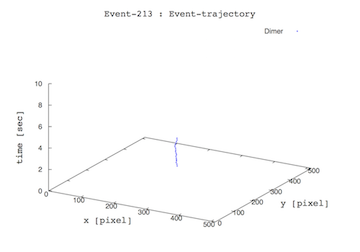}
        \hspace{0.4cm}
        \includegraphics[width=7.4cm]{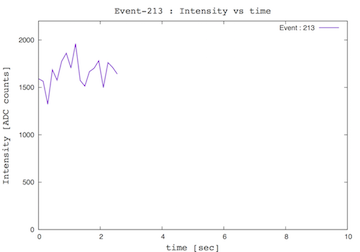}

  \centering
        \includegraphics[width=7.4cm]{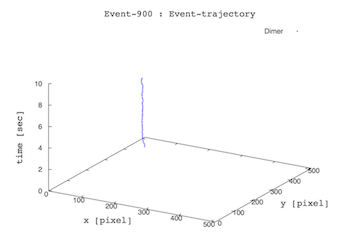}
        \hspace{0.4cm}
        \includegraphics[width=7.2cm]{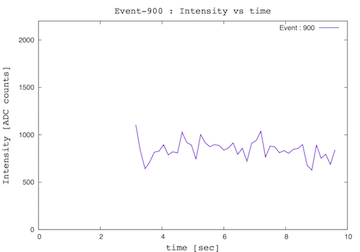}

  \caption{{\bf Examples of event-trajectories observed in the basal cell-region 1}. Left and right columns show event-trajectories and their intensity changes with time.}
  \label{fig;event_trajectory_tirfm_1}
\end{figure}

\begin{figure}[!h]
  \leftline{\bf \hspace{0.5cm} (c) False-events : Exchanges of monomer-dimer}
  \centering
        \includegraphics[width=7.4cm]{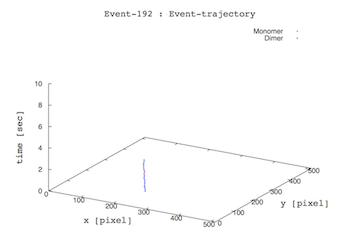}
        \hspace{0.4cm}
        \includegraphics[width=7.4cm]{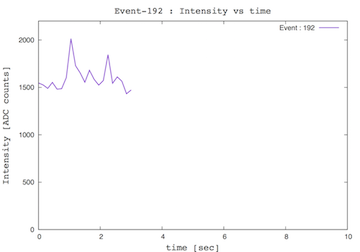}

  \centering
        \includegraphics[width=7.4cm]{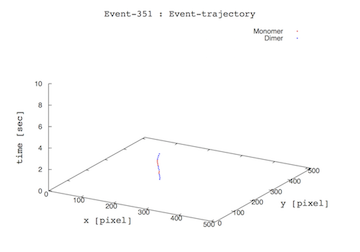}
        \hspace{0.4cm}
        \includegraphics[width=7.4cm]{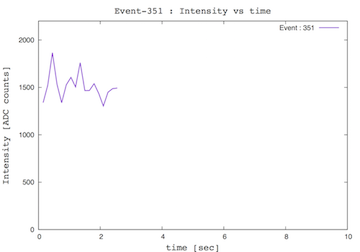}

  \leftline{\bf \hspace{0.5cm} (c) False-events : Exchanges of monomer-monoer (or dimer-dimer)}
  \centering
        \includegraphics[width=7.4cm]{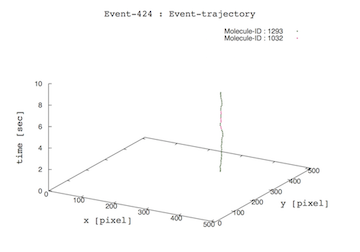}
        \hspace{0.4cm}
        \includegraphics[width=7.4cm]{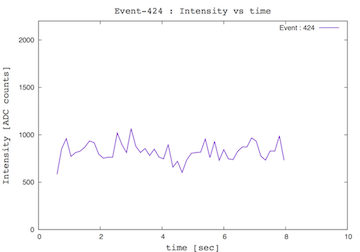}

  \centering
        \includegraphics[width=7.4cm]{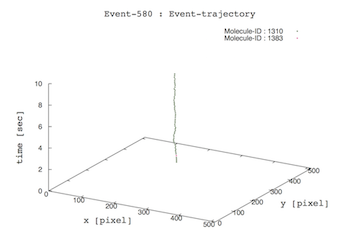}
        \hspace{0.4cm}
        \includegraphics[width=7.4cm]{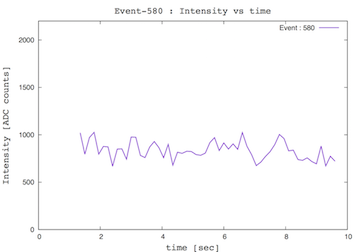}

  \caption{{\bf Examples of event-trajectories observed in the basal cell-region 2}. Green and pink spots represent molecular identification numbers assigned by Spatiocyte cell simulations.}
  \label{fig;event_trajectory_tirfm_2}
\end{figure}

\begin{figure}[!h]
  \leftline{\bf \hspace{0.5cm} (a) True-monomer events :}
  \centering
        \includegraphics[width=7.4cm]{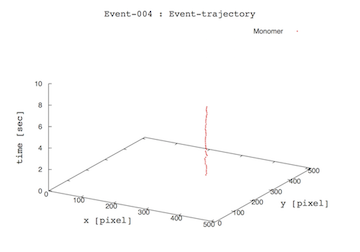}
        \hspace{0.4cm}
        \includegraphics[width=7.4cm]{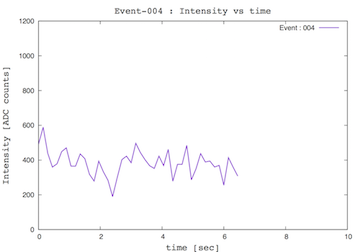}

  \centering
        \includegraphics[width=7.4cm]{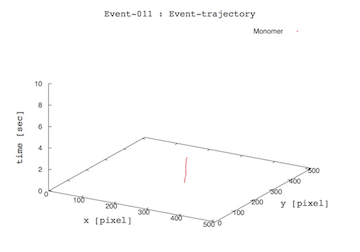}
        \hspace{0.4cm}
        \includegraphics[width=7.4cm]{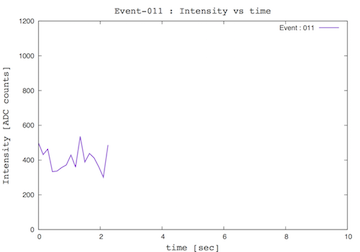}

  \leftline{\bf \hspace{0.5cm} (b) True-dimer events :}
  \centering
        \includegraphics[width=7.4cm]{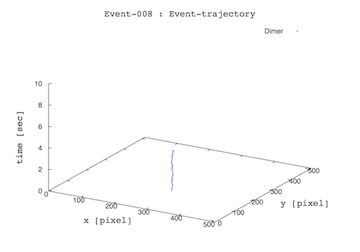}
        \hspace{0.4cm}
        \includegraphics[width=7.4cm]{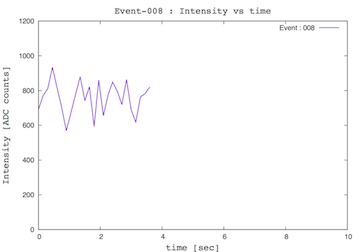}

  \centering
        \includegraphics[width=7.4cm]{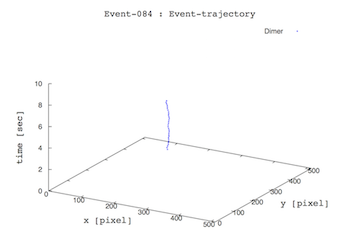}
        \hspace{0.4cm}
        \includegraphics[width=7.4cm]{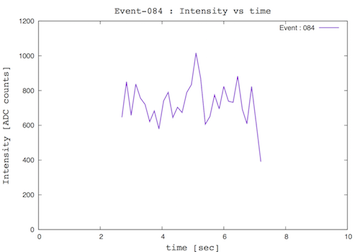}

  \caption{{\bf Examples of event-trajectories observed in the apical cell-region 1}. Left and right columns show event-trajectories and their intensity changes with time.}
  \label{fig;event_trajectory_epifm_1}
\end{figure}

\begin{figure}[!h]
  \leftline{\bf \hspace{0.5cm} (c) False-events : Exchanges of monomer-dimer}
  \centering
        \includegraphics[width=7.4cm]{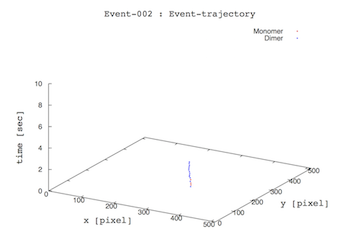}
        \hspace{0.4cm}
        \includegraphics[width=7.4cm]{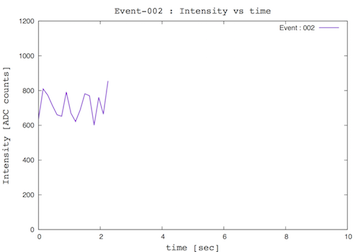}

  \centering
        \includegraphics[width=7.4cm]{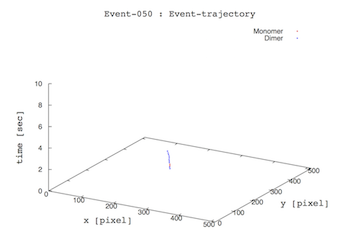}
        \hspace{0.4cm}
        \includegraphics[width=7.4cm]{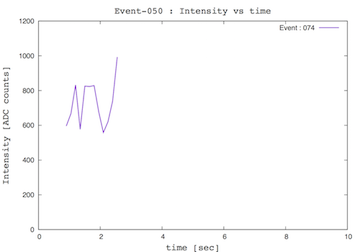}
        
  \leftline{\bf \hspace{0.5cm} (c) False-events : Exchanges of monomer-monomer (or dimer-dimer)}
  \centering
        \includegraphics[width=7.4cm]{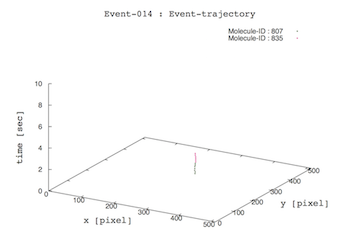}
        \hspace{0.4cm}
        \includegraphics[width=7.4cm]{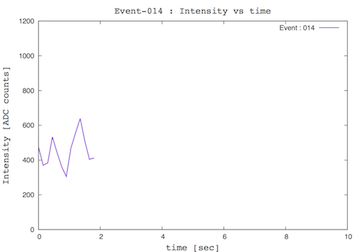}

  \centering
        \includegraphics[width=7.4cm]{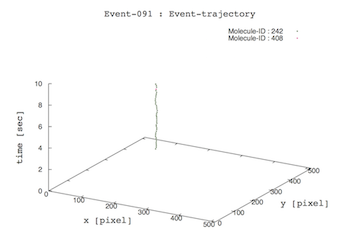}
        \hspace{0.4cm}
        \includegraphics[width=7.4cm]{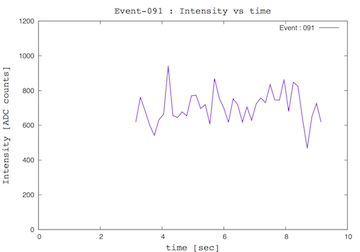}

  \caption{{\bf Examples of event-trajectories observed in the apical cell-region 2}. Green and pink spots represent molecular identification numbers assigned by Spatiocyte cell simulations.}
  \label{fig;event_trajectory_epifm_2}
\end{figure}

\paragraph{(3) Spot-tracking :}
In the simple cell-model, $118$ and $382$ events are reconstructed in the apical and basal cell-surfaces. These events can be categorized into three types; true-monomer, true-dimer, and false event-reconstruction. Table \ref{tab;events_frac_1} shows the number and fraction of each event types. Figures \ref{fig;event_trajectory_1} show space-time series of the observed and ground-truth reconstructed event-trajectories in the apical and basal cell-surfaces. Black spots represent observed states of detected spots. Red and blue spots represent the ground-truth of monomer and dimer. \\
\forceindent Figures \ref{fig;event_trajectory_2} represent all reconstructed trajectories for each event-types. Figures \ref{fig;event_trajectory_tirfm_1} to \ref{fig;event_trajectory_epifm_2} show example trajectories of each event-types and their reconstructed intensity (or spot pulse-height) changes in time; (a) True-monomer events are successfully reconstructed with lower-intensity. (b) True-dimer events are successfully reconstructed with higher-intensity. (c) Event-reconstruction failed. While tracking spots, true-monomer (or true-dimer) can be often exchanged with a neighbour molecule. For example, the third and fourth panels of Figure \ref{fig;event_trajectory_tirfm_2} and \ref{fig;event_trajectory_epifm_2} show exchanges of molecular-states and molecular identification-numbers assigned by Spatiocyte cell simulations. Such event-exchanges cannot be distinguished from other events. 

\begin{table}[!h]
\centering
\begin{tabular}{|c|c|c|c|c|}
\hline
%\multicolumn{2}{|c|}{\bf Area density of ErbB proteins} \\ \hline
Cell-surface & Total events & True-Monomer & True-Dimer & False-Reconstruction \\ \hline
Basal & $382$ & $175$ ($45.8\%$) & $106$ ($27.7\%$) & $101$ ($26.4\%$) \\ \hline
Apical & $118$ & $45$ ($38.1\%$) & $42$ ($35.6\%$) & $31$ ($26.3\%$) \\ \hline
\end{tabular}
\caption{{\bf The number and fraction of detected events}. The number and fraction of true-monomer, true-dimer, and false-reconstruction events to the total number of detected events.}
\label{tab;events_frac_1}
\end{table}

\newpage

\newpage

\newpage

\newpage

\newpage

\newpage

\begin{figure}[!h]
  \centering
        \includegraphics[width=7.4cm]{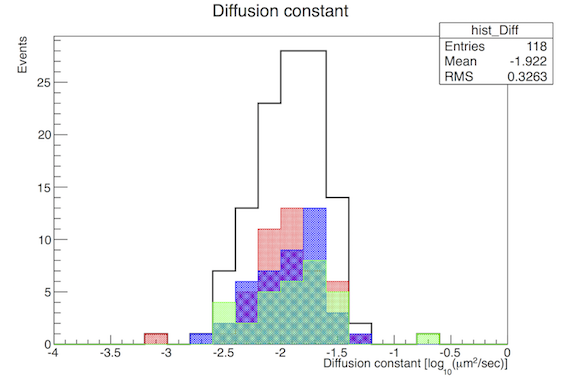}
        \hspace{0.4cm}
        \includegraphics[width=7.4cm]{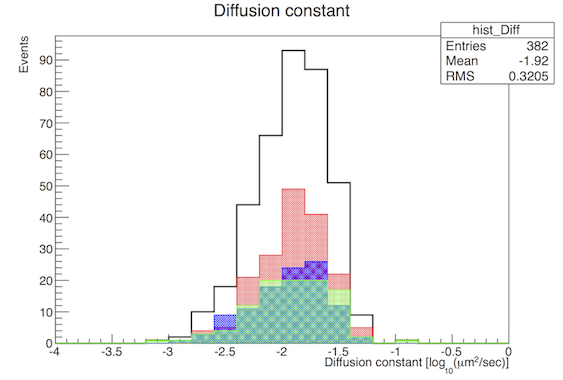}

  \centering
        \includegraphics[width=7.4cm]{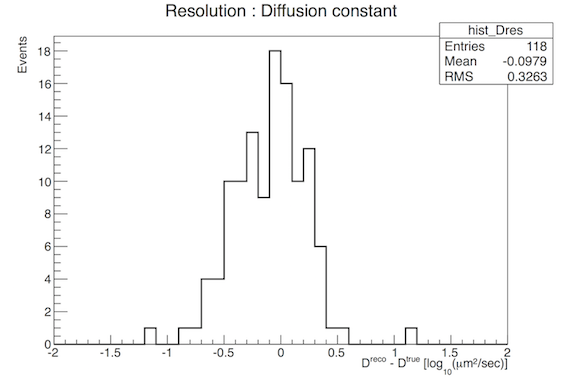}
        \hspace{0.4cm}
        \includegraphics[width=7.4cm]{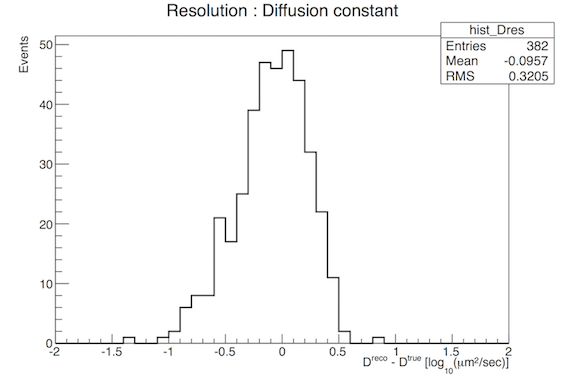}

  \centering
        \includegraphics[width=7.4cm]{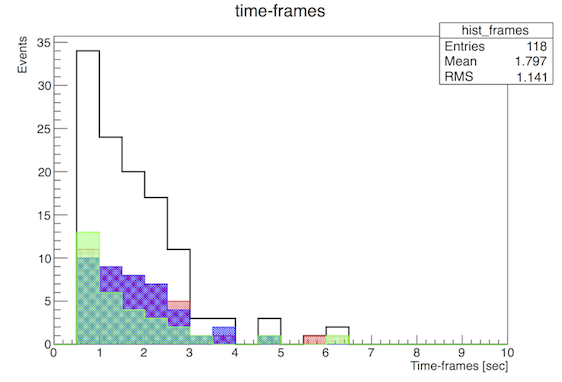}
        \hspace{0.4cm}
        \includegraphics[width=7.4cm]{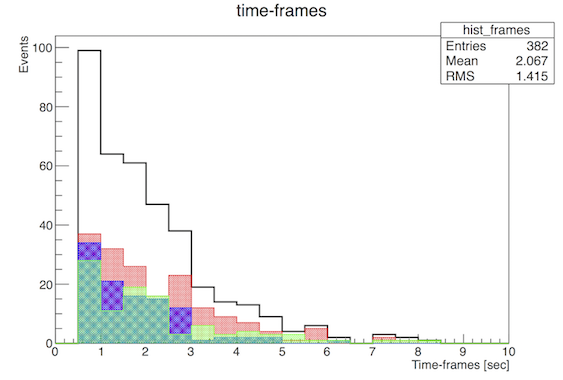}
 
  \centering
        \includegraphics[width=7.4cm]{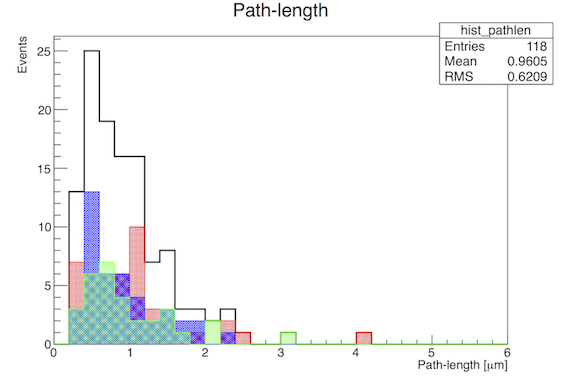}
        \hspace{0.4cm}
        \includegraphics[width=7.4cm]{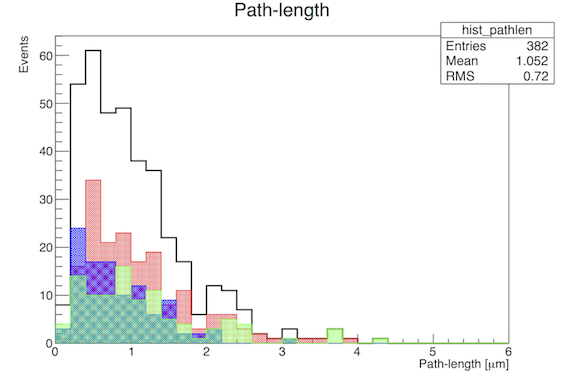}

  \centerline{\bf Apical cell-region \hspace{4.0cm} Basal cell-region}

  \caption{{\bf Event-properties 1}. Distributions of reconstructed diffusion constants, diffusional resolution (or diffusion reconstruction error), event time-frames and path-length are shown from top to bottom.}
  \label{fig;event_property_1}
\end{figure}

\begin{figure}[!h]
  \centering
        \includegraphics[width=7.4cm]{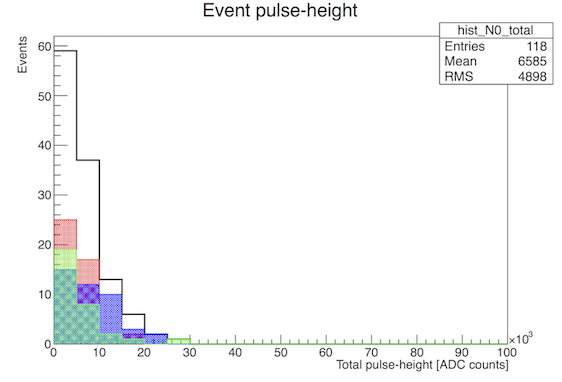}
        \hspace{0.4cm}
        \includegraphics[width=7.4cm]{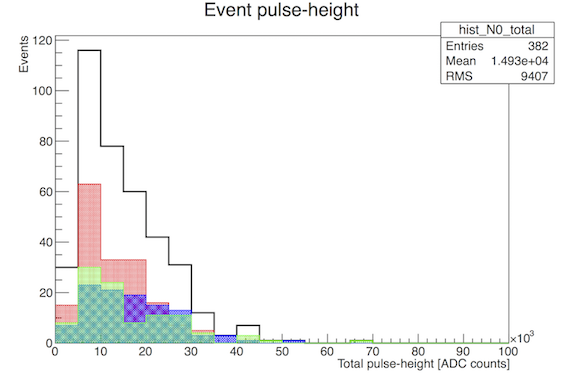}
 
  \centering
        \includegraphics[width=7.4cm]{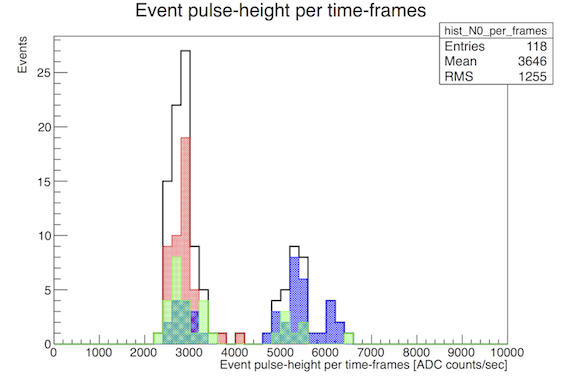}
        \hspace{0.4cm}
        \includegraphics[width=7.4cm]{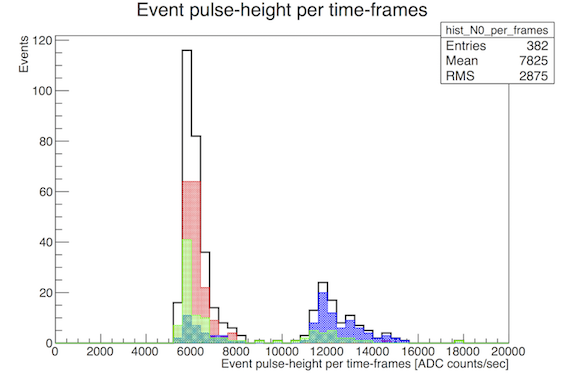}

  \centering
        \includegraphics[width=7.4cm]{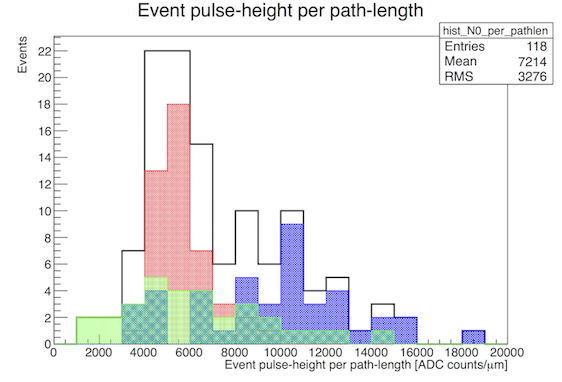}
        \hspace{0.4cm}
        \includegraphics[width=7.4cm]{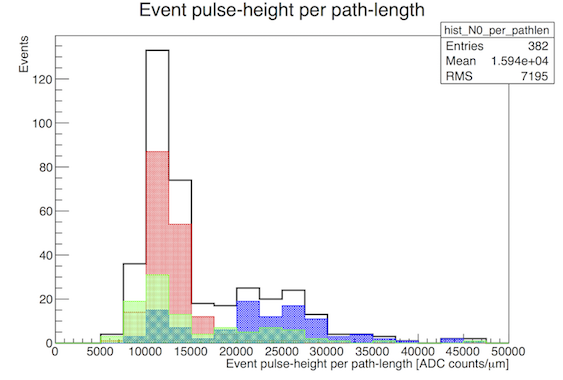}

  \centering
        \includegraphics[width=7.4cm]{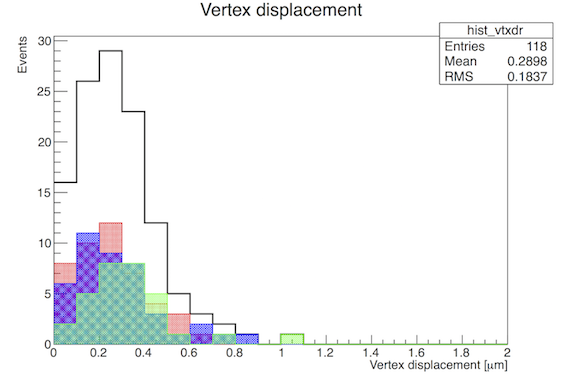}
        \hspace{0.4cm}
        \includegraphics[width=7.4cm]{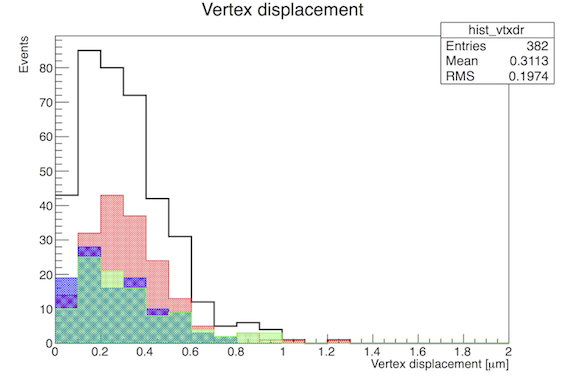}

  \centerline{\bf Apical cell-region \hspace{4.0cm} Basal cell-region}
  \caption{{\bf Event-properties 2}.  Distributions of event pulse-height, event pulse-height per time-frames, event pulse-height per path-length and event-vertex displacement are shown from top to bottom.}
  \label{fig;event_property_2}
\end{figure}

\begin{figure}[!h]
  \centering
        \includegraphics[width=7.4cm]{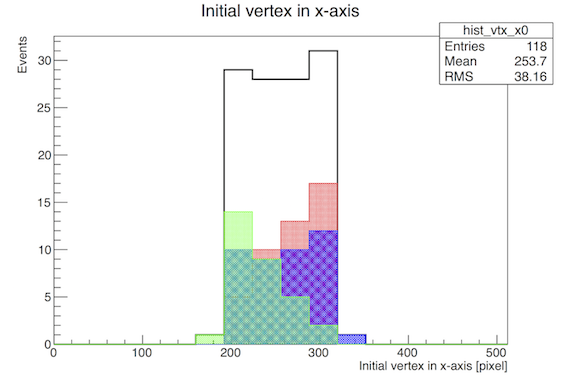}
        \hspace{0.4cm}
        \includegraphics[width=7.4cm]{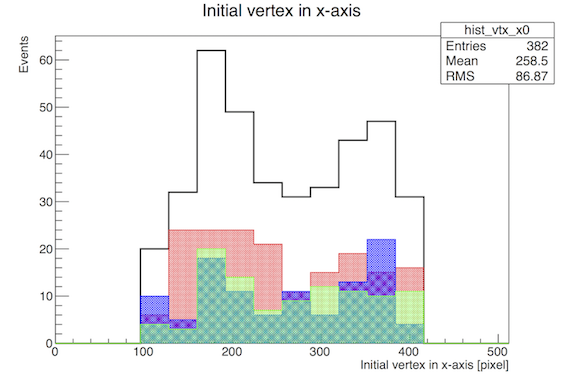}
 
  \centering
        \includegraphics[width=7.4cm]{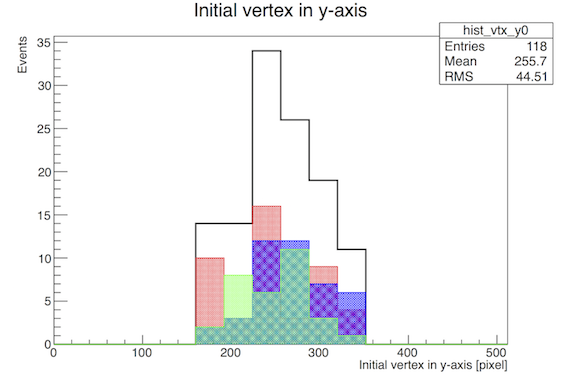}
        \hspace{0.4cm}
        \includegraphics[width=7.4cm]{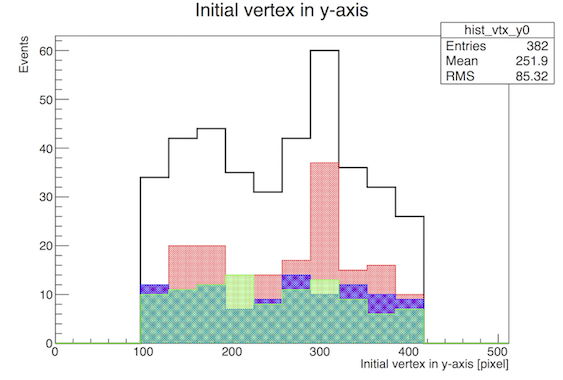}
 
  \centering
        \includegraphics[width=7.4cm]{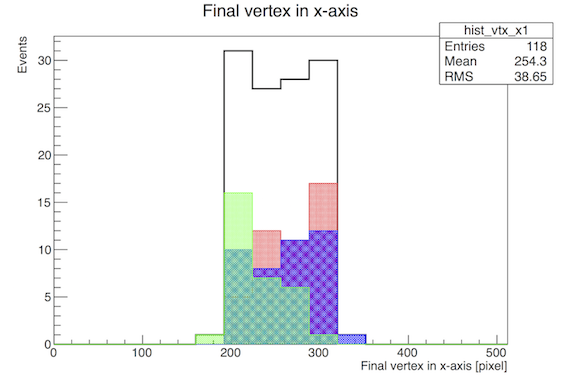}
        \hspace{0.4cm}
        \includegraphics[width=7.4cm]{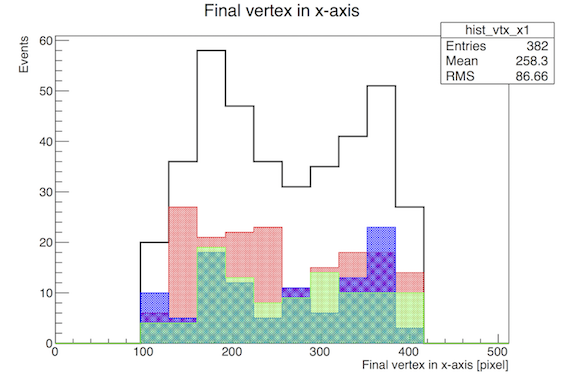}
 
  \centering
        \includegraphics[width=7.4cm]{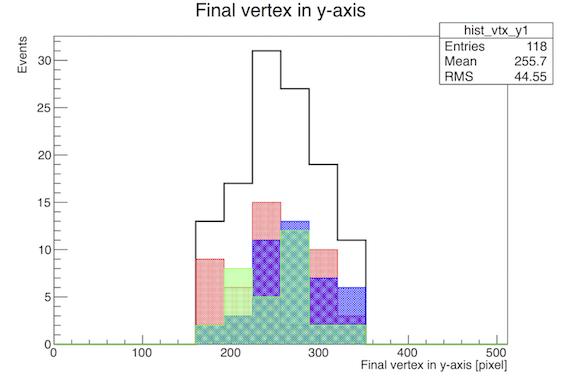}
        \hspace{0.4cm}
        \includegraphics[width=7.4cm]{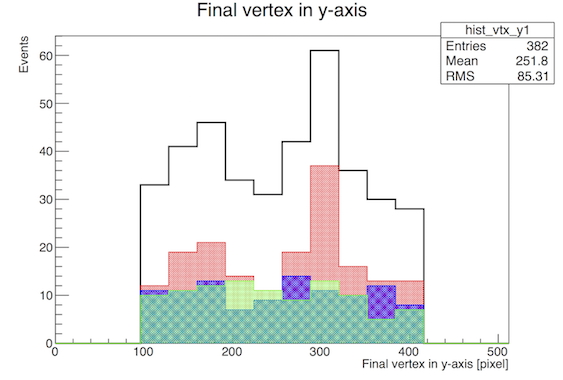}
 
  \centerline{\bf Apical cell-region \hspace{4.0cm} Basal cell-region}
  \caption{{\bf Event-properties 3}.  Distributions of initial and final event-vertex are shown from top to bottom.}
  \label{fig;event_property_3}
\end{figure}

\paragraph{(4) Event-property :}
Reconstructed track properties are presented as physical parameters: event-length, event-vertex, diffusion-coefficient, and event pulse-height. Distributions of each physical parameters are shown in Figures \ref{fig;event_property_1}, \ref{fig;event_property_2} and \ref{fig;event_property_3}. Red, blue and green filled histograms represent the ground-truth distribution of true-monomer, true-dimer and false event-reconstruction. Black solid lines represents the the observed distributions (distribution sum of each event-types). 

\begin{enumerate}
\item[(a)] Diffusion : The diffusion coefficient in lateral axes is reconstructed using short-range analysis method \cite{miyanaga2009}. The diffusion coefficient is determined by linear-fitting to mean squared displacement (MSD) for various time intervals ($\delta t$). The fitting function is written in the form of
\begin{eqnarray}
\left< \Delta r^2(\delta t) \right> & = & 4 D \delta t + a
\end{eqnarray}
where $D$ and $a$ represent diffusion coefficient and intersection in the MSD axis. 
\forceindent Distributions of reconstructed diffusion constant and its resolution ($D^{\ reco} - D^{\ true}$) are shown in the top panels of Figure \ref{fig;event_property_1}. In those Figures, we confirmed that peaks of the resolution distributions are located near zeros, and are roughly formed as Gaussian distributions in logarithmic scale. The RMS value represents the diffusional resolution to $0.32\ {\rm \log_{10}(\mu m^2/sec)}$. In addition, more cell-model samples are required to check the shape of the resolution distributions.
\item[(b)] Event-length : Time-frames and path-length are defined as the total time and distance that a molecule travels. Distributions of the time-frames and path-length are shown in the bottom panels of Figure \ref{fig;event_property_1}. 
\item[(c)] Event pulse-height : Distributions of event pulse-heights are shown in the top panels of Figure \ref{fig;event_property_2}. The event pulse-height is defined as the total spot pulse-height in a given trajectory, and written in the form of
\begin{eqnarray}
N_{evt} & = & \sum^{spots}_{j = 0} N_0 (\vec{r}_{j})
\end{eqnarray}
where $N_0$ is the spot pulse-height of $j$-th spot in a given trajectory. In addition, middle two panels of the Figures show distributions of event pulse-height per path-length and time-frame. Separation of true-monomer and true-dimer event ditributions can be clearly seen in those Figures.
\item[(d)] Event-vertex : Initial and final event-vertex distributions are shown in Figure \ref{fig;event_property_3}. Displacement distribution of the initial vertex to the final vertex is also shown in the bottom panels of Figure \ref{fig;event_property_3}. Initial and final x- and y-vertex distributions are expected to be flat in the apical and basal cell-regions. However, the vertex distributions are not flat in those Figures. More cell-model samples are required to verify the uniformity of each vertex distributions.
\end{enumerate}

\newpage

\newpage

\newpage

\newpage

\subsubsection{Simple ligand-receptor binding}
\paragraph{}
We construct a relatively simple cell-model of monomeric receptor binding to ligand on cellular membrane: ${\rm\bf L + r \leftrightarrow R}$. The monomers slowly diffuse at $0.015\ {\rm \mu m^2/sec}$. Monomer density and the dissociation constant are configured to $4.977\ {\rm \mu m^{-2}}$ and $3.63\ {\rm nM}$ ($k_1 = 0.00193\ {\rm nM^{-1}\ sec^{-1}},\ d_1 = 0.00700\ {\rm sec^{-1}}$). Cell-compartments are shown in Table \ref{tab;cell_compartments}. Experimental configuration is shown in Table \ref{tab;config_monomer}. 

\begin{table}[!h]
\centering
\begin{tabular}{|c|c|c|c|c|c|c|}
\hline
& Image acquisition & Time-lapse & Exposure-time & Frames & HRG-Ligand & Cell-samples \\ \hline
(A) & \multirow{2}{*}{$0 \sim 80\ {\rm min}$} & \multirow{2}{*}{$300\ {\rm sec}$} & \multirow{2}{*}{$0.150\ {\rm sec}$} & \multirow{2}{*}{$16$} & $1.0\ {\rm nM}$ & $10$ \\ \cline{1-1} \cline{6-7}
(B) & & & & & $0.300 \sim 400\ {\rm nM}$ & $120$\\ \hline
\end{tabular}
\caption{{\bf Experimental configuration}.}
\label{tab;config_monomer}
\end{table}

\leftline{\bf Experimental configuration A ($1.0\ {\rm nM}$ ligand input)}
\paragraph{}
Single-molecule imaging of the apical region of the ligand-receptor binding is simulated for the optical specification and its operating condition of the fluorescence microscopy simulation module shown in Table \ref{tab;specification3}. In particular, photobleaching is not included in the simulated single-molecule images. The results are as follows.

\begin{figure}[!h]
  \centering
        \includegraphics[width=6.6cm]{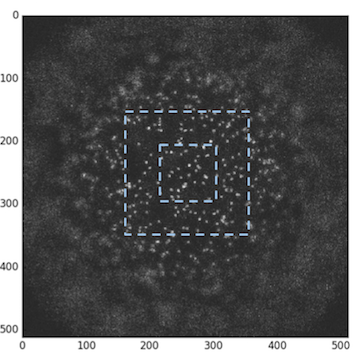}
        \hspace{1.2cm}
        \includegraphics[width=6.6cm]{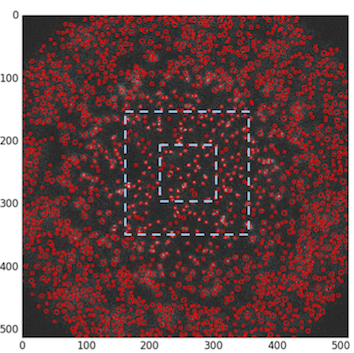}

  \caption{{\bf Single-molecule images of spot-detection on the apical cell-regions}. For  $1.0\ {\rm nM}$ ligand concentration, the single-molecule images before and after the spot-detection are shown in the left and the the right panels. The cut-1 and cut-2 are represented in inner and outer dashed boxes. Red circles represent the spots detected by the LoG method. Each image size is $512 \times 512\ {\rm pixels}$. Actual minimum and maximum values of the image intensity are $1,900$ and $2,500$ ADC counts. The image intensity is rescaled in the range of $0$ to $255$. The microscopy is configured to focus on the image center.}
  \label{fig;spot_images_2}

\vspace{0.8cm}

  \centering
        \includegraphics[width=7.4cm]{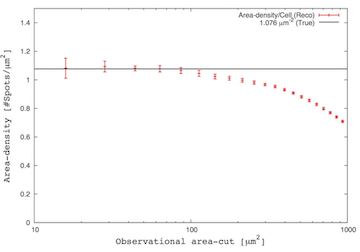}
        \hspace{0.4cm}
        \includegraphics[width=7.4cm]{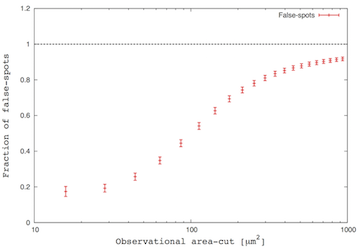}

  \centering
        \includegraphics[width=7.4cm]{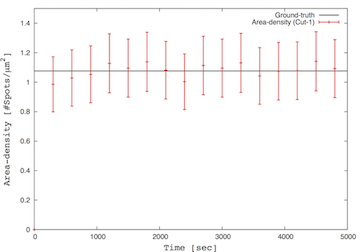}
        \hspace{0.4cm}
        \includegraphics[width=7.4cm]{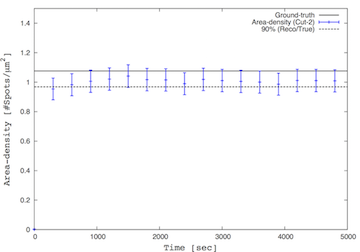}

  \caption{{\bf Reconstructed area-density, fractional occupancy of false-spots, and time-course data}.}
  \label{fig;reco_density_2}
\end{figure}

\paragraph{(1) Spot-detection :}
In this experimental configuration, the parameter values for the spot-detection algorithm are configured to $\sigma_{min} = 2.0\ {\rm pixels}$, $\sigma_{max} = 4.0\ {\rm pixels}$, $20$ intermediate values in the deviation range, threshold $= 15$, and overlap $= 0.5$. Figures \ref{fig;spot_images_2} show example images of spot-detection on the apical cell-regions for $100\ {\rm nM}$ ligand concentration. We assume two observational area-cuts to $80 \times 80\ {\rm pixels}$ (cut-1) and $200 \times 200\ {\rm pixels}$ (cut-2) with respect to the image center. 

\paragraph{}
Analysis results of various observational area-cuts are shown in Table \ref{tab;spots_monomer} and Figures \ref{fig;reco_density_2}; (a) the Table shows the number and fraction of simulated spots. The simulated spots are true-molecular spots and false-spots that arise from molecules and background noise. In particular, the false-spots are noise-like spots that can mimic the molecular spots. Relatively large fraction of the false-spots are captured with the cut-2. (b) Top left and right Figure panels show efficiency of area-density reconstruction and fractional occupancy of false-spots for various observational cuts. The area-cut is fixed at image center, and area-size varies from $10\ {\rm \mu m^2}$ to $1000\ {\rm \mu m^2}$. Ground-true spot area-density is set to $1.076\ {\rm spots/\mu m^2}$. The reconstruction efficiency is constant to $100\%$ below $100\ {\rm \mu m^2}$ area-cut. The efficiency is underestimated above the area-cut, due to including the defocused regions. (c) The left and the right bottom panels show time-course data for the cut-1 and cut-2. The reconstructed area-density varies in time, and saturated at $\sim 1.00\ {\rm spots/\mu m^2}$ for each area-cuts. 

\begin{figure}[!h]
  \centering
        \includegraphics[width=7.4cm]{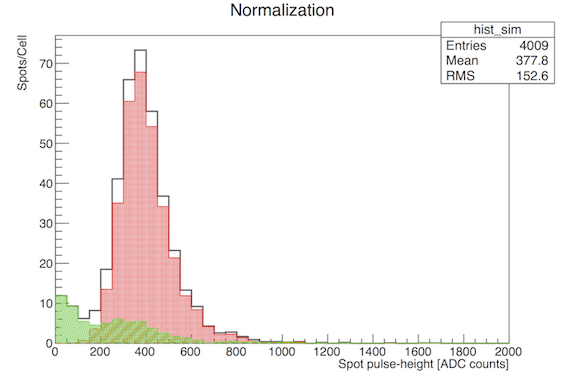}
        \hspace{0.4cm}
        \includegraphics[width=7.4cm]{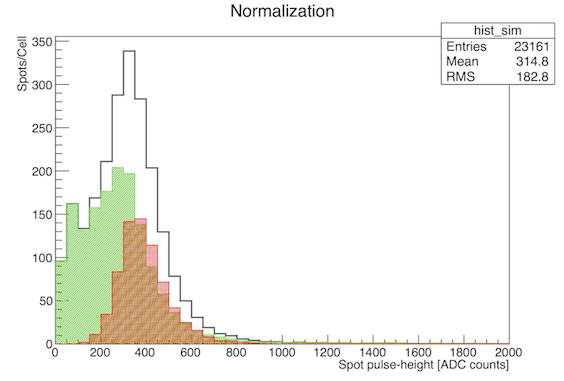}

  \centering
        \includegraphics[width=7.4cm]{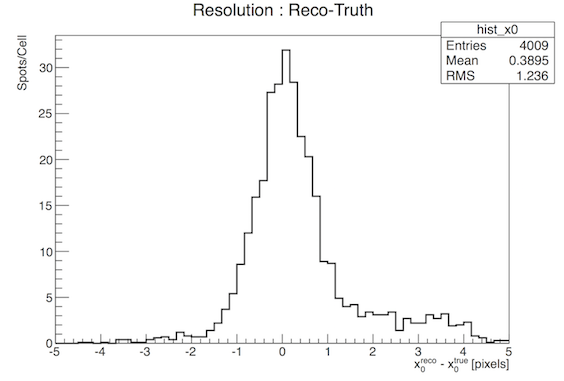}
        \hspace{0.4cm}
        \includegraphics[width=7.4cm]{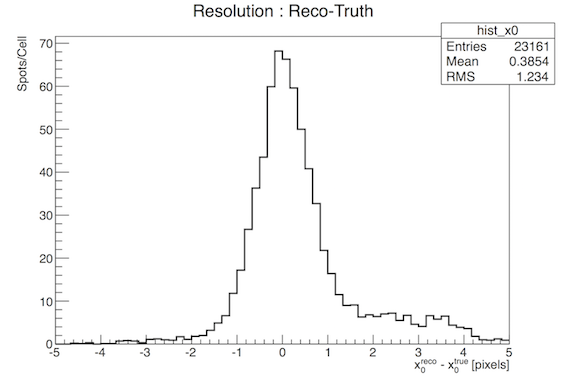}

  \centering
        \includegraphics[width=7.4cm]{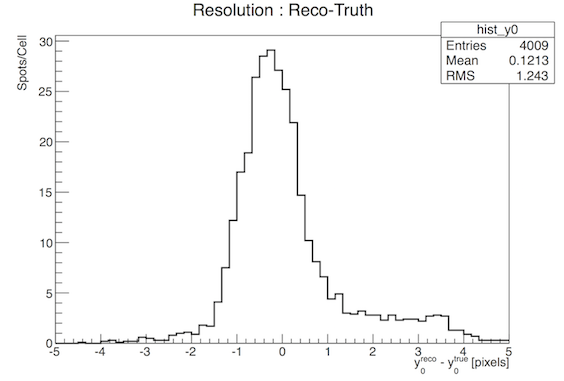}
        \hspace{0.4cm}
        \includegraphics[width=7.4cm]{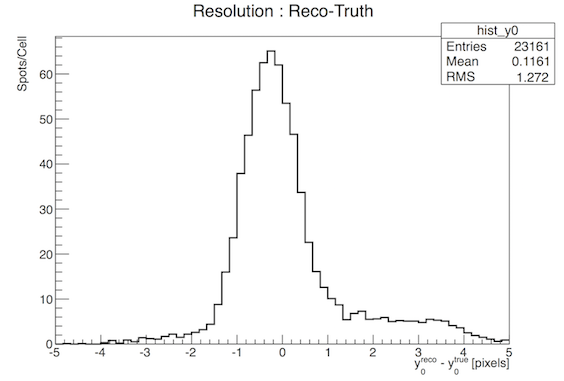}
 
  \centering
        \includegraphics[width=7.4cm]{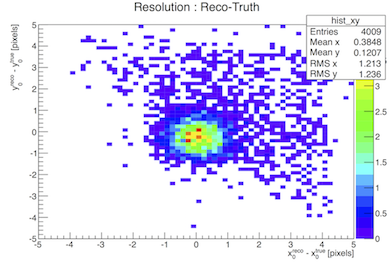}
        \hspace{0.4cm}
        \includegraphics[width=7.4cm]{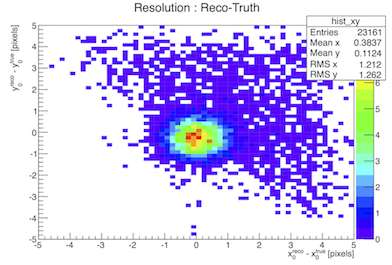}
 
  \centerline{\bf Cut-1 \hspace{7.0cm} Cut-2}
  \caption{{\bf Spot-properties 1}. The top panels show distributions of spot pulse-height. Distributions and correlation of positional resolution (or localization error) in x-y axes are shown in the bottom three panels.}
  \label{fig;spot_property_monomer_1}
\end{figure}

\begin{figure}[!h]
  \centering
        \includegraphics[width=7.4cm]{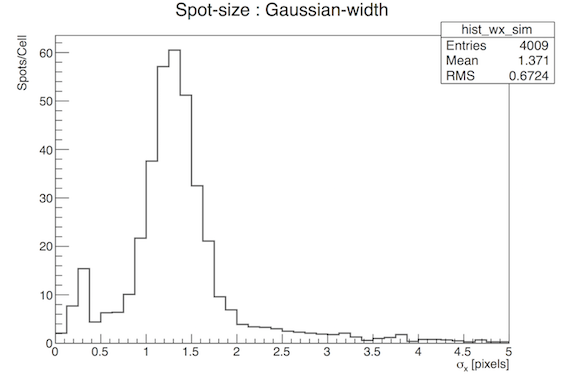}
        \hspace{0.4cm}
        \includegraphics[width=7.4cm]{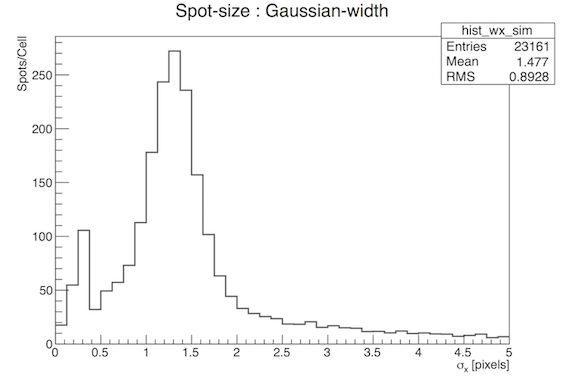}

  \centering
        \includegraphics[width=7.4cm]{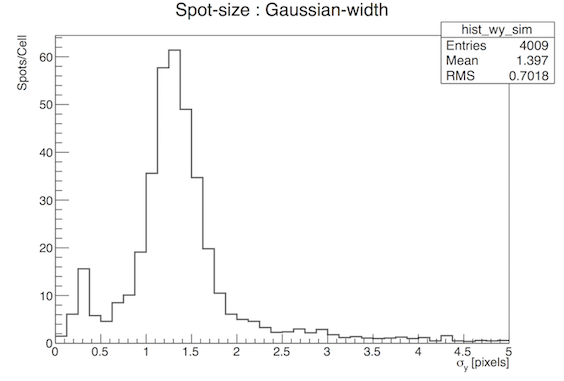}
        \hspace{0.4cm}
        \includegraphics[width=7.4cm]{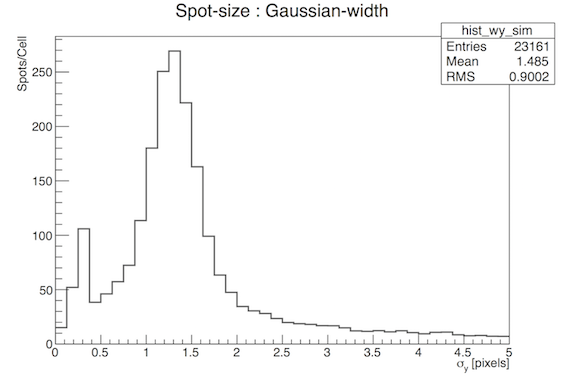}
 
  \centering
        \includegraphics[width=7.4cm]{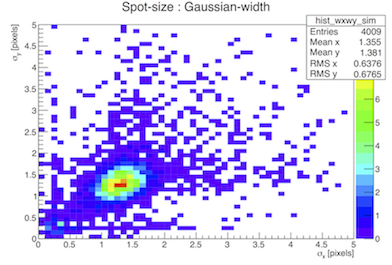}
        \hspace{0.4cm}
        \includegraphics[width=7.4cm]{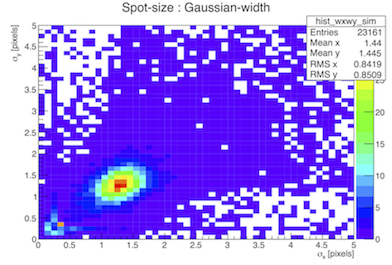} 

  \centering
        \includegraphics[width=7.4cm]{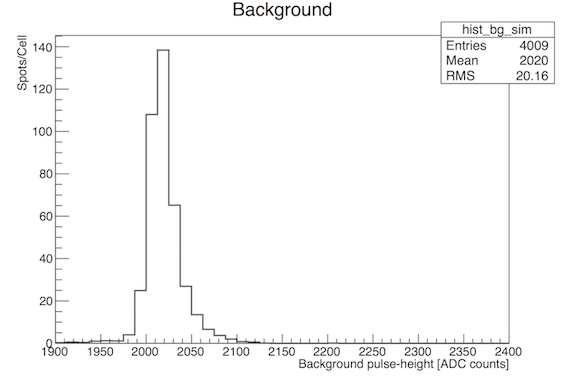}
        \hspace{0.4cm}
        \includegraphics[width=7.4cm]{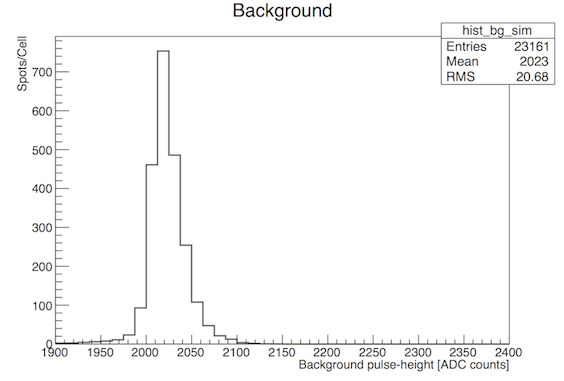} 

  \centerline{\bf Cut-1 \hspace{7.0cm} Cut-2}
  \caption{{\bf Spot-properties 2}. Distributions and correlations of reconstructed pulse-width (or Gaussian-witdth) in x-axis and y-axis are shown in the top three panels. Distribution of background pulse-height is shown in the bottom panels.}
  \label{fig;spot_property_monomer_2}
\end{figure}

\begin{figure}[!h]
  \centering
        \includegraphics[width=7.4cm]{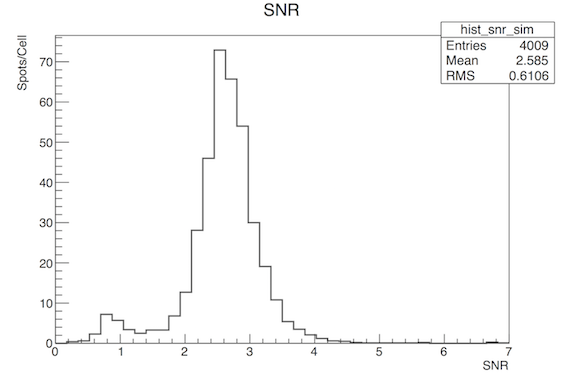}
        \hspace{0.4cm}
        \includegraphics[width=7.4cm]{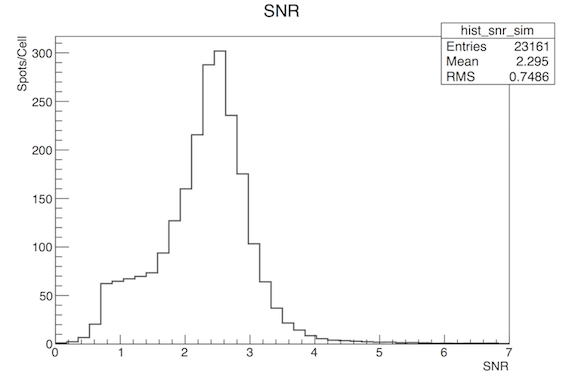} 

  \centering
        \includegraphics[width=7.4cm]{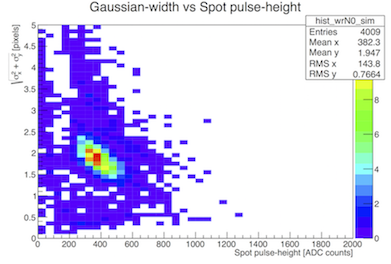}
        \hspace{0.4cm}
        \includegraphics[width=7.4cm]{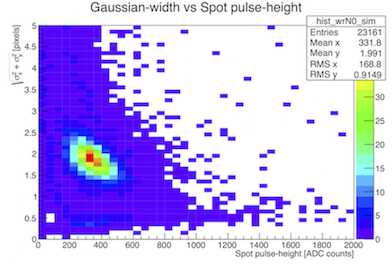} 

  \centering
        \includegraphics[width=7.4cm]{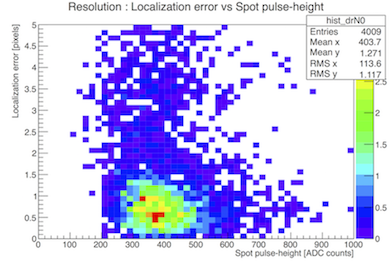}
        \hspace{0.4cm}
        \includegraphics[width=7.4cm]{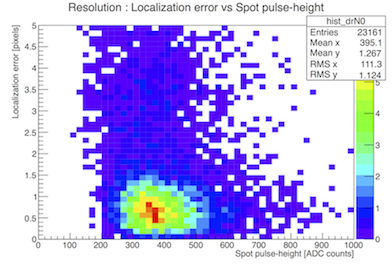} 

  \centerline{\bf Cut-1 \hspace{7.0cm} Cut-2}
  \caption{{\bf Spot-properties 3}. The top panels show SNR distribution. The bottom two panels show correlation distributions of spot pulse-height to spot-size and localization error.}
  \label{fig;spot_property_monomer_3}
\end{figure}

\paragraph{(2) Spot-property :}
Reconstructed spot-properties are presented as the Gaussian function of six parameters; the spot pulse-height (or normalization factor), central position, spot-size (or Gaussian-width) and background pulse-height. Distributions and correlations for each parameter is shown in Figures \ref{fig;spot_property_monomer_1} to \ref{fig;spot_property_monomer_3}. \\

\begin{table}[!h]
\centering
\begin{tabular}{|c|c|c|c|}
\hline
Spots/cell & Simulated & True-monomer & False-spot \\ \hline
Cut-1 & $400.9$  & $323.6$ ($80.7\%$) & $77.3$ ($19.3\%$) \\ \hline
Cut-2 & $2316.1$ & $709.6$ ($30.6\%$) & $1606.5$ ($69.4\%$) \\ \hline
\end{tabular}
\caption{{\bf The number and fraction of detected spots}.}
\label{tab;spots_monomer}
\end{table}

\newpage

\newpage

\newpage

\newpage

%\newpage

\leftline{\bf Experimental configuration B (various ligand inputs)}
\paragraph{}
In this experimental configuration, $10$ cell-model samples are prepared for $12$ ligand inputs ($0.3\ {\rm nM} \sim 400\ {\rm nM}$). Cell-compartments for each cell-model are listed in Table \ref{tab;cell_compartments}. We also assume two observational cut-1 and cut-2. Parameter values for the spot-detection algorithm are configured to $\sigma_{min} = 2.0\ {\rm pixels}$, $\sigma_{max} = 4.0\ {\rm pixels}$, $20$ intermediate values in the deviation range, threshold $= 15$, and overlap $= 0.5$. 

\begin{figure}[!h]
  \centering
        \includegraphics[width=7.4cm]{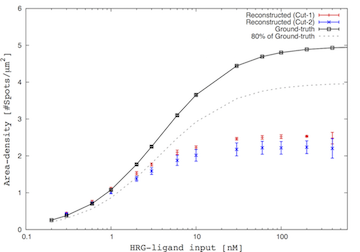}
        \hspace{0.4cm}
        \includegraphics[width=7.4cm]{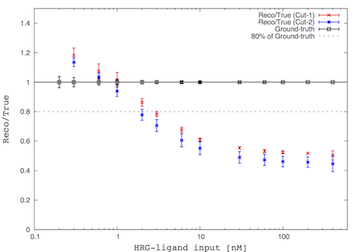}

  \vspace{0.1cm}
  \centering
        \includegraphics[width=7.4cm]{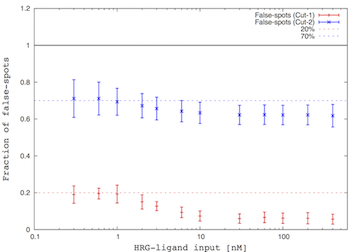}
        \hspace{0.4cm}
        \includegraphics[width=7.4cm]{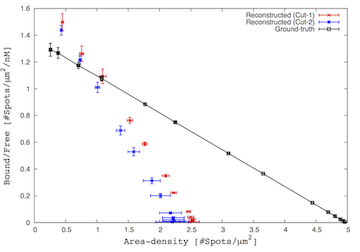}

  \caption{{\bf Equilibrium binding curve and Scatchard plot}. }
  \label{fig;monomer_results}

  \vspace{0.4cm}

  \centerline{\bf Cut-1 \hspace{7.0cm} Cut-2}
  \centering
        \includegraphics[width=7.4cm]{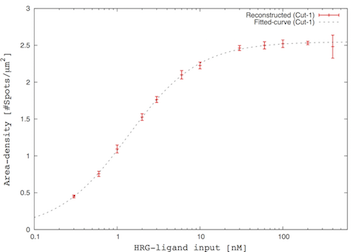}
        \hspace{0.4cm}
        \includegraphics[width=7.4cm]{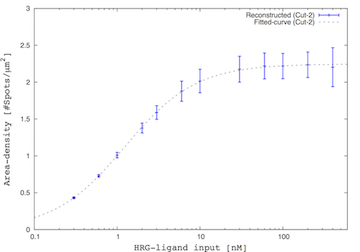}

  \vspace{0.1cm}
  \centering
        \includegraphics[width=7.4cm]{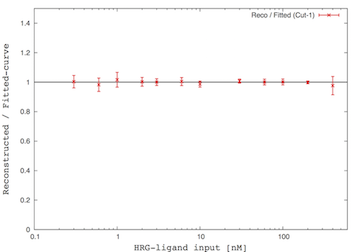}
        \hspace{0.4cm}
        \includegraphics[width=7.4cm]{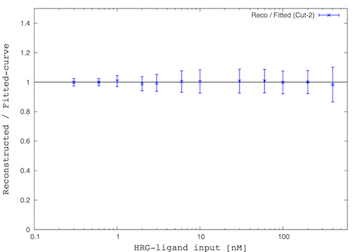}

  \caption{{\bf Fitting results}.}
  \label{fig;fitting_results}
\end{figure}

\paragraph{}
Analysis results are shown in Figures \ref{fig;monomer_results}. Red and blue crosses represent the reconstructed values obtained by the cut-1 and cut-2. Ground-truth and its $50\%$ recovery are presented with black solid and dashed lines. (1) The left and the right top panels show the equilibrium binding curve and its ratio of the reconstructed density to ground-truth. These plots clearly show that shape of the true binding curve is partially recovered in the reconstructed one. While the reconstruction efficiency is relatively high at low concentration, the efficiency is limited by up to $50\%$ at relatively higher ligand concentration input. (2) The bottom left panel shows the fraction of false-spots contaminated in all detected spots. Approximately $5\%$ of the detected spots are false-spots with the cut-1. Roughly, $65\%$ are contaminated by the cut-2. (3) The bottom right panel shows a Scatchard plot. The plot clearly shows that shape of true-Scatchard plot is not well-recovered in the reconstructed ones. The true-Scatchard plot is well-characterized as a straight line, representing no cooperativity. The shape of the reconstructed Scatchard plot are also represented with the straight line: No significant change was found in the simple binding system.

\paragraph{Hill coefficient :}
In a standard approach of biochemistry and biology, the Hill equation (1) can be fitted to the equilibrium binding curves to evaluate signal responses in simple ligand-receptor binding. The minimization function is given in the form of
\begin{eqnarray}
\chi^2 & = & \sum^{N bins}_{i = 0} \frac{\left(E_i - O_i\right)^2}{\sigma_i^2}
\end{eqnarray}
where $E_i$ and $O_i$ are the expected and the observed data at the i-th bin. $\sigma_i$ is statistical error at the i-th bin. The fitting results are shown in Table \ref{tab;monomer_fitting_results}. The top panels show the equilibrium binding curves for each cut. Bottom ones show the ratio of the reconstructed binding curves to the fitted ones. The Hill coefficients are nearly an unity: No significant change was found between the ground-true curve and reconstructed one, exhibiting no coopertivity in the simple binding system. 

%The Hill equation can be written in the form of
%\begin{eqnarray}
%B(L) = \frac{B_0 L^n}{K_A^n + L^n}
%\end{eqnarray}
%where $L$, $B_0$, $K_A$ and $n$ represents ligand concentration, maximum area-density of ligand binding, ligands occupying half of the binding sites and Hill coefficient. 

\begin{table}[!h]
\centering
\begin{tabular}{|c|c|c|c|c|}
\hline
& $B_0\ [{\rm spots/\mu m^2}]$ & $K_A\ [{\rm nM}]$ & $n$ & $\hat{\chi}_0^2$ \\ \hline
Ground-truth & \hspace{0.3cm} $4.977 \pm 0.000$ \hspace{0.3cm} & \hspace{0.2cm} $3.627 \pm 0.002$ \hspace{0.2cm} & \hspace{0.2cm} ${\bf\color{black} 1.001} \pm 0.001$ \hspace{0.2cm} & \hspace{0.1cm}  $0.066$ \hspace{0.1cm} \\ \hline
Reconstructed (Cut-1) & $2.549 \pm 0.007$ & $1.360 \pm 0.018$ & ${\bf\color{red} 1.023} \pm 0.011$ & $0.353$ \\ \hline
Reconstructed (Cut-2) & $2.247 \pm 0.010$ & $1.237 \pm 0.014$ & ${\bf\color{blue} 1.014} \pm 0.006$ & $0.125$\\ \hline
\end{tabular}
\caption{{\bf Results of fitting to the Hill equation}. The best fit values and uncertainties of each parameters are listed. $\hat{\chi}^2_0$ is the reduced minimum. If $n<1$, then the receptor system increases binding affinity of states and exhibits negative cooperativity. If $n>1$, then cooperativity is positive, decreasing the binding affinity of states. If $n=1$, then there is no cooperativity.}
\label{tab;monomer_fitting_results}
\end{table}

\begin{figure}[!h]
  \centerline{\bf Cut-1 \hspace{7.0cm} Cut-2}
  \centering
        \includegraphics[width=7.4cm]{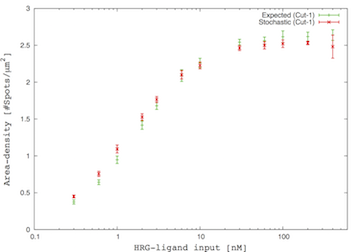}
        \hspace{0.4cm}
        \includegraphics[width=7.4cm]{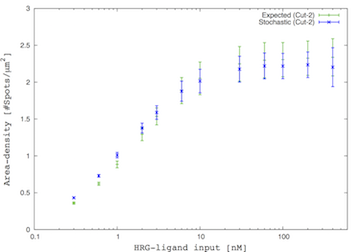}

  \centering
        \includegraphics[width=7.4cm]{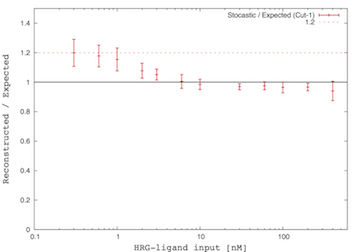}
        \hspace{0.4cm}
        \includegraphics[width=7.4cm]{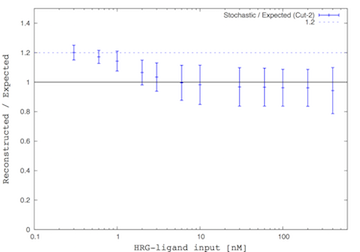}

  \centering
        \includegraphics[width=7.4cm]{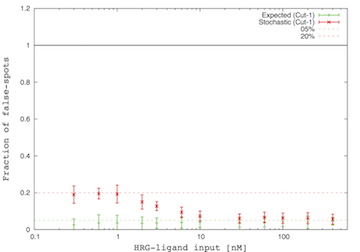}
        \hspace{0.4cm}
        \includegraphics[width=7.4cm]{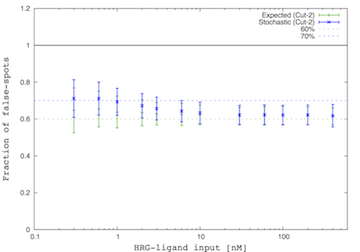}

  \centering
        \includegraphics[width=7.4cm]{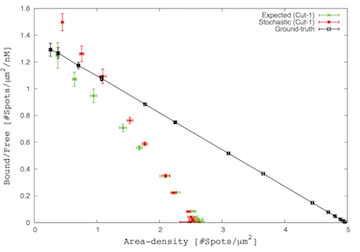}
        \hspace{0.4cm}
        \includegraphics[width=7.4cm]{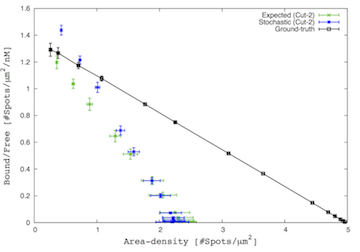}

  \caption{{\bf Stochasticity}. Comparison of the biological properties reconstructed from the expected and stochastic images.}
  \label{fig;stochasticity_1}
\end{figure}

\paragraph{Stochasticity :}
To see influence of stochastic photon-detection processes to the analytical procedure, we compared the biological properties reconstructed from the expected and stochastic images (see the SI section C.2.2 for more detailed explanation). Figure \ref{fig;stochasticity_1} shows the equilibrium binding curves and Scatchard plot for each area-cuts. Red and blue lines represent the reconstructed curves obtained by the cut-1 and cut-2. The reconstructed curves obtained from the stochastic images are presented with green lines. Ground-true Scatchard plots are presented with black solid lines. \\
\forceindent The top panels show the comparison of the equilibrium binding curves reconstructed from the expected and stochastic images. The restoration efficiency and defects of the analytical procedure are shown in the middle four panels. We computed the ratio of the reconstructed binding curves of the stochastic images to the expected ones. The 2nd top panels show the ratio curves, implying that the stochasticity can increase to approximately $20\%$ of the restoration efficiency of capturing more molecular-spots at the low concentration range ($< 6\ {\rm nM}$). The 2nd bottom panels show the fraction of false-spots contaminated in all detected spots. The left panel for the cut-1 represents that the stochasticity can increase to approximately $20\%$ of contamination of capturing the defected spots at the low concentration range. The right panel for the cut-2, however, shows no influence of the stochasticity to the false-spots contamination. Finally, the bottom panels show the comparison of the Scatchard plots reconstructed from the expected and stochastic images. Each reconstructed Scatchard plot is crossing over, do not align in parallel. Thus the stochasticity cannot conserve the shape and concavity of the binding curve and Scatchard plot. 

\newpage

\newpage

\newpage

\begin{figure}[!h]
  \centerline{\bf Cut-1 \hspace{7.0cm} Cut-2}
  \centering
        \includegraphics[width=7.4cm]{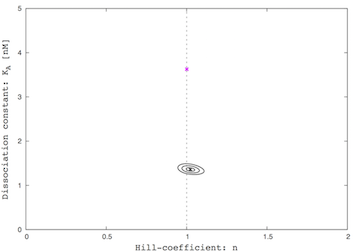}
        \hspace{0.4cm}
        \includegraphics[width=7.4cm]{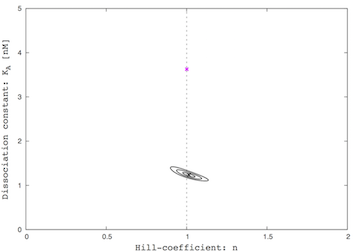}

  \vspace{0.4cm}
  \centering
        \includegraphics[width=7.4cm]{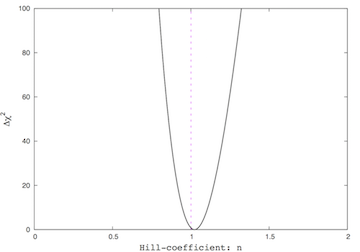}
        \hspace{0.4cm}
        \includegraphics[width=7.4cm]{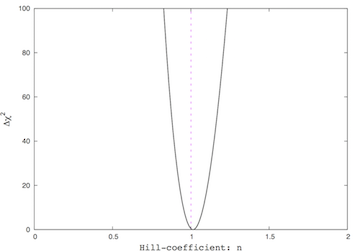}

  \vspace{0.4cm}
  \centering
        \includegraphics[width=7.4cm]{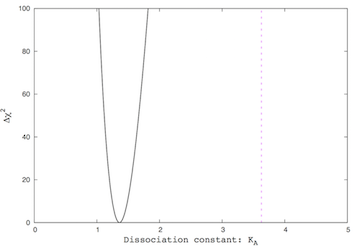}
        \hspace{0.4cm}
        \includegraphics[width=7.4cm]{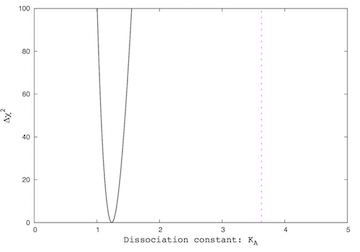}

  \caption{{\bf $\Delta \chi^2$ plots}. For each area-cuts, Figures show the $\Delta \chi^2$ plots of the Hill coefficient and the dissociation constant. $B_0$ is fixed to the best fit value.}
  \label{fig;monomer_dchi2_plots}
\end{figure}

\paragraph{Confidence interval :}
We estimate the uncertainties in the parameters to indicate numerically our confidence in our fitting results. $\Delta \chi^2$-plots for each area-cuts are shown in Figures \ref{fig;monomer_dchi2_plots} show. Black solid lines represent the $1\sigma$ ($68\%$), $2\sigma$ ($95\%$) and $3\sigma$  ($99\%$) confidence intervals. Black and pink points indicate the best fit and ground-truth. \\
\forceindent The top panels show the $\Delta \chi^2$ contour plots: the Hill coefficient ($n$) vs dissociation constant ($K_{A}$). The ground-truth (pink point) is located out of the $3\sigma$ confidence contour line, implying the restoration failure of the true parameter values. The middle four plots show the $\Delta \chi^2$ plots for the Hill coefficient ($n$) and dissociation constant ($K_{A}$).While the ground-true Hill coefficient is found within $3\sigma$ contour line, the true dissociation constant is clearly located out of the contour lines. Cooperativity is thus well-restored through the reconstruction procedure.

\newpage

\begin{wrapfigure}{r}{10.3cm}
\vspace{-10pt}
  \centering
    \includegraphics[width=10cm]{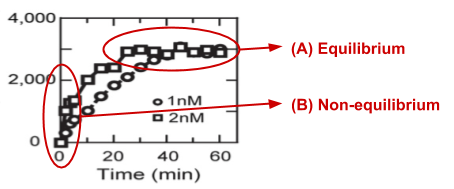}
    
  \caption{{\bf Experimental configurations} \cite{hiroshima2012}.}
  \label{fig;config_hiroshima2012}
\vspace{-10pt}
\end{wrapfigure}

\subsection{Dimer formation}
\paragraph{}
The reconstruction procedure can be also applied for Hiroshima's model of dimer formation. Model parameters and cell-compartments are shown in Table \ref{tab;hiroshima2012_dimer_model_parameters} and \ref{tab;cell_compartments}. In order to quantitatively evaluate the performance of the reconstruction procedure, two single-molecule experiments are configured at (A) the equilibrium region and (B) the nonequilibrium region \cite{hiroshima2012}. Details of each experimental configurations are shown in Figure \ref{fig;config_hiroshima2012} and Table \ref{tab;config_hiroshima2012}. \\

\begin{table}[!h]
\centering
\begin{tabular}{|c|c|c|c|c|c|c|}
\hline
 & Image acquisition & Time-lapse & Exposure-time & Frames & HRG-Ligand & Cell-samples \\ \hline
\multirow{3}{*}{(A)} & \multirow{3}{*}{$0 \sim 80\ {\rm min}$} & \multirow{3}{*}{$300\ {\rm sec}$} & \multirow{3}{*}{$0.150\ {\rm sec}$} & \multirow{3}{*}{$16$} & $1.0\ {\rm nM}$ & $10$ \\ \cline{6-7}
& & & & & $1.0\ {\rm pM} \sim 4.0\ {\rm nM}$ & $120$ \\ \cline{6-7}
& & & & & $0.2\ {\rm nM} \sim 4.0\ {\rm nM}$ & $60$ \\ \hline
%\cline{7-7}
%& & & & & & $2400$\\ \hline
%(A) & $0 \sim 80\ {\rm min}$ & $300.0\ {\rm sec}$ & $0.150\ {\rm sec}$ & $16$ & $1.0\ {\rm nM}$ & $10$  \\ \hline
(B) & $0 \sim 40\ {\rm sec}$ & $0.050\ {\rm sec}$ & $0.050\ {\rm sec}$ & $800$ & $2.0\ {\rm nM}$ & $10$  \\ \hline
\end{tabular}
\caption{{\bf Experimental configurations} \cite{hiroshima2012}.}
\label{tab;config_hiroshima2012}

\vspace{0.1cm}

\centering
\begin{tabular}{|p{4.0cm}|p{12cm}|}
\hline
\multicolumn{2}{|c|}{\bf Area-density of ErbB receptors} \\ \hline
Monomer & $0.564\ {\rm \mu m^{-2}}$ \\ \hline
Dimer & $2.207\ {\rm \mu m^{-2}}$ \\ \hline
\multicolumn{2}{|c|}{\bf Equilibrium constants and reaction rates} \\ \hline
$K_{1}=d_1/k_1$ & $3.63\ {\rm nM}$ \hspace{0.3cm} $(k_1 = 0.00193\ {\rm nM^{-1}\ sec^{-1}},\ d_1 = 0.00700\ {\rm sec^{-1}})$ \\ \hline
$K_{2}=d_2/k_2$ & $0.0155\ {\rm nM}$ \hspace{0.3cm} $(k_2 = 0.00255\ {\rm nM^{-1}\ sec^{-1}},\ d_2 = 3.95 \times 10^{-5}\ {\rm sec^{-1}})$ \\ \hline
$K_{3} = d_3/k_3$ & $0.553\ {\rm nM}$ \hspace{0.3cm} $(k_3 = 4.09\ {\rm nM^{-1}\ sec^{-1}},\ d_3 = 2.26\ {\rm sec^{-1}})$ \\ \hline
$K_{i} = d_i/k_i$ & $0.139 \hspace{0.3cm} (k_i = 4.51\ {\rm sec^{-1}},\ d_i = 0.629\ {\rm sec^{-1}})$ \\ \hline
$K_{4} = d_4/k_4$ & $26.316\ {\rm \mu m^{-2}}$ \hspace{0.3cm} (Assume $d_4 = 1.00\ {\rm sec^{-1}}$)\\ \hline
$K_{5} = d_5/k_5$ & $0.2247\ {\rm \mu m^{-2}}$ \hspace{0.3cm} (Assume $d_5 = 0.10\ {\rm sec^{-1}}$)\\ \hline
$K_{6} = d_6/k_6$ & $0.00238\ {\rm \mu m^{-2}}$ \hspace{0.3cm} (Assume $d_6 = 0.10\ {\rm sec^{-1}}$)\\ \hline
\end{tabular}
\caption{{\bf Model parameters for the dimer formation}. }
\label{tab;hiroshima2012_dimer_model_parameters}

\vspace{0.1cm}

\centering
\begin{tabular}{|c|c|c|c|c|}
\hline
& \hspace{0.1cm} VOXEL-RADIUS \hspace{0.1cm} & \hspace{0.1cm} LENGTH ($Y \times Z \times X$) \hspace{0.1cm} & \hspace{0.1cm} ORIGIN ($Y, Z, X$) \hspace{0.1cm} & \hspace{0.2cm} Surface-area \hspace{0.2cm} \\ \hline
(1) & $20\ {\rm nm}$ & $45 \times 35 \times 2$ ${\rm \mu m^3}$ & ($0, 0, -1$) & $2576.72$ ${\rm \mu m^2}$ \\ \hline
(2) & $20\ {\rm nm}$ & $40 \times 40 \times 2$ ${\rm \mu m^3}$ & ($0, 0, -1$) & $2615.06$ ${\rm \mu m^2}$ \\ \hline
(3) & $20\ {\rm nm}$ & $40 \times 30 \times 2$ ${\rm \mu m^3}$ & ($0, 0, -1$) & $1978.13$ ${\rm \mu m^2}$ \\ \hline
(4) & $20\ {\rm nm}$ & $40 \times 20 \times 2$ ${\rm \mu m^3}$ & ($0, 0, -1$) & $1346.36$ ${\rm \mu m^2}$ \\ \hline
(5) & $20\ {\rm nm}$ & $35 \times 40 \times 2$ ${\rm \mu m^3}$ & ($0, 0, -1$) & $2296.19$ ${\rm \mu m^2}$ \\ \hline
(6) & $20\ {\rm nm}$ & $35 \times 30 \times 2$ ${\rm \mu m^3}$ & ($0, 0, -1$) & $1736.84$ ${\rm \mu m^2}$ \\ \hline
(7) & $20\ {\rm nm}$ & $35 \times 20 \times 2$ ${\rm \mu m^3}$ & ($0, 0, -1$) & $1182.00$ ${\rm \mu m^2}$ \\ \hline
(8) & $20\ {\rm nm}$ & $25 \times 40 \times 2$ ${\rm \mu m^3}$ & ($0, 0, -1$) & $1661.28$ ${\rm \mu m^2}$ \\ \hline
(9) & $20\ {\rm nm}$ & $25 \times 30 \times 2$ ${\rm \mu m^3}$ & ($0, 0, -1$) & $1256.38$ ${\rm \mu m^2}$ \\ \hline
(10) & $20\ {\rm nm}$ & $25 \times 20 \times 2$ ${\rm \mu m^3}$ & ($0, 0, -1$) & $854.67$ ${\rm \mu m^2}$ \\ \hline
\end{tabular}
\caption{{\bf Cell-compartmental variables for 10 cell-samples}. GEOMETRY of each cell-model is configured to "Ellipsoid".}
\label{tab;cell_compartments}
\end{table}

\newpage

\subsubsection{Experimental configuration A (equilibrium region)}
\paragraph{}
Single-molecule imaging of apical regions of each model is simulated for the optical specification and operating condition of the fluorescence microscopy simulation module shown in Table \ref{tab;specification3}. In particular, photobleaching is not included in this experimental configuration. The results are shown as follows.

\paragraph{(1) Spot-detection :}
In this experimental configuration, the parameter values for the spot-detection algorithm are configured to $\sigma_{min} = 2.0\ {\rm pixels}$, $\sigma_{max} = 4.0\ {\rm pixels}$, $20$ intermediate values in the deviation range, threshold $= 15$, and overlap $= 0.5$. Figures \ref{fig;spot_images_3} show example images of spot-detection on the apical cell-regions. We assume two observational area-cuts (cut-1 and cut-2) to $80 \times 80\ {\rm pixels}$ ($28\ {\rm \mu m^2}$) and $200 \times 200\ {\rm pixels}$ ($176\ {\rm \mu m^2}$) represented with inner and outer dashed boxes. Red circles represent the spots detected by the LoG method. Each image size is $512 \times 512\ {\rm pixels}$. Actual minimum and maximum values of the image intensity are $1,900$ and $2,500$ ADC counts. The image intensity is rescaled in the range of $0$ to $255$. The microscopy is configured to focus on the image center. While molecular-spots are clear and properly picked up at the focused area, the spots are blured and not well detected near the curved-region of cellular membrane. 

\begin{figure}[!h]
  \centering
        \includegraphics[width=7.4cm]{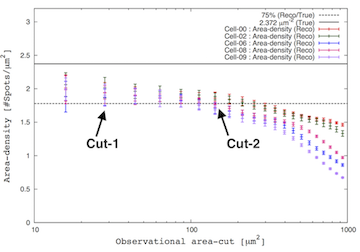}
        \hspace{0.4cm}
        \includegraphics[width=7.4cm]{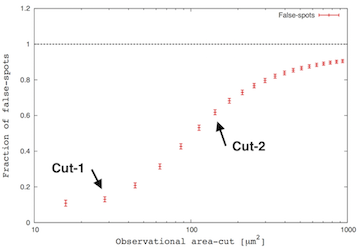}

  \centering
        \includegraphics[width=7.4cm]{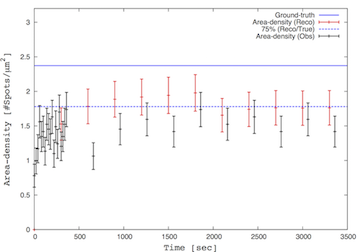}
        \hspace{0.4cm}
        \includegraphics[width=7.4cm]{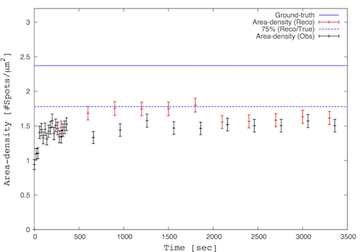}

  \caption{{\bf Reconstructed area-density, fractional occupancy of false-spots, and time-course data}.}
  \label{fig;reco_density_3}

  \centering
        \includegraphics[width=7.4cm]{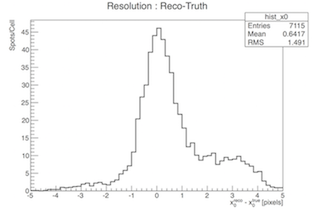}
        \hspace{0.4cm}
        \includegraphics[width=7.4cm]{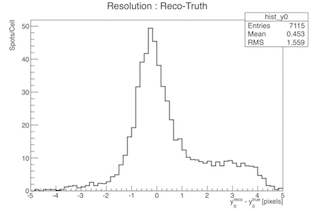}
 
  \centering
        \includegraphics[width=7.4cm]{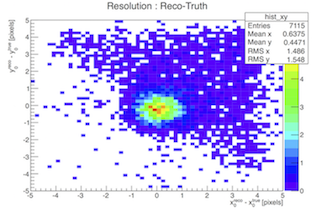}
        \hspace{0.4cm}
        \includegraphics[width=7.4cm]{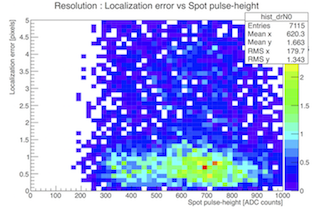}
 
  \caption{{\bf Spot-properties 1}. The $28\ {\rm \mu m^{-2}}$ area-cut is applied. Distributions and correlations of localization error ($\vec{r}^{\ reco}_0 - \vec{r}^{\ true}_0$) in  x-y axes are shown in the top and the bottom panels.}
  \label{fig;spot_property_dimer_A_1}
\end{figure}

\paragraph{}
Analysis results of various observational area-cuts are shown in Table \ref{tab;spots_dimer_A} and Figures \ref{fig;reco_density_3}; (a) the Table shows the number and fraction of observed and simulated spots. The simulated spots are true-molecular spots and false-spots that arise from molecules and background noise. In particular, the false-spots are noise-like spots that can mimic the molecular spots. A relatively large fraction of the false-spots are captured with the cut-2. (b) Top left and right Figure panels show efficiency of area-density reconstruction and fractional occupancy of false-spots for various observational cuts. The area-cut is fixed at the image center, and area-size varies from $10\ {\rm \mu m^2}$ to $1000\ {\rm \mu m^2}$. Ground-true spot area-density is set to $2.372\ {\rm spots/\mu m^2}$. While the reconstruction efficiency is constant at $75\%$ below $300\ {\rm \mu m^2}$ area-cut, the efficiency is underestimated above the area-cuts including the defocused regions. (c) The left and the right bottom panels show time-course data for the cut-1 and cut-2. The reconstructed area-density varies in time, and saturated at $\sim 1.7\ {\rm spots/\mu m^2}$ for each area-cuts. 

%\begin{table}[!h]
%\centering
%\begin{tabular}{|c|c|c|}
%\hline
% & \multicolumn{2}{c|}{Spots/frame/cell\ (Area-density)} \\ \hline 
%Area-cut & \hspace{0.3cm} $28\ {\rm \mu m^{2}}$ ($=\ 80 \times 80 \ {\rm pixels}$) \hspace{0.3cm} & \hspace{0.3cm} $397\ {\rm \mu m^{2}}$ ($=\ 300 \times 300 \ {\rm pixels}$) \hspace{0.3cm} \\ \hline
%Simulated & $44.05$ ($1.562\ {\rm \mu m^{-2}}$) & $91.74$ ($0.231\ {\rm \mu m^{-2}}$) \\ \hline
%\hspace{0.5cm} Hiroshima-2012 \hspace{0.5cm} & $16.64$ ($0.811\ {\rm \mu m^{-2}}$) & $170.55$ ($0.591\ {\rm \mu m^{-2}}$) \\ \hline
%\end{tabular}
%\caption{{\bf The number of detected spots per frame per cell}.}
%\label{tab;spots_dimer_A}
%\end{table}

\begin{table}[!h]
\centering
\begin{tabular}{|c|c|c|c|c|c|}
\hline
%\multicolumn{2}{|c|}{\bf Area density of ErbB proteins} \\ \hline
Spots/cell & Observed & Simulated & True-monomer & True-dimer & False-spot \\ \hline
Cut-1 & $425$ & $711.5$ & $237.2$ ($33.3\%$) & $382.4$ ($53.7\%$) & $91.9$ ($12.9\%$) \\ \hline
Cut-2 & $2685$ & $4068.6$ & $525.1$ ($12.9\%$) & $812.3$ ($20.0\%$) & $2731.2$ ($67.1\%$) \\ \hline
\end{tabular}
\caption{{\bf The number and fraction of detected spots}.}
\label{tab;spots_dimer_A}
\end{table}

\begin{figure}[!h]
  \centering
        \includegraphics[width=7.4cm]{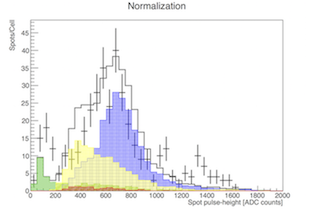}
        \hspace{0.4cm}
        \includegraphics[width=7.4cm]{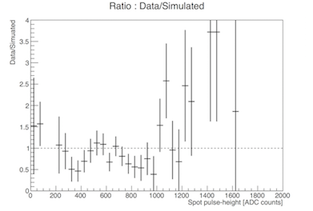}

  \centering
        \includegraphics[width=7.4cm]{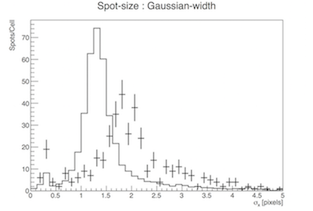}
        \hspace{0.4cm}
        \includegraphics[width=7.4cm]{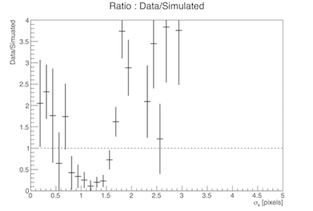}

  \centering
        \includegraphics[width=7.4cm]{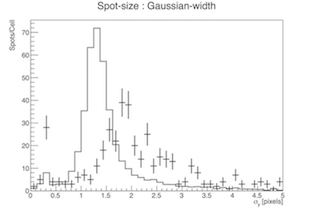}
        \hspace{0.4cm}
        \includegraphics[width=7.4cm]{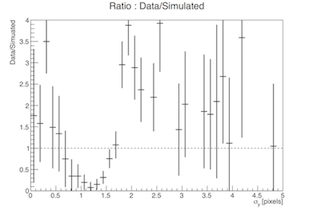}
 
  \centering
        \includegraphics[width=7.4cm]{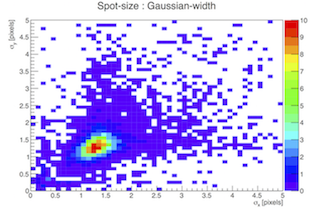}
        \hspace{0.4cm}
        \includegraphics[width=7.4cm]{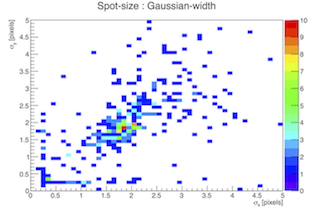}

  \caption{{\bf Comparison 1}. The cut-1 is applied. The simulated distributions of the spot pulse-height and spot-size are compared with the actual distributions obtained from Hiroshima's dataset. Black solid lines represent observed distributions of all reconstructed spots. Red, yellow, blue and green filled histograms represent ground-truth distribution of monomer (${\bf R}$), mono-dimer (${\bf rR\ {\rm or}\ r'R}$), and dimer (${\bf RR}$) spots. Crossed-lines represent Hiroshima's 2012 data.}
  \label{fig;spot_comparison_dimer_A_1}
\end{figure}

\begin{figure}[!h]
  \centering
        \includegraphics[width=7.4cm]{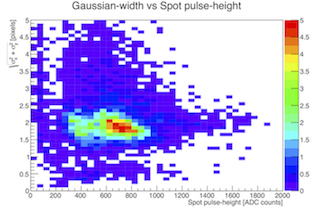}
        \hspace{0.4cm}
        \includegraphics[width=7.4cm]{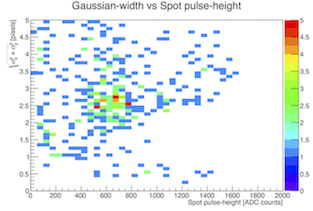}
        
  \centering
        \includegraphics[width=7.4cm]{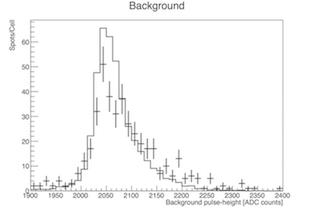}
        \hspace{0.4cm}
        \includegraphics[width=7.4cm]{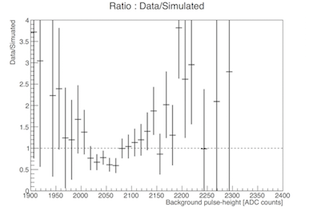}

  \centering
        \includegraphics[width=7.4cm]{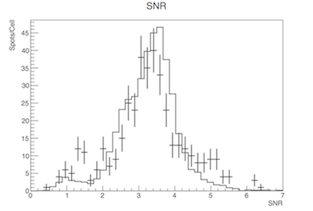}
        \hspace{0.4cm}
        \includegraphics[width=7.4cm]{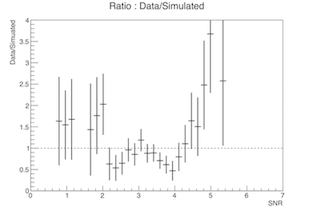}

  \caption{{\bf Comparison 2}. The cut-1 is applied. The top panels show that the simulated correlation (left) of the spot pulse-height to spot-size ($\sqrt{\sigma^2_{x} + \sigma^2_{y}}$) is compared with the actual correlation (right) obtained from Hiroshima's dataset. Distribution comparisons of background pulse-height and SNR are shown in bottom two panels.}
  \label{fig;spot_comparison_dimer_A_2}
\end{figure}

\begin{figure}[!h]
  \centering
        \includegraphics[width=7.4cm]{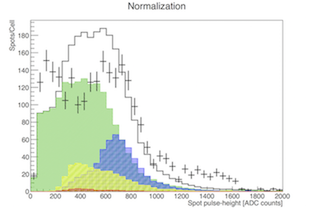}
        \hspace{0.4cm}
        \includegraphics[width=7.4cm]{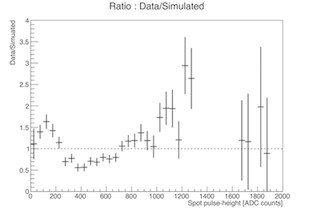}

  \centering
        \includegraphics[width=7.4cm]{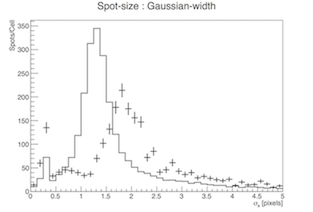}
        \hspace{0.4cm}
        \includegraphics[width=7.4cm]{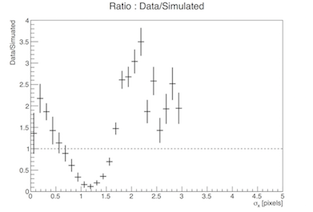}

  \centering
        \includegraphics[width=7.4cm]{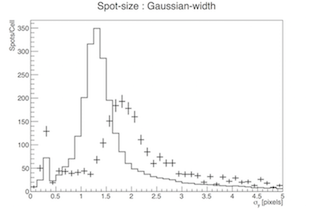}
        \hspace{0.4cm}
        \includegraphics[width=7.4cm]{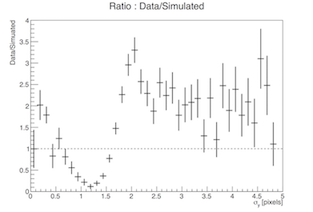}
 
  \centering
        \includegraphics[width=7.4cm]{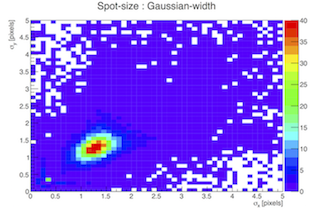}
        \hspace{0.4cm}
        \includegraphics[width=7.4cm]{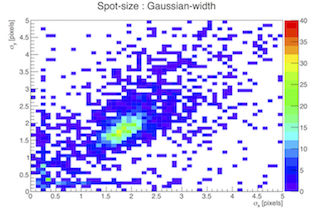}

  \caption{{\bf Comparison 3}. The cut-2 is applied. The simulated distributions of the spot pulse-height and spot-size are compared with the actual distributions obtained from Hiroshima's dataset. Black solid lines represent the observed distributions of all reconstructed spots. Red, yellow, blue and green filled histograms represent ground-truth distribution of monomer (${\bf R}$), mono-dimer (${\bf rR\ {\rm or}\ r'R}$), dimer (${\bf RR}$) spots. Crossed-lines represent Hiroshima's 2012 data.}
  \label{fig;spot_comparison_dimer_A_3}
\end{figure}

\begin{figure}[!h]
  \centering
        \includegraphics[width=7.4cm]{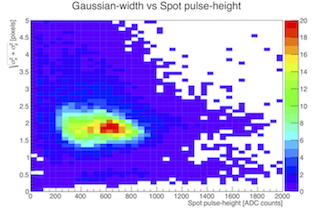}
        \hspace{0.4cm}
        \includegraphics[width=7.4cm]{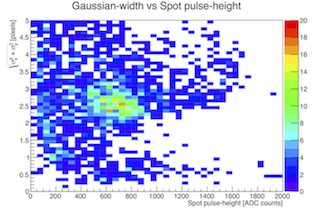}
        
  \centering
        \includegraphics[width=7.4cm]{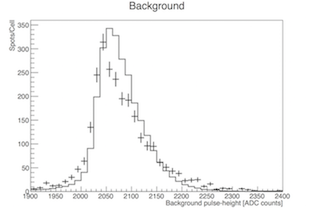}
        \hspace{0.4cm}
        \includegraphics[width=7.4cm]{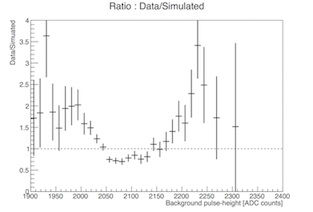}

  \centering
        \includegraphics[width=7.4cm]{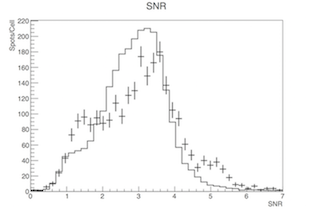}
        \hspace{0.4cm}
        \includegraphics[width=7.4cm]{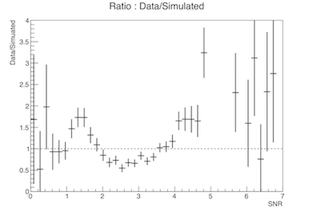}

  \caption{{\bf Comparison 4}. The cut-2 is applied. The top panels show that the simulated correlation (left) of the spot pulse-height to spot-size ($\sqrt{\sigma^2_{x} + \sigma^2_{y}}$) is compared with the actual correlation (right) obtained from Hiroshima's dataset. Distribution comparisons of background pulse-height and SNR are shown in bottom two panels.}
  \label{fig;spot_comparison_dimer_A_4}
\end{figure}

\paragraph{(2) Spot-property :}
Reconstructed spot-properties are presented as the Gaussian function of six parameters; spot pulse-height (or normalization factor), central position, spot-size (or Gaussian-width) and background pulse-height. Distributions and correlations of each parameter are shown in Figures \ref{fig;spot_property_dimer_A_1} to \ref{fig;spot_comparison_dimer_A_4}.

\paragraph{}
Positional resolutions (or localization error) of reconstructed spot-positions are shown in Figure \ref{fig;spot_property_dimer_A_1}. In those Figures, we confirmed that peaks of each resolution distribution are located near zeros, and are formed as nearly Gaussian functions. RMS value represents the positional resolution to $1.5\ {\rm pixels}$ ($100\ {\rm nm}$). However, the tail of each distribution appears to be asymmetric. One of the possible explanations of the asymmetry is because of the z-axis. In our analysis, we assumed that spots are characterized as a 2-D Gaussian function, ignoring the z-axis. 3-D Gaussian fitting may be able to resolve the asymmetry of each distribution. 

\paragraph{}
For each cut, we directly compare the simulated spectra to actual spectra in Figures \ref{fig;spot_comparison_dimer_A_1} to \ref{fig;spot_comparison_dimer_A_4}. The number of detected spots per frame per cell for each cuts are shown in Table \ref{tab;spots_dimer_A}. The left panel shows that the simulated distribution of reconstructed parameters is directly compared with Hiroshima's dataset. The right panel shows the ratio of the simulated spectra to the actual ones in each of bins. The errors are not only observed statistical errors, but also include the simulated sample statistical errors. All simulated spectra are normalized by the number of detected spots.

\begin{figure}[!h]
  \centering
        \includegraphics[width=6.6cm]{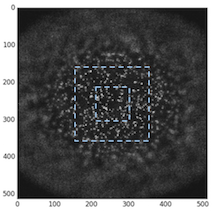}
        \hspace{1.2cm}
        \includegraphics[width=6.6cm]{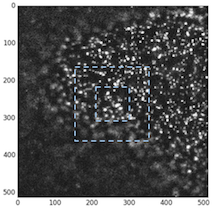}

  \vspace{0.1cm}
  \centering
        \includegraphics[width=6.6cm]{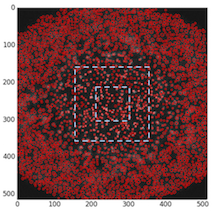}
        \hspace{1.2cm}
        \includegraphics[width=6.6cm]{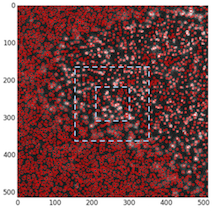}

  \centerline{\bf Cell-index $1$ \hspace{5.0cm} Hiroshima-2012}
  \caption{{\bf Single-molecule images of spot-detection on the apical cell-regions}. Single-molecule images before and after the spot-detection are shown in the top and the bottom panels.}
  \label{fig;spot_images_3}
\end{figure}

\newpage

\newpage

\newpage

\newpage

\newpage

\newpage

\subsubsection{Experimental configuration A (various ligand inputs)}
\paragraph{}
In this experimental configuration, $10$ cell-model samples are prepared for $12$ ligand inputs ($1.0\ {\rm pM} \sim 4.0\ {\rm nM}$). Cell-compartments for each cell-model are listed in Table \ref{tab;cell_compartments}. We also assume two observational cut-1 and cut-2. Parameter values for the spot-detection algorithm are configured to $\sigma_{min} = 2.0\ {\rm pixels}$, $\sigma_{max} = 4.0\ {\rm pixels}$, $20$ intermediate values in the deviation range, threshold $= 15$, and overlap $= 0.5$. 

\begin{figure}[!h]
  \centering
        \includegraphics[width=7.4cm]{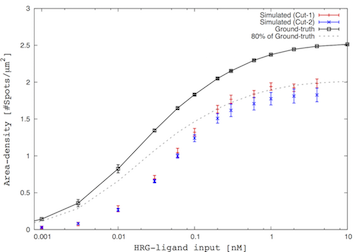}
        \hspace{0.4cm}
        \includegraphics[width=7.4cm]{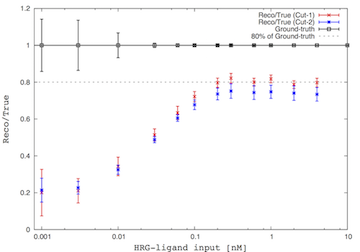}

  \vspace{0.4cm}
  
  \centering
        \includegraphics[width=7.4cm]{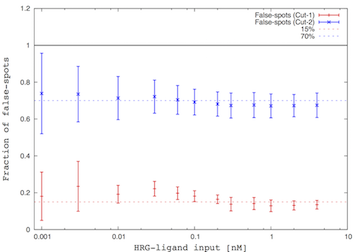}
        \hspace{0.4cm}
        \includegraphics[width=7.4cm]{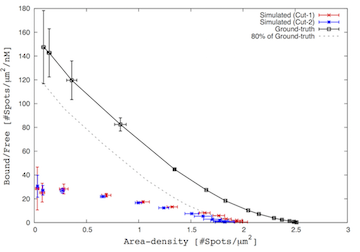}

  \caption{{\bf Equilibrium binding curve and Scatchard plot}. The left and the right panels in top row show that the reconstructed equilibrium curve and the ratio to true area-density in each HRG-ligand inputs. The bottom panels show the fraction of false-spots and Scachard plot. Red and blue crosses represent the reconstructed values obtained by the cut-1 and cut-2. Ground-truth and its $80\%$ recovery are presented with black solid and dashed lines.}
  \label{fig;dimer_results}
\end{figure}

\paragraph{}
Analysis results are shown in Figures \ref{fig;dimer_results}; (1) The left and the right top panels show the equilibrium binding curve and the ratio of the reconstructed density to ground-truth. These plots clearly show that shape of the true binding curve is partially recovered in the reconstructed one. While reconstruction efficiency is about $80\%$ and steady at relatively higher ligand concentration inputs ($> 0.2\ {\rm nM}$), the efficiency is significantly reduced at the lower ligand concentration region ($< 0.2\ {\rm nM}$). (2) The bottom left panel show fraction of false-spots contaminating all detected spots. Approximately $13\%$ of the detected spots are false-spots with the cut-1. Roughly, $70\%$ are contaminated by the cut-2. (3) The bottom right panel shows a Scatchard plot. The plot clearly shows that shape of true-Scatchard plot is not well-recovered in the reconstructed ones. The true-Scatchard plot is well-characterized as a concave-up curve. However, shape of the reconstructed plots is a concave-down curve or a straight line. Thus, measurement effect can significantly change apparent biological properties in single-molecule experiments.

\newpage

\paragraph{Hill coefficient :}
In a standard approach to biochemistry and biology, the Hill equation (1) can be fitted to the equilibrium binding curves to evaluate signal responses in a receptor system. The minimization function is given by the equation (13). The fitting results are shown in Table \ref{tab;dimer_fitting_results_1}. While model-truth of the receptor system exhibits the negative cooperativity, coopertivity is positive in the reconstructed curves. Restoration of the cooperative characteristics is thus failed in this measurements.

%The Hill equation can be written in the form of
%\begin{eqnarray}
%B(L) = \frac{B_0 L^n}{K_A^n + L^n}
%\end{eqnarray}
%where $L$, $B_0$, $K_A$ and $n$ represents ligand concentration, maximum area-density of ligand binding, ligands occupying half of the binding sites and Hill coefficient. 

\begin{table}[!h]
\centering
\begin{tabular}{|c|c|c|c|c|}
\hline
& $B_0\ [{\rm spots/\mu m^2}]$ & $K_A\ [{\rm nM}]$ & $n$ & $\hat{\chi}_0^2$ \\ \hline
Ground-truth & \hspace{0.3cm} $2.551 \pm 0.007$ \hspace{0.3cm} & \hspace{0.2cm} $0.027 \pm 0.001$ \hspace{0.2cm} & \hspace{0.2cm} ${\bf\color{black} 0.722} \pm 0.017$ \hspace{0.2cm} & \hspace{0.1cm}  $2.8831$ \hspace{0.1cm} \\ \hline
Reconstructed (Cut-1) & $1.989 \pm 0.014$ & $0.053 \pm 0.002$ & ${\bf\color{red} 1.109} \pm 0.026$ & $1.616$ \\ \hline
Reconstructed (Cut-2) & $1.850 \pm 0.010$ & $0.052 \pm 0.001$ & ${\bf\color{blue} 1.074} \pm 0.008$ & $0.352$\\ \hline
\end{tabular}
\caption{{\bf Results of fitting to the Hill equation}. The best fit values and uncertainties of each parameters are listed. $\hat{\chi}_0^2$ is the reduced minimum. If $n<1$, then the receptor system increases binding affinity of sites and exhibits negative cooperativity. If $n>1$, then cooperativity is positive, decreasing the binding affinity of sites. If $n=1$, then there is no cooperativity.}
\label{tab;dimer_fitting_results_1}
\end{table}

\begin{figure}[!h]
  \centerline{\bf Cut-1 \hspace{7.0cm} Cut-2}
  \centering
        \includegraphics[width=7.4cm]{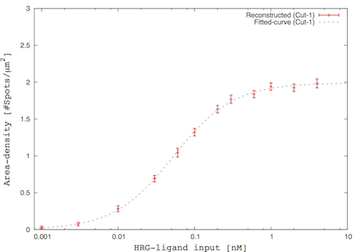}
        \hspace{0.4cm}
        \includegraphics[width=7.4cm]{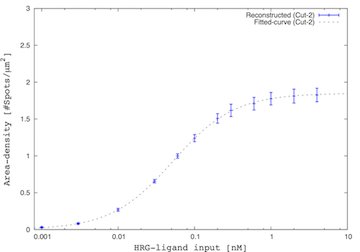}

  \vspace{0.4cm}
  \centering
        \includegraphics[width=7.4cm]{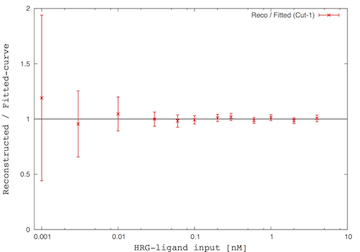}
        \hspace{0.4cm}
        \includegraphics[width=7.4cm]{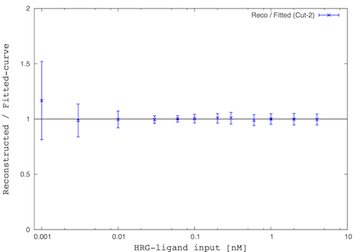}

  \caption{{\bf Fitting results}. The top panels show the equilibrium binding curves for each cut. The bottom panels show the ratio of the reconstructed binding curves to the fitted curves.}
  \label{fig;dimer_fitting_results_1}
\end{figure}

\newpage

\begin{figure}[!h]
  \centerline{\bf Cut-1 \hspace{7.0cm} Cut-2}
  \centering
        \includegraphics[width=7.4cm]{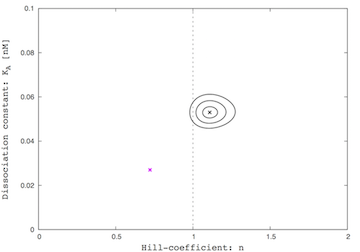}
        \hspace{0.4cm}
        \includegraphics[width=7.4cm]{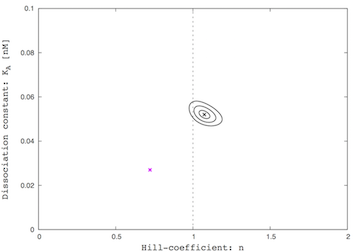}

  \vspace{0.4cm}
  \centering
        \includegraphics[width=7.4cm]{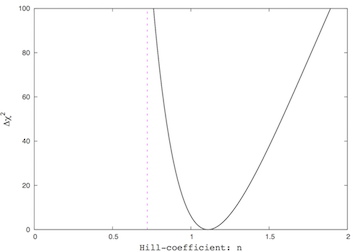}
        \hspace{0.4cm}
        \includegraphics[width=7.4cm]{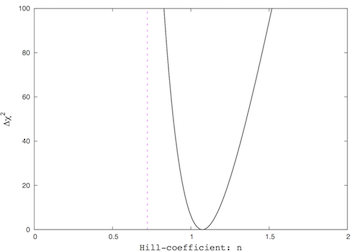}

  \vspace{0.4cm}
  \centering
        \includegraphics[width=7.4cm]{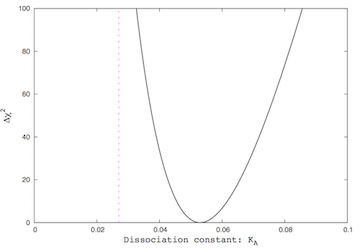}
        \hspace{0.4cm}
        \includegraphics[width=7.4cm]{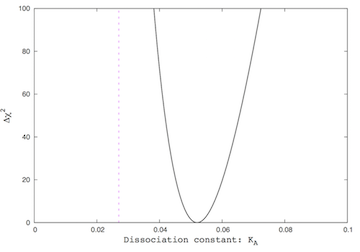}

  \caption{{\bf $\Delta \chi^2$ plots}. For each area-cuts, Figures show the $\Delta \chi^2$ plots of the Hill coefficient and the dissociation constant. $B_0$ is fixed to the best fit value.}
  \label{fig;dimer_dchi2_plots}
\end{figure}

\paragraph{Confidence interval :}
$\Delta \chi^2$-plots for each area-cuts are shown in Figures \ref{fig;dimer_dchi2_plots} show. Black solid lines represent the $1\sigma$ ($68\%$), $2\sigma$ ($95\%$) and $3\sigma$  ($99\%$) confidence intervals. Black and pink points indicate the best fit and ground-truth. \\
\forceindent The top panels show the $\Delta \chi^2$ contour plots: the Hill coefficient ($n$) vs dissociation constant ($K_{A}$). The ground-truth (pink point) is located out of the $3\sigma$ confidence contour line, implying the restoration failure of the true parameter values. The middle four plots show the $\Delta \chi^2$ plots for each model parameters: the ground-true Hill coefficient and the dissociation constant are clearly out of the contour lines. 

\newpage

\paragraph{Systematic sources :}
There are three major sources that can generate such systematic shifts. Details are discussed as follows.

\begin{figure}[!h]
  \centerline{\bf Cut-1 \hspace{7.0cm} Cut-2}
  \centering
        \includegraphics[width=7.4cm]{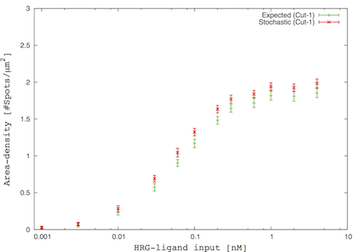}
        \hspace{0.4cm}
        \includegraphics[width=7.4cm]{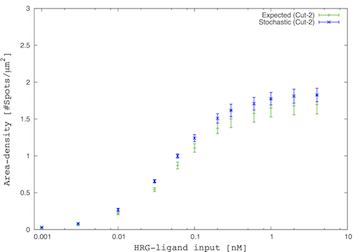}

  \centering
        \includegraphics[width=7.4cm]{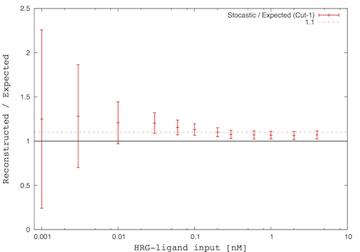}
        \hspace{0.4cm}
        \includegraphics[width=7.4cm]{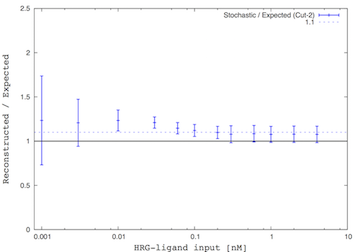}

  \centering
        \includegraphics[width=7.4cm]{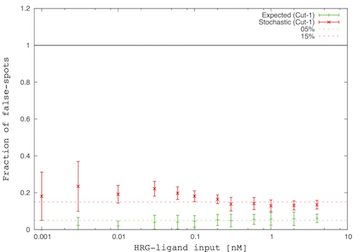}
        \hspace{0.4cm}
        \includegraphics[width=7.4cm]{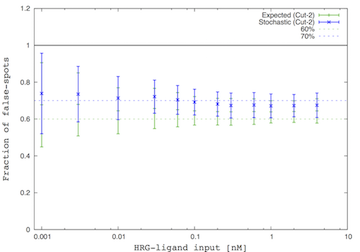}

  \centering
        \includegraphics[width=7.4cm]{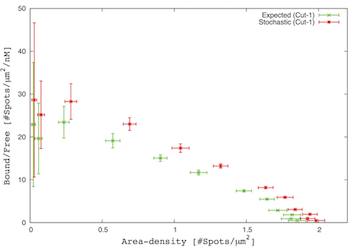}
        \hspace{0.4cm}
        \includegraphics[width=7.4cm]{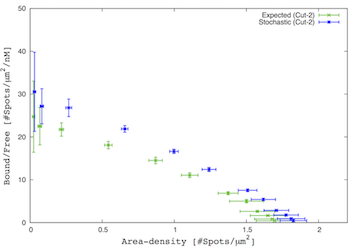}

  \caption{{\bf Stochasticity}. Comparison of the equilibrium binding curves and Scatchard plots reconstructed from the expected and stochastic images. Red and blue lines represent the reconstructed curves obtained by the cut-1 and cut-2. The reconstructed curves obtained from the stochastic images are presented with green lines.}
  \label{fig;stochasticity_2}
\end{figure}

\begin{figure}[!h]
  \leftline{\bf \hspace{0.5cm} Ligand : $4.000\ {\rm nM}$}
  \centering
        \includegraphics[width=7.3cm]{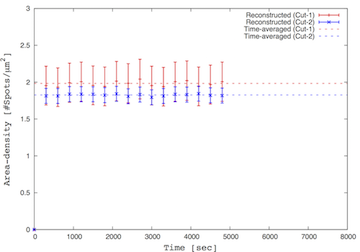}
        \hspace{0.4cm}
        \includegraphics[width=7.3cm]{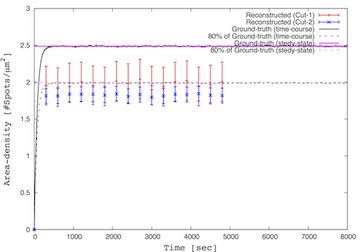}

  \leftline{\bf \hspace{0.5cm} Ligand : $1.000\ {\rm nM}$}
  \centering
        \includegraphics[width=7.3cm]{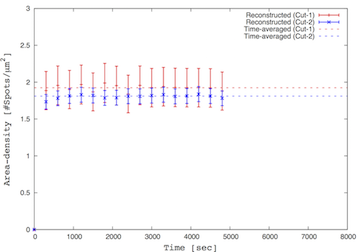}
        \hspace{0.4cm}
        \includegraphics[width=7.3cm]{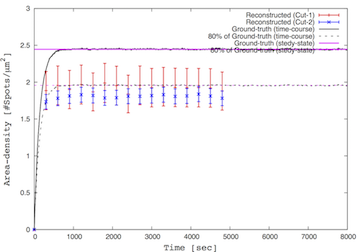}

  \leftline{\bf \hspace{0.5cm} Ligand : $2.000\ {\rm nM}$}
  \centering
        \includegraphics[width=7.3cm]{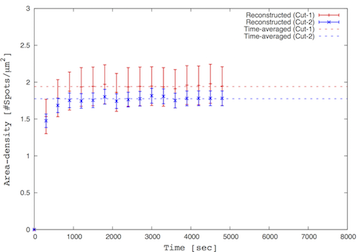}
        \hspace{0.4cm}
        \includegraphics[width=7.3cm]{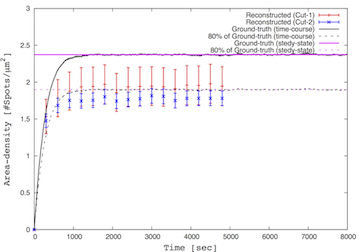}

  \caption{{\bf Time course data for the ligand concentration with inputs more than $0.2\ {\rm nM}$ (1)}. The left panels show that the reconstructed response of binding states in time compared with time-averaged values. Red and blue solid lines represent the reconstructed responses of binding state for each area-cuts. The time-averaged values are represented in red and blue dashed lines. The right panels show the comparison to the ground-true responses of binding states in time. Solid and dashed pink lines represent the ground-true equilibrium state and its $80\%$. The ground-true response is represented with the black solid line. The $80\%$ restoration of the true response is shown with the dashed black line.}
  \label{fig;time_course_1}
\end{figure}

\begin{figure}[!h]
  \leftline{\bf \hspace{0.5cm} Ligand : $0.600\ {\rm nM}$}
  \centering
        \includegraphics[width=7.3cm]{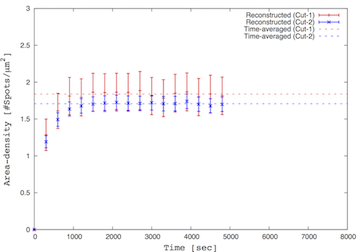}
        \hspace{0.4cm}
        \includegraphics[width=7.3cm]{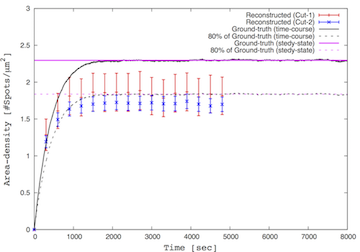}

  \leftline{\bf \hspace{0.5cm} Ligand : $0.300\ {\rm nM}$}
  \centering
        \includegraphics[width=7.3cm]{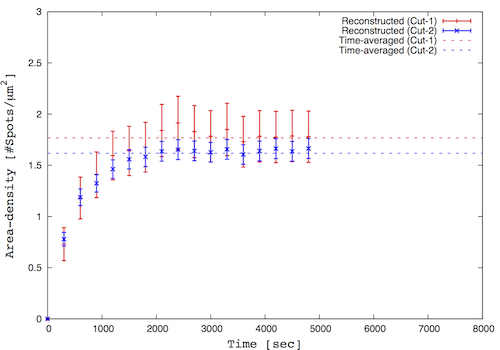}
        \hspace{0.4cm}
        \includegraphics[width=7.3cm]{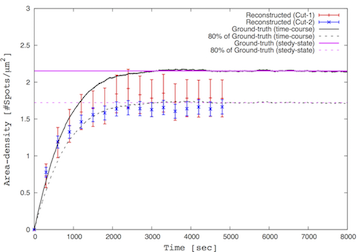}

  \leftline{\bf \hspace{0.5cm} Ligand : $0.200\ {\rm nM}$}
  \centering
        \includegraphics[width=7.3cm]{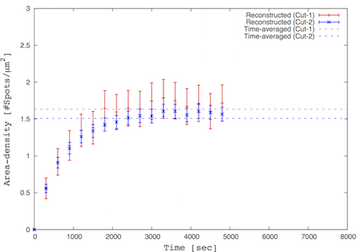}
        \hspace{0.4cm}
        \includegraphics[width=7.3cm]{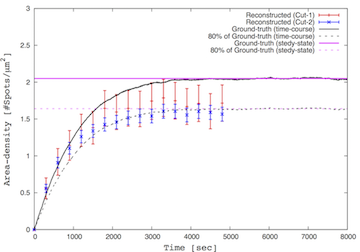}

  \caption{{\bf Time course data for the ligand concentration with inputs more than $0.2\ {\rm nM}$ (2)}. The left panels show that the reconstructed response of binding states in time compared with time-averaged values. Red and blue solid lines represent the reconstructed responses of binding state for each area-cuts. The time-averaged values are represented in red and blue dashed lines. The right panels show the comparison to the ground-true responses of binding states in time. Solid and dashed pink lines represent the ground-true equilibrium state and its $80\%$. The ground-true response is represented with the black solid line. The $80\%$ restoration of the true response is shown with the dashed black line.}
  \label{fig;time_course_2}
\end{figure}

\begin{figure}[!h]
  \leftline{\bf \hspace{0.5cm} Ligand : $0.100\ {\rm nM}$}
  \centering
        \includegraphics[width=7.3cm]{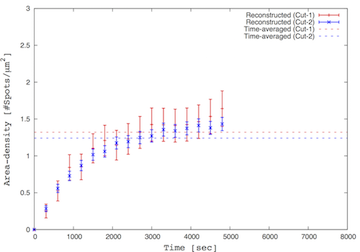}
        \hspace{0.4cm}
        \includegraphics[width=7.3cm]{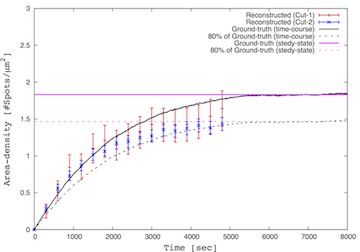}

  \leftline{\bf \hspace{0.5cm} Ligand : $0.060\ {\rm nM}$}
  \centering
        \includegraphics[width=7.3cm]{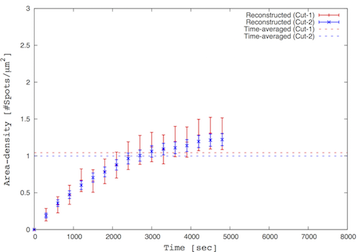}
        \hspace{0.4cm}
        \includegraphics[width=7.3cm]{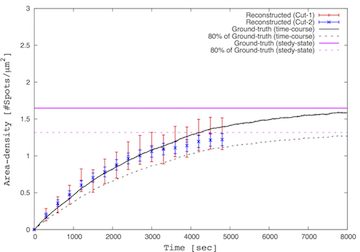}

  \leftline{\bf \hspace{0.5cm} Ligand : $0.030\ {\rm nM}$}
  \centering
        \includegraphics[width=7.3cm]{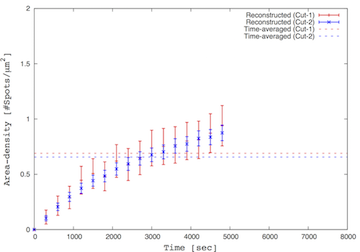}
        \hspace{0.4cm}
        \includegraphics[width=7.3cm]{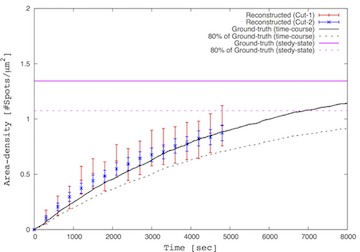}

  \caption{{\bf Time course data for the ligand concentration with inputs less than $0.2\ {\rm nM}$ (1)}. The left panels show that the reconstructed response of binding states in time compared with time-averaged values. Red and blue solid lines represent the reconstructed responses of the binding state for each area-cuts. The time-averaged values are represented in red and blue dashed lines. The right panels show the comparison to the ground-true responses of binding states in time. Solid and dashed pink lines represent the ground-true equilibrium state and its $80\%$. The ground-true response is represented with a black solid line. The $80\%$ restoration of the true response is shown with a dashed black line.}
  \label{fig;time_course_3}
\end{figure}

\begin{figure}[!h]
  \leftline{\bf \hspace{0.5cm} Ligand : $0.010\ {\rm nM}$}
  \centering
        \includegraphics[width=7.3cm]{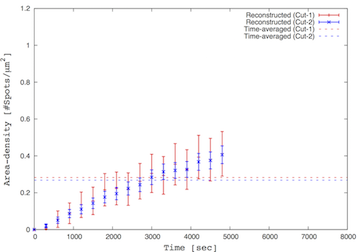}
        \hspace{0.4cm}
        \includegraphics[width=7.3cm]{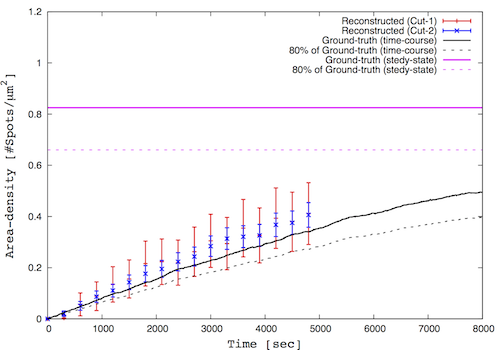}

  \leftline{\bf \hspace{0.5cm} Ligand : $0.003\ {\rm nM}$}
  \centering
        \includegraphics[width=7.3cm]{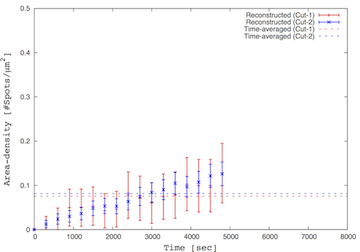}
        \hspace{0.4cm}
        \includegraphics[width=7.3cm]{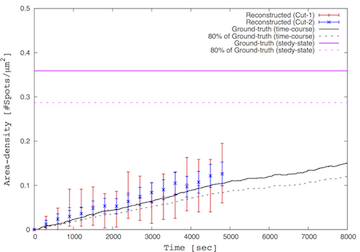}

  \leftline{\bf \hspace{0.5cm} Ligand : $0.001\ {\rm nM}$}
  \centering
        \includegraphics[width=7.3cm]{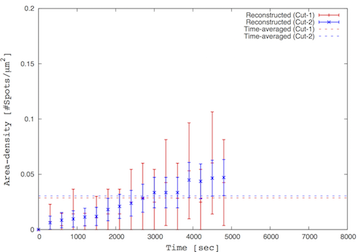}
        \hspace{0.4cm}
        \includegraphics[width=7.3cm]{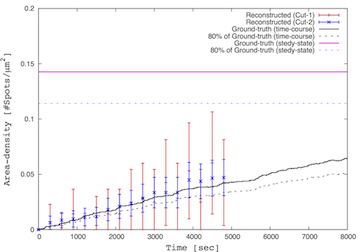}

  \caption{{\bf Time course data for the ligand concentration with inputs less than $0.2\ {\rm nM}$ (2)}. The left panels show that the reconstructed response of binding states in time compared with time-averaged values. Red and blue solid lines represent the reconstructed responses of the binding state for each area-cut. The time-averaged values are represented in red and blue dashed lines. The right panels show the comparison to the ground-true responses of binding states in time. Solid and dashed pink lines represent the ground-true equilibrium state and $80\%$ of the true equilibrium state. The ground-true response is represented with a black solid line. The $80\%$ restoration of the true response is shown with a dashed black line.}
  \label{fig;time_course_4}
\end{figure}

\begin{enumerate}
\item [(1)] Stochasticity: Stochastic process of photon-detection may influence the reconstruction procedures, changing the shape and concavity of the equilibrium binding curve and Scatchard plot. We turn off noise channels in the camera simulation module to generate expected images. Since intensity of the expected images is presented in the level of photoelectron unit, we convert the image intensity unit to the level of ADC counts for proper evaluation. We then applied the analytical procedure to the expected images in order to evaluate the systematic influence on the binding curve and Scatchard plot. \\

\begin{figure}[!h]
  \centering
        \includegraphics[width=6.6cm]{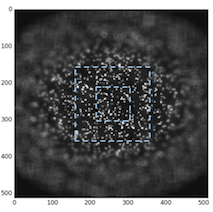}
        \hspace{1.2cm}
        \includegraphics[width=6.6cm]{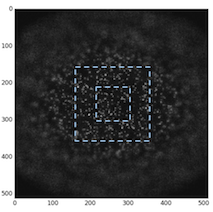}

  \vspace{0.1cm}
  \centering
        \includegraphics[width=6.6cm]{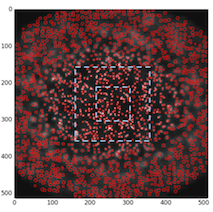}
        \hspace{1.2cm}
        \includegraphics[width=6.6cm]{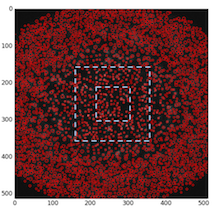}

  \centerline{\bf Expected image \hspace{5.0cm} Stochastic image}
  \caption{{\bf Comparison between expected and stochastic single-molecule images}. Single-molecule images before and after the spot-detection are shown in the top and the bottom panels. The observational cut-1 and cut-2 represented with inner and outer boxes. Red circles represent molecular-spots detected by using the LoG algorithm. Each image size is $512 \times 512\ {\rm pixels}$. Actual minimum and maximum values of the image intensity are $1,900$ and $3,200$ ADC counts. The image intensity is rescaled in the range of $0$ to $255$. The microscopy is configured to be focused on the image center. }
  \label{fig;spot_images_4}
\end{figure}

\newpage

\forceindent First, the expected images are compared with the stochastic images obtained by turning on the noise channels. Figure \ref{fig;spot_images_4} shows example images of spot-detection on the apical cell-surfaces. The red circles shown in the Figures are more sparsely distributed in the expected images, avoiding capture of the molecular-spots in the defocused region of the expected images. However, more spots are detected near the defocused region of the stochastic images.

\forceindent Figure \ref{fig;stochasticity_2} shows the equilibrium binding curves and Scatchard plot for each area-cuts. The top panels show the comparison of the equilibrium binding curves reconstructed from the expected and stochastic images. The restoration efficiency and defects of the analytical procedure are shown in the middle four panels. We computed the ratio of the reconstructed binding curves of the stochastic images to the expected images. The 2nd panels from the top show the ratio curves, implying that the stochasticity can increase approximately $10\%$ of the restoration efficiency in capturing more molecular-spots. The 2nd panels from the bottom show the fraction of false-spots contaminated in all detected spots, implying that the stochasticity can increase approximately $10\%$ of contamination in capturing the defected spots. Finally, the bottom panels show the comparison of the Scatchard plots reconstructed from the expected and stochastic images. The reconstructed Scatchard plots are aligned in parallel. Consequently, we found no significant change in the shape and concavity of the binding curve and Scatchard plot.

\item [(2)] Spot-detection algorithm: Figure \ref{fig;dimer_results} shows that the restoration efficiency of the LoG spot-detection method is limited by up to $\sim 80\%$. The fractional occupancy of the defected spots is steady at $\sim 13\%$ for various ligand concentrations. Such imperfect performance of the spot-detection algorithm may influence to our biological interpretation of cooperativity.

\forceindent Figures \ref{fig;time_course_1} and \ref{fig;time_course_2} show the time course data in the ligand concentration with inputs more than $0.2\ {\rm nM}$. In this concentration range, the reconstructed responses (represented by the red and blue crosses in the Figures) successfully converge to the $80\%$ of the true-equilibrium state (pink dashed lines), reaching to the $80\%$ restoration of the true-equilibrium state. As shown in Figure \ref{fig;dimer_results}B, the shape and concavity of the reconstructed binding curve and Scatchard plot are unchanged from the ground-truth in the concentration range, implying weak influence on the identification of cooperativity. In addition, more accurate algorithms \cite{ruusuvuori2010, smal2010, basset2015} may help improving the efficiency and defects, but those algorithms can only help shifting the entire spectra up and down: no significant change in the shape and concavity.

\item[(3)] Quasi-static response: The binding system often requires a longer image acquisition period to reach steady state, due to quasi-static response at very low concentration inputs, at $\sim 1\ {\rm pM}$ scale. This process is a slow transition of the binding system to equilibrium state. If the binding system is quite sensitive to such slow response, then the system cannot converge to the equilibrium state within the acquisition period. The quasi-static responses may accordingly become a critical issue for the biological interpretation of cooperativity.

\forceindent Figures \ref{fig;time_course_3} and \ref{fig;time_course_4} show the time course data of the ligand concentration with inputs less than $0.2\ {\rm nM}$. In this concentration range, the reconstructed responses still remain in a nonequilibrium state within the image acquisition period: $0$ to $5,000\ {\rm sec}$. Although the true-nonequilibrium binding  state (black lines) is well-restored through the reconstruction procedure, the reconstructed binding state (red and blue crosses) fails to converge to the true-equilibrium states (pink lines), generating a gap between the reconstructed binding state and the true-equilibrium state. Such gaps cannot be even detected during the image acquisition period, thus leading to the misidentification of cooperativity.
\end{enumerate}

\newpage

\newpage

\newpage

\newpage

\newpage

\newpage

\subsubsection{Experimental configuration A (high concentration range)}
\paragraph{}
In this experimental configuration, we used $10$ cell-samples for $6$ ligand inputs ($200\ {\rm pM} \sim 4.0\ {\rm nM}$).  Cell-compartments for each cell-model is listed in Table \ref{tab;cell_compartments}. We also assume two observational cuts: the cut-1 and the cut-2. Parameter values for the spot-detection algorithm are configured to $\sigma_{min} = 2.0\ {\rm pixels}$, $\sigma_{max} = 4.0\ {\rm pixels}$, $20$ intermediate values in the deviation range, threshold $= 15$, and overlap $= 0.5$. 

\begin{figure}[!h]
  \centerline{\bf Cut-1 \hspace{7.0cm} Cut-2}
  \centering
        \includegraphics[width=7.4cm]{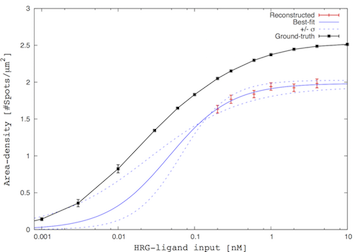}
        \hspace{0.4cm}
        \includegraphics[width=7.4cm]{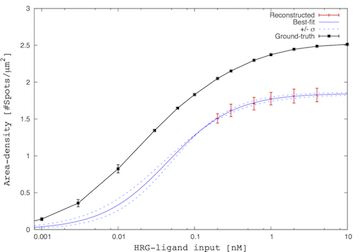}

%  \vspace{0.4cm}
%  \centering
%        \includegraphics[width=7.4cm]{figures_SI/reconstruction/dimer_model/config_A2/fitted_curve_ratio_cut1.png}
%        \hspace{0.4cm}
%        \includegraphics[width=7.4cm]{figures_SI/reconstruction/dimer_model/config_A2/fitted_curve_ratio_cut2.png}
%
  \vspace{0.4cm}
  \centering
        \includegraphics[width=7.4cm]{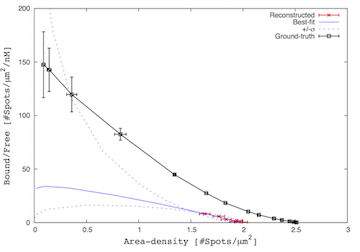}
        \hspace{0.4cm}
        \includegraphics[width=7.4cm]{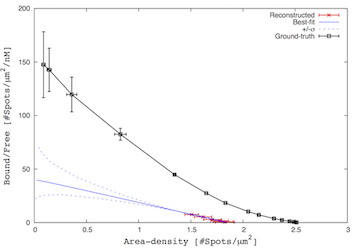}

  \caption{{\bf Equilibrium binding curves and Scatchard plot}. The top panels show the equilibrium binding curves for each cuts. Bottom ones show the Scatchard plots for each cuts. Blue solid and dashed lines represents the best fit curves and the curves $1\sigma$ shifted from the best fits.}
  \label{fig;dimer_fitting_results_2}
\end{figure}

\paragraph{Hill coefficient :}
The Hill equation (1) is fitted to the equilibrium binding curves to evaluate signal responses in the receptor system.  The minimization function is given by the equation (13). The fitting results are shown in Table \ref{tab;dimer_fitting_results_2} and Figures \ref{fig;dimer_fitting_results_2}. While model-truth of the receptor system exhibits the negative cooperativity, the Hill coefficients are largely fluctuated around an unity. The cooperative characteristics is thus indeterminable in this measurement.

%The Hill equation can be written in the form of
%\begin{eqnarray}
%B(L) = \frac{B_0 L^n}{K_A^n + L^n}
%\end{eqnarray}
%where $L$, $B_0$, $K_A$ and $n$ represents ligand concentration, maximum area-density of ligand binding, ligands occupying half of the binding sites and Hill coefficient. 

\begin{table}[!h]
\centering
\begin{tabular}{|c|c|c|c|c|}
\hline
& $B_0\ [{\rm spots/\mu m^2}]$ & $K_A\ [{\rm nM}]$ & $n$ & $\hat{\chi}_0^2$ \\ \hline
Ground-truth & \hspace{0.3cm} $2.551 \pm 0.007$ \hspace{0.3cm} & \hspace{0.2cm} $0.027 \pm 0.001$ \hspace{0.2cm} & \hspace{0.2cm} ${\bf\color{black} 0.722} \pm 0.017$ \hspace{0.2cm} & \hspace{0.1cm}  $2.8831$ \hspace{0.1cm} \\ \hline
Reconstructed (Cut-1) & $1.98332 \pm 0.04322$ & $0.0461509 \pm 0.02081$ & ${\bf 1.06812 \pm 0.3602}$ & $0.345343$ \\ \hline
Reconstructed (Cut-2) & $1.84792 \pm 0.01448$ & $0.0450399 \pm 0.006133$ & ${\bf 1.00477 \pm 0.1069}$ & $0.0128391$\\ \hline
\end{tabular}
\caption{{\bf Results of fitting to the Hill equation}. The best fit values and uncertainties of each parameters are listed. $\hat{\chi}^2_0$ is the reduced minimum.}
\label{tab;dimer_fitting_results_2}
\end{table}

% If $n<1$, then the receptor system increases binding affinity of sites and exhibits negative cooperativity. If $n>1$, then cooperativity is positive, decreasing the binding affinity of sites. If $n=1$, then no cooperativity.

\newpage

\begin{figure}[!h]
  \centerline{\bf Cut-1 \hspace{7.0cm} Cut-2}
  \centering
        \includegraphics[width=7.4cm]{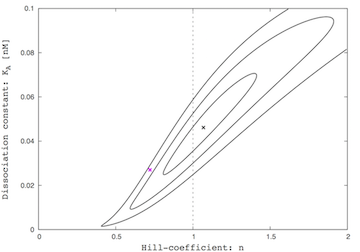}
        \hspace{0.4cm}
        \includegraphics[width=7.4cm]{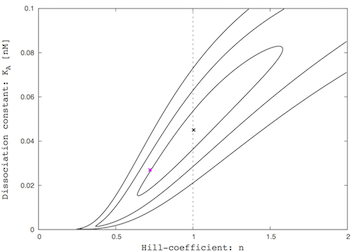}

  \vspace{0.4cm}
  \centering
        \includegraphics[width=7.4cm]{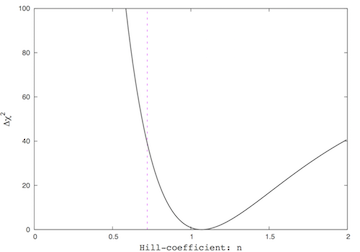}
        \hspace{0.4cm}
        \includegraphics[width=7.4cm]{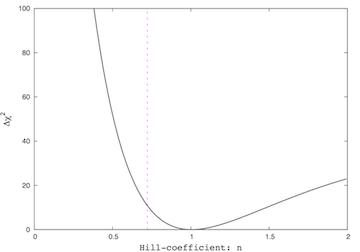}

  \vspace{0.4cm}
  \centering
        \includegraphics[width=7.4cm]{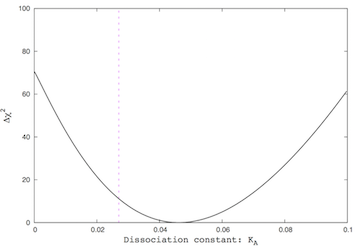}
        \hspace{0.4cm}
        \includegraphics[width=7.4cm]{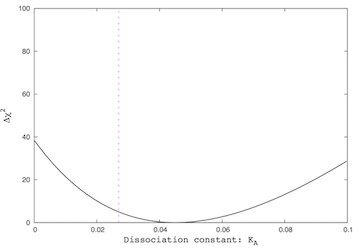}

  \caption{{\bf $\Delta \chi^2$ plots}. For each area-cuts, Figures show the $\Delta \chi^2$ plots of the Hill coefficient and the dissociation constant. $B_0$ is fixed to the best fit value.}
  \label{fig;dimer_dchi2_plots_2}
\end{figure}

\paragraph{Confidence interval :}
$\Delta \chi^2$-plots for each area-cuts are shown in Figures \ref{fig;dimer_dchi2_plots_2} show. Black solid lines represent the $1\sigma$ ($68\%$), $2\sigma$ ($95\%$) and $3\sigma$  ($99\%$) confidence intervals. Black and pink points indicate the best fit and ground-truth. \\
\forceindent The top panels show the $\Delta \chi^2$ contour plots: the Hill coefficient ($n$) vs dissociation constant ($K_{A}$). The ground-truth (pink point) is located within the $3\sigma$ confidence contour line, implying the successful restoration of the true parameter values. The middle four plots show the $\Delta \chi^2$ plots for the Hill coefficient and the dissociation constant.

\newpage

\subsubsection{Experimental configuration B (nonequilibrium region)}
\paragraph{}
Single-molecule imaging of apical regions of the dimer model is simulated for the optical specifications and operating conditions of the fluorescence microscopy simulation module shown in Table \ref{tab;specification3} and \ref{fig;config_hiroshima2012}. Photobleaching is included in the simulated single-molecule images. The results are shown as follows.

\paragraph{(1) Spot-detection :}
In this experimental configuration, the parameter values for the spot-detection algorithm are configured to $\sigma_{min} = 3.0\ {\rm pixels}$, $\sigma_{max} = 6.0\ {\rm pixels}$, $30$ intermediate values in the deviation range, threshold $= 30$, and overlap $= 0.0$. Figure \ref{fig;spot_images_3} shows example images of spot-detection on the apical cell-regions. Since the focal position is unknown and typically determined by human-eyes, we defined two observational area-cuts (cut-1 and cut-2) with $100 \times 300\ {\rm pixels}$ ($132\ {\rm \mu m^2}$) represented with dashed lines. While the cut-1 is placed at well-focused region of image center, the cut-2 is shifted out of the image center and located at the defocused region. Red circles represent the spots detected by the LoG method. Each image size is $512 \times 512\ {\rm pixels}$. Actual minimum and maximum values of the image intensity are $1,900$ and $2,600$ ADC counts. The image intensity is rescaled in the range of $0$ to $255$.

\begin{figure}[!h]
  \centering
        \includegraphics[width=7.4cm]{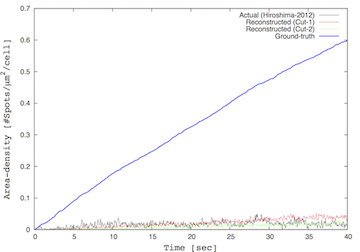}
        \hspace{0.4cm}
        \includegraphics[width=7.4cm]{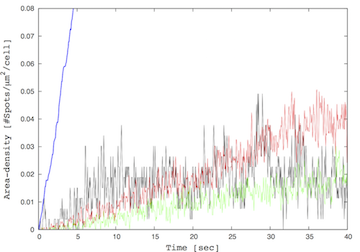}

  \centering
        \hspace{7.4cm}
        \hspace{0.4cm}
        \includegraphics[width=7.4cm]{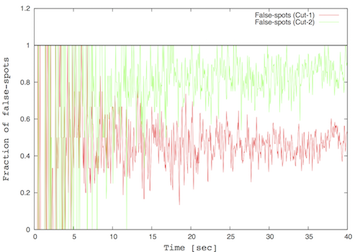}

  \caption{{\bf Reconstructed area-density}. The top panels show that the reconstructed density changes with time. Ground-truth is represented with the blue line. Black, red and green lines represent the observed and reconstructed area-density with the cut-1 and cut-2.}
  \label{fig;reco_density_4}

  \centering
        \includegraphics[width=7.4cm]{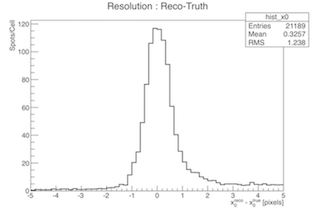}
        \hspace{0.4cm}
        \includegraphics[width=7.4cm]{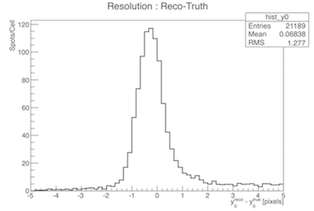}

  \centering
        \includegraphics[width=7.4cm]{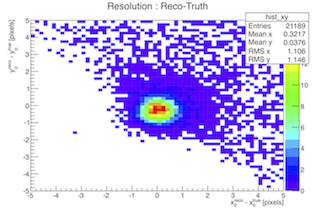}
        \hspace{0.4cm}
        \includegraphics[width=7.4cm]{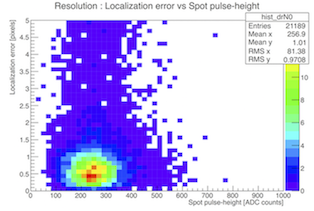}
 
  \caption{{\bf Spot-properties 1}.  The cut-1 is applied. Distributions and correlations of localization error ($\vec{r}^{\ reco}_0 - \vec{r}^{\ true}_0$) in  x-y axes are shown in the top and the bottom panels.}
  \label{fig;spot_property_dimer_B_1}
\end{figure}

\paragraph{}
Analysis results are shown in Table \ref{tab;spots_dimer_B} and Figures \ref{fig;reco_density_4}; (1) the Table shows the number and fraction of observed and simulated spots. The simulated spots are true-molecular spots and false-spots that arise from molecules and background noise. In particular, the false-spots are noise-like spots that can mimic the molecular spots. Quite a large fraction of the false-spots are captured with both cuts. (2) The left and the right panels of the Figure show binding curves and fractional occupancy of false-spots for the cut-1 and cut-2. Efficiency of the density reconstruction is approximately $10\ \%$ for observational cuts. While the reconstructed area-densities for cuts tend to be linearly increased in time, the observed density is relatively flat in time. In addition, the fractional occupancy of the false-spots is stable in time. 

\begin{table}[!h]
\centering
\begin{tabular}{|c|c|c|c|c|c|}
\hline
%\multicolumn{2}{|c|}{\bf Area density of ErbB proteins} \\ \hline
Spots/cell & Observed & Simulated & True-monomer & True-dimer & False-spot \\ \hline
Cut-1 & $1760$ & $2118.9$ & $284.6$ ($13.4\%$) & $854.3$ ($40.3\%$) & $980$ ($46.3\%$) \\ \hline
Cut-2 & $1760$ & $983.8$ & $42.8$ ($4.4\%$) & $130.8$ ($13.3\%$) & $810.2$ ($82.4\%$) \\ \hline
\end{tabular}
\caption{{\bf The number and fraction of detected spots}.}
\label{tab;spots_dimer_B}
\end{table}

\begin{figure}[!h]
  \centering
        \includegraphics[width=7.4cm]{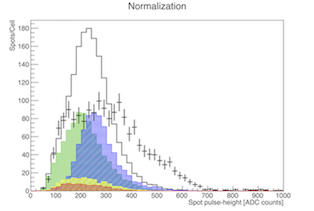}
        \hspace{0.4cm}
        \includegraphics[width=7.4cm]{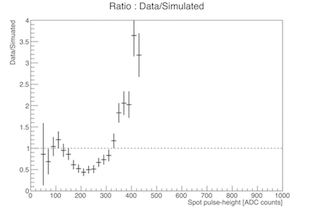}

  \centering
        \includegraphics[width=7.4cm]{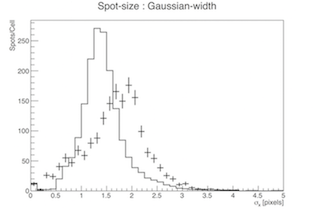}
        \hspace{0.4cm}
        \includegraphics[width=7.4cm]{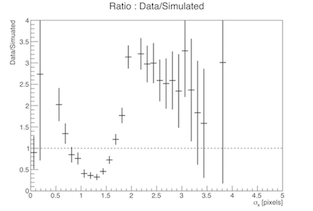}

  \centering
        \includegraphics[width=7.4cm]{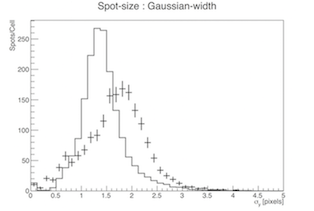}
        \hspace{0.4cm}
        \includegraphics[width=7.4cm]{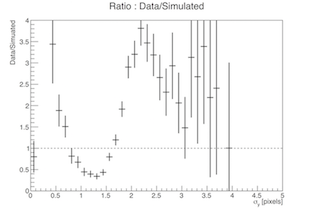}

  \centering
        \includegraphics[width=7.4cm]{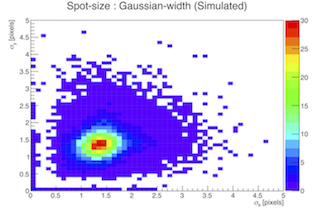}
        \hspace{0.4cm}
        \includegraphics[width=7.4cm]{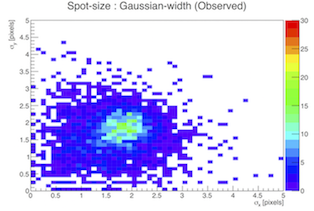}

  \caption{{\bf Comparison 1}. The cut-1 is applied. The simulated distributions of the spot pulse-height and spot-size are compared with the actual distributions obtained from Hiroshima's dataset. Black solid line represents observed distributions of all reconstructed spots. Red, blue and green filled histograms represent ground-truth distribution of monomer (${\bf R, rR\ {\rm or}\ r'R}$), dimer (${\bf RR}$) spots and defects. Crossed-lines represent Hiroshima's 2012 data.}
  \label{fig;spot_comparison_dimer_B_1}
\end{figure}

\begin{figure}[!h]
  \centering
        \includegraphics[width=7.4cm]{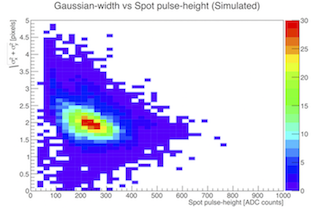} 
        \hspace{0.4cm}
        \includegraphics[width=7.4cm]{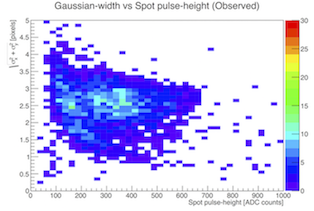} 
 
  \centering
        \includegraphics[width=7.4cm]{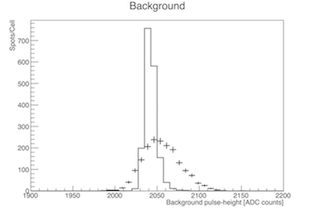}
        \hspace{0.4cm}
        \includegraphics[width=7.4cm]{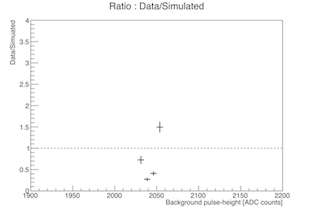}
 
  \centering
        \includegraphics[width=7.4cm]{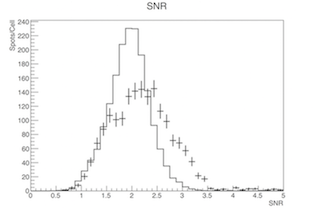}
        \hspace{0.4cm}
        \includegraphics[width=7.4cm]{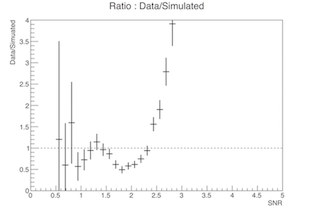}

  \caption{{\bf Comparison 2}. The cut-1 is applied. The top panels show that the simulated correlation (left) of the spot pulse-height to spot-size ($\sqrt{\sigma^2_{x} + \sigma^2_{y}}$) is compared with the actual correlation (right) obtained from Hiroshima's dataset. Comparisons of background pulse-height and SNR distributions are shown in the bottom two panels.}
  \label{fig;spot_comparison_dimer_B_2}
\end{figure}

\paragraph{(2) Spot-property :}
Reconstructed spot-properties are presented as the Gaussian function of six parameters; spot pulse-height (or normalization factor), central position, spot-size (or Gaussian-width) and background pulse-height. Distributions and correlations of each parameter are shown in Figures \ref{fig;spot_property_dimer_B_1} to \ref{fig;spot_comparison_dimer_B_2}. 

\paragraph{}
Positional resolution (or localization error) of reconstructed spot-positions are shown in Figure \ref{fig;spot_property_dimer_B_1}. In those Figures, we confirmed that the peaks of each resolution distributions are located near zeros, and are formed as nearly Gaussian distributions. The root mean squared (RMS) value represents the positional resolution to $1.5\ {\rm pixels}$ ($\sim 100\ {\rm nm}$). However, the tails of each distributions appear to be asymmetric. One of the possible explanations of the asymmetry relates to the z-axis. In our analysis, we assumed that spots are characterized by a 2-D Gaussian function, ignoring the z-axis. 3-D Gaussian fitting may be able to resolve the asymmetry of each distribution.

\paragraph{}
For each area-cuts, we directly compared the simulated spot-spectra to actual spectra in Figures \ref{fig;spot_comparison_dimer_B_1} and \ref{fig;spot_comparison_dimer_B_2}. Large discrepancies are clearly observed in all spectra. The left panel shows that the simulated distribution of reconstructed parameters is directly compared with Hiroshima's dataset. The right panel shows the ratio of the simulated spectra to the actual spectra in each of bins. The errors are not only observed statistical errors, but also include the simulated sample statistical errors. All simulated spectra are normalized by the number of detected spots. 

\begin{figure}[!h]
  \centering
        \includegraphics[width=6.6cm]{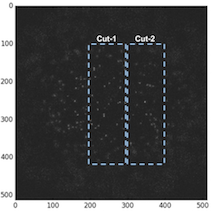}
        \hspace{1.2cm}
        \includegraphics[width=6.6cm]{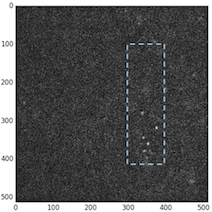}

  \vspace{0.1cm}
  \centering
        \includegraphics[width=6.6cm]{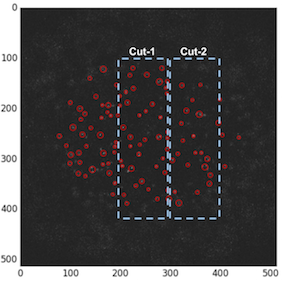}
        \hspace{1.2cm}
        \includegraphics[width=6.6cm]{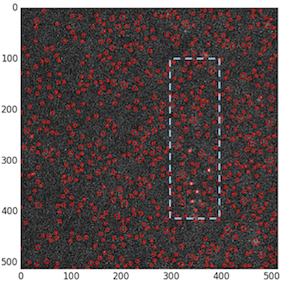}

  \centerline{\bf Cell-index $1$ \hspace{5.0cm} Hiroshima-2012}
  \caption{{\bf Single-molecule images and spot-detection on the apical cell-regions}. single-molecule images after time $30\ {\rm seconds}$ and the results of the spot-detection are shown in the top and the bottom panels.}
  \label{fig;spot_images_3}
\end{figure}

\newpage

\newpage

\newpage

\newpage

\begin{figure}[!h]
  \centering
        \includegraphics[width=7.4cm]{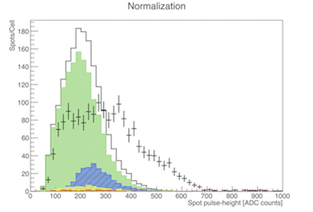}
        \hspace{0.4cm}
        \includegraphics[width=7.4cm]{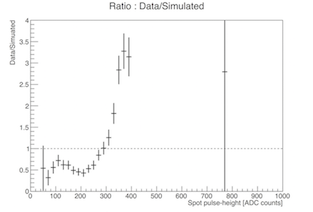}

  \centering
        \includegraphics[width=7.4cm]{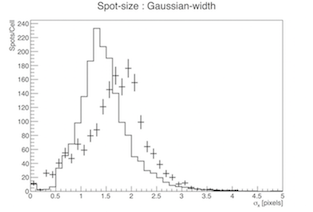}
        \hspace{0.4cm}
        \includegraphics[width=7.4cm]{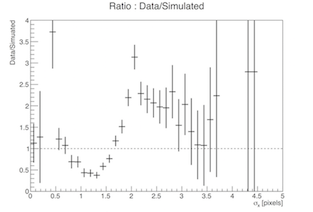}

  \centering
        \includegraphics[width=7.4cm]{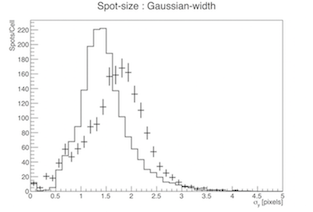}
        \hspace{0.4cm}
        \includegraphics[width=7.4cm]{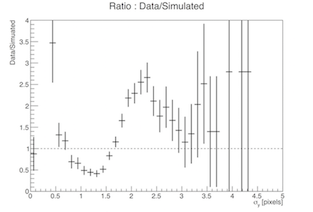}

  \centering
        \includegraphics[width=7.4cm]{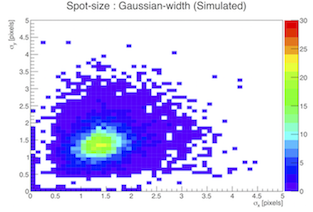}
        \hspace{0.4cm}
        \includegraphics[width=7.4cm]{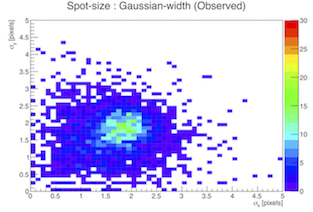}

  \caption{{\bf Comparison 3}. The cut-2 is applied. The simulated distributions of the spot pulse-height and spot-size are compared with the actual spot-distributions obtained from Hiroshima's dataset.}
  \label{fig;spot_comparison_dimer_B_3}
\end{figure}

\newpage

\begin{figure}[!h]
  \centering
        \includegraphics[width=7.4cm]{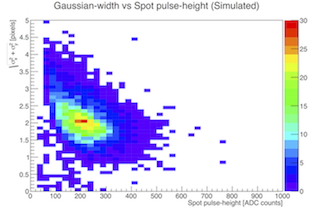} 
        \hspace{0.4cm}
        \includegraphics[width=7.4cm]{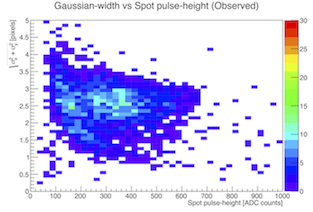} 
 
  \centering
        \includegraphics[width=7.4cm]{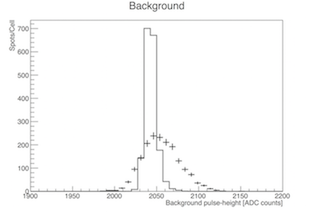}
        \hspace{0.4cm}
        \includegraphics[width=7.4cm]{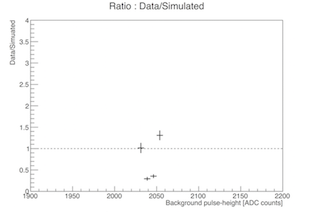}
 
  \centering
        \includegraphics[width=7.4cm]{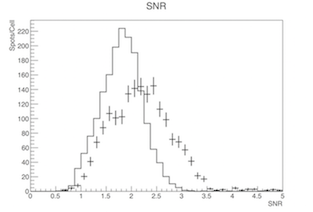}
        \hspace{0.4cm}
        \includegraphics[width=7.4cm]{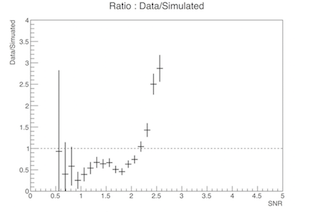}

  \caption{{\bf Comparison 3}. The cut-2 is applied. The top panels show the simulated correlation (left) of the spot pulse-height to spot-size ($\sqrt{\sigma^2_{x} + \sigma^2_{y}}$) is compared with the actual correlation (right) obtained from Hiroshima's dataset. Comparisons of background pulse-height and SNR distributions are shown in bottom two panels.}
  \label{fig;spot_comparison_dimer_B_4}
\end{figure}

\newpage

\paragraph{(3) Spot-tracking :}
To avoid linking spots having relatively larger size, we assume maximum threshold in the primary condition of linking two spots. An additional condition is the average size of the two spots must be less than $3.0\ {\rm pixels}$, and can be written in the form of
\begin{eqnarray}
\frac{\sigma_{r_{i,k}} + \sigma_{r_{j,k+1}}}{2} & < & 3.0\ {\rm pixels}
\end{eqnarray}
where $i$ and $j$ represent spot indexes at $k$-th and ($k+1$)-th image-frames.

\begin{figure}[!h]
  \leftline{\bf \hspace{0.5cm} State-transition events (Molecular interactions)}
  \centering
        \includegraphics[width=7.4cm]{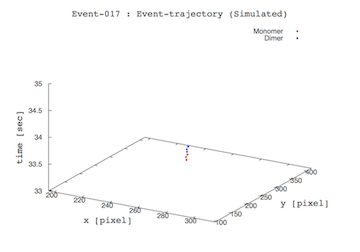}
        \hspace{0.4cm}
        \includegraphics[width=7.4cm]{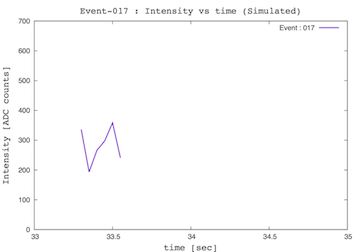}

  \centering
        \includegraphics[width=7.4cm]{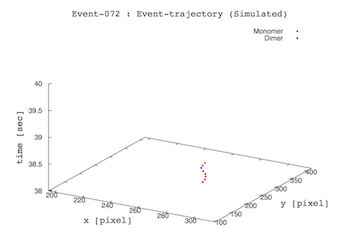}
        \hspace{0.4cm}
        \includegraphics[width=7.4cm]{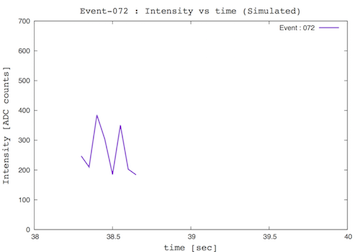}
        
  \leftline{\bf \hspace{0.5cm} Flase-reconstruction events (Molecular exchanges)}
  \centering
        \includegraphics[width=7.4cm]{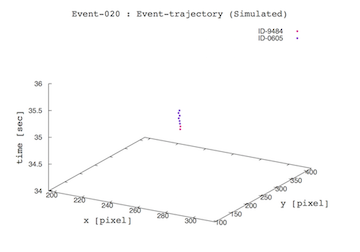}
        \hspace{0.4cm}
        \includegraphics[width=7.4cm]{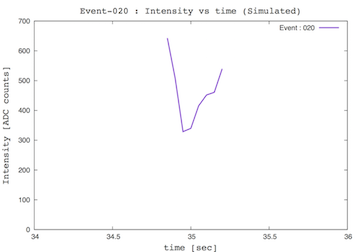}

  \centering
        \includegraphics[width=7.4cm]{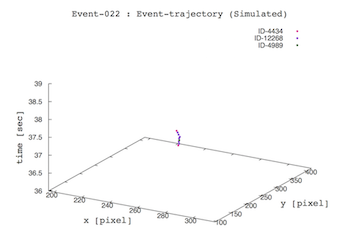}
        \hspace{0.4cm}
        \includegraphics[width=7.4cm]{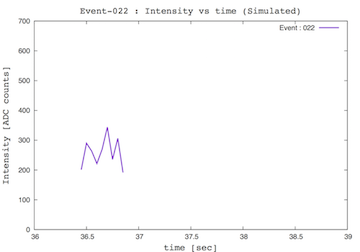}

  \caption{{\bf Examples of simulated event-trajectories 2}. The cut-1 is applied.}
  \label{fig;event_trajectory_sim_dimer_B_2}
\end{figure}

\paragraph{}
In the dimer model, $33$ and $12$ events per cell were reconstructed with each area-cuts. These events can be categorized into two types; (1) Ground-truth : true-monomer ({\bf R, rR, r'R}) and -dimer ({\bf RR}) events are the event-trajectories that are truly reconstructed with true-monomer and -dimer spots. State-transition events are the events that can interact between true-monomer and true-dimer events. While tracking spots, true-monomer (or true-dimer) can often interact with other-type of molecules. For example, the first and second panels of Figure \ref{fig;event_trajectory_sim_dimer_B_2} show interactions of molecular-states. Red and blue spots represent true-monomer and true-dimer. Such state-transition events cannot be distinguished from other events. (2) Failure : unfortunately $70$-$80\%$ of reconstructed event-trajectories are false-events. Event-trajectories are reconstructed under the assumption that the spot-tracking algorithm can successfully track identical molecules (or same molecular ID assigned by Spatiocyte). If the spot-tracking algorithm fails tracking the same molecule and captures other molecules traveling nearby, then the reconstructed event-trajectory is considered to be a false-event. Also, the reconstructed event-trajectory contaminated with non-molecular spots (or false-spots) is consider to be a false-event.

\begin{figure}[!h]
  \centering
        \includegraphics[width=7.4cm]{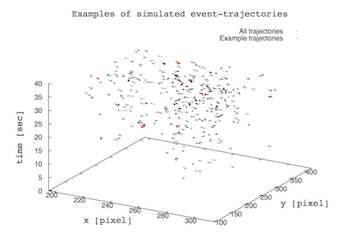}
        \hspace{0.4cm}
        \includegraphics[width=7.4cm]{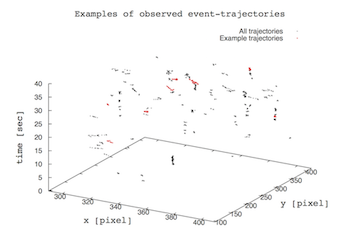}

  \caption{{\bf Simulated and observed event-trajectories}. The cut-1 is applied for the simulated event-trajectories. Red spots represent example event-trajectories shown in Figures \ref{fig;event_trajectory_obs_dimer_B_1} to \ref{fig;event_trajectory_sim_dimer_B_3}.}
  \label{fig;event_trajectory_dimer_B}
\end{figure}

\begin{figure}[!h]
  \centering
        \includegraphics[width=7.4cm]{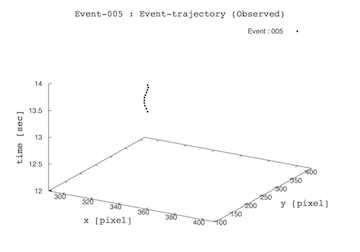}
        \hspace{0.4cm}
        \includegraphics[width=7.4cm]{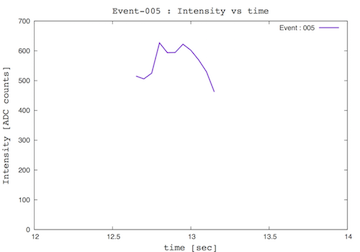}

  \centering
        \includegraphics[width=7.4cm]{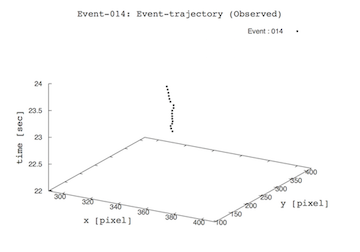}
        \hspace{0.4cm}
        \includegraphics[width=7.4cm]{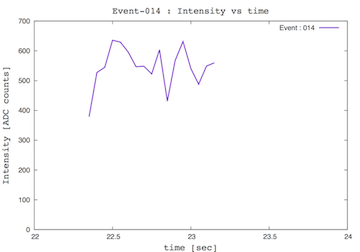}

  \centering
        \includegraphics[width=7.4cm]{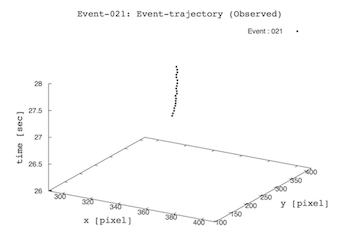}
        \hspace{0.4cm}
        \includegraphics[width=7.4cm]{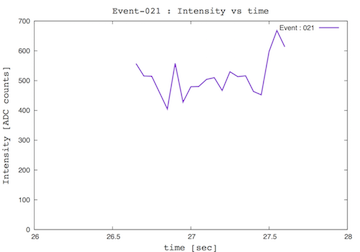}

  \centering
        \includegraphics[width=7.4cm]{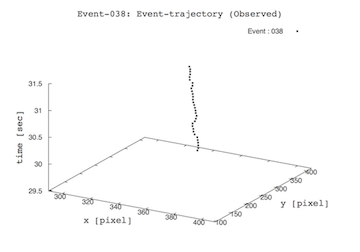}
        \hspace{0.4cm}
        \includegraphics[width=7.4cm]{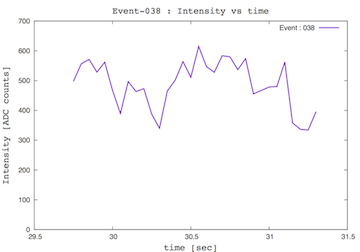}

  \caption{{\bf Examples of observed event-trajectories 1}.}
  \label{fig;event_trajectory_obs_dimer_B_1}
\end{figure}

\begin{figure}[!h]
  \centering
        \includegraphics[width=7.4cm]{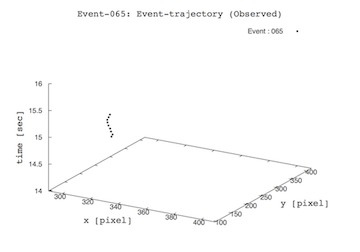}
        \hspace{0.4cm}
        \includegraphics[width=7.4cm]{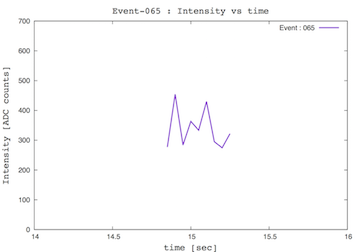}

  \centering
        \includegraphics[width=7.4cm]{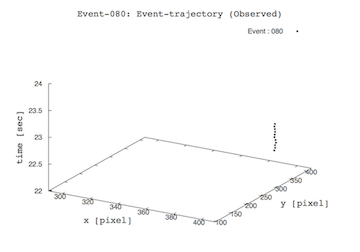}
        \hspace{0.4cm}
        \includegraphics[width=7.4cm]{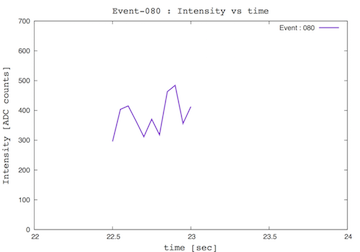}

  \centering
        \includegraphics[width=7.4cm]{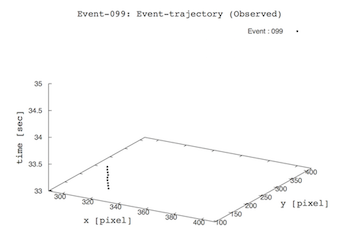}
        \hspace{0.4cm}
        \includegraphics[width=7.4cm]{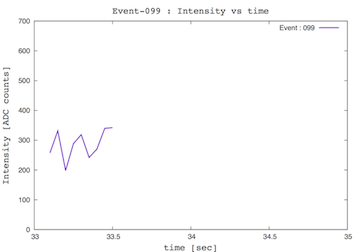}

  \centering
        \includegraphics[width=7.4cm]{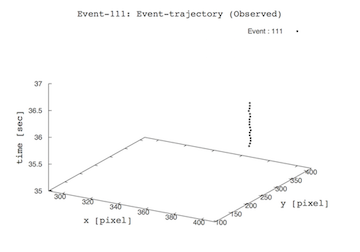}
        \hspace{0.4cm}
        \includegraphics[width=7.4cm]{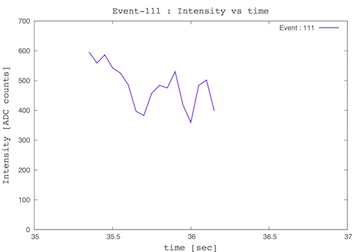}

  \caption{{\bf Examples of observed event-trajectories 2}.}
  \label{fig;event_trajectory_obs_dimer_B_2}
\end{figure}

\begin{figure}[!h]
  \leftline{\bf \hspace{0.5cm} Monomer event}
  \centering
        \includegraphics[width=7.4cm]{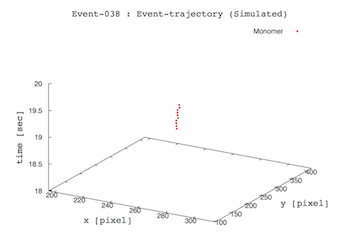}
        \hspace{0.4cm}
        \includegraphics[width=7.4cm]{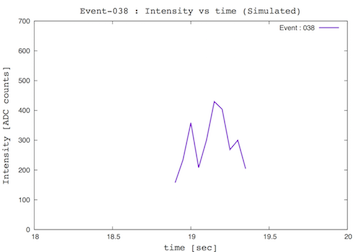}

  \centering
        \includegraphics[width=7.4cm]{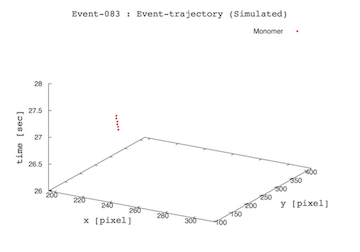}
        \hspace{0.4cm}
        \includegraphics[width=7.4cm]{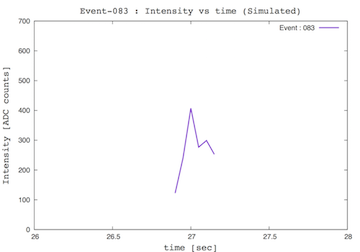}

  \leftline{\bf \hspace{0.5cm} Dimer event}
  \centering
        \includegraphics[width=7.4cm]{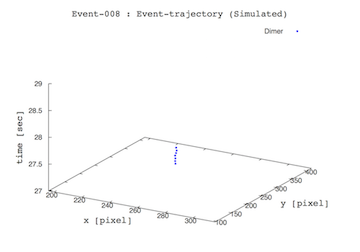}
        \hspace{0.4cm}
        \includegraphics[width=7.4cm]{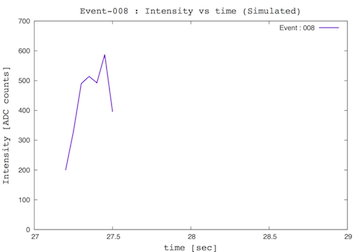}

  \centering
        \includegraphics[width=7.4cm]{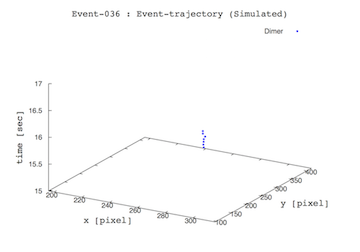}
        \hspace{0.4cm}
        \includegraphics[width=7.4cm]{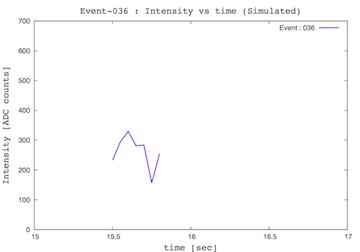}

  \caption{{\bf Examples of simulated event-trajectories 1}. The cut-1 is applied.}
  \label{fig;event_trajectory_sim_dimer_B_1}
\end{figure}

\begin{figure}[!h]
  \leftline{\bf \hspace{0.5cm} Flase-reconstruction events (False-spot)}
  \centering
        \includegraphics[width=7.4cm]{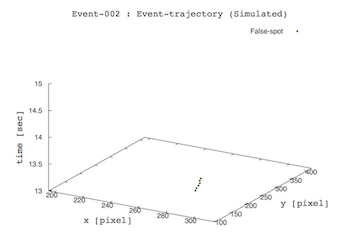}
        \hspace{0.4cm}
        \includegraphics[width=7.4cm]{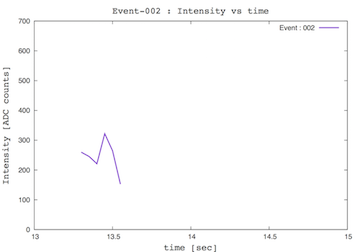}

  \centering
        \includegraphics[width=7.4cm]{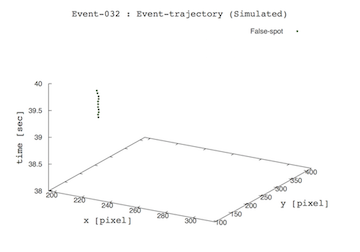}
        \hspace{0.4cm}
        \includegraphics[width=7.4cm]{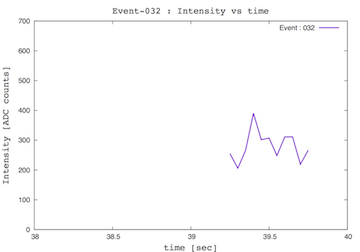}
        
  \centering
        \includegraphics[width=7.4cm]{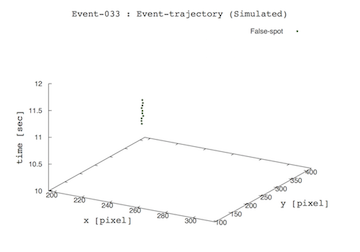}
        \hspace{0.4cm}
        \includegraphics[width=7.4cm]{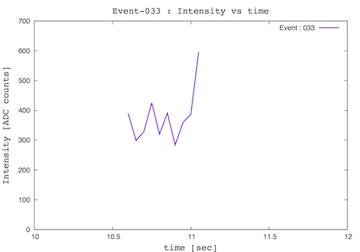}

  \centering
        \includegraphics[width=7.4cm]{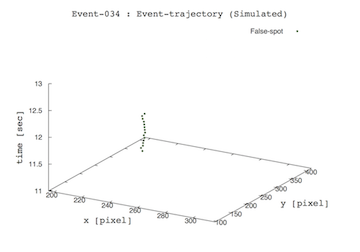}
        \hspace{0.4cm}
        \includegraphics[width=7.4cm]{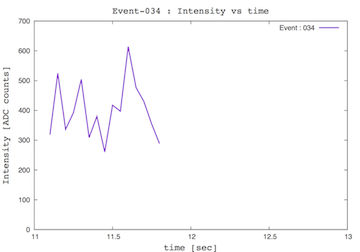}

  \caption{{\bf Examples of simulated event-trajectories 3}. The cut-1 is applied.}
  \label{fig;event_trajectory_sim_dimer_B_3}
\end{figure}

\paragraph{}
Properties of event-trajectories are shown in Table \ref{tab;events_dimer_B} and Figure \ref{fig;event_trajectory_dimer_B}. The Table shows the number and fraction of each event types. The Figures show space-time series of the observed and simulated event-trajectories. Black and red spots represent the spots selected by the cuts and example event-trajectories. Figures \ref{fig;event_trajectory_obs_dimer_B_1} to \ref{fig;event_trajectory_sim_dimer_B_3} show examples of the observed and simulated event-trajectories and their reconstructed intensity (or spot pulse-height) changes in time. 

\begin{table}[!h]
\centering
\begin{tabular}{|c|c|c|c|c|c|c|}
\hline
%\multicolumn{2}{|c|}{\bf Area density of ErbB proteins} \\ \hline
Events/cell & Observed & Simulated & True-monomer & True-dimer & State-transition & False-event \\ \hline
Cut-1 & $61.5$ & $33.3$ & $0.9$ ($2.7\%$) & $9.2$ ($27.6\%$) & $3.4$ ($10.2\%$) & $19.8$ ($68.5\%$) \\ \hline
Cut-2 & $61.5$ & $11.9$ & $0.4$ ($3.4\%$) & $1.1$ ($9.2\%$) & $0.9$ ($7.6\%$) & $9.5$ ($79.8\%$) \\ \hline
\end{tabular}
\caption{{\bf The number and fraction of detected events}.}
\label{tab;events_dimer_B}
\end{table}

\newpage

\newpage

\newpage

\newpage

\newpage

\newpage

\begin{figure}[!h]
  \centering
        \includegraphics[width=7.4cm]{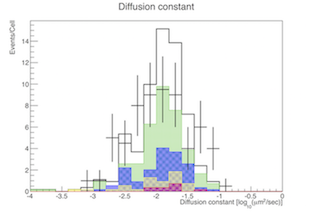}
        \hspace{0.4cm}
        \includegraphics[width=7.4cm]{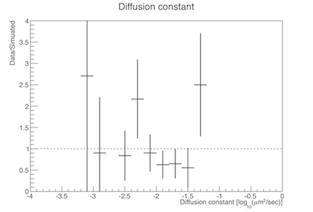}

  \centering
        \includegraphics[width=7.4cm]{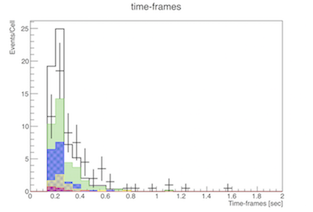}
        \hspace{0.4cm}
        \includegraphics[width=7.4cm]{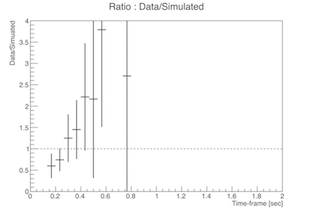}

  \centering
        \includegraphics[width=7.4cm]{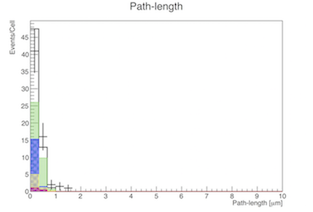}
        \hspace{0.4cm}
        \includegraphics[width=7.4cm]{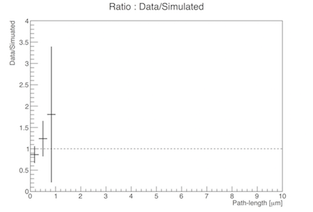}

  \caption{{\bf Event-properties 1}. The Cut-1 is applied. Distributions of diffusion constant, event time-frames and path-length are shown in the top, the middle and the bottom panels.}
  \label{fig;event_property_dimer_B_1}
\end{figure}

\begin{figure}[!h]
  \centering
        \includegraphics[width=7.4cm]{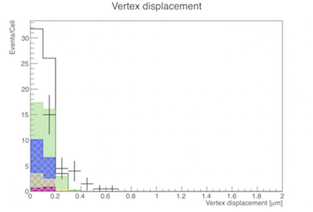}
        \hspace{0.4cm}
        \includegraphics[width=7.4cm]{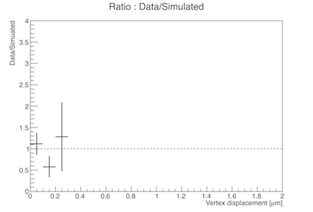}

  \centering
        \includegraphics[width=7.4cm]{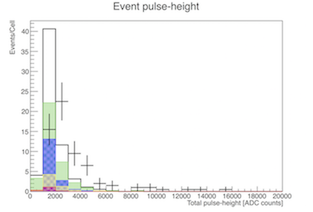}
        \hspace{0.4cm}
        \includegraphics[width=7.4cm]{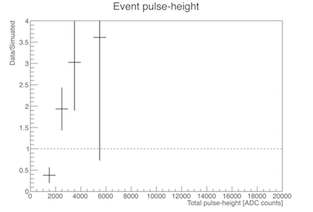}

  \centering
        \includegraphics[width=7.4cm]{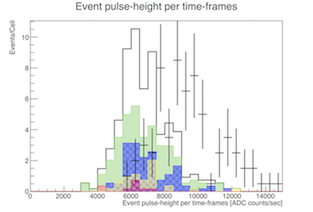}
        \hspace{0.4cm}
        \includegraphics[width=7.4cm]{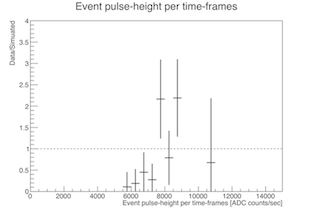}

  \centering
        \includegraphics[width=7.4cm]{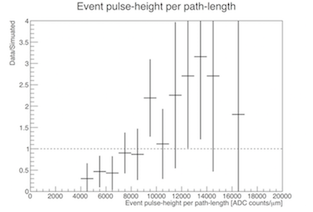}
        \hspace{0.4cm}
        \includegraphics[width=7.4cm]{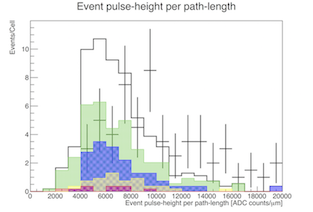}

  \caption{{\bf Event-properties 2}. The Cut-1 is applied. Distributions of event pulse-height, event pulse-height per time-frames, event pulse-height per path-length and event-vertex displacement are shown from top to bottom.}
  \label{fig;event_property_dimer_B_2}
\end{figure}

\begin{figure}[!h]
  \centering
        \includegraphics[width=7.4cm]{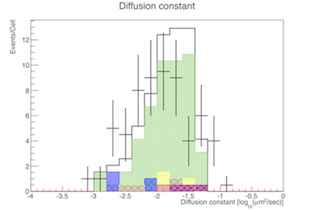}
        \hspace{0.4cm}
        \includegraphics[width=7.4cm]{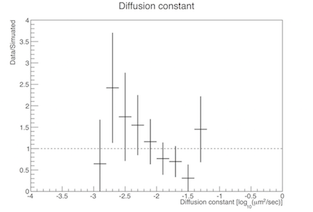}

  \centering
        \includegraphics[width=7.4cm]{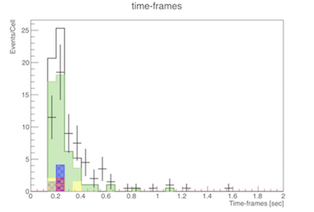}
        \hspace{0.4cm}
        \includegraphics[width=7.4cm]{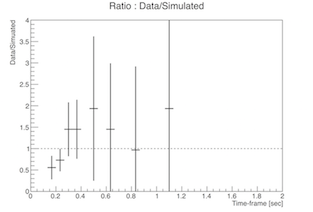}

  \centering
        \includegraphics[width=7.4cm]{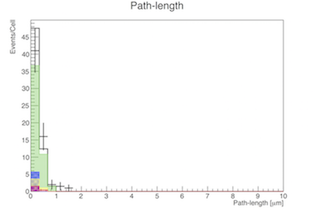}
        \hspace{0.4cm}
        \includegraphics[width=7.4cm]{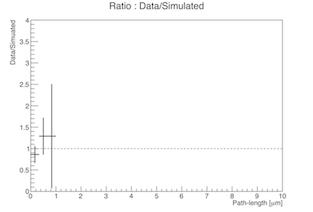}

  \centering
        \includegraphics[width=7.4cm]{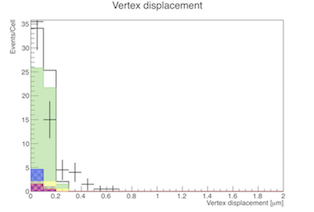}
        \hspace{0.4cm}
        \includegraphics[width=7.4cm]{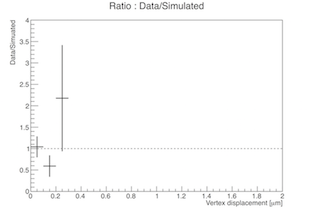}

  \caption{{\bf Event-properties 3}. The cut-2 is applied. Distributions of event time-frames, path-length and event-vertex displacement are shown from top to bottom.}
  \label{fig;event_property_dimer_B_3}
\end{figure}

\begin{figure}[!h]
  \centering
        \includegraphics[width=7.4cm]{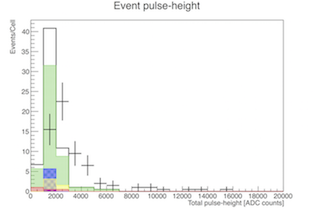}
        \hspace{0.4cm}
        \includegraphics[width=7.4cm]{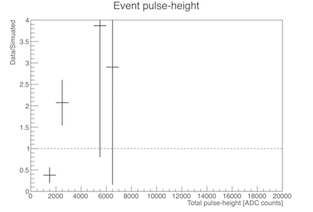}

  \centering
        \includegraphics[width=7.4cm]{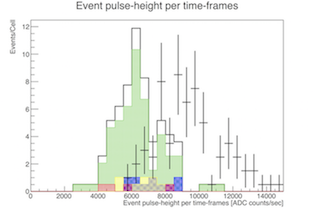}
        \hspace{0.4cm}
        \includegraphics[width=7.4cm]{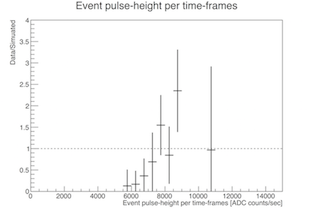}

  \centering
        \includegraphics[width=7.4cm]{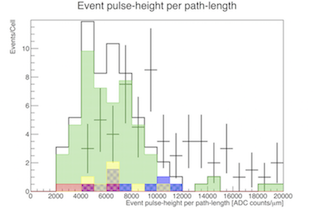}
        \hspace{0.4cm}
        \includegraphics[width=7.4cm]{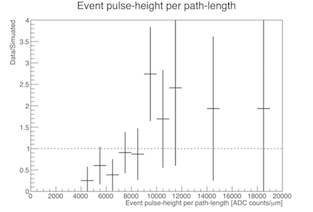}

  \caption{{\bf Event-properties 5}. The cut-2 is applied. Distributions of event pulse-height, event pulse-height per time-frames, and event pulse-height per path-length are shown from top to bottom.}
  \label{fig;event_property_dimer_B_5}
\end{figure}

\paragraph{(4) Event-property :}
For each area-cut, we directly compared the simulated event-spectra to actual spectra in Figures \ref{fig;event_property_dimer_B_1} to \ref{fig;event_property_dimer_B_5}. Red, blue, yellow and green filled histograms represent event-distribution of true-monomer, true-dimer, state-transition and false-reconstruction events. Black solid lines represent the sum of true-distributions. Black cross represents actual event-distributions. The left panel shows  the simulated distribution of reconstructed parameters directly compared with Hiroshima's dataset. The right panel shows the ratio of the simulated spectra to the actual spectra in each bin. The errors are not only observed statistical errors, but also include the simulated sample statistical errors. All simulated spectra are normalized by the number of detected spots. 

\newpage

\newpage

\newpage

\begin{thebibliography}{10}

\bibitem{king2009}
King RD et~al. (2009) {The automation of science.}
\newblock {\em Science (New York, N.Y.)} 324(5923):85--89.

\bibitem{marx2013}
Marx V (2013) {THE BIG CHALLENGES OF BIG DATA}.
\newblock {\em Nature} 498(7453):255--260.

\bibitem{jordan2015}
Jordan MI, Mitchell TM (2015) {Machine learning: Trends, perspectives, and
  prospects.}
\newblock {\em Science} 349(6245).

\bibitem{yachie2017}
Yachie N, {Robotic\ Biology\ Consortium}, Natsume T (2017) {Robotic crowd
  biology with Maholo LabDroids}.
\newblock {\em Nature} 35(4):3--5.

\bibitem{amrhein2019}
Amrhein V, Mcshane B (2019) {Retire statistical significance}.
\newblock {\em Nature} 567(7748):305--307.

\bibitem{geris2016}
Geris L, Gemez-Cabrero D (2016) {\em Uncertainty in Biology}.
\newblock (Springer).

\bibitem{taylor1997}
Taylor JR (1997) {\em An Introduction to Error Analysis.}
\newblock (University Science Book), 2nd edition.

\bibitem{bevington2003}
Bevington PR, Robinson DK (2003) {\em Data Reduction and Error Analysis.}
\newblock (McGrawHill), 3rd edition.

\bibitem{babtie2014}
Babtie AC, Kirk P, Stumpf MPH (2014) {Topological sensitivity analysis for
  systems biology}.
\newblock {\em Proc. Natl. Acad. Sci. U.S.A.} 111(52):18507--18512.

\bibitem{kirk2015}
Kirk PDW, Babtie AC, Stumpf MPH (2015) {Systems biology (un)certainties}.
\newblock {\em Science} 350(6259):386--388.

\bibitem{incerti2009}
Incerti S et~al. (2010) {The Geant4-DNA project}.
\newblock {\em Int. J. Model. Simul. Sci. Comput.} 1(2):157--178.

\bibitem{incerti2018}
Incerti S,  et~al. (2018) {Geant4-DNA example applications for track structure
  simulations in liquid water : A report from the Geant4-DNA Project}.
\newblock {\em Med. Phys.} 45(8):e722--e739.

\bibitem{agostinelli2003}
Agostinelli S, Allison J, Amako K (2003) {Geant4--a simulation toolkit}.
\newblock {\em Nucl. Instrum. Methods Phys. Res. A} 506(506):250--303.

\bibitem{allison2006}
Allison J,  et~al. (2006) {Geant4 developments and applications}.
\newblock {\em IEEE Trans. Nucl. Sci.} 53(1):270--278.

\bibitem{allison2016}
Allison J,  et~al. (2016) {Recent developments in GEANT4}.
\newblock {\em Nucl. Instrum. Methods Phys. Res. A} 835:186--225.

\bibitem{sbalzarini2013}
Sbalzarini IF (2013) {Modeling and simulation of biological systems from image
  data.}
\newblock {\em BioEssays : news and reviews in molecular, cellular and
  developmental biology} 35(5):482--90.

\bibitem{mai2013}
Mai J, Trump S, Lehmann I, Attinger S (2013) {Parameter Importance in FRAP
  Acquisition and Analysis: A Simulation Approach}.
\newblock {\em Biophys. J.} 104(9):2089--2097.

\bibitem{lakowicz2006}
Lakowicz J (2006) {\em Principles of Fluorescence Spectroscopy}.
\newblock (Springer), p. 954.

\bibitem{pawley2008}
Pawley J (2008) {\em Handbook of Biological Confocal Microscopy}.
\newblock (Springer), p. 988.

\bibitem{mansuripur2009}
Mansuripur M (2009) {\em Classical Optics and its Application}.
\newblock (Cambridge University Press), p. 714.

\bibitem{valeur2012}
Valeur B, Berberan-Santos M (2012) {\em Molecular Fluorescence}.
\newblock (Wiley), p. 592.

\bibitem{kaneko2006}
Kaneko K (2006) {\em Life: An introduction to complex systems biology}.
\newblock (Springer).

\bibitem{alon2007a}
Alon U (2007) {\em An introduction to systems biology --- Design principles of
  biological circuits.}
\newblock (Champman $\&$ Hall/CRC).

\bibitem{tomita1999}
Tomita M et~al. (1999) {E-CELL: Software environment for whole-cell
  simulation.}
\newblock {\em Bioinformatics} 15:72--84.

\bibitem{boulanger2009}
Boulanger J, Kervrann C, Bouthemy P (2009) {A simulation and estimation
  framework for intracellular dynamics and trafficking in video-microscopy and
  fluorescence imagery}.
\newblock {\em Med. Image Anal.} 13(1):132--142.

\bibitem{rezatofighi2013}
Rezatofighi SH et~al. (2013) {A framework for generating realistic synthetic
  sequences of total internal reflection fluorescence microscopy images}.
\newblock {\em Proceedings - International Symposium on Biomedical Imaging} pp.
  157--160.

\bibitem{angiolini2015}
Angiolini J, Plachta N, Mocskos E, Levi V (2015) {Exploring the Dynamics of
  Cell Processes through Simulations of Fluorescence Microscopy Experiments}.
\newblock {\em Biophys. J.} 108(11):2613--2618.

\bibitem{watabe2015}
Watabe M et~al. (2015) {A Computational Framework for Bioimaging Simulation}.
\newblock {\em PLOS One} 10(7):e0130089.

\bibitem{venkataramani2016}
Venkataramani V, Herrmannsd{\"{o}}rfer F, Heilemann M, Kuner T (2016) {SuReSim:
  simulating localization microscopy experiments from ground truth models}.
\newblock {\em Nat. Methods} 13(4).

\bibitem{linden2016}
Lind{\'{e}}n M, {\'{C}}uri{\'{c}} V, Boucharin A, Fange D, Elf J (2016)
  {Simulated single molecule microscopy with SMeagol}.
\newblock {\em Bioinformatics} p. btw109.

\bibitem{weigert2018}
Weigert M, Subramanian K, Bundschuh ST, Myers W, Kreysing M (2018) {Biobeam ---
  Multiplexed wave-optical simulations of light-sheet microscopy}.
\newblock {\em PLOS Comput. Biol.} 14(4):e1006079.

\bibitem{hiroshima2012}
Hiroshima M, Saeki Y, Okada-Hatakeyama M, Sako Y (2012) {Dynamically varying
  interactions between heregulin and ErbB proteins detected by single-molecule
  analysis in living cells.}
\newblock {\em Proc. Natl. Acad. Sci. U.S.A.} 109(35):13984--9.

\bibitem{arjunan2010}
Arjunan SNV, Tomita M (2010) {A new multicompartmental reaction-diffusion
  modeling method links transient membrane attachment of E. coli MinE to E-ring
  formation}.
\newblock {\em Syst. Synth. Biol.} 4:35--53.

\bibitem{takahashi2005}
Takahashi K, Arjunan S, Tomita M (2005) {Space in systems biology of signaling
  pathways -- towards intracellular molecular crowding in silico}.
\newblock {\em FEBS letters} 579:1783--1788.

\bibitem{vanderwalt2014}
van~der Walt S et~al. (2014) {scikit-image: image processing in Python.}
\newblock {\em PeerJ} 2:e453.

\bibitem{kitano2002_sci}
Kitano H (2002) {Systems biology: a brief overview.}
\newblock {\em Science (New York, N.Y.)} 295(5560):1662--4.

\bibitem{kitano2002_nat}
Kitano H (2002) {Computational systems biology.}
\newblock {\em Nature} 420(6912):206--10.

\bibitem{godfrey2003}
Godfrey-Smith P (2003) {\em Theory and Reality --- an introduction to the
  philosophy of science.}
\newblock (The University of Chicago Press).

\bibitem{zhao2011}
Zhao Q, Young IT, de~Jong JGS (2011) {Photon budget analysis for fluorescence
  lifetime imaging microscopy.}
\newblock {\em Journal of biomedical optics} 16(8):086007.

\bibitem{thompson2010}
Thompson Ma, Biteen JS, Lord SJ, Conley NR, Moerner WE (2010) {\em {Molecules
  and Methods for Super-Resolution Imaging}}.
\newblock (Elsevier Inc.) Vol.{} 475, 1 edition, pp. 27--59.

\bibitem{didier2005}
Didier P, Guidoni L, Bardou F (2005) {Infinite Average Lifetime of an Unstable
  Bright State in the Green Fluorescent Protein}.
\newblock {\em Physical Review Letters} 95(9):090602.

\bibitem{zondervan2004}
Zondervan R, Kulzer F, Kol'chenk Ma, Orrit M (2004) {Photobleaching of
  Rhodamine 6G in Poly(vinyl alcohol) at the Ensemble and Single-Molecule
  Levels}.
\newblock {\em The Journal of Physical Chemistry A} 108(10):1657--1665.

\bibitem{hoogenboom2007}
Hoogenboom JP, Hernando J, {Van Dijk} EMHP, {Van Hulst} NF,
  Garc{\'{i}}a-Paraj{\'{o}} MF (2007) {Power-law blinking in the fluorescence
  of single organic molecules}.
\newblock {\em ChemPhysChem} 8(6):823--833.

\bibitem{margolin2005}
Margolin G, Barkai E (2005) {Nonergodicity of blinking nanocrystals and other
  L{\'{e}}vy-walk processes}.
\newblock {\em Physical Review Letters} 94(8):1--4.

\bibitem{brokmann2003}
Brokmann X et~al. (2003) {Statistical Aging and Nonergodicity in the
  Fluorescence of Single Nanocrystals}.
\newblock {\em Physical Review Letters} 90(12):120601.

\bibitem{shimizu2001}
Shimizu K et~al. (2001) {Blinking statistics in single semiconductor
  nanocrystal quantum dots}.
\newblock {\em Physical Review B} 63(20):205316.

\bibitem{miyanaga2009}
Miyanaga Y, Matsuoka S, Ueda M (2009) {Single-Molecule Imaging Techniques to
  Visualize Chemotactic Signaling Events on the Membrane of Living
  Dictyostelium Cells}.
\newblock {\em Methods in Molecular Biology} 571:417--435.

\bibitem{axelrod2008}
Axelrod D (2008) {\em {Chapter 7: Total internal reflection fluorescence
  microscopy.}}
\newblock (Elsevier Inc.) Vol.{}~89, 1 edition, pp. 169--221.

\bibitem{wazawa2005}
Wazawa T, Ueda M, Rietdorf J (2005) {\em {Microscopy Techniques - Advances in
  Biochemical Engineering/Biotechnology}}.
\newblock (Springer Berlin / Heidelberg) Vol.{}~95, pp. 1297--1300.

\bibitem{axelrod2003}
Axelrod D (2003) {Total internal reflection fluorescence microscopy in cell
  biology.}
\newblock {\em Methods in enzymology} 361(2):1--33.

\bibitem{teramura2006}
Teramura Y et~al. (2006) {Single-molecule analysis of epidermal growth factor
  binding on the surface of living cells}.
\newblock {\em EMBO J.} 25(18):4215--4222.

\bibitem{uyemura2005}
Uyemura T, Takagi H, Yanagida T, Sako Y (2005) {Single-molecule analysis of
  epidermal growth factor signaling that leads to ultrasensitive calcium
  response.}
\newblock {\em Biophys. J.} 88(5):3720--30.

\bibitem{sako2002}
Sako Y, Uyemura T (2002) {Total Internal Reflection Fluorescence Microscopy for
  Single-molecule Imaging in Living Cells}.
\newblock {\em Cell Structure and Function} 27(5):357--365.

\bibitem{sako2000}
Sako Y, Minoghchi S, Yanagida T (2000) {Single-molecule imaging of EGFR
  signalling on the surface of living cells.}
\newblock {\em Nature Cell Biology} 2(3):168--172.

\bibitem{tokunaga1997}
Tokunaga M, Kitamura K, Saito K, Iwane AH, Yanagida T (1997) {Single molecule
  imaging of fluorophores and enzymatic reactions achieved by objective-type
  total internal reflection fluorescence microscopy.}
\newblock {\em Biochemical and biophysical research communications}
  235(1):47--53.

\bibitem{jena2008}
Jena BP (2008) {\em {Methods in Nano Cell Biology.}}
\newblock (Academic Press), 1 edition.

\bibitem{basset2015}
Basset A, Boulanger J, Salamero J, Bouthemy P, Kervrann C (2015) {Adaptive Spot
  Detection With Optimal Scale Selection in Fluorescence Microscopy Images}.
\newblock {\em IEEE Trans. Image Process.} 24(11):4512--4527.

\bibitem{ruusuvuori2010}
Ruusuvuori P et~al. (2010) {Evaluation of methods for detection of fluorescence
  labeled subcellular objects in microscope images}.
\newblock {\em BMC Bioinform.} 11:248.

\bibitem{smal2010}
Smal I, Loog M, Niessen W, Meijering E (2010) {Quantitative comparison of spot
  detection methods in fluorescence microscopy}.
\newblock {\em IEEE Trans. Med. Imaging} 29(2):282--301.

\end{thebibliography}
\end{document}